%% file: arxiv_volume_II.tex
\documentclass[oneside,a4paper,11pt,explicit]{book}
\def\ARXIVBUILD{1}
\input{II_Cosmology_Volume/volume_II.tex}
\usepackage[T1]{fontenc}
\usepackage[utf8]{inputenc}
\usepackage{textcomp}
\usepackage{vol2_kultem}

\usepackage{amsmath,amssymb}
\usepackage{booktabs}
\usepackage{tabularx}
\usepackage{array}
\usepackage{longtable}
\usepackage{lscape}
\usepackage{graphicx}
\graphicspath{{./}}
\usepackage[dvipsnames,svgnames,table]{xcolor}
\usepackage[most]{tcolorbox}
\usepackage{float}
\usepackage{hyperref}

\makeatletter
\renewcommand*\l@section{\@dottedtocline{1}{1.5em}{2.8em}}
\renewcommand*\l@subsection{\@dottedtocline{2}{3.8em}{3.7em}}
\renewcommand*\l@subsubsection{\@dottedtocline{3}{7.0em}{4.6em}}
\makeatother

\usetikzlibrary{arrows.meta,calc,positioning,shapes.geometric,
  decorations.pathmorphing,backgrounds,fit}
\numberwithin{equation}{chapter}

\definecolor{ink}{HTML}{151B23}
\definecolor{titlebg}{HTML}{100880}
\definecolor{softink}{HTML}{263441}
\definecolor{muted}{HTML}{5C6875}
\definecolor{paper}{HTML}{F6F7F2}
\definecolor{panel}{HTML}{FFFFFF}
\definecolor{line}{HTML}{D8DEE5}
\definecolor{teal}{HTML}{007C77}
\definecolor{blue}{HTML}{245AA6}
\definecolor{gold}{HTML}{B86B00}
\definecolor{rose}{HTML}{A7354D}
\definecolor{green}{HTML}{23724A}
\definecolor{violet}{HTML}{5A4CA0}

\newcommand{\doi}[1]{DOI:~\href{https://doi.org/#1}{#1}}
\newcommand{\facility}{3.5-meter Segmented-Mirror Robotic Space Telescope}

\newcommand{\fsig}{f\sigma_8}
\newcommand{\degti}{deg$^2$}
\newcommand{\kms}{km\,s$^{-1}$}
\newcommand{\fovsmall}{$10'\!\times10'$}
\newcommand{\fovlarge}{$30'\!\times30'$}
\newcolumntype{Y}{>{\raggedright\arraybackslash}X}

\newtcolorbox{leadbox}[2][]{
  enhanced, colback=paper, colframe=#2, boxrule=0.9pt, arc=2mm,
  left=2.2mm, right=2.2mm, top=1.8mm, bottom=1.8mm,
  fonttitle=\sffamily\bfseries, coltitle=white,
  attach boxed title to top left={xshift=2mm,yshift=-2mm},
  boxed title style={colback=#2,arc=1.2mm,boxrule=0pt}, #1 }
\newtcolorbox{metricbox}[1]{
  enhanced, colback=white, colframe=#1, boxrule=0.65pt, arc=1.3mm,
  left=1.8mm, right=1.8mm, top=1.6mm, bottom=1.6mm }

\makeatletter
\renewenvironment{thebibliography}[1]
  {\section*{References}\@mkboth{}{}%
   \list{\@biblabel{\@arabic\c@enumiv}}%
        {\settowidth\labelwidth{\@biblabel{#1}}%
         \leftmargin\labelwidth \advance\leftmargin\labelsep
         \usecounter{enumiv}\let\p@enumiv\@empty
         \renewcommand\theenumiv{\@arabic\c@enumiv}}%
   \small\sloppy\clubpenalty4000\widowpenalty4000\sfcode`\.\@m}
  {\def\@noitemerr{\@latex@warning{Empty `thebibliography' environment}}\endlist}
\makeatother

\input{vol2_front.tex}
\input{vol2_covers.tex}

\hypersetup{
  colorlinks=true,
  linkcolor=blue,
  citecolor=teal,
  urlcolor=gold,
  pdftitle={3.5-meter Segmented-Mirror Robotic Space Telescope -- \wpvolumelabel. \wpvolumetitle},
  pdfauthor={Juhan Kim et al.}}

\begin{document}
\frontmatter
\MakeFrontCover
\MakeColophon
\thispagestyle{fancy}
\vspace*{0.5em}
\noindent{\Large\bfseries Abstract}\par\smallskip
\noindent\wpabstract\par\medskip
\noindent\wpkeywords
\clearpage
\tableofcontents

\mainmatter
\renewcommand{\chaptername}{Volume}
\renewcommand{\thechapter}{\Roman{chapter}}
\setcounter{chapter}{\wpchapteroffset}
\MakeVolumeBody
\include{vol2_cosmology}

\clearpage
\definecolor{Blue1}{HTML}{1FABD5}
\definecolor{Blue2}{HTML}{1D8DB0}
\definecolor{Blue3}{HTML}{116E8A}
\backmatter
\MakeBackCover
\end{document}

%% file: II_Cosmology_Volume/volume_II.tex
\ifdefined\ARXIVBUILD
\else
  \csname input\endcsname{arxiv_volume_II.tex}
\fi

%% file: vol2_front.tex
\newcommand{\wpvolumelabel}{II}
\newcommand{\wpvolumetitle}{Key Scientific Mission: Wide-Field Cosmology and Galaxy Evolution}
\title{3.5-meter Segmented-Mirror Robotic Space Telescope}
\subtitle{Mission White Paper: \wpvolumelabel. \wpvolumetitle}
\newcommand{\wpauthors}{Juhan Kim$^{1}$, Yong-Woo Kang$^{2}$, Sang Hyun Lee$^{2,3}$, Jeong-Yeol Han$^{2,4}$, Sungwook E. Hong$^{2,4}$, Bongkon Moon$^{2,4}$, Donguk Song$^{2}$, Juhyung Kang$^{2}$, Myeong-Gu Park$^{5}$, Sang Chul Kim$^{2,4}$, Chung-Uk Lee$^{2}$, Sangmo Tony Sohn$^{6}$, Arman Shafieloo$^{2,4}$, David Parkinson$^{2,4}$, Hong Soo Park$^{2,4}$, Dohyeong Kim$^{7}$, Chan Park$^{2}$, Jungjoo Sohn$^{8}$, Young-Beom Jeon$^{2}$, Jong-Hak Woo$^{9}$, Hyung Mok Lee$^{9}$, Hong Bae Ann$^{7}$, Myungkook James Jee$^{10}$, Mansoo Choi$^{2}$, Changbom Park$^{1}$}
\date{2026}
\newcommand{\wpabstract}{%
\textbf{The \facility\ uses an image slicer for all spectroscopic observations. The planning baseline uses $R\simeq1000$ for the wide survey and retains selectable $R\simeq5000$ bands for precision line measurements.}
The central science case is a dense emission-line galaxy redshift survey for baryon acoustic oscillations and redshift-space distortions. Supernova and quasar programs exploit the stability, multiplexing, and repeatability of space operations. The supernova tier measures rest-frame $U$ and near-ultraviolet magnitudes that separate optical twins at subgroup precision to $z\simeq0.9$--$1.1$ in standard visits and to $z\simeq1.3$--$1.5$ in ten-hour stacks. Every wide-survey tile receives three spectroscopic orientations, and a joint scene reconstruction uses their different overlap geometries to recover the spectra. The flagship survey covers 100--300 deg$^2$ and targets $10^6$--$3\times10^6$ emission-line galaxies. A deep pencil-beam tier and a supernova time-domain tier complement the wide survey. The same observations provide a census of ultra-diffuse and low-surface-brightness galaxies, map intracluster light, and test cold, self-interacting, and fuzzy dark matter through dwarf-galaxy structure and low-mass halo abundance.
}
\newcommand{\wpkeywords}{\textbf{Keywords:} cosmology, galaxy evolution, emission-line galaxies, Type Ia supernovae, redshift-space distortions, low-surface-brightness galaxies}
\newcommand{\wpchapteroffset}{1}
\newcommand{\MakeVolumeBody}{%
  \definecolor{Blue1}{HTML}{9488AD}%
  \definecolor{Blue2}{HTML}{6F6295}%
  \definecolor{Blue3}{HTML}{514670}}

%% file: vol2_covers.tex
\newcommand{\MakeFrontCover}{%
  \begin{titlepage}
  \thispagestyle{empty}\sffamily\centering
  \noindent\colorbox{Blue3}{\parbox[t]{\dimexpr\textwidth-2\fboxsep\relax}{%
    \vspace{4mm}\centering
    {\color{white}\bfseries\fontsize{24}{29}\selectfont 3.5-meter Segmented-Mirror\\[1.5mm]
      Robotic Space Telescope}\\[3.5mm]
    {\color{white}\Large Mission White Paper}\\[1.5mm]
    {\color{white!88}\large \wpvolumelabel. \wpvolumetitle}%
    \vspace{4mm}}}
  \vfill
  \includegraphics[width=\textwidth]{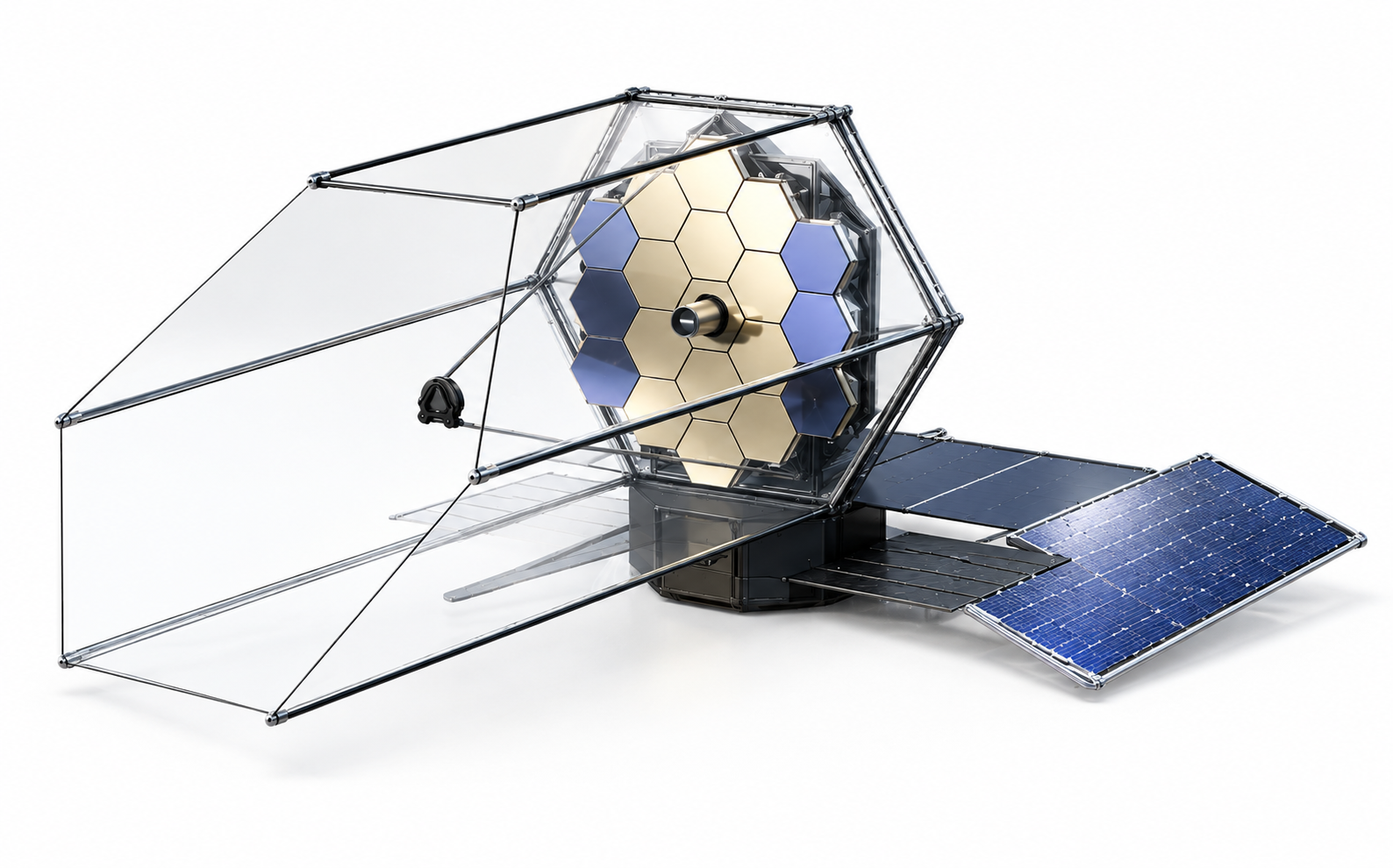}
  \vfill
  {\color{ink}\bfseries\normalsize \wpauthors\par}
  \vspace{3mm}
  {\color{muted}\small
    $^{1}$ Korea Institute for Advanced Study \textperiodcentered\
    $^{2}$ Korea Astronomy and Space Science Institute \\
    $^{3}$ University of Ulsan \textperiodcentered\
    $^{4}$ University of Science and Technology\\
    $^{5}$ Kyungpook National University \textperiodcentered\
    $^{6}$ Space Telescope Science Institute \textperiodcentered\
    $^{7}$ Pusan National University\\
    $^{8}$ Korea National University of Education \textperiodcentered\
    $^{9}$ Seoul National University \textperiodcentered\
    $^{10}$ Yonsei University\par}
  \vspace{5mm}
  {\color{Blue3}\bfseries\Large 2026}
  \vspace{4mm}
  \end{titlepage}}

\newcommand{\MakeColophon}{%
  \clearpage
  \thispagestyle{empty}\sffamily
  \null\vfill
  \noindent{\color{muted}\footnotesize
    {\color{ink}\bfseries 3.5-meter Segmented-Mirror Robotic Space Telescope}\\
    Mission White Paper: \wpvolumelabel. \wpvolumetitle\\[5mm]
    {\color{rose}\bfseries Release date September 17, 2026\\[5mm]}
    \textcopyright\ 2026 Korea Astronomy and Space Science Institute and the
    authors. All rights reserved.\\[2mm]
    Prepared by the mission study team at the Korea Astronomy and Space Science
    Institute, University of Ulsan, University of Science and Technology, the
    Space Telescope Science Institute, the Korea Institute for Advanced Study,
    Pusan National University, Korea National University of Education, Seoul
    National University, Kyungpook National University, and Yonsei University.\\[5mm]
    {\color{ink}\bfseries Suggested citation:} Kim, J., Kang, Y.-W., Lee, S.-H.,
    et al.\ (2026), \textit{3.5-meter Segmented-Mirror Robotic Space Telescope:
    Mission White Paper, \wpvolumelabel. \wpvolumetitle}.\\[2mm]
    {\color{ink}\bfseries Corresponding author:} Sang Hyun Lee
    \textperiodcentered\ \texttt{shlee@kasi.re.kr}.\par}
  \vspace{8mm}
  \clearpage}

\newcommand{\MakeBackCover}{%
  \clearpage
  \ifodd\value{page} \hbox{}\thispagestyle{empty}\newpage \fi
  \thispagestyle{empty}\null
  \clearpage
  \thispagestyle{empty}\sffamily\centering
  \null\vspace*{\stretch{1}}
  \includegraphics[width=\textwidth]{vol2_3.5ST_GPT.png}
  \vspace*{\stretch{2.2}}
  \noindent\colorbox{Blue3}{\parbox[t]{\dimexpr\textwidth-2\fboxsep\relax}{%
    \vspace{3mm}\centering
    {\color{white}\bfseries\Large 3.5-meter Segmented-Mirror Robotic Space Telescope}\\[1.5mm]
    {\color{white!88}\normalsize Mission White Paper: \wpvolumelabel. \wpvolumetitle}%
    \vspace{3mm}}}
  \vspace{3mm}
  \clearpage}

%% file: vol2_cosmology.tex
\chapter{Wide-Field Cosmology and Galaxy Evolution}\label{app:cosmology}

\vspace{2mm}
\begin{center}
\resizebox{\textwidth}{!}{%
\begin{tikzpicture}[font=\sffamily]
  \foreach \x/\title/\val/\col/\vfont in {
    0/{Resolving-power trade}/{$R=1000/5000$}/teal/\large,
    4.4/{Point-source resolution}/{300 / 60 \kms}/blue/\large,
    8.8/{Field of view}/{10'--30' square}/gold/\normalsize,
    13.2/{Flagship sample}/{$10^6$--$3\times10^6$ ELGs}/green/\small} {
    \fill[\col!10,draw=\col,rounded corners=2pt] (\x,0) rectangle ++(4.1,1.35);
    \node[anchor=west,color=muted,font=\scriptsize] at (\x+0.25,0.98) {\title};
    \node[anchor=west,color=ink,font=\bfseries\vfont] at (\x+0.25,0.42) {\val};
  }
\end{tikzpicture}}
\end{center}

\section{Instrument Baseline}

This volume develops the cosmology program within the broader context of
recent large-aperture space-observatory concepts \cite{roy2026II,wevers2026II}.
Its contribution is defined by a wide-field, image-sliced emission-line
survey that measures galaxy redshifts at scale. The central performance
test is the recovered BAO and RSD covariance after spectral overlap,
redshift failures, survey geometry, and calibration uncertainty have been
propagated through the catalog. This measurement-centered definition keeps
the cosmology case distinct from a general architecture description.

The image slicer forms part of the adopted spectrograph architecture. The slicer reformats a wide entrance region into a pseudo-slit without imposing the light loss of a narrow physical slit. The planning baseline assigns $R\simeq1000$ to the wide ELG survey because this choice preserves field and survey efficiency. Selectable bands at $R\simeq5000$ remain available for compact objects, precision line profiles, and tests of line identification. The cosmology study must compare both choices within the same image-slicer architecture. The comparison depends on redshift success, catastrophic line misidentification, spectral overlap, detector format, survey area, $nP$, and the resulting BAO and RSD precision. The numerical yield estimates below use $R=1000$ as the wide-survey reference rather than treating it as a final hardware commitment.

For a point source,
\begin{equation}
  \Delta v_{\rm FWHM}\simeq \frac{c}{R}
  \simeq 300\ {\rm km\,s^{-1}}\left(\frac{1000}{R}\right).
\end{equation}
The corresponding point-source widths are about $300$\,\kms\ at $R=1000$ and $60$\,\kms\ at $R=5000$. Both choices measure ELG redshifts for the same statistical clustering analysis. RSD constrains $\fsig(z)$ through the anisotropic galaxy power spectrum $P(k,\mu)$ or correlation function. The $R=1000$ design requires fewer detector samples and reduces overlap between neighbouring spectra. The $R=5000$ design separates the [O\,\textsc{ii}] doublet more securely and reduces catastrophic confusion among single-line identifications. The higher resolving power also measures line profiles and stellar absorption features more accurately. The preferred cosmology specification is the choice that yields the smaller final BAO and RSD covariance after completeness, purity, redshift uncertainty, and survey area enter the forecast.

The wavelength range sets the most restrictive requirement on spectral length. A constant-resolution spectrum over 0.3--3\,$\mu$m requires
\begin{equation}
  N_{\rm res}=R\ln\left(\frac{3}{0.3}\right)
  \simeq
  \begin{cases}
  2303 & (R=1000),\\
  11513 & (R=5000),
  \end{cases}
\end{equation}
or about 4600 and 23,000 pixels at two pixels per resolution element. Neither design records the full wavelength decade as one trace. The baseline uses two band-limited arms and three orientations per tile. The image slicer remains present at both resolving powers.

\begin{leadbox}[title=Baseline Configuration]{blue}
\begin{tabularx}{\textwidth}{@{}p{4.2cm}Y@{}}
\toprule
\textbf{Element} & \textbf{Adopted baseline for the proposal}\\
\midrule
Telescope & \facility\ with a diffraction-limited segmented aperture and robotic survey operations.\\
Spectroscopy & Image-sliced grism or prism channels. The wide-survey baseline is $R_{\rm point}\simeq1000$ with selectable $R_{\rm point}\simeq5000$ bands for precision spectroscopy.\\
Wavelength strategy & 0.3--1.0\,$\mu$m optical channel and a near-IR channel with science throughput to at least 2.70\,$\mu$m. The operational band-edge goal is 3.0\,$\mu$m. A reduction to 2.5\,$\mu$m is the formal engineering off-ramp if thermal background, cooling, detector, mass, power, or cost constraints prevent the longer band. Each channel is divided into filters to limit spectrum length.\\
Field of view & \fovsmall\ threshold and \fovlarge\ goal. Phase A must determine the recorded field at each candidate resolving power from the detector format and spectral length.\\
Roll strategy & Three image-sliced orientations per science tile near $0^\circ$, $45^\circ$, and $90^\circ$ at the selected resolving power.\\
Direct imaging & Required for source detection, morphology priors, wavelength zero-points, and spectral-overlap modeling.\\
\bottomrule
\end{tabularx}
\end{leadbox}

\subsection{Two-Arm Image-Sliced Spectrograph}

The spectrograph uses an image slicer and two wavelength arms. A dichroic sends 0.3--1.0\,$\mu$m light to the optical arm and 1.0--3.0\,$\mu$m light to the near-infrared arm. Band-limiting filters and a grism or prism prevent either arm from producing one trace across the full wavelength decade. The baseline operates both arms simultaneously. Sequential operation would require a second visit with the same orientation and pointing. The near-infrared arm must retain calibrated science throughput to at least $2.70\,\mu$m. The $3.0\,\mu$m value is an operational band-edge goal that provides margin beyond the science threshold. The $2.5\,\mu$m value is a formal off-ramp rather than the preferred design. The additional margin keeps the $z=3$ H$\alpha$+[N\,II]+[S\,II] complex away from filter transitions, detector quantum-efficiency roll-off, and wavelength-calibration edge losses. Section~\ref{sec:etc} compares all three red cutoffs and therefore supplies the engineering trade needed for the final detector and cooling decision.

The filter plan follows two properties of grism dispersion. A grism disperses at nearly constant $\Delta\lambda$ per pixel rather than constant $R=\lambda/\Delta\lambda$. The resolving power rises approximately linearly across one setting. The quoted $R=1000$ and $R=5000$ values refer to the centre of the corresponding setting. A grating also produces overlapping orders because second-order light at $\lambda$ lands at the same detector position as first-order light at $2\lambda$. At most one octave, $\lambda_{\max}\leq2\lambda_{\min}$, therefore remains free of cross-order contamination. The survey observes each arm through two band-limiting filters that obey the octave limit. The optical arm covers $0.36$--$0.70$ and $0.50$--$1.0\,\mu$m. The near-infrared arm covers $1.0$--$1.75$ and $1.6$--$3.0\,\mu$m. HST, JWST, Euclid, and Roman slitless instruments use the same order-sorting principle. The two settings per channel define the exposure times and survey schedule in the proposal. The $R=1000$ reference traces occupy about $1090$--$1330$ pixels at two pixels per resolution element. The $R=5000$ design requires about five times as many spectral samples over the same setting. A fixed detector format must absorb the increase through detector area, wavelength coverage, spatial field, or a combination of the three. The trade belongs to the comparison between resolving powers rather than to the adoption of the image slicer.

The image slicer acts as a wide entrance slit. The slicer accepts the same source area at both resolving powers and assigns each spatial element to a spectral trace. Increasing $R$ redistributes the same source and sky photons among more spectral samples. The signal-to-noise integrated over the same physical emission line or final wavelength interval is therefore independent of $R$ in the photon-limited limit. Binning an $R=5000$ spectrum to the wavelength bins used at $R=1000$ recovers the same signal-to-noise. One unbinned $R=5000$ sample has a lower value because the sample spans a narrower wavelength interval. The redshift fit combines all samples across the line profile and therefore retains the information in the complete line. Read noise, dark current, scattered light, spectral cross-talk, and extraction losses introduce a residual dependence when the higher resolving power illuminates more pixels.

The image-slicer geometry also limits the source width projected into the dispersion direction and reduces blending between different slices. The remaining cosmology trade does not follow from a fundamental photon penalty at $R=5000$. The $R=1000$ design uses shorter traces and reduces detector-format pressure. The $R=5000$ design separates neighbouring lines and lowers the catastrophic redshift rate. Phase A simulations must propagate both effects into the secure ELG density $n(z)$ and the usable survey area. The final comparison must report BAO and RSD covariance rather than S/N alone. Broad supernova features and reionization damping wings need no $R=5000$ resolution while individual Ly$\alpha$ forest absorbers require $R\gtrsim30{,}000$ and remain outside both choices.

\begin{leadbox}[title=Cosmology Selection Rule]{green}
\textbf{Use $R\simeq1000$ for the wide survey and reserve $R\simeq5000$ for selectable bands unless the matched forecast supports a wider high-resolution deployment.}
The $R\simeq5000$ spectrum retains the integrated photon-limited S/N after binning and adds information that separates the [O\,\textsc{ii}] doublet and rejects incorrect line identities. The $R\simeq1000$ design preserves shorter traces and wider survey efficiency. The deciding quantity is the final BAO and RSD covariance from the recovered catalog after secure-redshift density, usable area, and contamination enter the forecast.
\end{leadbox}

Robotic servicing may later add an $R\gtrsim15000$ high-dispersion module for
specialized follow-up. That module is an extension path and is excluded from
the baseline five-year yield, detector budget, and observing-time allocation.

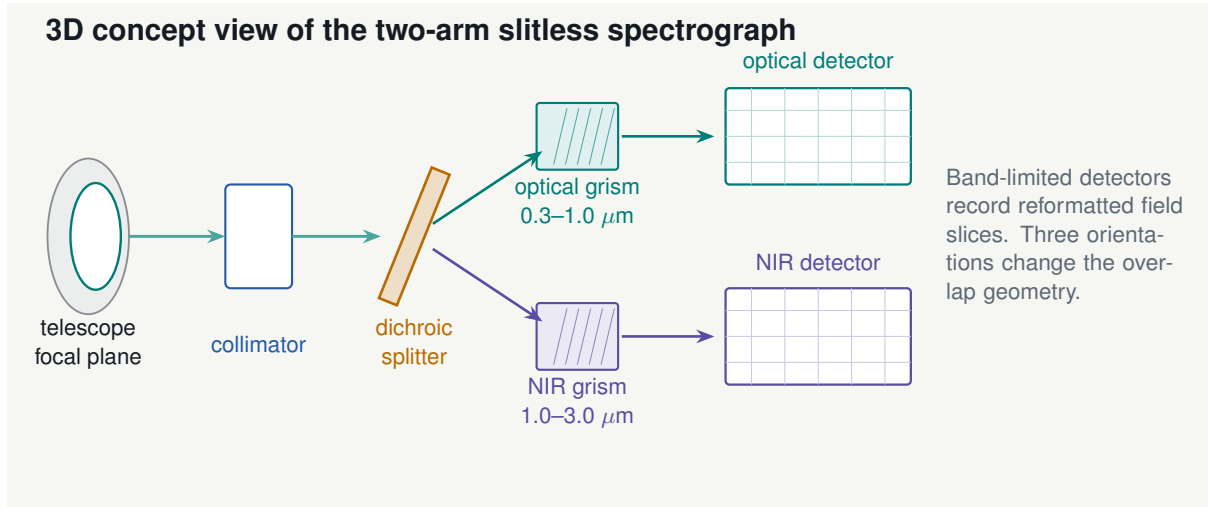
\begin{figure}[htbp]
\centering
\begin{tikzpicture}[font=\sffamily\footnotesize,scale=0.98, every node/.style={transform shape}]
  \fill[paper] (-0.4,-0.45) rectangle (15.8,6.35);
  \node[anchor=west,font=\bfseries\large,color=ink] at (0,5.95) {3D concept view of the two-arm slitless spectrograph};

  \coordinate (tel) at (0.7,3.2);
  \coordinate (coll) at (3.0,3.2);
  \coordinate (dich) at (5.1,3.2);
  \coordinate (optg) at (7.3,4.55);
  \coordinate (nrg) at (7.3,1.85);
  \coordinate (optd) at (10.55,4.55);
  \coordinate (nrd) at (10.55,1.85);

  \draw[fill=ink!8,draw=ink!50,line width=0.7pt] (tel) ellipse (0.55 and 1.05);
  \draw[fill=white,draw=teal,line width=0.9pt] ($(tel)+(0.1,0)$) ellipse (0.34 and 0.72);
  \node[color=ink,align=center] at (0.7,1.75) {telescope\\focal plane};

  \draw[fill=white,draw=blue,line width=0.8pt,rounded corners=2pt] ($(coll)+(-0.45,-0.7)$) rectangle ($(coll)+(0.45,0.7)$);
  \node[color=blue,align=center] at (3.0,1.75) {collimator};

  \draw[fill=gold!22,draw=gold,line width=1.0pt,rotate around={-22:(dich)}] ($(dich)+(-0.13,-0.95)$) rectangle ($(dich)+(0.13,0.95)$);
  \node[color=gold,align=center] at (5.1,1.75) {dichroic\\splitter};

  \foreach \p/\name/\col/\txt in {optg/optical grism/teal/{0.3--1.0 $\mu$m},nrg/NIR grism/violet/{1.0--3.0 $\mu$m}} {
    \draw[fill=\col!12,draw=\col,line width=0.8pt,rounded corners=1.2pt] ($(\p)+(-0.55,-0.45)$) rectangle ($(\p)+(0.55,0.45)$);
    \foreach \dx in {-0.32,-0.16,0,0.16,0.32}
      \draw[\col!70,line width=0.35pt] ($(\p)+(\dx,-0.38)$) -- ($(\p)+(\dx+0.18,0.38)$);
    \node[color=\col,align=center] at ($(\p)+(0,-0.9)$) {\name\\\txt};
  }

  \foreach \p/\name/\col in {optd/optical detector/teal,nrd/NIR detector/violet} {
    \draw[fill=white,draw=\col,line width=0.85pt,rounded corners=1.5pt] ($(\p)+(-1.25,-0.65)$) rectangle ($(\p)+(1.25,0.65)$);
    \foreach \x in {-0.9,-0.45,0,0.45,0.9}
      \draw[\col!30] ($(\p)+(\x,-0.65)$) -- ($(\p)+(\x,0.65)$);
    \foreach \y in {-0.35,0,0.35}
      \draw[\col!30] ($(\p)+(-1.25,\y)$) -- ($(\p)+(1.25,\y)$);
    \node[color=\col,align=center] at ($(\p)+(0,1.0)$) {\name};
  }

  \draw[-{Stealth[length=2.4mm]},line width=1.1pt,color=white!30!teal] (1.25,3.2) -- (2.55,3.2);
  \draw[-{Stealth[length=2.4mm]},line width=1.1pt,color=white!30!teal] (3.45,3.2) -- (4.65,3.2);
  \draw[-{Stealth[length=2.4mm]},line width=1.0pt,color=teal] (5.35,3.37) -- (6.83,4.35);
  \draw[-{Stealth[length=2.4mm]},line width=1.0pt,color=violet] (5.35,3.03) -- (6.83,2.05);
  \draw[-{Stealth[length=2.4mm]},line width=1.0pt,color=teal] (7.9,4.55) -- (9.15,4.55);
  \draw[-{Stealth[length=2.4mm]},line width=1.0pt,color=violet] (7.9,1.85) -- (9.15,1.85);

  \node[anchor=west,text width=3.2cm,align=left,color=muted] at (12.15,3.2)
  {{\hyphenpenalty=10000 Band-limited \mbox{detectors} record reformatted field slices. Three orientations change the overlap geometry.}};
\end{tikzpicture}
\caption{Self-generated 3D-style spectrograph schematic. An image slicer reformats the entrance field. A dichroic then splits the beam into optical and near-infrared arms with separate band-limited dispersers and detectors. The image slicer belongs to both candidate resolving-power designs.}
\label{fig:spectrograph}
\end{figure}

\subsection{Crowded-Field Slitless Deblending}\label{sec:deblending}

Slitless spectroscopy in crowded extragalactic fields is a deliberate design choice for the wide survey. A single dispersion direction is insufficient because neighbouring galaxy spectra overlap according to wavelength, morphology, and orientation. The baseline extraction therefore fits a \textbf{joint scene model over multiple orientations} rather than extracting each one-dimensional spectrum independently.

For each survey tile, the telescope obtains direct imaging and spectra at orientations separated by approximately $45^\circ$ and $90^\circ$. The $90^\circ$ observation supplies a nearly orthogonal dispersion direction. The $45^\circ$ observation places pairs that overlap in both orthogonal views into a third projection. The slitless extraction fit uses the following joint forward model.
\begin{equation}
  \mathbf{d}_{r}=\mathbf{A}_{r}(\mathbf{x},\mathbf{m},\lambda)\,\mathbf{s}+\mathbf{n}_{r},
  \qquad r\in\{0^\circ,45^\circ,90^\circ\},
  \label{eq:scene}
\end{equation} 
Here $\mathbf{d}_{r}$ denotes the detector image at orientation $r$ and $\mathbf{s}$ contains the source spectra. The projection operator $\mathbf{A}_{r}$ follows from the direct-image catalogue, source morphology, spectral traces, point-spread function, and wavelength calibration. A regularized least-squares fit recovers all spectra from the overdetermined system. Sparse matrix operations evaluate the fit without forming an explicit inverse. The direct-image morphology together with smoothness, non-negativity, and learned morphology priors constrains fully blended projections. The redshift catalogue records the covariance of the recovered spectra and the estimated contamination. Sources that remain degenerate after the three-orientation fit receive a low-confidence redshift flag or are excluded from the cosmology selection.

Three orientations define the current planning baseline for the wide survey. Two nearly orthogonal orientations provide the minimum geometrical constraint but leave some source pairs, detector defects, and extended-source self-blends degenerate. A third orientation separated by approximately $45^\circ$ supplies an independent overlap geometry. Injection-recovery tests must determine whether the gain in secure redshift density justifies the one-third reduction in survey area relative to two orientations. Additional orientations are reserved for reference fields, ultra-deep calibration fields, and unusually crowded targets.

\begin{table}[htbp]
\centering
\sffamily\small
\begin{tabularx}{\textwidth}{@{}c p{3.2cm} Y Y@{}}
\toprule
\textbf{Rolls} & \textbf{Survey role} & \textbf{Deblending value} & \textbf{Cost judgment}\\
\midrule
1 & Not acceptable for cosmology & A single view cannot distinguish many overlapping spectra from genuine lines. The configuration supplies no independent geometrical check. & Lowest cost with a high risk of redshift failure and contamination.\\
2 & Minimum operational mode & Near-orthogonal pair resolves many overlaps and morphology projections. & Useful fallback, but still vulnerable to pairs blended in both directions.\\
3 & Baseline for wide BAO/RSD & Supplies a third overlap geometry and permits tests in which one orientation is omitted. & Adopted planning point pending quantitative injection-recovery tests.\\
4+ & Calibration/deep fields & Constrains rare degeneracies and tests the stability of the spectral extraction. & Assigned only when the calibration gain exceeds the corresponding loss of survey area.\\
\bottomrule
\end{tabularx}
\caption{Cost-benefit logic for the number of slitless roll orientations. The proposal adopts three rolls for the wide cosmology survey and uses additional rolls only where calibration value exceeds lost survey area.}
\label{tab:rollcost}
\end{table}

The baseline spectral extraction combines calibrated detector forward modelling, simultaneous fits to all catalogued sources, and regularized sparse linear inversion across every orientation and dither. HST slitless software defines the extraction geometry from direct-image positions, source morphology, trace solutions, flat-field cubes, and contamination models \cite{axe2009}. The LINEAR formalism assembles the transformations between direct-image sources and dispersed pixels into a sparse linear system and solves the system with LSQR \cite{linear2018}. Crowded-field three-dimensional spectroscopy applies the same physical approach by fitting many sources simultaneously with a wavelength-dependent point-spread function and an external high-resolution catalogue prior \cite{kamann2013,husser2016}. Machine-learning classifiers support rather than replace the detector-level fit. The classifiers identify uncatalogued neighbours, define trace masks, reject artifacts, flag highly contaminated pixels, and test catalogue completeness. Astronomical Mask R-CNN applications demonstrate convolutional and instance-segmentation methods for those tasks \cite{burke2019}. Bayesian variational inference estimates posterior distributions for source counts, positions, and fluxes \cite{liu2021}. The posterior distributions propagate into contamination and redshift-quality estimates. Non-negative matrix factorization has reduced calibration crosstalk in the Keck OSIRIS integral-field spectrograph and supplies a relevant test for calibration components \cite{horstman2022}. Generative models, including diffusion models, remain outside the baseline cosmology extraction unless the models reproduce the detector pixels through the calibrated forward operator and satisfy injection-recovery bias limits. Acceptance depends on recovered line-flux bias, redshift failure rate, contamination covariance, and injection-recovery completeness as functions of neighbour separation, magnitude contrast, morphology, and orientation geometry. Visual cleanliness alone does not establish cosmology-grade performance.

Four method families provide useful components for the calibrated forward model. Constrained matrix factorization forms the first family. SCARLET represents each source with a small number of components whose spectral energy distributions multiply non-parametric spatial profiles. Non-negativity, radial monotonicity, and symmetry regularize the source separation \cite{scarlet2018}. SCARLET is a semi-parametric optimization method and serves as the production image deblender for the Rubin Observatory data-processing system. Deep generative priors form the second family. Variational autoencoders have been applied directly to astronomical deblending \cite{arcelin2021}, while deep generative galaxy models supply morphology priors and realistic simulated populations \cite{lanusse2021}. Normalizing flows, score-based models, and diffusion models belong to the same family. The cited studies have not yet demonstrated the line-flux bias control required for slitless cosmology. Instance-segmentation networks such as Mask R-CNN form the third family and identify overlapping sources at the pixel level \cite{burke2019}. Probabilistic catalogue inference forms the fourth family and estimates posterior distributions for source positions and fluxes \cite{liu2021}. Cosmology-grade acceptance depends on line-flux bias rather than image-plane separation alone. The baseline therefore restricts learned models to morphology priors, source masks, and quality tests within the calibrated detector fit. Every learned prior must reproduce the detector data through the same forward operator and satisfy injection-recovery limits before contributing to the cosmology sample.

Euclid and Roman use different source-separation strategies that define relevant precedents for the proposed survey. The Euclid Morphology Challenge compared SourceXtractor++ with four other profile-fitting methods on about 1.5 million simulated galaxies. The tests forecast robust structural measurements for at least 400 million survey galaxies and quantify the loss of accuracy caused by non-analytic morphologies \cite{sourcext2020,euclidmorph2023}. Euclid's SIR spectroscopic reduction uses a photometric source catalogue to extract NISP spectra and model cross-contamination \cite{euclidsir2026}. Roman applies pyLINEAR to grism spectra. The sparse inversion assembles every transformation between direct-image sources and dispersed pixels into one system \cite{linear2018,wang2021}. Roman supernova simulations extend the method across multiple orientations and reconstruct a three-dimensional spectral scene that separates host-galaxy light from the supernova \cite{astraatmadja2026}. Forward-model emulators such as ESpRESSO generate realistic Roman slitless scenes for training and validation \cite{espresso2024}. Rubin imaging uses the constrained SCARLET factorization for source separation \cite{scarlet2018}. Because the proposed survey combines slitless spectra with multiple orientations, the Roman and HST detector forward model in Eq.~\eqref{eq:scene} defines the baseline extraction. Euclid profile fits, Rubin source models, and generative morphology models supply constrained priors and validation tests.

\begin{figure}[htbp]
\centering
\includegraphics[width=\textwidth]{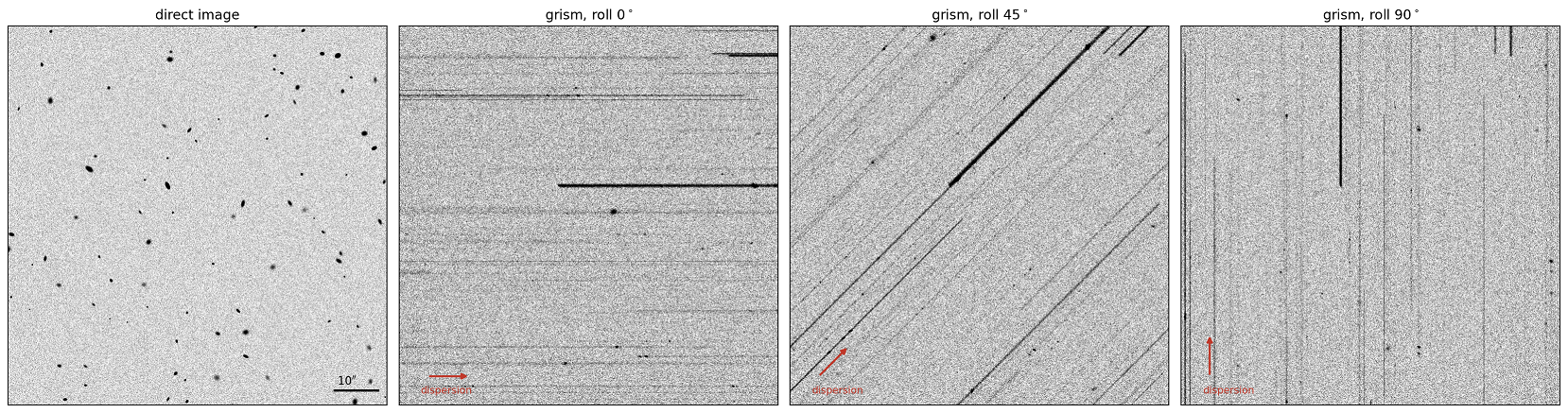}
\caption{Unsliced control simulation of crowded-field slitless exposures with $0.11''$ pixels and the $1.0$--$1.75\,\mu$m band-limited setting at $R=1000$. The constant dispersion gives $1091$-pixel first-order traces. Sources follow deep near-infrared counts, $40$ per arcmin$^2$ over $18.5<m_{\rm AB}<24.5$, with FSPS galaxy template spectra at $0.4<z<2.2$. \emph{Top left} shows the direct image of the displayed $88''\times88''$ field. The other panels show the same field dispersed at three survey orientations. Traces from sources outside the crop enter the displayed region as on a real detector. Every source produces a long one-sided trace and emission lines appear as compact knots. Neighbouring spectra overlap heavily at one orientation. The rotated views place source pairs into different overlap configurations. The control case demonstrates the confusion that the adopted image slicer and the joint scene reconstruction of Eq.~\eqref{eq:scene} must reduce. The panel illustrates overlap geometry rather than completed recovery statistics. Section~\ref{sec:e2eval} defines the required survey-scale tests of purity, completeness, and flux bias.}
\label{fig:crowded}
\end{figure}

The three survey orientations produce distinct overlap geometries. The adopted image slicer divides the focal-plane region into strips and reformats the strips into a pseudo-slit. The same geometry applies to both candidate resolving powers. Sources in different slices do not overlap. Sources that share a slice remain subject to spectral blending. The restricted adjacency reduces the number of coupled sources relative to the unsliced control in Figure~\ref{fig:crowded}. Flight instruments including the JWST NIRSpec integral-field unit use image slicers \cite{boker2022}. The flight heritage establishes the optical principle. The proposed wide-field implementation still requires an optical demonstration.

Detector format and survey grasp constrain both resolving-power designs. The $R=5000$ design requires five times as many spectral samples over the same band as the $R=1000$ design. The increased trace length may reduce the spatial field or require more detectors. The $R=1000$ reference calculation currently supplies the quoted survey yield because the proposal has not completed a matched $R=5000$ detector-layout and recovery simulation. The reference status does not select the flight resolution. Phase A must determine the slice width and field for both choices from the same end-to-end optical model. The study must measure throughput, cross-talk, wavelength stability, redshift completeness, catastrophic failures, and blind recovery of injected sources before the science team recommends one resolving power.

\begin{figure}[htbp]
\centering
\includegraphics[width=\textwidth]{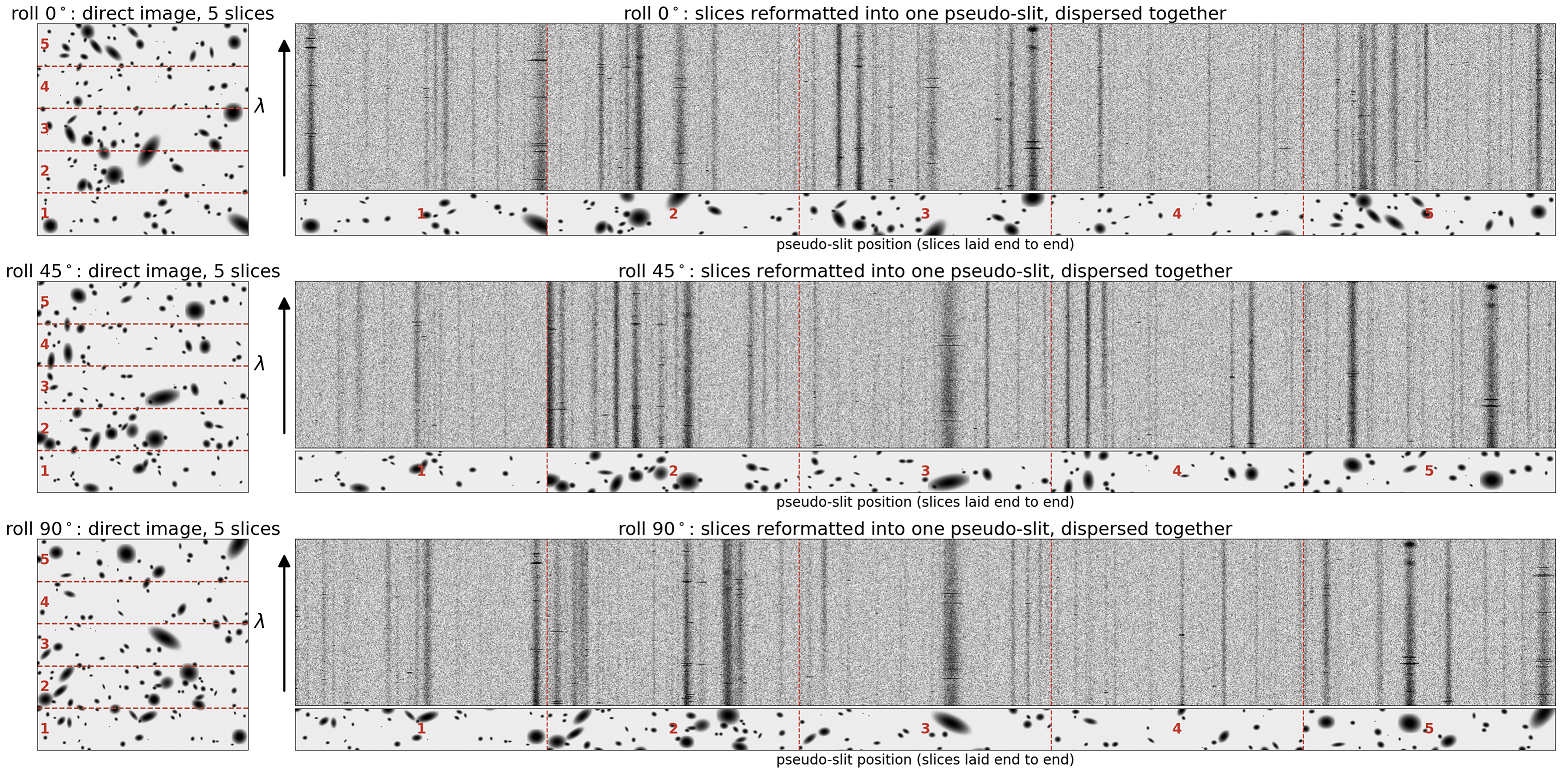}
\caption{Image-slicer geometry applied to a scene whose source positions, brightnesses, sizes, ellipticities, and position angles come from a Subaru HSC-SSP deep image of the SXDS/UDS/XMM-LSS field \cite{aihara2018}. The rendering uses the proposed $0.11''$ pixels and diffraction-limited point-spread function. Each row shows one orientation. The left panel divides the reconstructed image into five horizontal slices. The right panel reformats the slices into one pseudo-slit and disperses the result. The demonstration uses $R=1000$ to keep the image scale directly comparable with Figure~\ref{fig:crowded}. The same slicing principle applies at $R=5000$. Dashed lines mark the segment assigned to each slice. Sources in different slices do not overlap. The five-slice layout illustrates the mapping and does not define the final optical design.}
\label{fig:slicer}
\end{figure}

\subsection{Effective Resolution for Galaxies}\label{sec:effres}

Galaxy morphology along the dispersion axis reduces the effective resolving power according to
\begin{equation}
  R_{\rm eff}\simeq R_{\rm point}
  \frac{\theta_{\rm PSF}}{(\theta_{\rm PSF}^2+\theta_{\rm src,\parallel}^2)^{1/2}} .
  \label{eq:reff1000}
\end{equation}
At 1\,$\mu$m, a 3.5 m diffraction-limited telescope has a point-spread-function FWHM near $0.06''$. Without spatial reformatting, $R_{\rm point}=1000$ would give $R_{\rm eff}\simeq510$ for a $0.10''$ emitting region and $R_{\rm eff}\simeq290$ for a $0.20''$ emitting region. The corresponding values at $R_{\rm point}=5000$ would reach about $2550$ and $1450$. The adopted image slicer limits $\theta_{\rm src,\parallel}$ through the slice width at either resolving power. The recovered $R_{\rm eff}$ depends on slice width, source centering, optical aberrations, and line-emitting morphology. Phase A must measure the line-spread function with extended-source scenes at both candidate resolutions rather than infer the performance from point sources alone.

Extended galaxies also create an internal deblending problem. Different regions of the same galaxy project to different detector pixels along the dispersion direction, which mixes spatial structure with wavelength. The calibrated forward model fits this morphology-induced self-blending together with neighbouring-source contamination. The direct image supplies a wavelength-zero-point prior, segmentation map, Sersic or multi-component morphology, and clump masks. For galaxies whose emission-line morphology differs from the broadband continuum, the joint fit represents the galaxy with a small number of internal components,
\begin{equation}
  \mathbf{d}_{r,g} =
  \sum_{c=1}^{N_c}
  \mathbf{A}_{r,g,c}(\mathbf{x}_{g,c},\mathbf{m}_{g,c},\lambda)\,
  \mathbf{s}_{g,c}
  +\mathbf{n}_{r,g},
  \label{eq:scenecomp}
\end{equation}
where $g$ labels the galaxy and $c$ labels internal components such as a disk, bulge, or star-forming knot. Equation~\eqref{eq:scenecomp} expands the scene model in Eq.~\eqref{eq:scene} rather than defining an independent model. The per-object spectrum $\mathbf{s}$ for galaxy $g$ becomes a sum over internal spectra $\mathbf{s}_{g,c}$, each with a projection operator $\mathbf{A}_{r,g,c}$. A single linear system simultaneously fits contamination between galaxies and self-blending within each galaxy. The component expansion is the extended-source analogue of the LINEAR sparse forward model and the aXe morphology-based extraction geometry \cite{axe2009,linear2018}. The CSST slitless emulator similarly treats galaxy size, Sersic index, position angle, and axis ratio as determinants of slitless self-broadening \cite{cess2024}. The science catalogue therefore reports $R_{\rm eff}$, line-flux covariance, morphology flags, and a self-blending quality flag. Sources whose inferred redshift or line flux shifts beyond the validated tolerance when the morphology prior, component number, or omitted orientation is varied remain available for ancillary science but are excluded from the cosmology-grade BAO and RSD sample.

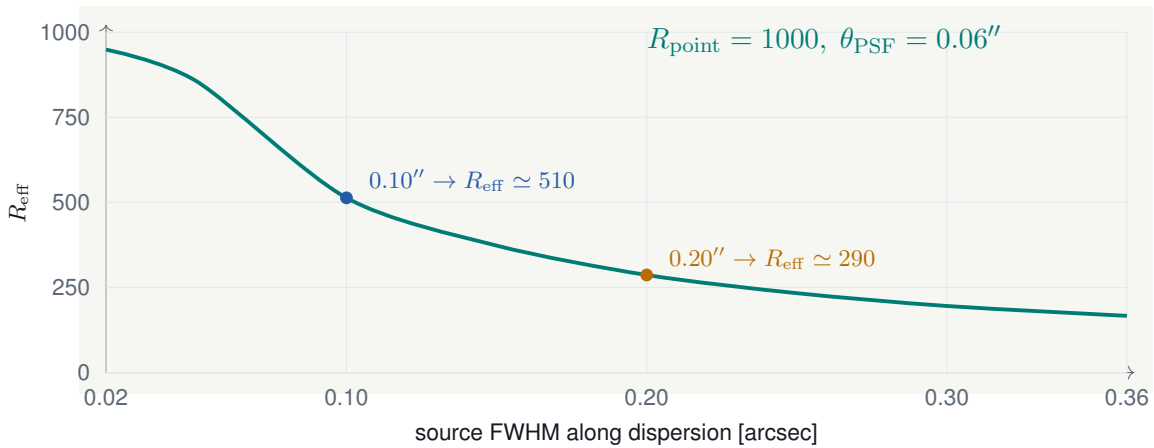
\begin{figure}[htbp]
\centering
\begin{tikzpicture}[font=\sffamily\footnotesize]
  \def\w{13.5}
  \def\h{4.5}
  \fill[paper] (-0.35,-0.28) rectangle (\w+0.35,\h+0.35);
  \draw[->,color=muted] (0,0) -- (\w+0.1,0);
  \draw[->,color=muted] (0,0) -- (0,\h+0.1);
  \foreach \y/\lab in {0/0,1.125/250,2.25/500,3.375/750,4.5/1000} {
    \draw[line!70] (0,\y) -- (\w,\y);
    \node[anchor=east,color=muted] at (-0.08,\y) {\lab};
  }
  \foreach \x/\lab in {0/0.02,3.18/0.10,7.15/0.20,11.12/0.30,13.5/0.36} {
    \draw[line!70] (\x,0) -- (\x,\h);
    \node[anchor=north,color=muted] at (\x,-0.08) {\lab};
  }
  \draw[teal,line width=1.45pt,smooth]
    plot coordinates {(0,4.27) (1.2,3.85) (3.18,2.31) (5.16,1.68) (7.15,1.29) (9.13,1.05) (11.12,0.88) (13.5,0.75)};
  \fill[blue] (3.18,2.31) circle (2.4pt);
  \fill[gold] (7.15,1.29) circle (2.4pt);
  \node[anchor=west,color=blue] at (3.35,2.55) {$0.10''\rightarrow R_{\rm eff}\simeq510$};
  \node[anchor=west,color=gold] at (7.32,1.52) {$0.20''\rightarrow R_{\rm eff}\simeq290$};
  \node[anchor=south,color=teal,font=\bfseries] at (9.5,4.05) {$R_{\rm point}=1000,\ \theta_{\rm PSF}=0.06''$};
  \node[anchor=north,color=ink] at (0.5*\w,-0.55) {source FWHM along dispersion [arcsec]};
  \node[rotate=90,anchor=south,color=ink] at (-0.9,0.5*\h) {$R_{\rm eff}$};
\end{tikzpicture}
\caption{At $R=1000$, morphology sets the effective resolution for galaxies. The relevant design goal is robust redshift recovery, not galaxy internal kinematics.}
\label{fig:reff1000}
\end{figure}

\subsection{Field of View as a Science Requirement}

The instantaneous field of view determines whether the observatory completes a wide cosmological survey within the mission lifetime. The two candidate field sizes differ by a factor of nine in area according to
\begin{align}
  A(10'\times10') &= 100\ {\rm arcmin^2}=0.0278\ {\rm deg^2},\\
  A(30'\times30') &= 900\ {\rm arcmin^2}=0.2500\ {\rm deg^2}.
\end{align}
For a 100 \degti\ survey, the threshold field requires roughly 3,600 pointings before the required three orientations and dithers while the maximum field requires 400. For 300 \degti, the corresponding counts are 10,800 and 1,200. Three slitless orientations multiply the raw spectroscopic visit count by three. The \fovsmall\ mode is appropriate for deep calibration fields and time-domain fields. The wide cosmology survey requires the \fovlarge\ mode.

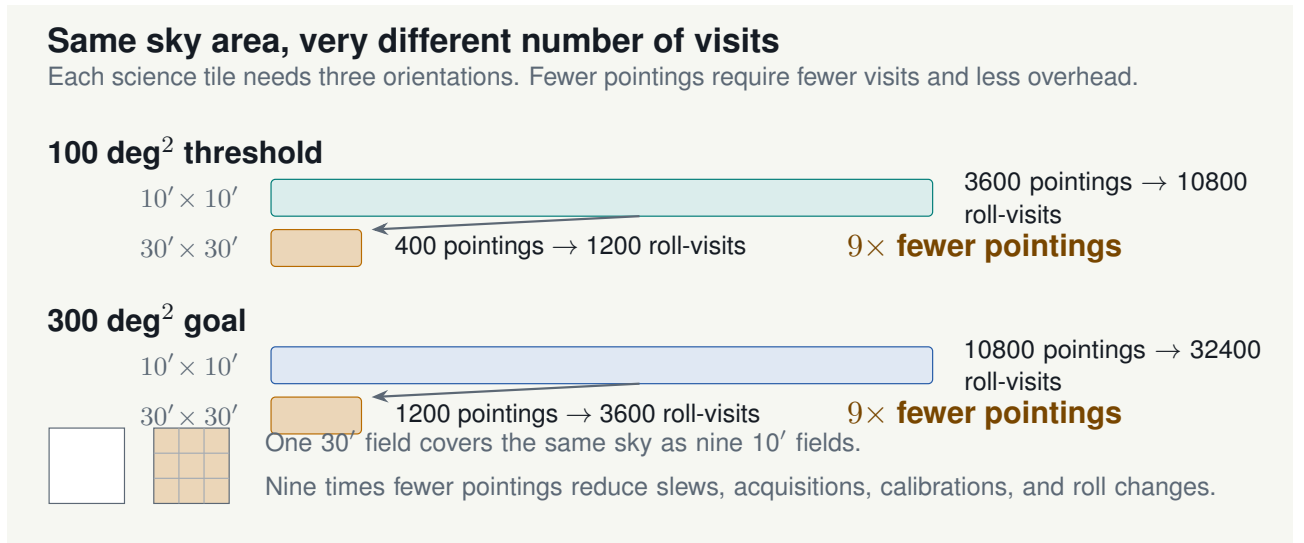
\begin{figure}[htbp]
\centering
\resizebox{\textwidth}{!}{%
\begin{tikzpicture}[font=\sffamily\footnotesize]
  \fill[paper] (-0.35,-0.45) rectangle (15.0,6.0);
  \node[anchor=west,font=\bfseries\large,color=ink] at (0,5.55) {Same sky area, very different number of visits};
  \node[anchor=west,color=muted] at (0,5.12) {Each science tile needs three orientations. Fewer pointings require fewer visits and less overhead.};

  \foreach \y/\survey/\smallp/\smallv/\largep/\largev/\col in {
    3.45/{100 deg$^2$ threshold}/3600/10800/400/1200/teal,
    1.45/{300 deg$^2$ goal}/10800/32400/1200/3600/blue} {
    \node[anchor=west,font=\bfseries,color=ink] at (0,\y+0.78) {\survey};
    \node[anchor=east,color=muted] at (2.55,\y+0.25) {\fovsmall};
    \node[anchor=east,color=muted] at (2.55,\y-0.35) {\fovlarge};

    \fill[\col!14,draw=\col,rounded corners=1.5pt] (2.8,\y+0.03) rectangle (10.7,\y+0.47);
    \fill[gold!28,draw=gold,rounded corners=1.5pt] (2.8,\y-0.57) rectangle (3.88,\y-0.13);

    \node[anchor=west,color=ink,text width=3.8cm] at (10.95,\y+0.25) {\smallp\ pointings $\rightarrow$ \smallv\ roll-visits};
    \node[anchor=west,color=ink] at (4.15,\y-0.35) {\largep\ pointings $\rightarrow$ \largev\ roll-visits};
    \node[anchor=west,font=\bfseries,color=gold!65!black] at (9.55,\y-0.35) {$9\times$ fewer pointings};

    \draw[-{Stealth[length=2mm]},color=muted,line width=0.75pt] (7.2,\y+0.03) -- (4.0,\y-0.13);
  }

  \begin{scope}[xshift=0.15cm,yshift=0.05cm]
    \draw[draw=muted,fill=white] (0,0) rectangle (0.9,0.9);
    \draw[draw=muted,fill=gold!28] (1.25,0) rectangle (2.15,0.9);
    \foreach \xx in {1.55,1.85} \draw[muted!55] (\xx,0) -- (\xx,0.9);
    \foreach \yy in {0.30,0.60} \draw[muted!55] (1.25,\yy) -- (2.15,\yy);
    \node[anchor=west,text width=11.8cm,align=left,color=muted] at (2.45,0.72)
      {One 30$'$ field covers the same sky as nine 10$'$ fields.};
    \node[anchor=west,text width=11.8cm,align=left,color=muted] at (2.45,0.18)
      {{\hyphenpenalty=10000 Nine times fewer pointings reduce slews, acquisitions, calibrations, and roll changes.}};
  \end{scope}
\end{tikzpicture}}
\caption{Field-of-view scaling for the wide RSD survey. The \fovlarge\ mode covers the same sky with one ninth as many pointings as \fovsmall. After the required three roll orientations, a 100 deg$^2$ survey needs about 1200 rather than 10,800 slitless visits.}
\label{fig:fov}
\end{figure}

\subsection{Sensitivity and the Exposure-Time Calculator}\label{sec:etc}

The sensitivity of the image-sliced two-arm spectrograph and the survey planning depth adopted in Section~\ref{sec:elgdepth} follow from a detector-level photon budget. The calculation separates the absolute sensitivity estimate from the comparison of resolving power. For an emission line of flux $F_{\rm line}$ observed at wavelength $\lambda$, a total exposure $t$ produces
\begin{equation}
  S=F_{\rm line}\,A_{\rm tel}\,\eta\,t\,\frac{\lambda}{hc},
  \label{eq:etcsig}
\end{equation}
where $A_{\rm tel}$ is the effective collecting area and $\eta$ is the end-to-end throughput including optics, disperser, and detector quantum efficiency. The factor $\lambda/hc$ converts line energy to photon number. The image slicer assigns a spatial element and a wavelength interval $\delta\lambda_{\rm pix}$ to each detector sample. The diffuse background collected in one sample per second is
\begin{equation}
  B_{\rm pix}=i_{\lambda,{\rm sky}}\,A_{\rm tel}\,\eta\,\Omega_{\rm pix}\,\delta\lambda_{\rm pix}\,\frac{\lambda}{hc},
  \label{eq:etcbg}
\end{equation}
where $i_{\lambda,{\rm sky}}$ is the sky surface brightness per unit wavelength and $\Omega_{\rm pix}$ is the solid angle assigned to one spatial sample. Let $\Delta\lambda_{\rm sci}$ denote the physical wavelength interval used for a line-flux or redshift measurement. The number of spectral samples is $N_\lambda=\Delta\lambda_{\rm sci}/\delta\lambda_{\rm pix}$. Summing the sky over the full spatial extraction $\Omega_{\rm ext}$ gives
\begin{equation}
 B_{\rm sci}=i_{\lambda,{\rm sky}}A_{\rm tel}\eta\Omega_{\rm ext}
 \Delta\lambda_{\rm sci}\frac{\lambda}{hc}.
 \label{eq:etcslicerbg}
\end{equation}
The factors $\delta\lambda_{\rm pix}$ and $N_\lambda$ cancel. Equation~\eqref{eq:etcslicerbg} therefore contains no explicit resolving power. Source shot noise, integrated sky emission, dark current $D$, and read noise $\sigma_{\rm read}$ over $n_{\rm det}$ illuminated detector samples give the total variance
\begin{equation}
  N^2=S+B_{\rm sci}t+D\,n_{\rm det}t+n_{\rm read}\,n_{\rm det}\,\sigma_{\rm read}^2,
  \label{eq:etcnoise}
\end{equation}
The $5\sigma$ line-flux limit is the flux for which $S=5N$. The sky term dominates under the space-based near-infrared background when detector noise remains subdominant. Setting $S=5N$ with $F_{\rm line}=F_{5\sigma}$ gives
\[
  F_{5\sigma}\,A_{\rm tel}\,\eta\,t\,\frac{\lambda}{hc}=5\,(B_{\rm sci}t)^{1/2},
\]
Solving for the flux yields
\begin{equation}
  F_{5\sigma}=5\left(\frac{i_{\lambda,{\rm sky}}\Omega_{\rm ext}
  \Delta\lambda_{\rm sci}}
  {A_{\rm tel}\eta(\lambda/hc)t}\right)^{1/2}.
  \label{eq:etc}
\end{equation}
The background-limited depth scales as $t^{-1/2}$. More importantly, the same $\Delta\lambda_{\rm sci}$ gives the same photon-limited S/N at $R=1000$ and $R=5000$. The image slicer acts as a wide slit and retains the same source area. Higher dispersion only redistributes the photons among more detector samples.

For continuum flux density $f_\lambda$ integrated over the fixed science interval $\Delta\lambda_{\rm sci}$, the background-limited signal-to-noise is
\begin{equation}
 \left(\frac{S}{N}\right)_{\rm cont}
 \simeq
 \frac{f_\lambda\,[A_{\rm tel}\eta t(\lambda/hc)\Delta\lambda_{\rm sci}]^{1/2}}
 {(i_{\lambda,{\rm sky}}\Omega_{\rm ext})^{1/2}}
 \propto R^{0},
 \label{eq:snrcontR}
\end{equation}
because neither the source photons nor the integrated sky photons depend on the number of spectral samples. A quoted S/N per native resolution element uses $\Delta\lambda_{\rm sci}=\lambda/R$ and scales as $R^{-1/2}$. The native-bin convention does not describe the S/N of the same physical feature. Binning the $R=5000$ spectrum to the science interval used at $R=1000$ recovers the same photon-limited S/N.

An emission line integrated over a fixed physical profile follows
\begin{equation}
 \left(\frac{S}{N}\right)_{\rm line}
 \bigg|_{\Delta\lambda_{\rm sci}\ {\rm fixed}}
 \propto R^0,
 \label{eq:snrlineR}
\end{equation}
when the same line interval is used at both resolutions. The $R=5000$ design does not gain total line photons. The design instead resolves line structure and improves the likelihood of the correct line identity. Read noise, dark current, scattered light, spectral cross-talk, and extraction losses produce a practical penalty when the higher dispersion uses more detector samples. Equations~\eqref{eq:snrcontR} and~\eqref{eq:snrlineR} state the comparison relevant to cosmological redshift recovery.

Table~\ref{tab:etc} lists the inputs required for the image-sliced calculation. The mean zodiacal sky brightness near $1.6\,\mu$m follows the standard reference model \cite{leinert1998}. The collecting area follows from the $3.5$\,m aperture and the detector terms represent near-infrared HgCdTe arrays. The slice geometry must supply $\Omega_{\rm ext}$ and $n_{\rm det}$ before Eq.~\eqref{eq:etc} produces a flight sensitivity. The current proposal therefore retains $1.0\times10^{-16}\,{\rm erg\,s^{-1}\,cm^{-2}}$ at $0.75$\,hr as a conservative planning reference derived from comparable space surveys. The reference depth applies to both candidate resolving powers in the photon-limited comparison. A matched image-slicer ETC must quantify the residual detector penalty at $R=5000$.

The ETC must evaluate the red channel at three explicit cutoffs. The $2.5\,\mu$m case is the formal engineering off-ramp. The $2.70\,\mu$m case is the minimum calibrated science-throughput requirement. The $3.0\,\mu$m case is the operational band-edge goal. The comparison must report wavelength-dependent throughput, thermal background, detector noise, line-flux depth, and secure-redshift fraction for each cutoff. It must also propagate the cutoff into the accessible H$\alpha$ and [S\,II] volume rather than comparing only a single representative wavelength. These three cases separate a science requirement from the additional margin that the instrument team may retain if the cooling and detector design permit it.

\begin{table}[htbp]
\centering
\sffamily\small
\begin{tabularx}{\textwidth}{@{}l p{3.1cm} X@{}}
\toprule
\textbf{Parameter} & \textbf{Baseline value} & \textbf{Note}\\
\midrule
Effective area $A_{\rm tel}$ & $8\times10^{4}$ cm$^{2}$ & $3.5$\,m aperture with central obstruction\\
Throughput $\eta$ & $0.30$ & optics, disperser, and detector quantum efficiency\\
Wavelength $\lambda$ & $1.6\,\mu$m & representative near-infrared line. Repeat the ETC for red cutoffs of $2.5$, $2.70$, and $3.0\,\mu$m\\
Sky brightness $\bar i_{\rm sky}$ & $9.6\times10^{-19}$ erg s$^{-1}$ cm$^{-2}$ \AA$^{-1}$ arcsec$^{-2}$ & reddened zodiacal light at $1.6\,\mu$m, $\simeq21.6$ AB arcsec$^{-2}$ \cite{leinert1998}\\
Extraction solid angle $\Omega_{\rm ext}$ & Set by the slice geometry & same accepted source region in both resolving-power calculations\\
Science interval $\Delta\lambda_{\rm sci}$ & Set by the fitted line profile & identical physical interval for the $R=1000$ and $R=5000$ comparison\\
Detector samples $n_{\rm det}$ & Set by slice width and dispersion & increases with $R$ and controls detector-noise corrections\\
Dark current $D$ & $0.01$ e$^-$ s$^{-1}$ pix$^{-1}$ & near-infrared HgCdTe baseline\\
Read noise $\sigma_{\rm read}$ & $10$ e$^-$ pix$^{-1}$ & per read, three reads\\
\bottomrule
\end{tabularx}
\caption{Baseline detector inputs for the image-sliced exposure-time calculation. The comparison between $R=1000$ and $R=5000$ uses the same source aperture and physical line interval. The photon-limited S/N is therefore independent of resolving power. Detector terms retain a dependence through the number of illuminated samples.}
\label{tab:etc}
\end{table}

\begin{table}[htbp]
\centering
\sffamily\small
\begin{tabularx}{\textwidth}{@{}p{2.5cm}p{3.6cm}YY@{}}
\toprule
\textbf{Resolution} & \textbf{Cosmology benefit} & \textbf{Cosmology cost} & \textbf{S/N at fixed line interval}\\
\midrule
$R=1000$ & Shorter spectra and lower detector-format pressure & More blended line complexes and a higher risk of single-line misidentification & Equal in the photon-limited limit\\
$R=5000$ & Secure [O\,\textsc{ii}] separation, lower catastrophic-redshift fraction, and more precise line centroids & Five times more spectral samples and a possible loss of field or wavelength coverage & Equal in the photon-limited limit
\end{tabularx}
\caption{Cosmology trade between the two candidate resolving powers. Both designs use the adopted image slicer. The slicer accepts the same wide entrance region and preserves the integrated photon-limited S/N. The final selection requires matched detector layouts and injection tests that measure secure redshift density, usable area, and BAO and RSD covariance.}
\label{tab:etcmodes}
\end{table}

The numerical calculator\footnote{Open-source at \url{https://github.com/kjhan0606/3.5mST}.} tracks the photon budget wavelength by wavelength. The calculator integrates the zodiacal spectrum, Galactic cirrus, and telescope thermal emission over the selected extraction interval. The calculation also includes wavelength-dependent throughput, quantum efficiency, optimal-extraction efficiency, detector noise, and cosmic-ray losses \cite{ienaka2013,jwstcrdocs}. The present calculation returns the absolute sensitivity reference and permits comparisons with JWST, Roman, and Euclid. The next instrument configuration must include the adopted slice geometry and evaluate $R=1000$ and $R=5000$ through the same source aperture. Only the number of detector samples differs between the two cases. The matched calculation will test the photon-limited $R^0$ scaling and measure departures caused by detector noise, cross-talk, scattered light, and extraction losses. Figure~\ref{fig:etcgui} shows the present graphical interface.

The baseline sensitivities quoted in this key science do not apply a cosmic-ray loss correction. The planned observing sequence uses multiple exposures. The resulting efficiency loss will enter the final exposure-time budget after the detector readout pattern, ramp-fitting performance, and in-orbit cosmic-ray rate are fixed. Cosmic-ray losses therefore remain outside the present planning depth and science forecasts.

The ETC parameter files are versioned planning inputs rather than a flight calibration. Phase A must deliver one instrument description for each grism, image-slicer, and imaging setting. Each description must contain the dichroic, filter, disperser, mirror, and detector quantum-efficiency curves on a common wavelength grid. The image-slicer description must also contain the slice geometry, field mapping, throughput, cross-talk, scattered-light model, and line-spread function. Every description must specify detector temperature, readout pattern, dark-current distribution, bad-pixel fraction, full well, read-noise covariance, telescope emissivity, and the thermal model seen by the near-infrared focal plane. Each simulated detector scene records the ETC inputs used for its realization. A forecast qualifies for acceptance testing only after measured component curves replace the planning curves and their uncertainties propagate through spectral extraction.

JWST, Roman, and Euclid provide sensitivity comparisons for similar aperture classes or observing modes. Table~\ref{tab:jwstval} evaluates their documented imaging and spectroscopic reference cases and compares the ETC estimates with the published sensitivities.

\begin{table}[htbp]
\centering
\sffamily\footnotesize
\setlength{\tabcolsep}{4pt}
\begin{tabularx}{\textwidth}{@{}l Y c c c@{}}
\toprule
\textbf{Mission case} & \textbf{Published sensitivity} & \textbf{Exposure} & \textbf{ETC estimate} & \textbf{Estimate/reference}\\
\midrule
\multicolumn{5}{@{}l}{\textbf{JWST} \cite{jdoxnircamsens,jdoxnirspecsens}}\\
NIRCam F200W (imaging) & AB\,29.0 point, S/N$=10$ & 10\,ks & S/N$=8.7$ & $0.87\times$ (13\% low)\\
NIRCam F444W (imaging) & AB\,27.9 point, S/N$=10$ & 10\,ks & S/N$=11.0$ & $1.10\times$ (10\% high)\\
NIRSpec R=1000 (line) & $5.7\times10^{-19}$ @\,2\,$\mu$m, S/N$=10$ & 100\,ks & S/N$=7.8$ & $0.78\times$ (22\% low)\\
\multicolumn{5}{@{}l}{\textbf{Roman / Euclid grism, H$\alpha$ at 1.6\,$\mu$m}}\\
Roman HLSS grism \cite{wang2021} & $1.0\times10^{-16}$ erg\,s$^{-1}$\,cm$^{-2}$, $6.5\sigma$, $0.3''$ src & $1000$\,s & $7.3\times10^{-17}$ & $1.4\times$ deeper\\
Euclid Wide NISP grism \cite{euclid2011} & $3.5\times10^{-16}$ erg\,s$^{-1}$\,cm$^{-2}$, $3.5\sigma$, $0.5''$ src & $2240$\,s & $8.7\times10^{-17}$ & $4.0\times$ deeper\\
\multicolumn{5}{@{}l}{\textbf{Roman / Euclid imaging}}\\
Roman WFI F158 (Wide Tier) \cite{romandocs} & $26.2$ AB, $5\sigma$ point source & $642$\,s & $26.07$ AB & $0.13$\,mag shallower\\
Euclid NISP $H$ (design floor) \cite{scaramella2022} & $24.0$ AB, $5\sigma$ point source & $1344$\,s & $24.83$ AB & $0.83$\,mag deeper\\
\bottomrule
\end{tabularx}
\caption{ETC comparison with documented JWST, Roman, and Euclid sensitivities. The published-sensitivity column lists each mission's reference limit for the stated source, exposure, and configuration. The JWST values include the NIRCam point-source depths at AB $29.0$ in F200W and $24.5$\,nJy in F444W together with the NIRSpec $R=1000$ line-flux limit from the STScI JDox sensitivity pages \cite{jdoxnircamsens,jdoxnirspecsens}. The Roman and Euclid rows use published grism survey limits \cite{wang2021,euclid2011}. The Roman imaging row gives the Wide Tier F158 $5\sigma$ point-source depth for the full $3\times2$ dither sequence with $107$\,s per dither \cite{romandocs}. The Euclid imaging row gives the survey design floor for the NISP $Y_E$, $J_E$, and $H_E$ bands rather than the estimated achieved depth of $24.5$\,AB. The $1344$\,s exposure follows from four reference dithers with $3\times112$\,s of NISP imaging per dither \cite{scaramella2022}. The ETC column uses the same aperture, detector parameters, and target as each reference. The JWST estimates agree within approximately $10$--$25\%$. The idealized Roman and Euclid grism estimates are $1.4$--$4.0$ times deeper than the published survey limits because the photon budget omits spectral-extraction losses and wide-survey self-contamination. Different significance thresholds and source sizes explain most of the difference between the Roman and Euclid ratios. Pixel scale alone shifts the estimated depth by only a few percent. Roman assumes $6.5\sigma$ and a $0.3''$ source while Euclid assumes $3.5\sigma$ and a $0.5''$ source. The imaging estimates agree within one magnitude.}
\label{tab:jwstval}
\label{tab:romaneuclidval}
\end{table}

Published Roman and Euclid survey limits define the flagship ELG depth in Eq.~\eqref{eq:f5sigma}. The idealized ETC does not define that depth because Table~\ref{tab:romaneuclidval} places the ETC estimates $1.4$--$4.0$ times deeper than the slitless survey limits achieved by the two missions. The supernova rest-frame near-ultraviolet reach in Section~\ref{sec:snia}, the void dark-galaxy star-formation-rate threshold in Section~\ref{sec:dwarfvoid}, and the Lyman-$\alpha$ column-density threshold in Section~\ref{sec:lyaforest} instead use the uncalibrated slitless ETC estimates. The corresponding reaches are idealized upper bounds rather than validated survey depths. Imaging estimates require a smaller correction. The imaging rows in Table~\ref{tab:romaneuclidval} agree with the Roman and Euclid published depths within one magnitude. The low-surface-brightness, intracluster-light, and dwarf and void galaxy depths in Sections~\ref{sec:lsb} and~\ref{sec:dwarfvoid} therefore have a smaller and better quantified systematic margin.

\begin{figure}[htbp]
\centering
\includegraphics[width=\textwidth]{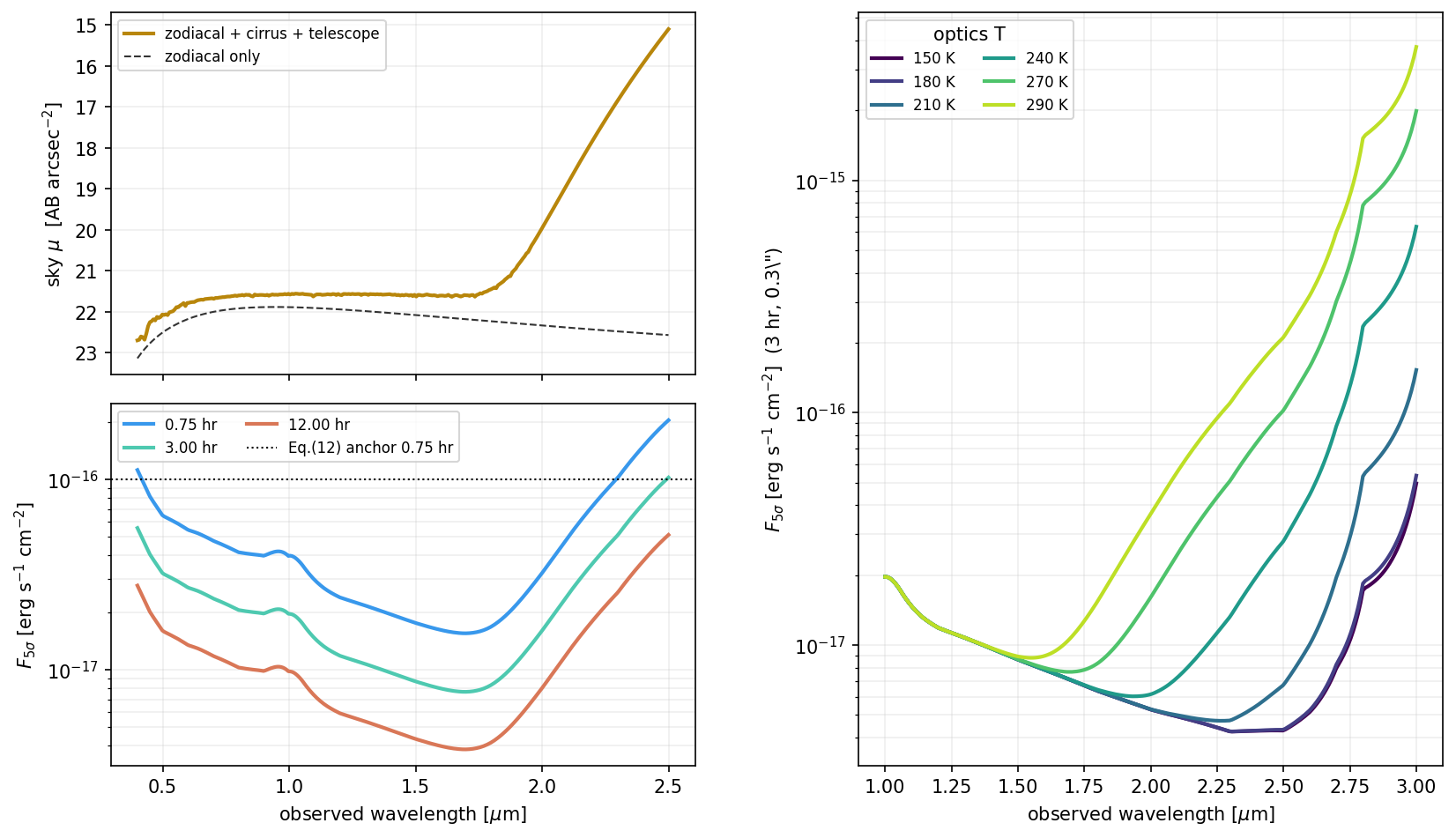}
\caption{Full numerical exposure-time reference for the spectroscopic survey. \emph{Top left} shows the modelled zodiacal, galactic-cirrus, and telescope thermal backgrounds. \emph{Bottom left} gives the $5\sigma$ line-flux depth $F_{5\sigma}(\lambda)$ for three exposures and marks the conservative $10^{-16}$ planning reference of Eq.~\eqref{eq:f5sigma}. \emph{Right} gives the telescope-temperature trade in a $3$\,hr exposure. The present curves establish the wavelength dependence and cooling requirement. The matched $R=1000$ and $R=5000$ calculation must use the same image-slicer aperture and physical line interval.}
\label{fig:etc}
\label{fig:etccool}
\end{figure}

\begin{figure}[htbp]
\centering
\includegraphics[width=0.72\textwidth]{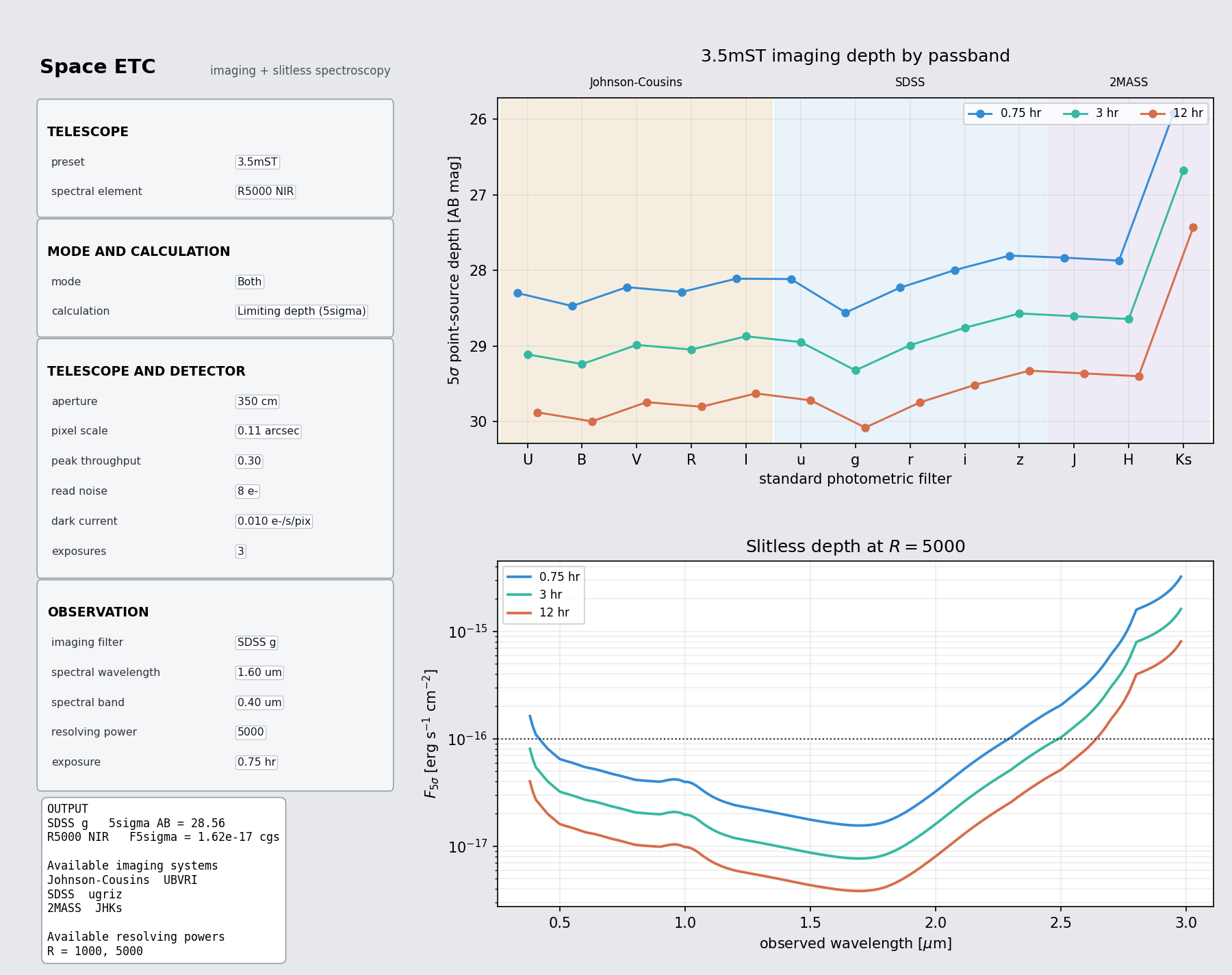}
\caption{The open-source exposure-time calculator and its graphical output (\url{https://github.com/kjhan0606/3.5mST}). A telescope preset and an observing setting define the aperture and detector parameters. The imaging panel shows the $5\sigma$ point-source depth for Johnson--Cousins $UBVRI$, SDSS $ugriz$, and 2MASS $JHK_{\rm s}$ photometry at three exposure times. Each passband uses its pivot wavelength and rectangular-equivalent width. The lower panel gives the line-flux depth across the full wavelength range. The resolving-power selector includes $R=1000$ and $R=5000$. The current curves define the wavelength dependence of the sensitivity. The matched flight comparison must use the same image-slicer geometry and physical line interval at both resolving powers as specified in Table~\ref{tab:etcmodes}.}
\label{fig:etcgui}
\end{figure}

Telescope temperature sets a restrictive red wavelength limit. Below approximately $2\,\mu$m, zodiacal emission dominates the background and the depth depends weakly on optics temperature. Telescope thermal emission rises steeply at longer wavelengths and overtakes zodiacal emission at a crossover that shifts blueward as the optics warm (Figure~\ref{fig:etccool}). For passively warm optics near $270$\,K, the crossover occurs near $1.9\,\mu$m and the $5\sigma$ depth at $2.7\,\mu$m becomes more than an order of magnitude shallower. Cooling the optics to $\lesssim200$\,K moves the thermal-background crossover beyond $2.5\,\mu$m and keeps the full $1$--$3\,\mu$m channel background limited. Many [O\,II] and [O\,III] redshifts lie below $2\,\mu$m. However, the H$\alpha$ control sample required to avoid a tracer-selection discontinuity across $2<z<3$ lies at $2.0$--$2.63\,\mu$m. The red [S\,II] component reaches $2.69\,\mu$m at $z=3$. The red wavelength range therefore supports flagship selection calibration rather than detached companion science. The $2.70\,\mu$m science-throughput threshold and the $3.0\,\mu$m band-edge goal directly set the near-infrared cooling requirement.

\begin{figure}[htbp]
\centering
\includegraphics[width=\textwidth]{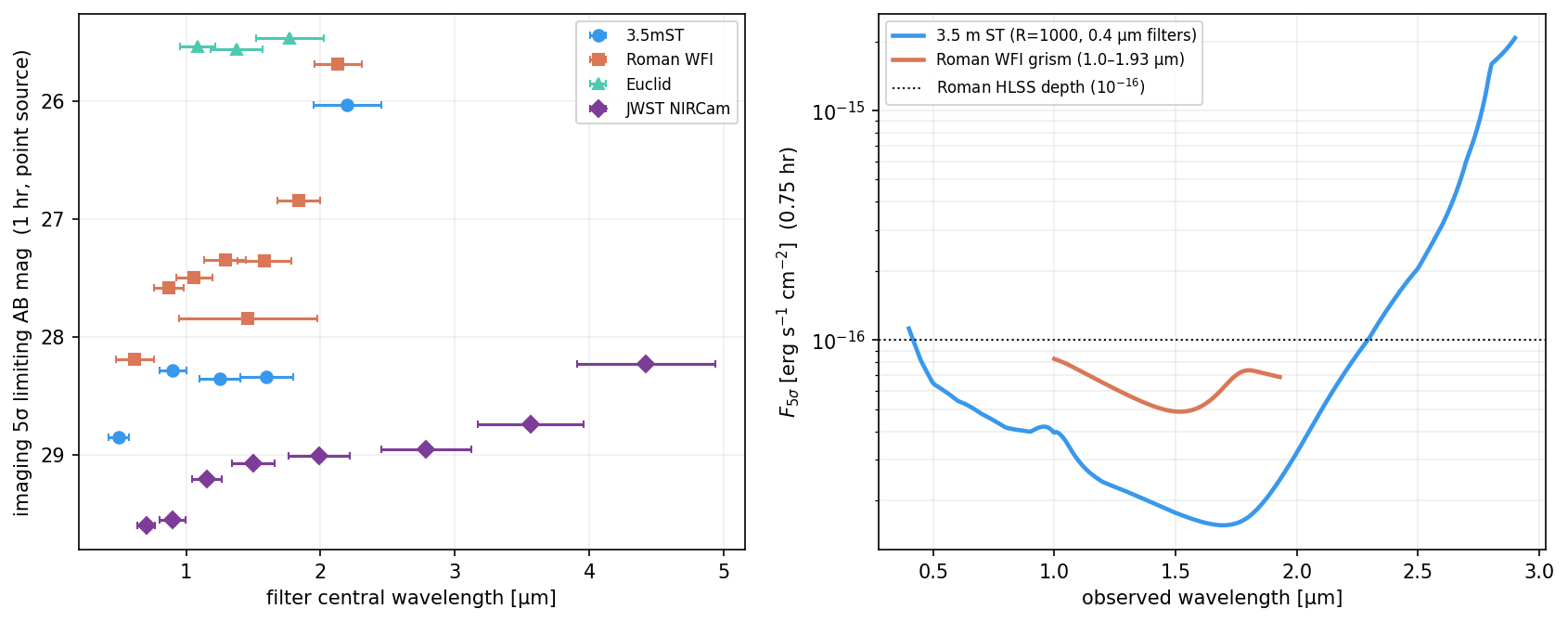}
\caption{The left panel gives the imaging $5\sigma$ point-source depth in a $1$\,hr exposure for each mission in its own filters. The Roman sequence includes F213 and therefore reaches $2.30\,\mu$m in imaging. The right panel gives the $5\sigma$ line-flux reference in a $0.75$\,hr exposure for the $3.5$\,m concept and Roman G150. Roman G150 ends at $1.93\,\mu$m. The Roman HLSS planning depth is marked. The plotted $3.5$\,m curve uses the current $R=1000$ reference calculation with band-limiting $0.4\,\mu$m filters. The $R=5000$ comparison must adopt the same image-slicer aperture and line-integration interval.}
\label{fig:compareimg}
\label{fig:comparespec}
\end{figure}

The same calculator permits a direct comparison with existing missions. Table~\ref{tab:compareimg} gives the imaging $5\sigma$ point-source depth at common wavelengths for the $3.5$\,m concept, Roman, Euclid, and JWST. Figure~\ref{fig:compareimg} plots the depth in each mission's filters. Figure~\ref{fig:comparespec} compares the slitless line-flux depth with Roman. The $3.5$\,m concept lies between Roman and JWST in imaging depth. The band-limiting $0.4\,\mu$m filters reduce the dispersed sky relative to the full Roman G150 band and produce a deeper ideal line-flux limit in the overlap region. Warm $270$\,K optics become thermal limited beyond $\sim2\,\mu$m. Cooling to $\lesssim200$\,K removes most of the penalty through the science band (Figure~\ref{fig:etccool}).

Roman wavelength limits depend on observing mode. WFI imaging spans $0.5$--$2.3\,\mu$m and the F213 imaging filter spans $1.95$--$2.30\,\mu$m. Roman slitless spectroscopy does not extend to $2.3\,\mu$m. G150 spans $1.00$--$1.93\,\mu$m at $R\simeq451$ and P127 spans $0.75$--$1.80\,\mu$m at $R\simeq80$--$180$ \cite{romanwfi2025}. Table~\ref{tab:romanmodecompare} separates the imaging and spectroscopic limits. The distinction matters because the red F213 edge contributes imaging depth but does not retain H$\alpha$ or [O\,\textsc{iii}] in a Roman spectrum beyond the G150 cutoff.

\begin{table}[htbp]
\centering
\sffamily\small
\begin{tabularx}{\textwidth}{@{}p{2.5cm}p{3.1cm}p{2.7cm}Y@{}}
\toprule
\textbf{Facility mode} & \textbf{Wavelength range} & \textbf{Spectral resolution} & \textbf{Role in the comparison}\\
\midrule
3.5\,m image sliced & $0.36$--$3.00\,\mu$m in band-limited settings & $R_{\rm point}=1000$ or $5000$ & Resolving-power trade for the same ELG survey architecture\\
Roman WFI imaging & $0.48$--$2.30\,\mu$m overall. F213 spans $1.95$--$2.30\,\mu$m & Not applicable & Wide imaging through eight filters\\
Roman WFI G150 & $1.00$--$1.93\,\mu$m & $R\simeq451$ & Wide slitless spectroscopy\\
Roman WFI P127 & $0.75$--$1.80\,\mu$m & $R\simeq80$--$180$ & Low-resolution slitless spectroscopy
\end{tabularx}
\caption{Wavelength and resolution comparison with Roman WFI. Roman imaging reaches $2.30\,\mu$m through F213 while Roman spectroscopy ends at $1.93\,\mu$m for G150 and $1.80\,\mu$m for P127. The Roman values come from the STScI WFI quick reference and the Roman technical-information release cited there \cite{romanwfi2025}. The $3.5$\,m concept adopts image slicing and retains $R=1000$ and $R=5000$ as candidate specifications.}
\label{tab:romanmodecompare}
\end{table}

\begin{table}[htbp]
\centering
\sffamily\footnotesize
\setlength{\tabcolsep}{4pt}
\renewcommand{\arraystretch}{0.92}
\begin{tabularx}{\textwidth}{@{}c Y Y Y Y !{\hskip 8pt} c Y Y Y Y@{}}
\toprule
\textbf{$\lambda$ [$\mu$m]} & \textbf{3.5\,m} & \textbf{Roman} & \textbf{Euclid} & \textbf{JWST} & \textbf{$\lambda$ [$\mu$m]} & \textbf{3.5\,m} & \textbf{Roman} & \textbf{Euclid} & \textbf{JWST}\\
\midrule
\multicolumn{5}{@{}l}{\textbf{1\,hr}} & \multicolumn{5}{@{\hskip8pt}l}{\textbf{24\,hr}}\\
0.6 & 29.01 & 28.23 & -- & -- & 0.6 & 30.75 & 29.98 & -- & --\\
1.0 & 28.28 & 27.63 & 25.59 & 29.62 & 1.0 & 30.02 & 29.38 & 27.35 & 31.39\\
1.5 & 28.25 & 27.21 & 25.30 & 29.03 & 1.5 & 29.99 & 28.98 & 27.07 & 30.82\\
2.0 & 27.02 & 26.03 & -- & 28.73 & 2.0 & 28.75 & 27.77 & -- & 30.54\\
2.5 & 24.49 & -- & -- & 28.41 & 2.5 & 26.21 & -- & -- & 30.19\\
3.5 & -- & -- & -- & 28.17 & 3.5 & -- & -- & -- & 29.95\\
\multicolumn{5}{@{}l}{\textbf{5\,hr}} & \multicolumn{5}{@{\hskip8pt}l}{\textbf{48\,hr}}\\
0.6 & 29.89 & 29.12 & -- & -- & 0.6 & 31.12 & 30.36 & -- & --\\
1.0 & 29.16 & 28.53 & 26.50 & 30.53 & 1.0 & 30.40 & 29.76 & 27.73 & 31.77\\
1.5 & 29.13 & 28.12 & 26.21 & 29.96 & 1.5 & 30.36 & 29.36 & 27.44 & 31.19\\
2.0 & 27.90 & 26.92 & -- & 29.68 & 2.0 & 29.13 & 28.15 & -- & 30.91\\
2.5 & 25.36 & -- & -- & 29.33 & 2.5 & 26.59 & -- & -- & 30.57\\
3.5 & -- & -- & -- & 29.09 & 3.5 & -- & -- & -- & 30.33\\
\multicolumn{5}{@{}l}{\textbf{10\,hr}} & \multicolumn{5}{@{\hskip8pt}l}{}\\
0.6 & 30.27 & 29.50 & -- & -- & & & & & \\
1.0 & 29.54 & 28.91 & 26.88 & 30.91 & & & & & \\
1.5 & 29.51 & 28.50 & 26.59 & 30.34 & & & & & \\
2.0 & 28.28 & 27.29 & -- & 30.06 & & & & & \\
2.5 & 25.74 & -- & -- & 29.71 & & & & & \\
3.5 & -- & -- & -- & 29.47 & & & & & \\
\bottomrule
\end{tabularx}
\caption{Imaging $5\sigma$ point-source limiting AB magnitude through a $0.3\,\mu$m band in exposures of $1$, $5$, and $10$\,hr (left) and $24$ and $48$\,hr (right) for the four telescopes evaluated with the same calculator from their real apertures and detector parameters. Dashes mark wavelengths outside a mission's band. JWST is deepest (largest, cold aperture), followed by the $3.5$\,m concept, Roman, and Euclid. The $3.5$\,m entry at $2.5\,\mu$m is depressed by warm-optics thermal background, which cooling removes.}
\label{tab:compareimg}
\end{table}

\clearpage
\section{Flagship ELG BAO and RSD Survey}

\subsection{Science Objective}

The flagship program is a band-limited slitless survey of emission-line galaxies over 100--300 \degti. BAO measures the expansion history while RSD measures the growth of structure through $\fsig(z)$ in broad bins across $1\lesssim z\lesssim3$. The primary tracers are H$\alpha$, [O\,III], H$\beta$, and [O\,II] selected by line flux, colour, and direct-image morphology.

The all-line survey targets approximately $10^6$ secure redshifts over the 100 \degti threshold area and up to a few $10^6$ over the 300 \degti goal area. The target is not a DESI [O\,II]-only surface density and does not follow from aperture alone. The predicted ELG count depends on the distribution of spectral line strengths and the redshift-dependent luminosity functions of H$\alpha$, [O\,III], H$\beta$, and [O\,II] after application of the slitless selection function. A direct calculation with DESI DR1 \texttt{ELG\_LOPnotqso} gives a conservative [O\,II] reference of $\simeq3.4\times10^2$ deg$^{-2}$ at $F_{\rm [O\,II]}\ge10^{-16}$ erg s$^{-1}$ cm$^{-2}$ over $0.6<z<1.6$ and $\simeq2.2\times10^2$ deg$^{-2}$ over $1.0<z<1.6$. The headline count uses the wide-tier line-flux threshold of $10^{-16}$\,erg\,s$^{-1}$\,cm$^{-2}$ in Table~\ref{tab:wedding} and includes H$\alpha$, [O\,III], H$\beta$, and [O\,II]. The medium-tier threshold near $5\times10^{-17}$\,erg\,s$^{-1}$\,cm$^{-2}$ instead calibrates the redshift selection function and the high-redshift luminosity function against simulated detector images.

BAO provides a standard ruler through the sound horizon $r_d$. The survey measures transverse and radial BAO combinations,
\begin{equation}
  \theta_{\rm BAO}(z)\simeq \frac{r_d}{D_M(z)}, \qquad
  \Delta z_{\rm BAO}(z)\simeq \frac{H(z)r_d}{c},
\end{equation}
and therefore constrains $D_M(z)/r_d$ and $H(z)r_d$. RSD adds the growth observable $\fsig(z)$ from the anisotropic clustering amplitude, where
\begin{equation}
  f(z)\equiv\frac{d\ln D}{d\ln a}
\end{equation}
is the logarithmic growth rate of the linear growth factor $D(a)$ with scale factor $a$. The wide tier drives the BAO/RSD statistical power. The medium and deep tiers calibrate the redshift selection function that otherwise becomes the dominant systematic.

\subsection{Wavelength Cutoff Requirement}\label{sec:wavecut}

The red cutoff follows from the stated cosmology experiment rather than from a general preference for infrared coverage. A survey advertised over $1<z<3$ must retain a consistent tracer definition throughout that volume. For a line with rest wavelength $\lambda_0$, a red cutoff $\lambda_{\rm cut}$ gives
\begin{equation}
  z_{\max}=\frac{\lambda_{\rm cut}}{\lambda_0}-1.
  \label{eq:zmaxcut}
\end{equation}
Table~\ref{tab:wavecut} applies Equation~\eqref{eq:zmaxcut} to the principal ELG lines. The rest wavelengths are taken from the NIST Atomic Spectra Database \cite{nistlines}.

\begin{table}[htbp]
\centering
\sffamily\small
\setlength{\tabcolsep}{3.2pt}
\begin{tabularx}{\textwidth}{@{}c c c c c Y@{}}
\toprule
\textbf{Red cutoff} & \textbf{$z_{\max}$(H$\alpha$)} & \textbf{$z_{\max}$([S\,II])} & \textbf{$z_{\max}$([O\,III])} & \textbf{$z_{\max}$([O\,II])} & \textbf{Consequence for the $1<z<3$ survey}\\
\midrule
1.30\,$\mu$m & 0.98 & 0.93 & 1.60 & 2.49 & H$\alpha$ is absent from the flagship volume. [O\,III] is lost above $z=1.60$, and [O\,II] does not reach $z=3$.\\
2.00\,$\mu$m & 2.05 & 1.97 & 2.99 & 4.37 & Oxygen lines reach approximately $z=3$, but H$\alpha$ disappears above $z=2.05$. The highest-redshift bins become oxygen selected.\\
2.70\,$\mu$m & 3.11 & 3.01 & 4.39 & 6.24 & H$\alpha$ and the [S\,II] doublet remain observable through $z=3$. The rest-optical line complex is retained across the full flagship volume.\\
3.00\,$\mu$m & 3.57 & 3.46 & 4.99 & 7.05 & The adopted band edge provides an 11\% guard band beyond the science threshold. H$\alpha$ also remains available to $z=3.57$.\\
\bottomrule
\end{tabularx}
\caption{Maximum observable redshift as a function of the near-infrared cutoff, computed with Equation~\eqref{eq:zmaxcut}. The calculation uses H$\alpha$ $6563$\,\AA, the red [S\,II] component at $6731$\,\AA, [O\,III] $5007$\,\AA, and [O\,II] $3727$\,\AA. Values are rounded to two decimal places. A Roman-like $1.93\,\mu$m grism cutoff lies close to the $2.00\,\mu$m row and gives $z_{\max}=1.94$ for H$\alpha$ and $2.85$ for [O\,III] \cite{romandocs,wang2021}.}
\label{tab:wavecut}
\end{table}

The number of lines allowed by wavelength coverage is a geometric upper bound. A secure redshift also requires sufficient line flux after the wavelength-dependent background, throughput, source size, and spectral overlap are included. Table~\ref{tab:z3lineplacement} shows the observed locations of the full diagnostic set at the upper boundary of the flagship survey.

\begin{table}[htbp]
\centering
\sffamily\small
\begin{tabularx}{0.74\textwidth}{@{}Y c@{}}
\toprule
\textbf{Emission line} & \textbf{Observed wavelength at $z=3$}\\
\midrule
\multicolumn{2}{@{}l}{\textbf{Rest-UV supplementary diagnostic}}\\
Ly$\alpha$ $1215.67$\,\AA & $0.486\,\mu$m\\
\addlinespace[0.35em]
\multicolumn{2}{@{}l}{\textbf{Rest-optical systemic-redshift lines}}\\
{[O\,II]} $3727$\,\AA & $1.491\,\mu$m\\
H$\beta$ $4861$\,\AA & $1.944\,\mu$m\\
{[O\,III]} $4959$\,\AA & $1.984\,\mu$m\\
{[O\,III]} $5007$\,\AA & $2.003\,\mu$m\\
H$\alpha$ $6563$\,\AA & $2.625\,\mu$m\\
{[N\,II]} $6584$\,\AA & $2.634\,\mu$m\\
{[S\,II]} $6716$\,\AA & $2.686\,\mu$m\\
{[S\,II]} $6731$\,\AA & $2.692\,\mu$m\\
\bottomrule
\end{tabularx}
\caption{Observed wavelengths of the principal ELG diagnostic lines at $z=3$, computed from $\lambda_{\rm obs}=\lambda_0(1+z)$. Rest wavelengths are from the NIST Atomic Spectra Database \cite{nistlines}. Ly$\alpha$ enters the $0.3\,\mu$m blue edge at $z=1.47$ and appears at $0.486\,\mu$m by $z=3$. Resonant transfer through neutral hydrogen modifies Ly$\alpha$ visibility and may displace the line centroid from the systemic redshift. Ly$\alpha$ is therefore a supplementary diagnostic rather than a guaranteed systemic-redshift line \cite{dijkstra2014}. A $2.0\,\mu$m cutoff contains [O\,II], H$\beta$, and the blue [O\,III] component at this redshift. Wavelength coverage alone does not guarantee detection of the weaker H$\beta$ line. Calibrated throughput to $2.70\,\mu$m retains the independent H$\alpha$+[N\,II]+[S\,II] complex.}
\label{tab:z3lineplacement}
\end{table}

\newpage
The four cutoff cases have distinct scientific consequences.
\begin{itemize}[leftmargin=1.7em,itemsep=0.45em,topsep=0.4em]
\item \textbf{$1.30\,\mu$m cutoff.} The cutoff cannot execute the stated BAO and RSD survey. [O\,III] leaves the band above $z=1.60$ and H$\beta$ leaves above $z=1.67$. Only [O\,II] among the principal rest-optical lines remains between $z\simeq1.67$ and $2.49$. No principal rest-optical line remains over $2.49<z<3$. Ly$\alpha$ remains available as a supplementary rest-UV line but cannot define the systemic redshift by itself.
\item \textbf{$2.00\,\mu$m cutoff.} [O\,II] and H$\beta$ are formally inside the band at $z=3$, while [O\,III] $5007$ lies immediately beyond the nominal edge at $2.003\,\mu$m. H$\beta$ is weaker than H$\alpha$ and [O\,III] in many ELGs. Wavelength coverage alone therefore does not ensure a two-line redshift. H$\alpha$ also disappears above $z=2.05$, which makes the highest-redshift sample predominantly oxygen selected.
\item \textbf{$2.70\,\mu$m science threshold.} All lines in Table~\ref{tab:z3lineplacement} remain in band through $z=3$. The H$\alpha$+[N\,II]+[S\,II] complex supplies an independent strong feature for redshift confirmation and provides line ratios that diagnose active nuclei and variations in nebular conditions.
\item \textbf{$3.00\,\mu$m operational band edge.} The additional $0.30\,\mu$m keeps the $2.70\,\mu$m science boundary away from filter transitions, detector quantum-efficiency roll-off, wavelength-calibration edge losses, and flight margin. The same band edge extends H$\alpha$ to $z=3.57$ and [O\,III] to $z=4.99$.
\end{itemize}

The $2.0\,\mu$m transition removes targets and shifts the tracer population toward oxygen-line emitters. Above $z\simeq2.05$ the catalogue switches from a mixed Balmer and oxygen-line selection to a predominantly oxygen-line selection. Galaxy bias, redshift success, and completeness consequently acquire a selection discontinuity. H$\alpha$ traces recombination powered by massive stars while [O\,III] and [O\,II] depend strongly on metallicity and ionization state \cite{kennicutt1998,khostovan2015,saito2020}. The discontinuity occurs in the same redshift range where the mission interprets anisotropic clustering as a growth measurement. Retaining the Balmer-selected population also preserves the effective number density $n(z)$ and the effective clustering volume defined in Equation~\eqref{eq:veff}. Loss of one tracer has a larger statistical cost as $nP$ approaches unity because the weight scales as $[nP/(1+nP)]^2$.

\begin{leadbox}[title=Cosmology Wavelength Requirement]{green}
The broad wavelength coverage is required to secure multiple-line redshifts across the full survey volume. Calibrated throughput to $2.70\,\mu$m retains the H$\alpha$+[N\,II]+[S\,II] complex at $z=3$ while the $3.0\,\mu$m operational band edge keeps the complex away from throughput roll-off and calibration losses. The baseline requirement is set by secure multiple-line redshifts through $z=3$. The additional reach beyond $z=3$ provides a secondary high-redshift science return. Phase A shall compare $1.30$, $2.00$, $2.70$, and $3.00\,\mu$m cutoffs with the same luminosity functions, ETC, deblending model, and survey time. The decision data products are $n(z)$, secure-redshift fraction, line-confusion rate, $nP$, $V_{\rm eff}$, and forecast uncertainties in $D_M/r_d$, $Hr_d$, and $f\sigma_8$.
\end{leadbox}

\subsection{Survey Precedent and Redshift Precision}

Roman's High Latitude Spectroscopic Survey provides the closest design precedent. The large-area slitless survey has $R\simeq435$--865, forecasts millions of H$\alpha$ and [O\,III] redshifts, and targets BAO and RSD cosmology \cite{wang2021}. The $R=1000$ candidate retains line-separation margin while keeping spectra shorter. The $R=5000$ candidate resolves more line structure and reduces ambiguous identifications. The design goal is a calibrated high-purity redshift sample whose radial smearing remains small compared with the BAO scale after quality cuts. Internal kinematics provide an additional return rather than the primary cosmology criterion.

Wang et al.'s reference HLSS design covers $2000$\,deg$^2$ to a $10^{-16}\,{\rm erg\,s^{-1}\,cm^{-2}}$, $6.5\sigma$ flux limit. The forecast contains approximately $10^7$ H$\alpha$ redshifts at $1\lesssim z\lesssim2$ and $2\times10^6$ [O\,III] redshifts at $2\lesssim z\lesssim3$. The resulting fractional precision is approximately $1$--$2\%$ for BAO distance and $6$--$7\%$ for the growth rate in each redshift bin \cite{wang2021}. The proposed $100$--$300$\,deg$^2$ footprint and $10^6$--$3\times10^6$ redshift yield occupy a smaller statistical volume. Sample size alone does not surpass Roman HLSS. Three complementary properties motivate the proposed survey. The source selection includes H$\alpha$, [O\,III], H$\beta$, and [O\,II] rather than the two lines in the Roman forecast. An independent three-orientation detector fit permits a comparison of extraction systematics with Roman. The footprint and cadence support joint multi-tracer measurements with Roman, DESI, and HETDEX-class surveys as described in Section~\ref{sec:anchorsynergy}. The scientific gain therefore comes from the combined tracer sample and independent tests of survey systematics rather than from a stand-alone volume larger than Roman HLSS.

\begin{leadbox}[title=RSD Survey Definition]{green}
\begin{tabularx}{\textwidth}{@{}p{4.3cm}Y@{}}
\toprule
\textbf{Requirement} & \textbf{Adopted value}\\
\midrule
Field of view & \fovlarge\ required for the wide survey. \fovsmall\ reserved for deep fields and technology validation.\\
Survey area & 100 \degti threshold. 300 \degti goal.\\
Redshift sample & $10^6$ threshold. $3\times10^6$ goal secure ELG redshifts.\\
Spectrograph & Image slicer with $R_{\rm point}=1000$ and $5000$ under trade. Current yield values use the $R=1000$ reference calculation.\\
Redshift quality & completeness and purity calibrated as functions of flux, size, line ratio, wavelength, field density, and roll angle.\\
Cosmology target & 3--5\% $\fsig$ precision in broad redshift bins, after end-to-end forecast validation.\\
\bottomrule
\end{tabularx}
\end{leadbox}

\subsection{Representative ELG Spectrum and Detectable Lines}

The target population is a star-forming ELG with blue continuum, nebular recombination lines, and collisionally excited oxygen lines. The most important redshift features are [O\,II] $\lambda3727$, H$\beta$ $\lambda4861$, [O\,III] $\lambda4959,\lambda5007$, H$\alpha$ $\lambda6563$, and the neighboring [N\,II]/[S\,II] complex. The $R=1000$ candidate separates the [O\,III] doublet for compact sources over the full cosmological range. The $R=5000$ candidate provides more secure [O\,II] separation and reduces confusion within the H$\alpha$+[N\,II] complex. The improvement must be measured as a reduction in catastrophic redshifts rather than assumed from resolving power alone.

Table~\ref{tab:lines} maps the line set onto the proposed two-arm wavelength coverage and gives the redshift interval over which each diagnostic falls in the optical and near-infrared arms. Multiple lines normally constrain a redshift across $1<z<3$, which reduces dependence on a single ambiguous feature.

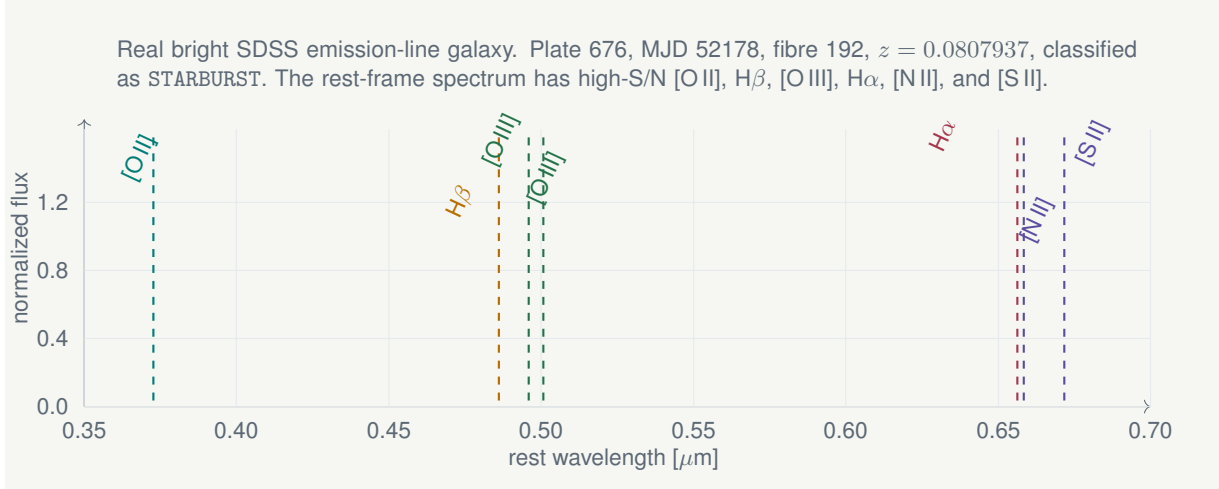
\begin{figure}[htbp]
\centering
\begin{tikzpicture}[font=\sffamily\footnotesize]
  \fill[paper] (-0.35,-0.35) rectangle (15.7,6.12);
  \node[anchor=west,font=\bfseries\large,color=ink] at (0,6.78) {Observed bright low-redshift ELG spectrum shifted to rest frame};
  \draw[->,color=muted] (0.7,0.75) -- (14.8,0.75);
  \draw[->,color=muted] (0.7,0.75) -- (0.7,4.55);
  \node[anchor=north,color=muted] at (7.7,0.32) {rest wavelength [$\mu$m]};
  \node[rotate=90,anchor=south,color=muted] at (0.08,2.85) {normalized flux};
  \foreach \x/\lab in {0.70/0.35,2.71/0.40,4.73/0.45,6.74/0.50,8.76/0.55,10.77/0.60,12.79/0.65,14.80/0.70} {
    \draw[line!65] (\x,0.75) -- (\x,4.42);
    \node[anchor=north,color=muted] at (\x,0.68) {\lab};
  }
  \foreach \y/\lab in {0.75/0.0,1.65/0.4,2.55/0.8,3.45/1.2} {
    \draw[line!65] (0.7,\y) -- (14.8,\y);
    \node[anchor=east,color=muted] at (0.62,\y) {\lab};
  }
  \begin{scope}[shift={(-13.400,0.82)},x=40.286cm,y=2.28cm]
    \draw[blue,line width=0.72pt] plot file {vol2_sdss_elg_rest_spectrum.dat};
  \end{scope}
  \foreach \lam/\lab/\col/\yy/\rot/\dx in {
    0.3727/{[O\,II]}/teal/3.92/65/0.06,
    0.4861/{H$\beta$}/gold/3.34/65/-0.26,
    0.4959/{[O\,III]}/green/4.16/65/-0.12,
    0.5007/{[O\,III]}/green/3.66/65/0.28,
    0.6563/{H$\alpha$}/rose/4.24/65/-0.78,
    0.6584/{[N\,II]}/violet/3.10/65/0.42,
    0.6717/{[S\,II]}/violet/4.12/65/0.62} {
    \pgfmathsetmacro{\xx}{0.7 + (\lam-0.35)*40.286}
    \draw[\col,line width=0.8pt,dashed] (\xx,0.83) -- (\xx,4.36);
    \node[anchor=south,rotate=\rot,color=\col] at (\xx+\dx,\yy) {\lab};
  }
  \node[anchor=west,text width=13.5cm,color=muted] at (1.0,5.26)
  {Real bright SDSS emission-line galaxy. Plate 676, MJD 52178, fibre 192, $z=0.0807937$, classified as \texttt{STARBURST}. The rest-frame spectrum has high-S/N [O\,II], H$\beta$, [O\,III], H$\alpha$, [N\,II], and [S\,II].};
\end{tikzpicture}
\caption{Observed bright low-redshift emission-line galaxy spectrum from SDSS DR17. The nearby starburst ELG demonstrates the high-S/N line set used for slitless redshift identification. Bright lines include [O\,II], H$\beta$, [O\,III], H$\alpha$, [N\,II], and [S\,II]. The data are from SDSS DR17 spectrum spec-0676-52178-0192 \cite{sdssdr17}.}
\label{fig:elgspectrum}
\end{figure}

\begin{table}[htbp]
\centering
\sffamily\small
\begin{tabularx}{\textwidth}{@{}p{2.8cm}p{2.1cm}p{2.8cm}p{2.8cm}Y@{}}
\toprule
\textbf{Line} & \textbf{Rest $\lambda$} & \textbf{0.3--1.0 $\mu$m arm} & \textbf{1.0--3.0 $\mu$m arm} & \textbf{Cosmology use}\\
\midrule
{[O\,II]} & 0.3727 $\mu$m & $0<z<1.68$ & $1.68<z<7.05$ & Key line for $z\gtrsim1.7$ when H$\alpha$ leaves the band. Doublet and SED priors improve purity.\\
H$\beta$ & 0.4861 $\mu$m & $0<z<1.06$ & $1.06<z<5.17$ & Secondary confirmation line with [O\,III].\\
{[O\,III]} & 0.4959, 0.5007 $\mu$m & $0<z<1.00$ & $1.00<z<4.99$ & High-EW tracer for $1<z<3$. Doublet separation improves redshift purity.\\
H$\alpha$ & 0.6563 $\mu$m & $0<z<0.52$ & $0.52<z<3.57$ & Primary low-to-mid-$z$ BAO tracer and strongest line for many star-forming galaxies.\\
{[S\,II]} & 0.6716, 0.6731 $\mu$m & $0<z<0.49$ & $0.49<z<3.47$ & Template constraint near H$\alpha$ and diagnostic in bright calibration spectra.\\
\bottomrule
\end{tabularx}
\caption{Observed redshift windows for the proposed two-arm spectrograph. The ELG BAO core is $1\lesssim z\lesssim3$, where H$\alpha$, [O\,III], and [O\,II] provide overlapping redshift confirmation paths.}
\label{tab:lines}
\end{table}

\begin{figure}[htbp]
\centering
\includegraphics[width=0.98\textwidth,height=0.50\textheight,keepaspectratio]{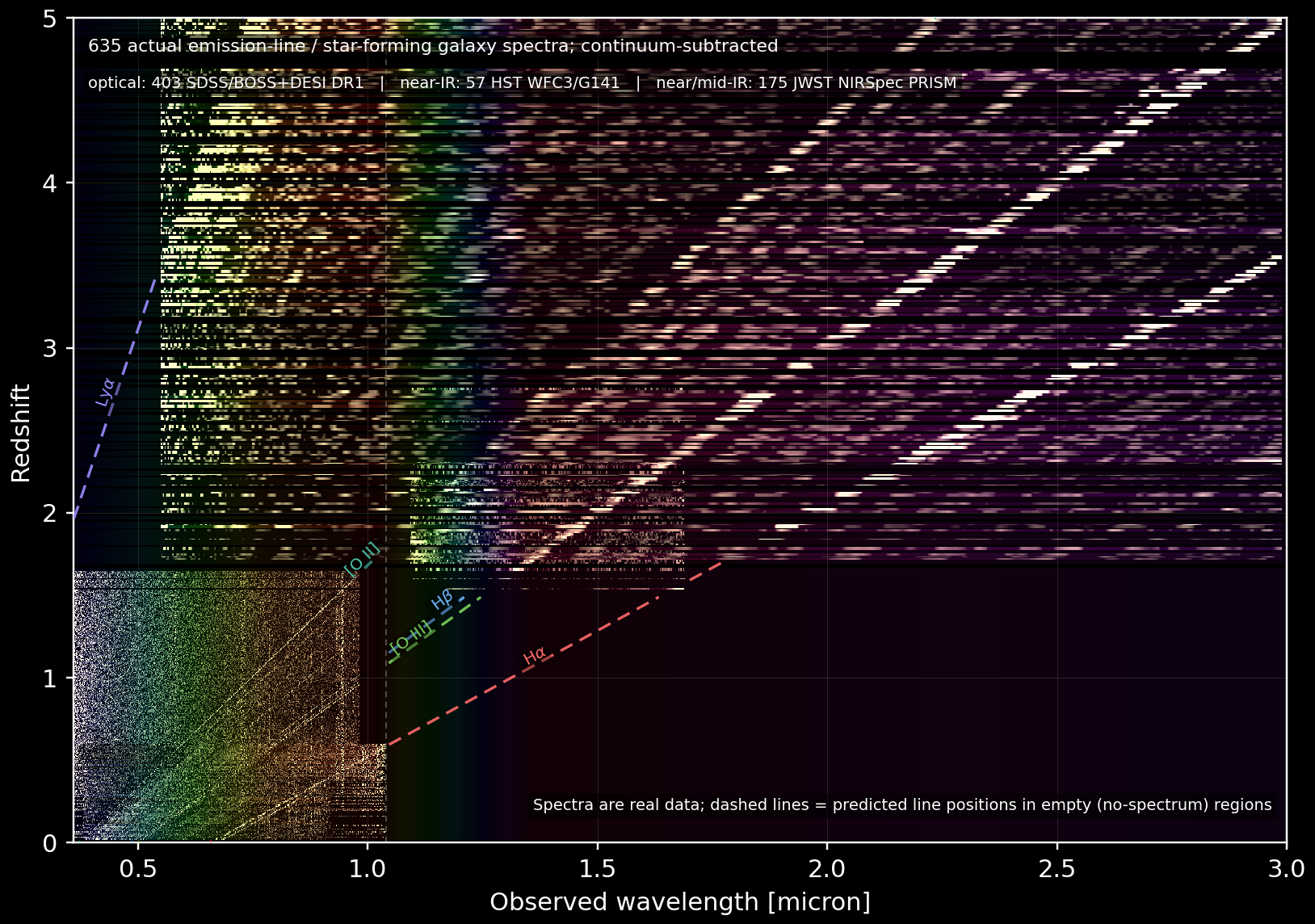}
\caption{Actual ELG/star-forming galaxy spectral stack in observed wavelength and redshift, in the compact ``spectral barcode'' style of the 2dF QSO atlas, spanning the full $0.36$--$3.0\,\mu$m proposed band and $0<z<5$, with the dashed line near $1.04\,\mu$m marking the optical/near-IR arm split. Four real datasets are stacked at their catalog redshifts, 110 SDSS/BOSS and 293 DESI DR1 ELGs ($z\lesssim1.65$, via SPARCL) in the optical arm, 57 HST WFC3/G141 3D-HST COSMOS grism spectra ($1.5<z<2.75$) in the near-IR, and 175 JWST/NIRSpec PRISM star-forming galaxies ($1.7<z<5.0$, DAWN JWST Archive) at high redshift, $635$ spectra in total. Each row is continuum-subtracted and scaled to emphasize line residuals. The continuous diagonal tracks ([O\,II], H$\beta$+[O\,III], H$\alpha$) run unbroken from the optical block through the JWST band, demonstrating that the line set the survey relies on stays observable across $1\lesssim z\lesssim3$ and beyond. No spectra are synthetic. Dashed labelled lines trace the predicted $\lambda_{\rm obs}=\lambda_{\rm rest}(1+z)$ track of each diagnostic line only in the empty, no-data regions, without overplotting real data \cite{raichoor2022,desidr1,sparcl2024,brammer2012,momcheva2016,degraaff2024,heintz2023}.}
\label{fig:elgobswavez}
\end{figure}

\subsection{ELG Luminosity, Distance, and Depth}\label{sec:elgdepth}

The planning model uses a flat $\Lambda$CDM cosmology with $H_0=70$ km s$^{-1}$ Mpc$^{-1}$ and $\Omega_m=0.3$. For continuum magnitude estimates,
\begin{equation}
  m_{\rm AB}=M_{\rm AB}+\mu(z)+K(z), \qquad
  \mu(z)=5\log_{10}\!\left[\frac{D_L(z)}{\rm Mpc}\right]+25,
\end{equation}
where $K(z)$ is the bandpass-dependent $K$-correction. For emission-line selection,
\begin{equation}
  F_{\rm line}=\frac{L_{\rm line}}{4\pi D_L^2}.
\end{equation}
Representative star-forming ELGs at $1<z<3$ occupy roughly $M_{\rm AB}\simeq-19$ to $-22.5$, with the highest-yield BAO sample concentrated near $M_{\rm AB}\simeq-20.5$ to $-21.5$ and line luminosities $L_{\rm H\alpha}$ or $L_{\rm [O\,III]}\sim10^{42}$--$10^{43}$ erg s$^{-1}$. Table~\ref{tab:elgdepth} gives planning values before detailed luminosity-function integration and ETC validation. The table relates absolute magnitude, observed continuum magnitude, emission-line flux, and the tier-dependent line-luminosity depth from Eq.~\eqref{eq:f5sigma} on a common redshift grid.

\begin{table}[htbp]
\centering
\sffamily\small
\setlength{\tabcolsep}{4.5pt}
\resizebox{\textwidth}{!}{%
\begin{tabular}{@{}c c c c c c !{\hskip 8pt} c c c@{}}
\toprule
& & & & & & \multicolumn{3}{c@{}}{\textbf{Line-luminosity limit $L_{5\sigma}$ [erg s$^{-1}$]}}\\
\cmidrule(l){7-9}
\textbf{$z$} & \textbf{$\mu(z)$} & \textbf{$m(M=-20)$} & \textbf{$m(M=-21)$} & \textbf{$F(10^{42})$} & \textbf{$F(3\times10^{42})$} & \textbf{Wide} & \textbf{Medium} & \textbf{Deep}\\
 & mag & mag & mag & erg s$^{-1}$ cm$^{-2}$ & erg s$^{-1}$ cm$^{-2}$ & & &\\
\midrule
1.0 & 44.10 & 24.1 & 23.1 & $1.9\times10^{-16}$ & $5.7\times10^{-16}$ & $5.2\times10^{41}$ & $2.6\times10^{41}$ & $1.3\times10^{41}$\\
1.5 & 45.19 & 25.2 & 24.2 & $7.0\times10^{-17}$ & $2.1\times10^{-16}$ & $1.4\times10^{42}$ & $7.1\times10^{41}$ & $3.6\times10^{41}$\\
2.0 & 45.96 & 26.0 & 25.0 & $3.5\times10^{-17}$ & $1.0\times10^{-16}$ & $2.9\times10^{42}$ & $1.4\times10^{42}$ & $7.2\times10^{41}$\\
2.5 & 46.55 & 26.5 & 25.5 & $2.0\times10^{-17}$ & $6.0\times10^{-17}$ & $5.0\times10^{42}$ & $2.5\times10^{42}$ & $1.2\times10^{42}$\\
3.0 & 47.03 & 27.0 & 26.0 & $1.3\times10^{-17}$ & $3.9\times10^{-17}$ & $7.8\times10^{42}$ & $3.9\times10^{42}$ & $1.9\times10^{42}$\\
\bottomrule
\end{tabular}}
\caption{The left columns relate ELG absolute magnitude, apparent magnitude, and line flux on the same redshift grid used for the flux-to-luminosity conversion. Continuum magnitudes omit $K(z)$, which ranges from several tenths to approximately one magnitude depending on rest-frame wavelength and galaxy SED. The right columns give the line-luminosity limit $L_{5\sigma}$ implied by the wide, medium, and deep planning depths in Eq.~\eqref{eq:f5sigma}. An ELG with $3\times10^{42}$ erg s$^{-1}$ is detectable in the wide tier to $z\simeq2$, in the medium tier to $z\simeq2.5$, and in the deep tier beyond $z\simeq3$ depending on line identity and background.}
\label{tab:elgdepth}
\label{tab:llumlim}
\end{table}

For survey planning, the line-flux limit is parameterized as
\begin{equation}
  F_{5\sigma}(t_{\rm tot})\simeq
  1.0\times10^{-16}
  \left(\frac{0.75\ {\rm hr}}{t_{\rm tot}}\right)^{1/2}
  {\rm erg\ s^{-1}\ cm^{-2}},
  \label{eq:f5sigma}
\end{equation}
where $t_{\rm tot}$ is the summed spectroscopic exposure over the required three orientations in a given band. The $t_{\rm tot}^{-1/2}$ dependence follows from background-limited statistics. For a faint line integrated over a fixed number of resolution-element pixels, the accumulated sky signal grows as $t_{\rm tot}$ and the corresponding Poisson noise grows as $t_{\rm tot}^{1/2}$ while the line signal grows as $t_{\rm tot}$. Therefore, ${\rm S/N}\propto t_{\rm tot}^{1/2}$ at fixed flux and $F_{5\sigma}\propto t_{\rm tot}^{-1/2}$. The quantity $F_{5\sigma}$ denotes the line flux whose integrated signal equals five times the sky-background noise. Detector read noise and source shot noise are subdominant at the planning depth. Published $5$--$6.5\sigma$ line-flux limits from Roman HLSS and Euclid calibrate the normalization of $1.0\times10^{-16}\,{\rm erg\,s^{-1}\,cm^{-2}}$ at $t_{\rm tot}=0.75$ hr \cite{wang2021,euclid2011}. Equation~\eqref{eq:f5sigma} then gives $F_{5\sigma}\simeq1.0\times10^{-16}$, $5\times10^{-17}$, and $2.5\times10^{-17}\,{\rm erg\,s^{-1}\,cm^{-2}}$ at $t_{\rm tot}=0.75$, 3, and 12 hr, respectively. For the adopted flat $\Lambda$CDM cosmology, $D_L(z=2)\simeq15.5$ Gpc and $L_{\rm line}=4\pi D_L^2 F_{5\sigma}$. The three exposure times therefore correspond to $L_{\rm line}\simeq2.9\times10^{42}$, $1.4\times10^{42}$, and $7\times10^{41}$ erg s$^{-1}$ at $z\simeq2$. Section~\ref{sec:etc} derives the instrumental sensitivity and compares the result with Roman, Euclid, and JWST.

For an isolated line, the planning signal-to-noise ratio is
\begin{equation}
  {\rm S/N}_{\rm line}\simeq 5\,\frac{F_{\rm line}}{F_{5\sigma}(t_{\rm tot})}\,C_{\rm blend},
  \label{eq:snline}
\end{equation}
where $C_{\rm blend}\simeq0.7$--1 represents loss from morphology, background, and residual overlap after the three-orientation deblending fit. Table~\ref{tab:snrelg} gives the pre-blending S/N for $C_{\rm blend}=1$. The fitted $C_{\rm blend}$ and the spectral covariance reduce the object-level S/N.

\begin{table}[htbp]
\centering
\sffamily\small
\begin{tabularx}{\textwidth}{@{}p{2.3cm}p{3.3cm}c c c c@{}}
\toprule
 & & \multicolumn{4}{c}{\textbf{Approximate emission-line S/N}}\\
\cmidrule(lr){3-6}
\textbf{Tier} & \textbf{$F_{5\sigma}$ \ (exposure $t_{\rm tot}$)} & \multicolumn{4}{c}{at line flux $F$ [erg s$^{-1}$ cm$^{-2}$]:}\\
 & [erg s$^{-1}$ cm$^{-2}$] & $3\times10^{-17}$ & $10^{-16}$ & $3\times10^{-16}$ & $10^{-15}$\\
\midrule
Wide & $1.0\times10^{-16}$ \ (0.75 hr) & 1.5 & 5 & 15 & 50\\
Medium & $5.0\times10^{-17}$ \ (3 hr) & 3 & 10 & 30 & 100\\
Deep pencil & $2.5\times10^{-17}$ \ (12 hr) & 6 & 20 & 60 & 200\\
Ultra-deep & $1.3\times10^{-17}$ \ (48 hr) & 12 & 38 & 115 & 385\\
\bottomrule
\end{tabularx}
\caption{Approximate emission-line S/N before applying the object-specific blending/morphology factor. The second column lists, for each survey tier, the line-flux depth $F_{5\sigma}$ together (in parentheses) with the corresponding single-region total exposure time $t_{\rm tot}$, i.e. the three-orientation summed science exposure per band, related by $F_{5\sigma}\propto t_{\rm tot}^{-1/2}$ and matching the tiers of Table~\ref{tab:wedding}. All line fluxes ($F_{5\sigma}$ and the $F$ values in the column headers) are in erg s$^{-1}$ cm$^{-2}$. Secure cosmology redshifts require either S/N $\gtrsim7$--10 in a single line with strong priors, or multi-line/template confirmation at lower individual-line S/N.}
\label{tab:snrelg}
\end{table}

Table~\ref{tab:llumlim} gives the same depth model in luminosity rather than flux. Each entry is the line luminosity detected at S/N=5 before the object-specific blending and morphology correction. For a source with luminosity $L_{\rm line}$ at the listed redshift, the planning S/N is approximately
\begin{equation}
  {\rm S/N}_{\rm line}\simeq
  5\,\frac{L_{\rm line}}{L_{5\sigma}(z,t_{\rm tot})}\,C_{\rm blend}.
\end{equation}
A source equal to a Table~\ref{tab:llumlim} entry has S/N$\simeq5$ for $C_{\rm blend}=1$ and S/N$\simeq3.5$--5 for the nominal $C_{\rm blend}=0.7$--1 range. Doubling the source luminosity doubles the S/N.

\subsubsection{Line-Luminosity-Function Count Model}

The conversion from survey depth to an expected ELG catalog size requires an explicit line-luminosity-function model. For a line $i$ with rest wavelength $\lambda_i$ and luminosity function $\Phi_i(L,z)$, the 5$\sigma$ flux limit gives a redshift-dependent luminosity threshold
\begin{equation}
  L_{{\rm lim},i}(z)=4\pi D_L^2(z)F_{{\rm lim},i}(z),
\end{equation}
modified in practice by throughput, background, morphology, blending, and wavelength-dependent contamination. The differential flux distribution entering the exposure-time calculation is
\begin{equation}
  \frac{d^2N_i}{dF\,dz\,d\Omega}
  =
  4\pi D_L^2(z)\,
  \Phi_i\!\left(4\pi D_L^2F,z\right)
  \frac{dV_c}{dz\,d\Omega},
\end{equation}
where $\Phi_i$ is defined per unit luminosity. If the LF is written per dex in luminosity, the equivalent expression includes the factor $d\log L/dF=(F\ln10)^{-1}$. The count forecast therefore integrates the observed line-strength distribution above the detection threshold in each redshift slice. An ETC depth alone does not determine the count.

For analytic planning, a Schechter function represents the line LF \cite{schechter1976,saito2020,mehta2015},
\begin{equation}
  \Phi_i(L,z)dL =
  \phi_i^\star(z)
  \left(\frac{L}{L_i^\star(z)}\right)^{\alpha_i(z)}
  \exp\!\left[-\frac{L}{L_i^\star(z)}\right]
  \frac{dL}{L_i^\star(z)} ,
\end{equation}
where the redshift evolution is part of the model, not a post-processing correction. A convenient planning parameterization is
\begin{equation}
  \begin{aligned}
  \log_{10}L_i^\star(z) &=
  \log_{10}L_{i,0}^\star+
  Q_{L,i}\log_{10}\!\left(\frac{1+z}{1+z_0}\right),\\
  \log_{10}\phi_i^\star(z) &=
  \log_{10}\phi_{i,0}^\star+
  Q_{\phi,i}\log_{10}\!\left(\frac{1+z}{1+z_0}\right),\\
  \alpha_i(z) &= \alpha_{i,0}+Q_{\alpha,i}(z-z_0),
  \end{aligned}
  \label{eq:lf_evolution_planning}
\end{equation}
with separate coefficients for H$\alpha$, [O\,III], H$\beta$, and [O\,II]. Equation~\eqref{eq:lf_evolution_planning} interpolates the redshift-evolving Schechter LF fits of Saito et al. \cite{saito2020} and the single-bin WISP [O\,III] fits of Mehta et al. \cite{mehta2015}. For the Saito et al. H$\alpha$ and [O\,II] rows in Table~\ref{tab:lfparams_eq16}, the adopted correspondence is $z_0=0$, $Q_{L,i}=\beta$, $Q_{\alpha,i}=0$, and $Q_{\phi,i}=\gamma$ below $z_{\rm pivot}$. Above $z_{\rm pivot}$, the source model uses the broken normalization slope $Q_{\phi,i}=-\epsilon$ with the continuity factor given below \cite{saito2020}. The mission analysis fits the parameters in redshift bins and uses smooth evolution only to interpolate between calibrated bins. The measured line-flux distribution is a direct projection of the evolving LF,
\begin{equation}
  \frac{d^2N_i}{dz\,d\Omega}(>F_{\rm lim}) =
  \frac{dV_c}{dz\,d\Omega}
  \int_{4\pi D_L^2(z)F_{\rm lim}}^\infty
  \Phi_i(L,z)\,C_i(L,z)\,dL ,
\end{equation}
where $C_i(L,z)$ is the completeness function of line $i$, the probability between zero and one that a galaxy of line luminosity $L$ at redshift $z$ yields a secure detection of that line in the survey data. In the ideal case $C_i=1$ the line-selected surface density in a redshift bin reduces to the standard luminosity-function projection over comoving volume \cite{schechter1976,saito2020,mehta2015},
\begin{equation}
  \Sigma_i(z_1,z_2;F_{\rm lim}) =
  \int_{z_1}^{z_2}
  \frac{dV_c}{dz\,d\Omega}
  \int_{L_{{\rm lim},i}(z)}^\infty
  \Phi_i(L,z)\,dL\,dz .
  \label{eq:lf_surface_density}
\end{equation}
Equivalently, for a pure Schechter LF this inner integral is $\phi_i^\star\Gamma(\alpha_i+1,L_{\rm lim}/L_i^\star)$, but numerical integration is preferred because ELG faint-end slopes, LF evolution, and completeness functions are not perfectly described by a single Schechter law.
Table~\ref{tab:lfparams_eq16} lists the explicit literature parameter values adopted as starting inputs for Eq.~\eqref{eq:lf_evolution_planning} and for the Eq.~\eqref{eq:lf_surface_density} surface-density integral. Saito et al. use
$L^\star(z)=L^\star_{*,0}(1+z)^\beta$ and
$\Phi^\star(z)=\Phi^\star_{*,0}(1+z)^\gamma$ below $z_{\rm pivot}$, with
$\Phi^\star(z)=\Phi^\star_{*,0}(1+z_{\rm pivot})^{\gamma+\epsilon}(1+z)^{-\epsilon}$ above $z_{\rm pivot}$ \cite{saito2020}. The planning table does not assign an independent LF to H$\beta$. The line-ratio model instead propagates H$\beta$ as a secondary confirmation line. The Saito et al. prescription uses H$\alpha$/H$\beta=2.9$ and [O\,III]/H$\beta=4.1$ \cite{saito2020}.

\begin{table}[htbp]
\centering
\sffamily\scriptsize
\setlength{\tabcolsep}{3pt}
\begin{tabularx}{\textwidth}{@{}p{1.25cm}p{1.45cm}c c c c c c Y@{}}
\toprule
\textbf{Line} & \textbf{$z$ range} & \textbf{$\log_{10}\Phi^\star_{*,0}$} & \textbf{$\gamma$} & \textbf{$\epsilon$} & \textbf{$z_{\rm pivot}$} & \textbf{$\log_{10}L^\star_{*,0}$} & \textbf{$\alpha$} & \textbf{Source / note}\\
 & & Mpc$^{-3}$ & & & & erg s$^{-1}$ & & \\
\midrule
H$\alpha$ & $0.3<z<2.0$ & $-2.92\pm0.03$ & $1.30\pm0.12$ & $1.86\pm1.40$ & $1.53\pm0.12$ & $41.59\pm0.03$ & $-1.35$ & Saito et al. with $\beta=1.91\pm0.08$.\\
{[O\,II]} & $0.3<z<2.5$ & $-1.89\pm0.04$ & $-1.96\pm0.18$ & $-2.48\pm0.21$ & $1.00\pm0.02$ & $40.73\pm0.02$ & $-1.25$ & Saito et al. with $\beta=2.61\pm0.08$.\\
{[O\,III]} & $0.8<z<1.2$ & $-3.17^{+0.27}_{-0.39}$ & \multicolumn{3}{c}{single-bin Schechter LF} & $42.21^{+0.22}_{-0.18}$ & $-1.42^{+0.23}_{-0.43}$ & Mehta et al. WISP [O\,III] LF.\\
{[O\,III]} & $1.85<z<2.2$ & $-2.69^{+0.31}_{-0.51}$ & \multicolumn{3}{c}{single-bin Schechter LF} & $42.55^{+0.28}_{-0.19}$ & $-1.57^{+0.28}_{-0.77}$ & Mehta et al. WISP [O\,III] LF.\\
\bottomrule
\end{tabularx}
\caption{Literature LF parameters for the redshift evolution in Eq.~\eqref{eq:lf_evolution_planning} and the surface-density integral in Eq.~\eqref{eq:lf_surface_density}. The values initialize the exposure-time and number-count forecasts. The mission forecast refits the parameters with the medium and deep calibration tiers and includes completeness $C_i(L,z)$.}
\label{tab:lfparams_eq16}
\end{table}

The secure-redshift catalogue is the union of several line selections rather than the sum of independent H$\alpha$, [O\,III], H$\beta$, and [O\,II] counts. The forecast therefore integrates over galaxy properties $\boldsymbol{\theta}$ including line ratios, equivalent widths, continuum magnitude, half-light radius, dust attenuation, and local spectral crowding.
\begin{equation}
  N_{\rm secure} =
  \Omega_{\rm surv}
  \int dz\,\frac{dV_c}{dz\,d\Omega}
  \int d\boldsymbol{\theta}\,
  n(\boldsymbol{\theta}|z)\,
  P_{\rm secure}(\boldsymbol{\theta},z).
\end{equation}
The function $n(\boldsymbol{\theta}|z)$ is the comoving number density of galaxies per unit property interval at redshift $z$. The function generalizes the luminosity function in Eq.~\eqref{eq:lf_surface_density} by replacing luminosity with the full property vector $\boldsymbol{\theta}$. The normalization $\int d\boldsymbol{\theta}\,n(\boldsymbol{\theta}|z)$ recovers the total comoving number density at $z$. The factor $P_{\rm secure}(\boldsymbol{\theta},z)$ is the fraction of galaxies with the specified properties for which spectral extraction returns a cosmology-grade redshift. The expression is a selection-weighted population count rather than a Bayesian inference. The conditional notation $(\,\cdot\,|z)$ denotes evaluation at redshift $z$ rather than a posterior. The redshift-success probability is approximately
\begin{equation}
  P_{\rm secure}=1-\prod_i
  \left[1-C_i(F_i,z,\boldsymbol{\theta})\right],
\end{equation}
where the $C_i$ are correlated because the same galaxy supplies multiple lines. For slitless data, $C_i$ includes line S/N, direct-image continuum priors, source size along the dispersion direction, orientation-dependent overlap probability, line confusion, and the probability that a single-line solution satisfies the cosmology-grade criteria. The medium and deep tiers measure the LF tails, line-ratio distribution, and incompleteness required to convert the wide-tier target density into an unbiased clustering selection function.

The mission requires the following LF calibration set.

\begin{itemize}
  \item H$\alpha$ LF over $0.5\lesssim z\lesssim3.5$ because H$\alpha$ is the strongest line for many star-forming galaxies, anchoring the low-to-mid redshift BAO sample.
  \item [O\,III] and H$\beta$ LFs over $1\lesssim z\lesssim3$ because high-equivalent-width [O\,III] emitters dominate part of the near-IR selection, providing strong multi-line confirmation.
  \item [O\,II] LF over $1\lesssim z\lesssim3$ because [O\,II] remains in band after H$\alpha$ shifts toward the long-wavelength edge, supplying continuity with DESI/eBOSS.
  \item Joint distributions of line ratios, equivalent width, dust, size, and continuum magnitude because the probability of detecting at least one secure redshift feature depends on correlated galaxy properties, not on four independent one-line LFs.
\end{itemize}
The DESI-derived [O\,II] calculation below therefore serves only as a low-redshift empirical reference. The full 3.5 m all-line survey yield must combine redshift-dependent luminosity functions with the slitless selection model and must be validated with simulated detector images and deeper calibration fields.

\subsubsection{DESI [O\,II] Luminosity-Function Reference}

The first redshift-dependent number-count estimate is derived directly from the local DESI DR1 LSS \texttt{ELG\_\allowbreak LOPnotqso} FITS table rather than from a fitted analytic Schechter function. DESI ELG target selection and validation show that the Main ELG sample is selected by a $g$-band cut plus a $(g-r)$--$(r-z)$ color box, with two disjoint subsamples of about 1940 and 460 targets deg$^{-2}$, and that reliable ELG redshifts are assessed using [O\,II] flux information \cite{raichoor2022}. We therefore set the effective area of the downloaded table by
\begin{equation}
  \Omega_{\rm eff}=
  \frac{N_{\rm target}}{2400\ {\rm deg}^{-2}}
  =
  \frac{971427}{2400}
  \simeq405\ {\rm deg}^{2},
\end{equation}
We select secure [O\,II] galaxies with \texttt{ZWARN=0}, finite positive redshift, positive \texttt{OII\_FLUX}, and
\begin{equation}
  {\rm S/N}_{\rm [O\,II]}=
  {\tt OII\_FLUX}\,({\tt OII\_FLUX\_IVAR})^{1/2}>5 .
\end{equation}
The line luminosity is computed as
\begin{equation}
  L_{\rm [O\,II]}=4\pi D_L^2(z)F_{\rm [O\,II]},
\end{equation}
using the same flat $\Lambda$CDM cosmology as the rest of this section. In a redshift bin $z_1<z<z_2$, the binned observed luminosity function is
\begin{equation}
  \widehat{\Phi}(\log L_k,z_j)=
  \frac{N_k}
  {\Omega_{\rm eff}\,V_{\rm c,deg^2}(z_1,z_2)\,\Delta\log L\,\overline{C}_k},
\end{equation}
where $V_{\rm c,deg^2}$ is the comoving volume per square degree in the redshift bin and $\overline{C}_k$ is the mean completeness correction in luminosity bin $k$. The conservative DESI table below sets $\overline{C}_k=1$. The medium-depth counts therefore represent lower limits rather than a complete faint-end extrapolation. Summing the measured LF bins above the luminosity threshold gives the predicted count distribution in one field of view,
\begin{equation}
  \Delta N_{{\rm FOV},j}(>F_{\rm lim}) =
  A_{\rm FOV}\,
  V_{\rm c,deg^2}(z_j)
  \sum_{\log L_k>\log L_{\rm lim}(z_j)}
  \widehat{\Phi}(\log L_k,z_j)\,
  \Delta\log L\,
  \overline{P}_{{\rm secure},k,j},
\end{equation}
with $A_{\rm FOV}=0.25$ deg$^2$ for one \fovlarge\ frame. The count equation converts an observed LF into the number of galaxies expected in one pointing. The equivalent observable surface density for a survey flux limit $F_{\rm lim}$ is
\begin{equation}
  \Sigma(>F_{\rm lim},z_1,z_2)=
  \int_{\log L_{\rm lim}}^\infty
  \widehat{\Phi}(\log L,z)\,V_{\rm c,deg^2}\,d\log L
  =
  \frac{N(F_{\rm [O\,II]}>F_{\rm lim})}{\Omega_{\rm eff}},
\end{equation}
with $L_{\rm lim}=4\pi D_L^2(z_{\rm mid})F_{\rm lim}$. A single \fovlarge\ frame has area 0.25 deg$^2$ and therefore contains $N_{\rm frame}=0.25\Sigma$. Table~\ref{tab:desiLFcounts} summarizes the DESI-derived reference based only on [O\,II]. The table provides a conservative reference for the proposed all-line slitless survey rather than the full H$\alpha$+[O\,III]+[O\,II] yield. DESI does not provide the same direct [O\,II] calibration above $z=1.6$. External [O\,II] and H$\alpha$ luminosity functions and number counts must calibrate the high-redshift extension \cite{comparat2015,saito2020,mehta2015}.

\begin{table}[htbp]
\centering
\sffamily\scriptsize
\begin{tabularx}{\textwidth}{@{}c c c c c c c@{}}
\toprule
\textbf{$z$ bin} & \textbf{$\log L_{\rm lim,wide}$} & \textbf{$n_{\rm DESI}$} & \textbf{$\Sigma_{\rm wide}$} & \textbf{$\Sigma_{\rm med}$} & \textbf{$N_{\rm frame,wide}$} & \textbf{$N_{\rm frame,med}$}\\
 & $\log_{10}({\rm erg\ s}^{-1})$ & $10^{-5}$ Mpc$^{-3}$ & deg$^{-2}$ & deg$^{-2}$ & ELGs per 0.25 deg$^2$ & ELGs per 0.25 deg$^2$\\
 & $F_{\rm lim}=10^{-16}$ & $>L_{\rm lim,wide}$ & $F_{\rm lim}=10^{-16}$ & $F_{\rm lim}=5\times10^{-17}$ & $F_{\rm lim}=10^{-16}$ & $F_{\rm lim}=5\times10^{-17}$\\
\midrule
0.6--0.8 & 41.34 & 2.34 & 25.9 & 35.4 & 6.5 & 8.8\\
0.8--1.0 & 41.60 & 6.47 & 94.6 & 118.4 & 23.6 & 29.6\\
1.0--1.2 & 41.82 & 5.35 & 93.6 & 119.2 & 23.4 & 29.8\\
1.2--1.4 & 42.00 & 3.95 & 77.9 & 104.7 & 19.5 & 26.2\\
1.4--1.6 & 42.15 & 2.45 & 52.4 & 69.4 & 13.1 & 17.4\\
\bottomrule
\end{tabularx}
\caption{DESI-derived [O\,II] number-count reference by redshift. The table is computed from the local DESI DR1 \texttt{ELG\_LOPnotqso} catalog using \texttt{ZWARN=0} and [O\,II] S/N$>5$. The effective area is inferred from the DESI Main ELG target density, $1940+460\simeq2400$ deg$^{-2}$, and the 971,427 targets in the local table \cite{raichoor2022}. The quoted $n_{\rm DESI}$ is the integrated observed number density above the wide-tier luminosity limit, not a Schechter-fit $\phi^\star$. Medium counts are lower limits for a truly deeper survey because the DESI target catalog is not complete at arbitrarily faint [O\,II] flux.}
\label{tab:desiLFcounts}
\end{table}

Summing the observed-LF redshift bins in Table~\ref{tab:desiLFcounts} gives the expected [O\,II]-only count per \fovlarge\ pointing shown in Table~\ref{tab:desiFOVsum}. The values do not constitute the final all-line ELG yield. They provide a conservative empirical floor that maps the measured DESI [O\,II] LF into one 0.25 deg$^2$ field at the adopted flux thresholds.

\begin{table}[htbp]
\centering
\sffamily\small
\begin{tabularx}{\textwidth}{@{}c c c c c@{}}
\toprule
\textbf{Redshift range} & \textbf{$\Sigma_{\rm wide}$} & \textbf{$N_{\rm FOV,wide}$} & \textbf{$\Sigma_{\rm med}$} & \textbf{$N_{\rm FOV,med}$}\\
 & deg$^{-2}$ & ELGs per 0.25 deg$^2$ & deg$^{-2}$ & ELGs per 0.25 deg$^2$\\
\midrule
0.6--1.0 & 120.5 & 30.1 & 153.8 & 38.5\\
1.0--1.6 & 223.9 & 56.0 & 293.3 & 73.3\\
0.6--1.6 & 344.4 & 86.1 & 447.1 & 111.8\\
\bottomrule
\end{tabularx}
\caption{Cumulative field-of-view counts obtained by inserting the observed DESI [O\,II] LF into the FOV count equation. The wide threshold is $F_{\rm lim}=10^{-16}$ erg s$^{-1}$ cm$^{-2}$ and the medium threshold is $F_{\rm lim}=5\times10^{-17}$ erg s$^{-1}$ cm$^{-2}$. A larger all-line space-survey count requires support from the H$\alpha$, [O\,III], H$\beta$, and high-redshift [O\,II] luminosity functions together with the slitless redshift-success model.}
\label{tab:desiFOVsum}
\end{table}

The corresponding continuum planning depths, for binned low-resolution spectral elements or matched direct imaging used as extraction priors, are approximately
\begin{equation}
  m_{\rm lim}(t_{\rm tot})\simeq24.3+
  1.25\log_{10}\!\left(\frac{t_{\rm tot}}{0.75\ {\rm hr}}\right).
\end{equation}
The relation gives $m_{\rm AB}\simeq24.3$, 25.1, 25.8, and 26.6 mag for $t_{\rm tot}=0.75$, 3, 12, and 48 hr, respectively. The line-selected ELG catalogue therefore reaches lower star-formation activity than the continuum reaches in stellar-mass completeness. The medium and deep tiers calibrate the resulting selection function.

\begin{figure}[htbp]
\centering
\includegraphics[width=0.78\textwidth]{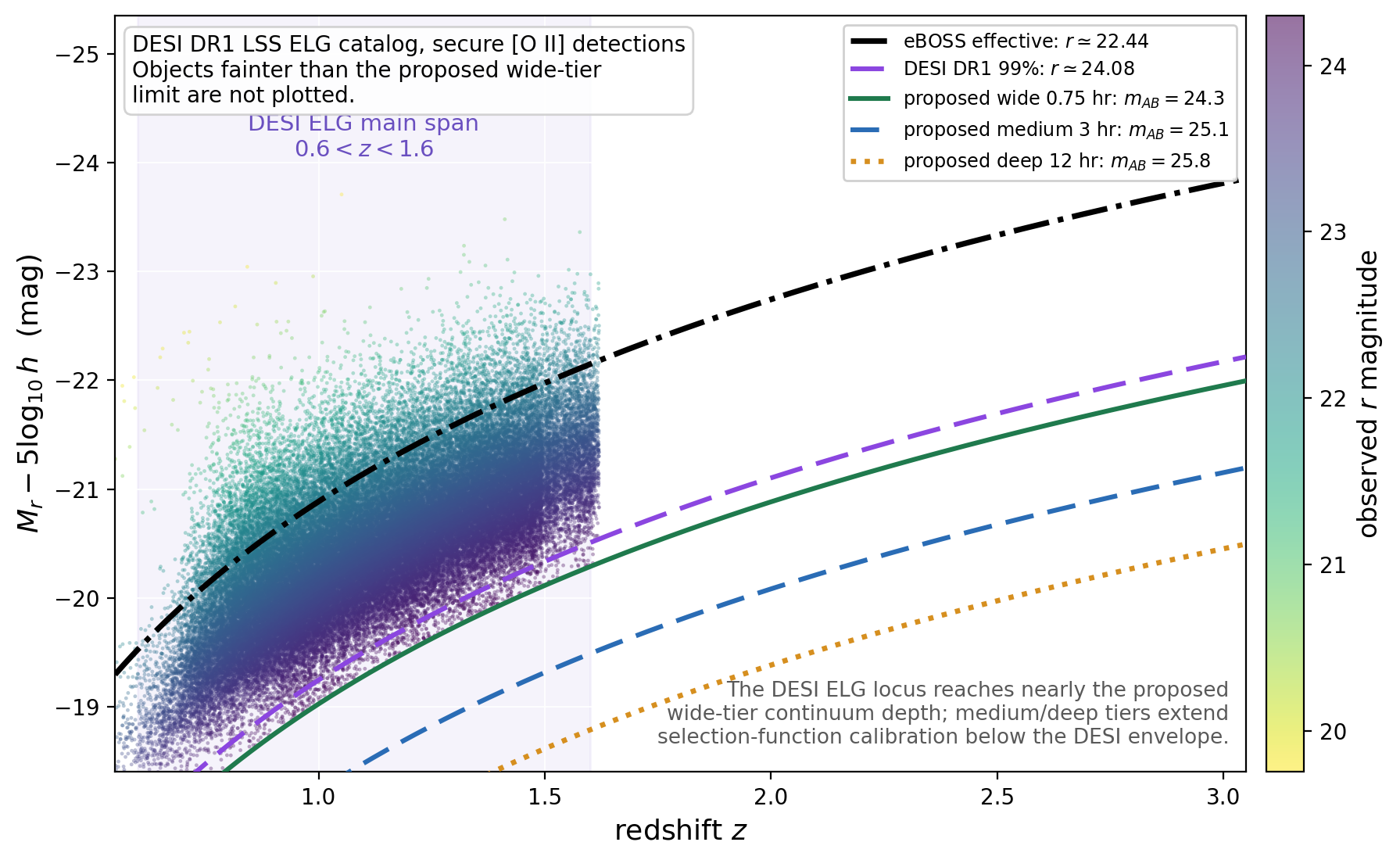}
\caption{Observed-frame ELG absolute magnitude from the DESI DR1 LSS ELG catalogue as a function of redshift. Points are secure DESI ELGs with [O\,II] detections and $m_r\le24.3$. The black curve gives the effective eBOSS depth and the purple curve gives the DESI DR1 99th-percentile $r$-band depth of the plotted sample. Green, blue, and orange curves give the faintest $M_r-5\log_{10}h$ at the proposed wide, medium, and deep continuum limits. The planning comparison uses observed $r$ band and omits $K$-corrections \cite{raichoor2022}.}
\label{fig:elgmagdepth}
\end{figure}

Figure~\ref{fig:elgmagdepth} places the continuum limits on an observed ELG catalogue. We use the full DESI DR1 LSS \texttt{ELG\_LOPnotqso} catalogue, select galaxy spectra with \texttt{ZWARN=0} and [O\,II] signal-to-noise greater than 5, and compute an observed-frame $r$-band magnitude from the Legacy Surveys fluxes,
\begin{equation}
  M_r-5\log_{10}h
  =
  m_r-
  \left[
  5\log_{10}\!\left(\frac{D_L}{h^{-1}{\rm Mpc}}\right)+25
  \right],
\end{equation}
without applying a $K$-correction. The black curve marks the effective eBOSS $r$-band limit of $r\simeq22.44$ and provides a historical comparison. The purple curve marks the DESI DR1 99th-percentile depth of $r\simeq24.08$ for the plotted secure ELG sample. DESI target selection combines a $g$-band magnitude cut with a $(g-r)$ versus $(r-z)$ colour box rather than applying a pure $r$-band cut. The plotted depth curves indicate observed-frame catalogue truncation and do not represent formal single-band selection boundaries \cite{raichoor2022}. The catalogue points are restricted to the adopted wide-tier continuum limit of $m_r\le24.3$. DESI covers the key $0.6<z<1.6$ interval and nearly reaches the proposed wide-tier continuum depth. The proposed space survey extends the selection to $1\lesssim z\lesssim3$ with deeper medium and deep tiers and near-infrared emission lines.

\begin{figure}[htbp]
\centering
\includegraphics[width=0.78\textwidth]{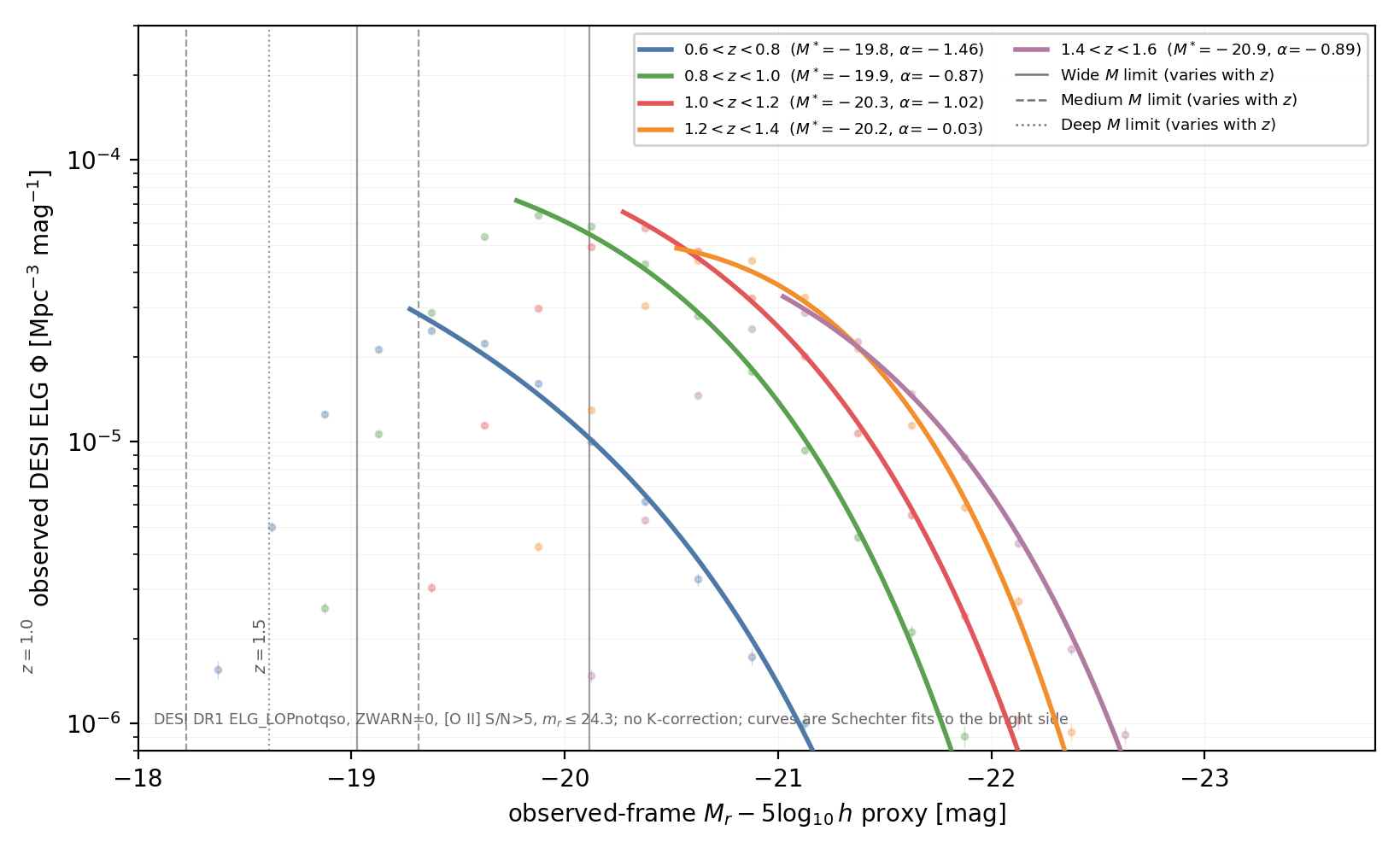}
\caption{Observed-frame DESI ELG absolute-magnitude luminosity function in redshift bins from the secure \texttt{ELG\_LOPnotqso} sample used in Figure~\ref{fig:elgmagdepth}. Points give the binned luminosity function. Solid curves give Schechter fits ($\Phi^\star,M^\star,\alpha$) to the well-sampled bright side of each bin and exclude magnitudes beyond the faint-end turnover imposed by DESI target selection. The curves must not be extrapolated through the incompleteness boundary. Vertical markers show the proposed continuum limits at representative redshifts. The luminosity functions use DESI-observed quantities without rest-frame $K$-corrections \cite{raichoor2022}.}
\label{fig:desiMlf}
\end{figure}

To convert Figure~\ref{fig:elgmagdepth} into a redshift-dependent count estimate, we bin the same secure DESI sample in $M_r-5\log_{10}h$ and redshift. The observed-frame continuum luminosity function is
\begin{equation}
  \widehat{\Phi}_M(M_k,z_j)=
  \frac{N_k}
  {\Omega_{\rm eff}\,V_{\rm c,deg^2}(z_1,z_2)\,\Delta M},
\end{equation}
with the same $\Omega_{\rm eff}$ and cosmology used for the [O\,II] LF calibration above. Figure~\ref{fig:desiMlf} shows the resulting evolution over $0.6<z<1.6$. The turnover at faint $M_r$ marks the DESI target-selection and redshift-success boundary rather than a turnover in the galaxy LF. The figure therefore constrains the planning depth but does not by itself support extrapolation to medium-tier or deep-tier yields.

Table~\ref{tab:desiMcounts} lists the corresponding DESI-observed count integral and answers a narrower question than the line-flux table. For sources represented in the DESI ELG selection the table gives the number per \fovlarge\ frame above the proposed continuum depth. The nearly identical wide, medium, and deep counts diagnose truncation of the DESI parent catalogue near the wide-tier depth. The similarity does not imply that additional depth is unimportant. DESI alone does not measure the incremental population expected at $m_{\rm AB}\simeq25.1$--26.6.

\begin{table}[htbp]
\centering
\sffamily\scriptsize
\begin{tabularx}{\textwidth}{@{}c c c c c c c@{}}
\toprule
\textbf{$z$ bin} &
\textbf{$M_{\rm lim,wide}$} & \textbf{$N_{\rm frame,wide}$} &
\textbf{$M_{\rm lim,med}$} & \textbf{$N_{\rm frame,med}$} &
\textbf{$M_{\rm lim,deep}$} & \textbf{$N_{\rm frame,deep}$}\\
 & \multicolumn{1}{c}{$m_r=24.3$} & \multicolumn{1}{c}{0.25 deg$^2$} &
   \multicolumn{1}{c}{$m_r=25.1$} & \multicolumn{1}{c}{0.25 deg$^2$} &
   \multicolumn{1}{c}{$m_r=25.8$} & \multicolumn{1}{c}{0.25 deg$^2$}\\
\midrule
0.6--0.8 & -18.07 & 8.9 & -17.27 & 8.9 & -16.57 & 8.9\\
0.8--1.0 & -18.74 & 29.6 & -17.94 & 29.7 & -17.24 & 29.7\\
1.0--1.2 & -19.28 & 29.6 & -18.48 & 29.7 & -17.78 & 29.7\\
1.2--1.4 & -19.73 & 26.0 & -18.93 & 26.1 & -18.23 & 26.1\\
1.4--1.6 & -20.11 & 17.2 & -19.31 & 17.3 & -18.61 & 17.3\\
\bottomrule
\end{tabularx}
\caption{DESI-observed continuum count integral corresponding to Figures~\ref{fig:elgmagdepth} and \ref{fig:desiMlf}. The count is $N_{\rm frame}=0.25\,\Sigma$ after integrating the binned DESI observed-frame luminosity function brighter than the listed $M_r-5\log_{10}h$ limit. Because the local DESI catalogue was restricted to $m_r\le24.3$, the medium and deep columns are lower limits and identify the magnitude range where external deep surveys must supply the missing faint population.}
\label{tab:desiMcounts}
\end{table}

External calibration is therefore required. DESI gives the wide-area normalization and the $0.6<z<1.6$ colour-selected ELG reference. DEEP2, VVDS, and COSMOS-based emission-line catalogues supply the fainter-continuum and higher-redshift information needed to model the DESI truncation boundary, colour incompleteness, and faint-end LF slope. The zCOSMOS group catalogue adds environmental and membership information \cite{newman2013deep2,lefevre2013vvds,saito2020,knobel2012zcosmos}. Published [O\,II], H$\beta$+[O\,III], and H$\alpha$ luminosity functions show strong redshift evolution in both $L^\star$ and normalization. The final yield model must jointly fit DESI and the deeper samples rather than extrapolate the DESI observed-frame luminosity function alone \cite{comparat2015,comparat2016,khostovan2015,mehta2015}.

\subsection{Wedding-Cake Survey Strategy}

The BAO and RSD program uses a wedding-cake survey. The wide tier supplies cosmological volume. The medium tier measures redshift success for fainter ELGs. The deep pencil beams constrain luminosity-function tails, completeness, and spectral deblending under high source density.

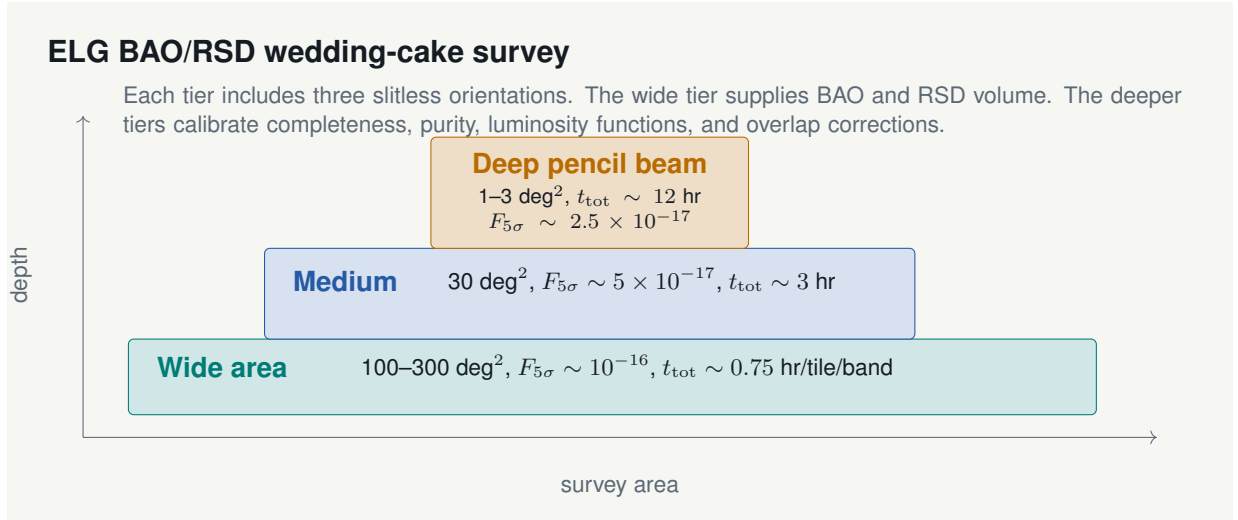
\begin{figure}[htbp]
\centering
\begin{tikzpicture}[font=\sffamily\footnotesize]
  \fill[paper] (-0.4,-0.35) rectangle (15.8,6.5);
  \node[anchor=west,font=\bfseries\large,color=ink] at (0,5.82) {ELG BAO/RSD wedding-cake survey};
  \draw[->,color=muted] (0.6,0.75) -- (14.8,0.75);
  \draw[->,color=muted] (0.6,0.75) -- (0.6,5.0);
  \node[anchor=north,color=muted] at (7.7,0.32) {survey area};
  \node[rotate=90,anchor=south,color=muted] at (0.05,2.9) {depth};

  \fill[teal!18,draw=teal,rounded corners=2pt] (1.2,1.05) rectangle (14.0,2.05);
  \fill[blue!18,draw=blue,rounded corners=2pt] (3.0,2.05) rectangle (11.6,3.25);
  \fill[gold!22,draw=gold,rounded corners=2pt] (5.2,3.25) rectangle (9.4,4.72);

  \node[anchor=west,font=\bfseries,color=teal] at (1.45,1.68) {Wide area};
  \node[anchor=west,color=ink] at (4.15,1.68) {100--300 deg$^2$, $F_{5\sigma}\sim10^{-16}$, $t_{\rm tot}\sim0.75$ hr/tile/band};
  \node[anchor=west,font=\bfseries,color=blue] at (3.25,2.82) {Medium};
  \node[anchor=west,color=ink] at (5.3,2.82) {30 deg$^2$, $F_{5\sigma}\sim5\times10^{-17}$, $t_{\rm tot}\sim3$ hr};
  \node[font=\bfseries,color=gold] at (7.3,4.34) {Deep pencil beam};
  \node[align=center,color=ink,font=\scriptsize,text width=3.9cm] at (7.3,3.78)
    {1--3 deg$^2$, $t_{\rm tot}\sim12$ hr\\ $F_{5\sigma}\sim2.5\times10^{-17}$};

  \node[anchor=west,text width=14.0cm,color=muted] at (1.0,5.06)
  {Each tier includes three slitless orientations. The wide tier supplies BAO and RSD volume. The deeper tiers calibrate completeness, purity, luminosity functions, and overlap corrections.};
\end{tikzpicture}
\caption{The ELG program combines several depths. Deeper pencil beams calibrate the selection function of the wide BAO and RSD survey. The quoted line-flux depths $F_{5\sigma}$ are in erg s$^{-1}$ cm$^{-2}$.}
\label{fig:weddingcake}
\end{figure}

Table~\ref{tab:wedding} turns the wedding-cake strategy in Figure~\ref{fig:weddingcake} into exposure and depth requirements. The wide tier supplies the BAO/RSD volume, while the medium, deep, and ultra-deep tiers are included to measure redshift failures, overlap losses, and luminosity-function tails that are not calibrated from the wide survey alone.

\begin{table}[htbp]
\centering
\sffamily\small
\begin{tabularx}{\textwidth}{@{}p{2.5cm}p{2.4cm}p{2.8cm}p{3.1cm}Y@{}}
\toprule
\textbf{Tier} & \textbf{Area} & \textbf{3-orientation exposure} & \textbf{Line-flux limit $F_{5\sigma}$} & \textbf{Primary role}\\
 & deg$^2$ & science time per band & erg s$^{-1}$ cm$^{-2}$ (smaller $=$ deeper) & \\
\midrule
Wide BAO/RSD & 100 \degti threshold. 300 \degti goal & $3\times900$ s per band & $1.0\times10^{-16}$ & Cosmological volume with $10^6$--$3\times10^6$ secure ELG redshifts.\\
Medium calibration & 30 \degti & $3\times3600$ s per band & $5.0\times10^{-17}$ & Redshift failures, faint ELGs, line-ratio priors, survey transfer functions.\\
Deep pencil beam & 1--3 \degti & $3\times4$ hr per band & $2.5\times10^{-17}$ & Luminosity-function tail, overlap calibration, completeness to $z\simeq3$.\\
Ultra-deep validation & 0.1--0.3 \degti & $3\times16$ hr per band & $1.3\times10^{-17}$ & External spectroscopic reference fields and extraction-stability tests.\\
\bottomrule
\end{tabularx}
\caption{Illustrative ELG survey tiers. Exposure times give total science exposure per filter and include the required three orientations. Slew, readout, dither, calibration, and scheduling overheads are excluded. A smaller $F_{5\sigma}$ denotes a fainter detectable line and therefore a greater depth. In the background-limited regime, $F_{5\sigma}\propto t_{\rm tot}^{-1/2}$. The wide tier reaches $1.0\times10^{-16}$ in $3\times900$ s while the ultra-deep tier reaches $1.3\times10^{-17}$ in $3\times16$ hr.}
\label{tab:wedding}
\end{table}

\subsection{BAO, \texorpdfstring{$H_0$}{H0}, and Growth Precision}

The ELG BAO measurement is designed as a precision distance-ratio experiment. In each redshift bin the measured parameters are
\begin{equation}
  \alpha_\perp(z)=
  \frac{[D_M(z)/r_d]}{[D_M(z)/r_d]_{\rm fid}},\qquad
  \alpha_\parallel(z)=
  \frac{[H(z)r_d]_{\rm fid}}{H(z)r_d},
\end{equation}
where $D_M$ is the transverse comoving distance and $r_d$ is the sound horizon at the drag epoch. The survey alone does not determine an absolute $H_0$. An absolute value requires calibration of $r_d$ from the CMB, BBN, or another early-Universe constraint. Without an external calibration, the late-time observable is the inverse-distance-ladder combination $H_0r_d$.

The growth measurement comes from the redshift-space galaxy power spectrum,
\begin{equation}
  P_s(k,\mu,z)\simeq
  \left[b(z)+f(z)\mu^2\right]^2 P_m(k,z)
  \exp\!\left[-k^2\mu^2\sigma_r^2(z)\right],
\end{equation}
with $\sigma_r=c\,\sigma_z/H(z)$ the radial redshift-smearing scale. The principal growth observable is $\fsig(z)$. For compact reporting we also quote a growth-amplitude summary,
\begin{equation}
  S_8^{\rm grow}\equiv \sigma_8(\Omega_m/0.3)^{1/2},
\end{equation}
after marginalizing over galaxy bias in the ELG clustering model. The statistical power scales approximately as
\begin{equation}
  V_{\rm eff}(k,\mu)=
  \int \left[\frac{n(z)P(k,\mu,z)}{1+n(z)P(k,\mu,z)}\right]^2 dV,
  \label{eq:veff}
\end{equation}
so the wide tier sets the cosmological precision while the medium and deep tiers reduce redshift-failure and line-identification systematics.

In configuration space the same information is measured from the two-dimensional galaxy correlation function, using the minimum-variance Landy--Szalay estimator \cite{landyszalay1993},
\begin{equation}
  \xi(\sigma,\pi)=\frac{DD(\sigma,\pi)-2DR(\sigma,\pi)+RR(\sigma,\pi)}
  {RR(\sigma,\pi)},
  \label{eq:xiestimator}
\end{equation}
where $\sigma$ is the transverse separation and $\pi$ is the line-of-sight separation. The BAO feature appears as an annular excess at $r_{\rm BAO}\simeq100$--$110\,h^{-1}$ Mpc. The ring is nearly circular in real space. Peculiar velocities and coherent infall distort the redshift-space contours and produce anisotropy that is fitted jointly with the BAO scale. The transverse and radial BAO positions give
\begin{equation}
  \sigma_{\rm BAO}\propto D_M(z)/r_d,\qquad
  \pi_{\rm BAO}\propto [H(z)r_d]^{-1},
\end{equation}
while the quadrupole-to-monopole structure of $\xi(\sigma,\pi)$ constrains $\fsig(z)$. The ELG survey is therefore designed to recover the full two-dimensional correlation function, not only a spherically averaged BAO peak.

\begin{figure}[htbp]
\centering
\includegraphics[width=0.80\textwidth]{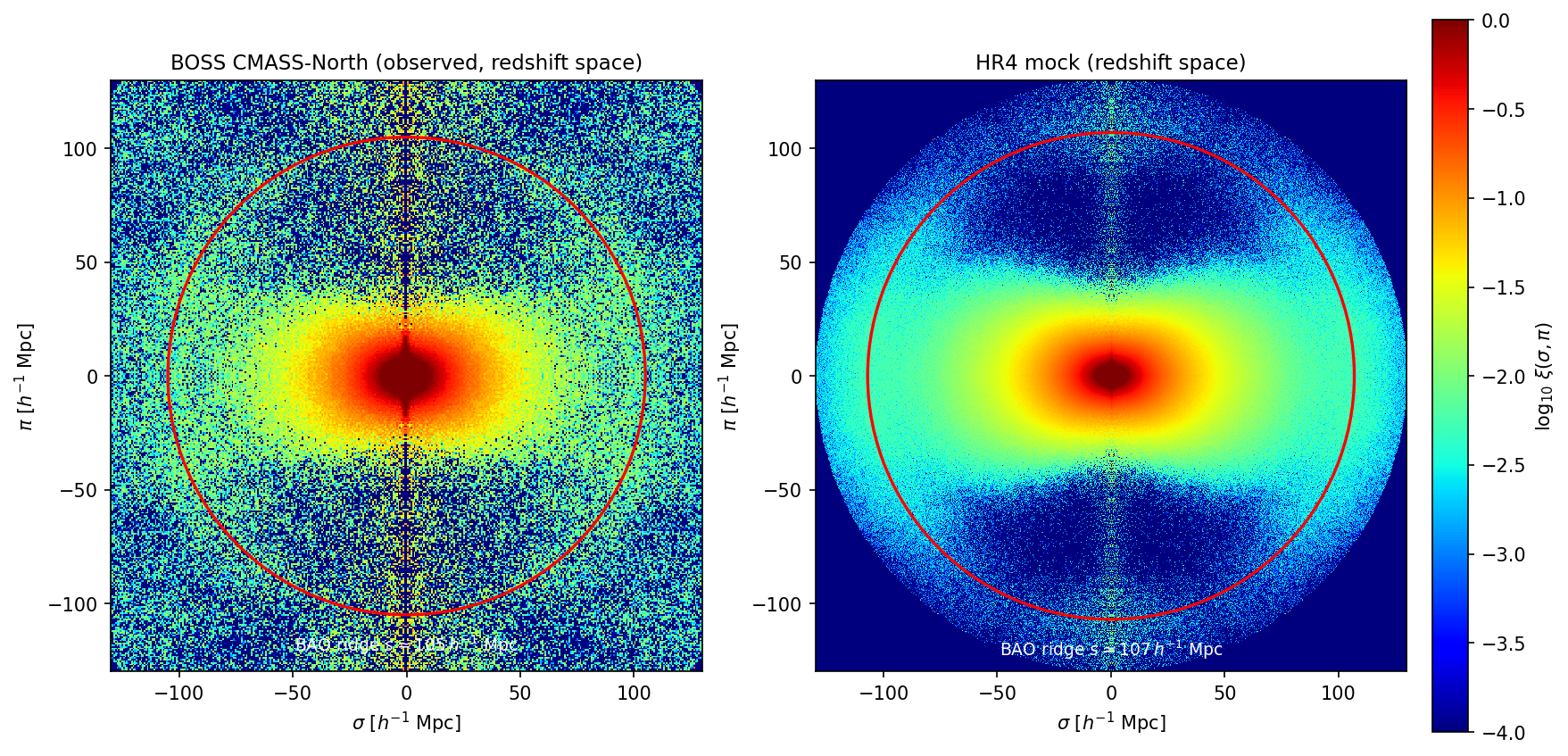}
\caption{Two-dimensional galaxy correlation function $\xi(\sigma,\pi)$ in redshift space. Both panels extend to a common $\pm130\,h^{-1}$\,Mpc radius and use the same $\log_{10}|\xi|$ colour scale. The left panel contains 568{,}776 galaxies from the observed BOSS CMASS-North catalogue and 5{,}000{,}000 random points. The Landy--Szalay estimator in Eq.~\eqref{eq:xiestimator} gives a fitted BAO ridge at $s\simeq105\,h^{-1}$\,Mpc. The right panel gives the author's pair-count measurement of the Horizon Run 4 mock catalogue \cite{hr4} and a fitted BAO ridge at $s\simeq107\,h^{-1}$\,Mpc. The mock reproduces the observed redshift-space anisotropy, including small-scale fingers of god along $\pi$ and large-scale Kaiser compression. The same clustering estimators define the ELG forecast.}
\label{fig:hr4xi}
\end{figure}

The clustering analysis fits $\xi(\sigma,\pi)$ in redshift slices using mock-derived covariance matrices and survey random catalogues. Free parameters include the dilation factors $(\alpha_\perp,\alpha_\parallel)$, broadband nuisance terms, galaxy bias, redshift smearing, and RSD parameters. The recovered $(D_M/r_d,Hr_d,\fsig)$ measurements constrain flat $\Lambda$CDM, $w_0w_a$ dark energy, and modified-growth models. The sequence connects the observed ELG redshift catalogue to $H_0$, the expansion history, and the growth factor.

The percentage precisions in Table~\ref{tab:baoforecast} are planning scalings, not the output of a full Fisher matrix. The threshold case is normalized to a conservative BAO/RSD performance target for a 100 deg$^2$, $N_{\rm ELG}\sim10^6$ high-purity redshift survey. The target is 1.5--2.5\% per broad redshift bin in the BAO dilation parameters and 4--6\% per bin in $\fsig$. The goal case is then scaled with effective survey volume,
\begin{equation}
  \sigma_{\rm stat,goal}\simeq
  \sigma_{\rm stat,thr}
  \left(\frac{V_{\rm eff,thr}}{V_{\rm eff,goal}}\right)^{1/2},
\end{equation}
where $V_{\rm eff}$ is the quantity defined above. If the usable area and secure-redshift count both increase from 100 deg$^2$ and $\sim10^6$ to 300 deg$^2$ and $\sim3\times10^6$, the ideal statistical improvement is close to $3^{1/2}$. Hence, a 1.5--2.5\% threshold BAO-bin precision becomes about 0.9--1.5\% and a 4--6\% threshold $\fsig$ precision becomes about 2.5--4\%. The combined-distance entries are slightly tighter because they represent the multi-bin distance-scale constraint after combining transverse and radial BAO information while retaining a floor for redshift failures, line confusion, and survey-window calibration. The quoted $H_0$ values are not direct late-time measurements of $H_0$ alone. They are obtained by adding the combined BAO distance precision and an assumed external sound-horizon calibration in quadrature,
\begin{equation}
  \left(\frac{\sigma_{H_0}}{H_0}\right)^2
  \simeq
  \left(\frac{\sigma_{\rm BAO}}{\alpha}\right)^2+
  \left(\frac{\sigma_{r_d}}{r_d}\right)^2 ,
\end{equation}
with $\sigma_{r_d}/r_d$ taken to be of order 1--2\% for the proposal-level estimate. The final row retains the same wide-area statistical BAO scale but lowers the systematic allowance after applying the medium and deep calibration tiers. The reduced systematic allowance improves the $H_0$ and growth floors without assuming a larger wide-survey volume.

\begin{table}[htbp]
\centering
\sffamily\small
\begin{tabularx}{\textwidth}{@{}p{3.0cm}p{2.2cm}p{2.4cm}p{2.8cm}Y@{}}
\toprule
\textbf{Survey configuration} & \textbf{Secure ELGs and required survey period} & \textbf{BAO distance precision} & \textbf{$H_0$ precision} & \textbf{Growth precision}\\
area in deg$^2$, redshift dimensionless & counts, calendar yr, exposure hr & fractional \% & fractional \% & fractional \%\\
\midrule
Threshold wide survey. 100 \degti\ at $1<z<3$ & $\sim10^6$. Calendar time of 1.5--2.5 yr with approximately 1200--1800 hr of science exposure including three orientations & $1.5$--$2.5\%$ per broad bin in $D_M/r_d$ and $H r_d$. $1.5$--$2.0\%$ combined distance scale & $2.5$--$3.5\%$ with an external $r_d$ prior. Otherwise $H_0r_d$ only & $\sigma(\fsig)=4$--$6\%$ per bin. $S_8^{\rm grow}$ to $\sim4\%$ after bias marginalization\\
Goal wide survey. 300 \degti\ at $1<z<3$ & $\sim3\times10^6$. Calendar time of 3--4 yr with approximately 2000--3000 hr of science exposure including three orientations & $0.9$--$1.5\%$ per broad bin. $0.9$--$1.2\%$ combined distance scale & $1.5$--$2.2\%$ with an external $r_d$ prior. $H_0r_d$ at approximately the same fractional precision & $\sigma(\fsig)=2.5$--$4\%$ per bin. $S_8^{\rm grow}$ to $\sim2.5$--$3.5\%$\\
Goal plus calibrated wedding-cake tiers & $\sim3\times10^6$ wide sources plus fainter calibration samples. Total ELG program of 4--5 yr with approximately 3000--4200 hr of science exposure & Same statistical BAO scale as the goal wide survey with reduced redshift-failure and purity bias & $1.3$--$2.0\%$ if selection-function systematics remain subdominant and $r_d$ is externally calibrated & $2$--$3\%$ growth-amplitude floor targeted by overlap calibration, mock recovery, and bias modelling\\
\bottomrule
\end{tabularx}
\caption{Planning-level cosmological precision and required survey period for the ELG BAO and RSD program. The values are proposal forecasts rather than final Fisher-matrix requirements. The forecasts assume high-purity redshifts, three-orientation overlap correction, realistic survey masks, and external calibration of $r_d$ when quoting absolute $H_0$.}
\label{tab:baoforecast}
\end{table}

The science requirement extends beyond the detection of many emission lines. Spectral extraction must recover the BAO scale and anisotropic RSD signal without imposing an orientation-dependent, morphology-dependent, or overlap-dependent selection function. The baseline cosmology sample requires redshift purity above $98\%$, a catastrophic line-confusion rate below $1\%$, and a redshift-success model whose spatial variation induces less than $0.3\%$ bias in the BAO distance scale. Blinded simulated detector images will test the requirements. Repeated fits using two orientations will reserve the third orientation as independent validation data.

\subsection{Parameter Estimation from Redshifts to Cosmological Constraints}

Four analysis stages convert the redshift catalogue into posterior constraints on $H_0r_d$, $\Omega_m$, $w_0$, $w_a$, and $S_8^{\rm grow}$. Template fitting and line classification first assign a redshift and quality flag to each deblended spectrum. Configuration-space and Fourier-space estimators then measure two-point clustering. Survey simulations determine the covariance of the clustering statistics. A cosmological likelihood finally compares the measured statistics with theoretical predictions. The detector-image validation in Section~\ref{sec:e2eval} applies the same sequence to simulated observations before analysis of flight data. The planning precision in Table~\ref{tab:baoforecast} rests on a Gaussian likelihood sampled with affine-invariant ensemble MCMC. CosmoMC and Cobaya extend the likelihood to joint multi-probe inference. Nested sampling is reserved for Bayesian evidence comparisons. A neural emulator evaluates the nonlinear matter power spectrum rapidly enough for the larger number of likelihood evaluations required by nested sampling. Simulation-based inference provides an independent test and does not set the planning precision.

\subsubsection{Redshift and Quality Classification}

Each deblended one-dimensional spectrum from the scene reconstruction in Section~\ref{sec:effres} is fitted against an ELG template library that spans the H$\alpha$/[O\,III]/[O\,II] flux-ratio sequence. The $\chi^2$ fit follows the template classification and redshift method used for BOSS and eBOSS spectra \cite{bolton2012}. A supervised neural classifier independently examines DESI spectra, SDSS spectra, and injected slitless observations from Section~\ref{sec:e2eval}. QuasarNET provides the precedent for locating emission lines with a convolutional network \cite{quasarnet2018}. The template fit returns the redshift and formal uncertainty. The classifier identifies catastrophic single-line degeneracies that remain unresolved by the $\chi^2$ grid. The classification uses the multi-line pattern, the morphology prior, and the redundancy among the three observed orientations represented by Eq.~\eqref{eq:scene}. The template and classifier results jointly determine the redshift-quality flag defined in Section~\ref{sec:effres} and select the cosmology sample with purity above $98\%$.

\subsubsection{Two-Point Clustering Estimators in Configuration and Fourier Space}

The correlation function of Eq.~\eqref{eq:xiestimator} is measured with random catalogs that encode the survey mask, the overlap losses between the three position angles, and the redshift-dependent selection function. The same information is measured independently in Fourier space. The FKP-weighted density field
\begin{equation}
  F(\mathbf{r}) = w(\mathbf{r})\big[n_g(\mathbf{r})-\alpha_{\rm rand}\,n_s(\mathbf{r})\big],
  \qquad
  w(\mathbf{r}) = \frac{1}{1+n(\mathbf{r})P_0},
  \label{eq:fkp}
\end{equation}
where $n_g$ and $n_s$ denote the galaxy and synthetic-random densities and $\alpha_{\rm rand}$ gives their relative normalization \cite{fkp1994}. The Yamamoto line-of-sight estimator measures the minimum-variance power-spectrum multipoles $P_0(k)$, $P_2(k)$, and $P_4(k)$ with fast Fourier transforms \cite{yamamoto2006,hand2017nbodykit}. The power-spectrum multipoles and correlation-function wedges constrain the same $(\alpha_\perp,\alpha_\parallel,\fsig)$ parameters. Agreement between configuration-space and Fourier-space results tests residual window-function errors, redshift-failure gradients, and orientation-dependent selection effects.

\subsubsection{Covariance Matrices}\label{sec:covariance}

The likelihood requires a covariance matrix for the compressed BAO and RSD parameters in each redshift bin. The baseline covariance comes from mock galaxy catalogues constructed from N-body simulations with the resolution and volume of Horizon Run 4 \cite{hr4}. An ELG halo-occupation model matched to the luminosity functions in Section~\ref{sec:elgdepth} populates the simulated haloes. The same selection function, masks, and redshift-quality cuts are applied to simulated and observed catalogues. Approximate mock catalogues calibrated against the full N-body realizations provide enough independent realizations for stable covariance inversion. Jackknife resampling of the survey footprint supplies a data-derived covariance estimate for comparison before cosmological inference.

\subsubsection{Likelihood, MCMC, and Simulation-Based Inference}

The baseline likelihood is Gaussian in the compressed parameters,
\begin{equation}
  -2\ln\mathcal{L} =
  \big[\mathbf{p}_{\rm obs}-\mathbf{p}_{\rm model}(\theta)\big]^{T}
  \mathbf{C}^{-1}
  \big[\mathbf{p}_{\rm obs}-\mathbf{p}_{\rm model}(\theta)\big],
  \label{eq:gausslike}
\end{equation}
with $\mathbf{p}=(\alpha_\perp,\alpha_\parallel,\fsig)$ per redshift bin, $\mathbf{C}$ the covariance of Section~\ref{sec:covariance}, and $\theta$ the cosmological and nuisance parameters. For the ELG-only fit the posterior is sampled with an affine-invariant ensemble MCMC sampler \cite{emcee2013}. For the joint fit that adds Planck/ACT CMB priors and the internal SN\,Ia and DSPL strong-lensing likelihoods of the companion programs above, the same posterior is sampled with the CosmoMC and Cobaya cosmological-parameter-estimation codes \cite{lewisbridle2002,torradolewis2021},
\begin{equation}
  \mathcal{P}(\theta\mid D) \;\propto\;
  \mathcal{L}_{\rm ELG}(D_{\rm ELG}\mid\theta)\,
  \mathcal{L}_{\rm SN}(D_{\rm SN}\mid\theta)\,
  \mathcal{L}_{\rm lens}(D_{\rm lens}\mid\theta)\,
  \mathcal{L}_{\rm CMB}(D_{\rm CMB}\mid\theta)\,
  \Pi(\theta),
  \label{eq:jointpost}
\end{equation}
with $\Pi(\theta)$ the prior on $\theta$. The joint posterior constrains $H_0$, $\Omega_m$, and $(w_0,w_a)$ together with the supernova program. Bayesian comparison of flat $\Lambda$CDM with dynamic dark energy uses nested sampling to determine the evidence \cite{polychord2015}. A joint multi-probe calculation requires approximately $10^6$--$10^7$ likelihood evaluations. A neural emulator trained on Boltzmann-code and perturbation-theory calculations supplies the nonlinear matter power spectrum in Eq.~\eqref{eq:gausslike} \cite{cosmopower2022,euclidemulator2021}. The emulator reduces each theoretical power-spectrum evaluation to approximately a millisecond.

Simulation-based inference provides a second test of the Gaussian result. A neural density estimator learns the conditional distribution of the full mock $P(k,\mu)$ or $\xi(\sigma,\pi)$ measurements given cosmological parameters \cite{cranmer2020}. The estimator does not assume a Gaussian likelihood or an analytic covariance. The planning precision in Table~\ref{tab:baoforecast} does not depend on this calculation. The simulation-based analysis instead tests whether non-Gaussian small-scale information, including mildly nonlinear fingers of god, gives a consistent growth rate.

Before unblinding, a fixed undisclosed offset is applied to $(\alpha_\perp,\alpha_\parallel,\fsig)$ in the catalogue following the blind-analysis practice of BOSS, eBOSS, and DESI. The offset is removed only after the spectral selection, covariance prescription, and likelihood choices are fixed. The procedure extends the held-out-orientation test into the cosmological analysis.

\subsection{Validation with Simulated Detector Images}\label{sec:e2eval}

The present detector-scene validation performs a blind compact-line search from direct-image positions. The search examines the full first-order trace at all three orientations without using the true wavelength or line flux. The search records H$\alpha$, [O\,III], and [O\,II] single-line hypotheses. The candidate list is fixed before comparison with the simulated truth catalogue and gives a provisional selection table in redshift bins. The compact-line search is a diagnostic. The search neither fits neighbouring spectra jointly nor resolves single-line degeneracies and therefore does not enter the science forecast.

The validation renderer obtains the direct-image count rate, slitless continuum, line-electron yield, zodiacal background, dark current, and read noise from the tracked ETC. The current calculation uses the $1.0$--$1.75\,\mu$m setting and the same $900$ s exposure for every orientation and dither. The compact-line diagnostic does not replace the joint scene extraction required for an acceptance result.

The redshift-binned table provides the interface between image-level recovery and the yield integral of Section~\ref{sec:elgdepth}. A Phase A acceptance realization must use a calibrated deep-field catalogue, the calibrated ETC package, and the flight sparse multi-roll forward-model extractor. The resulting $P_{\rm secure}(F,z,\boldsymbol{\theta})$ replaces the planning selection factor in each tier yield and defines the same selection in the clustering random catalogue.

Phase A must demonstrate survey performance with simulated detector images rather than analytic exposure-time arguments alone. Survey-scale tests have not yet established the required purity, completeness, or flux-bias limits. Figure~\ref{fig:crowded} illustrates the deblending geometry on one tile and Figure~\ref{fig:slicer} illustrates the adopted image-slicer mapping. Neither figure supplies recovery statistics. The computational cost of the joint sparse inversion also remains unmeasured for approximately $10^6$--$3\times10^6$ sources across $10^4$ orientation visits. The Phase A suite must inject realistic galaxies into detector scenes and recover redshifts with the flight spectral-extraction method. The following quantities are verification requirements rather than demonstrated performance.

The first validation realization uses \texttt{detector\_scene\_validation.py}. The input catalogue contains position, redshift, continuum magnitude, size, axis ratio, line equivalent width, and line flux. Independent noisy direct images and dispersed exposures at $0^\circ$, $45^\circ$, and $90^\circ$ include an imperfect source catalogue with missed objects, centroid errors, and spurious detections. The present forced-line measurement tests detector geometry and the association between input and recovered sources. The calculation is not a redshift measurement and does not enter a science forecast. The next stage applies the calibrated sparse multi-orientation forward model in Eq.~\ref{eq:scene} without true wavelengths or line fluxes as priors.

A verification realization uses a $1536\times1536$ pixel tile, $1091$ pixel traces, three orientations, and two dithers per orientation. The realization produces truth and direct-image catalogues, detector images, and recovery tables binned by magnitude and nearest-neighbour separation. The outputs test source association across the complete observing pattern. The analytic source population and forced-line measurement do not support a survey-performance claim. The Phase A acceptance realization must use a calibrated deep-field catalogue, measured detector calibration files, and a blind redshift search with the flight forward model.

A second verification realization projects a 3D-HST COSMOS catalogue into the same detector geometry \cite{skelton2014,momcheva2016}. The input contains 907 sources to $F160W_{\rm AB}=26$ with measured positions, magnitudes, half-light radii, axis ratios, position angles, reference redshifts, and fitted line properties. Use of reference redshifts and an ETC conversion from line flux to detector electrons restricts the realization to tests of source density and data products. The realization does not test blind redshift recovery or survey sensitivity. Phase A must replace the simplified inputs with a calibrated detector model and blind flight spectral extraction.

\begin{itemize}
  \item redshift completeness, purity, and catastrophic-failure rates.
  \item single-line misidentification rates for H$\alpha$, [O\,III], and [O\,II].
  \item morphology-induced wavelength shifts and line broadening.
  \item overlap statistics for \fovsmall\ and \fovlarge\ fields.
  \item sensitivity to orientation, dithering, and direct-image priors.
  \item random catalogues, masks, and survey-window functions for clustering.
  \item blinded recovery of the BAO scale and $\fsig(z)$ from simulated detector images.
  \item processor time and memory required for the joint scene inversion in Eq.~\eqref{eq:scene} per tile and for the full survey.
\end{itemize}

\subsection{Synergy with Legacy Deep Fields}\label{sec:deepfields}

\begin{figure}[htbp]
\centering
\includegraphics[width=0.92\textwidth]{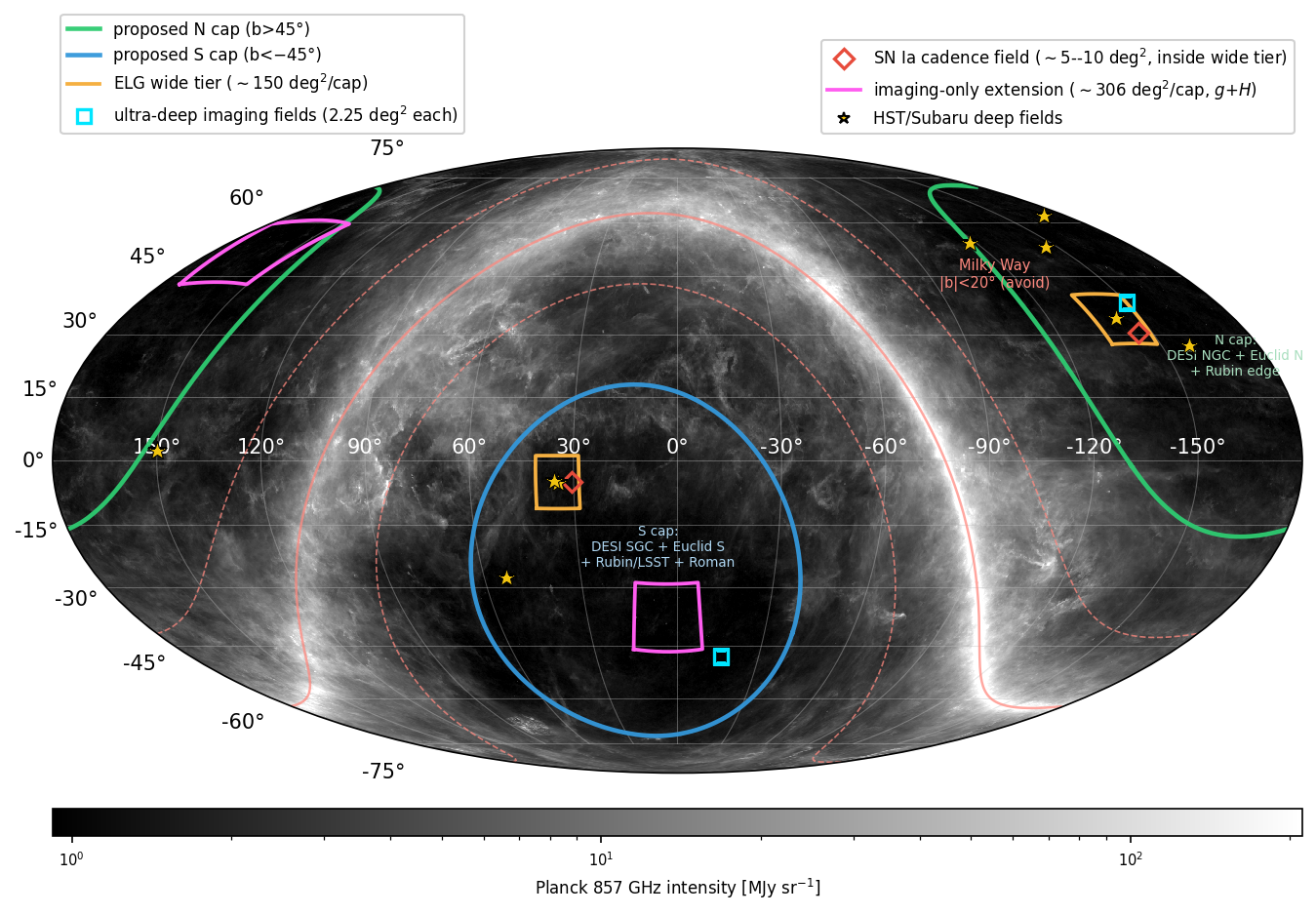}
\caption{Proposed ELG survey fields on an equatorial Mollweide projection over the Planck 857\,GHz thermal-dust map \cite{planck2018hfi}. The map has a native resolution of approximately $5'$ and was obtained through CDS \textit{hips2fits}. Green and blue outlines show the northern and southern Galactic caps at $|b|\gtrsim45^\circ$. Red curves delimit the Milky Way zone at $|b|<20^\circ$. The caps overlap the DESI, Euclid, Rubin/LSST, and Roman footprints. Orange open boxes show the two ELG wide-tier footprints at their true area of approximately 150\,deg$^2$ each. Cyan squares show the two 2.25\,deg$^2$ ultra-deep imaging fields near the Planck minima in Figure~\ref{fig:imgfields}. Red diamonds show the 5--10\,deg$^2$ SN Ia cadence fields nested within the wide tiers. Magenta outlines show the approximately 306\,deg$^2$ imaging extensions in each cap. Gold stars identify legacy HST and Subaru deep fields. Figure~\ref{fig:deepzoom} enlarges the nested survey fields.}
\label{fig:skymap}
\end{figure}

\begin{figure}[htbp]
\centering
\includegraphics[width=0.98\textwidth]{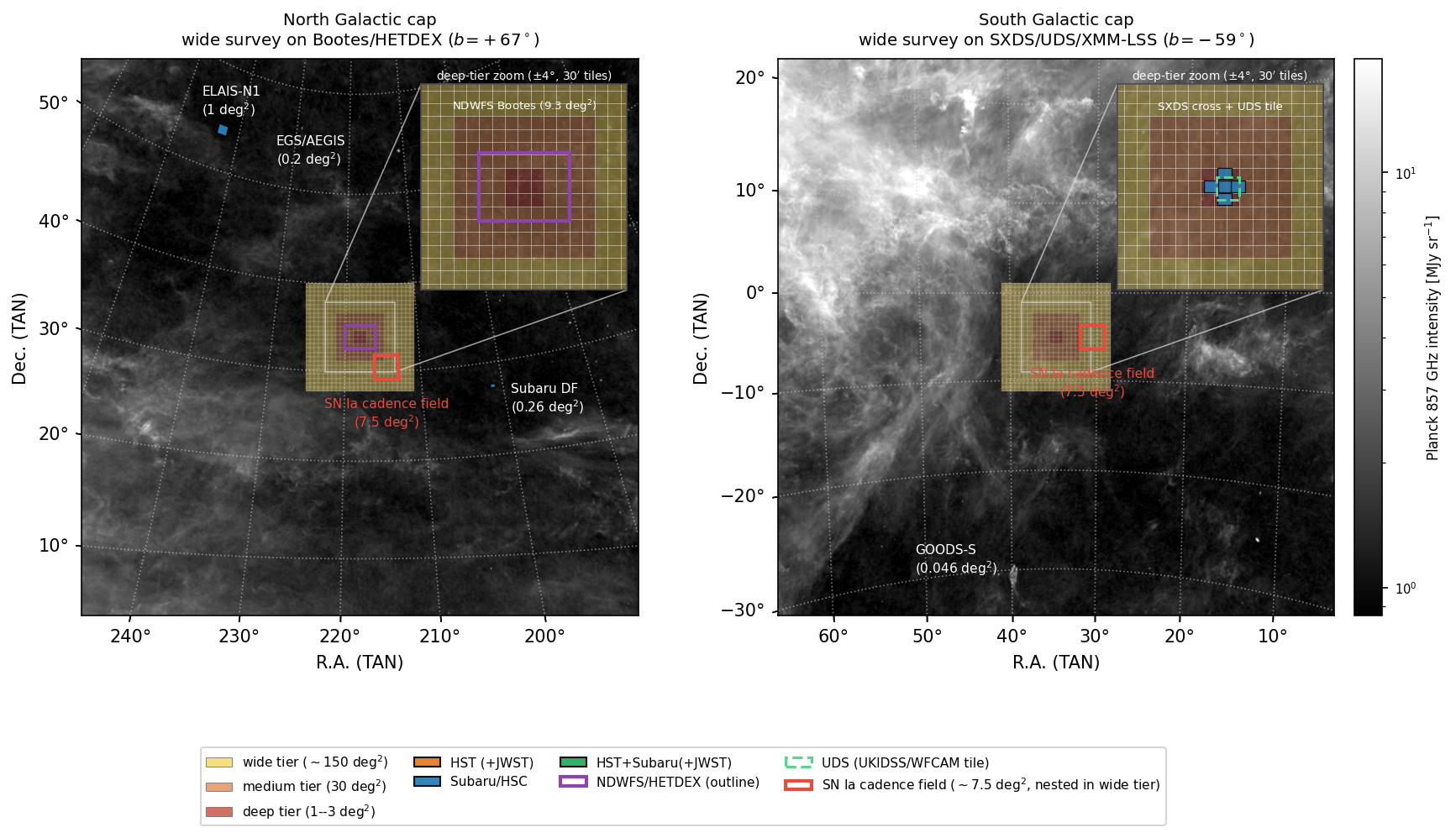}
\caption{Nested wedding-cake tiers of the proposed ELG survey, one Galactic cap per panel, in gnomonic (TAN) projection on the Planck 857\,GHz thermal-dust map \cite{planck2018hfi} (native $\sim$5$'$ resolution, CDS \textit{hips2fits}). Both reference fields, the NDWFS/HETDEX Bo\"otes field ($b\simeq+67^\circ$) and the SXDS/UDS/XMM-LSS field ($b\simeq-59^\circ$), sit in dark low-cirrus windows. The bright complex at the upper left of the southern panel, toward Taurus, illustrates the foreground the field selection avoids. Neighbouring legacy fields (ELAIS-N1, the Extended Groth Strip, the Subaru Deep Field, GOODS-S) are shown for context. Each $\pm4^\circ$ inset zooms on the reference and shows the true footprint shape of the legacy fields together with the wedding-cake nesting. The wide tier ($\sim$150 deg$^2$, gold) encloses the medium tier (30 deg$^2$, orange), which encloses the deep tier (1--3 deg$^2$, red). Each tier is a square paving of $30'\times30'$ tiles rather than a circle. The red outline inside each wide tier is the SN Ia time-domain cadence field ($\sim$7.5\,deg$^2$), nested inside rather than external to the wide tier and sited away from the visible cirrus structure. Its hosts therefore get real slitless redshifts and environment from the flagship survey. The remaining legacy fields of the caps are marked on the all-sky map of Figure~\ref{fig:skymap}.}
\label{fig:deepzoom}
\end{figure}

The two high-Galactic-latitude caps avoid the Milky Way and include major legacy deep fields (Figure~\ref{fig:skymap}). The northern cap contains COSMOS, GOODS-N, the Extended Groth Strip, and the Subaru Deep Field. The southern cap contains GOODS-S, the Hubble Ultra Deep Field, UDS/SXDS, and XMM-LSS. Decades of HST and Subaru imaging and extensive ground-based and space-based spectroscopy cover the fields. JWST adds CEERS in the Extended Groth Strip, JADES in GOODS-S, and COSMOS-Web. The wide survey therefore gains deep ancillary data within its footprint. The $100$--$300$ deg$^2$ wide tier and the nested medium and deep tiers centre on the NDWFS/HETDEX Bo\"otes field at $b\simeq+67^\circ$ in the north and the SXDS/UDS/XMM-LSS field at $b\simeq-59^\circ$ in the south (Figure~\ref{fig:deepzoom}). High absolute Galactic latitude limits extinction and cirrus. HETDEX provides the Bo\"otes field with a blind emission-line comparison sample while other legacy fields in both caps provide additional deep and calibration pointings.

The overlap primarily calibrates the emission-line selection function. Deep multi-band photometry and spectroscopic redshift catalogues from zCOSMOS, VVDS, VUDS, DEEP2, MOSDEF, and JWST NIRSpec reach well below the ELG continuum limit. The data provide reference redshifts, line fluxes, and line ratios. Matching slitless spectra to the reference samples directly measures redshift success, interloper fractions, line misidentification, and the effective selection function.

Deep HST and JWST imaging also supplies the morphological priors required for crowded-field deblending. Band-limited slitless self-broadening depends on galaxy size, S\'ersic index, and orientation. High-resolution measurements of those quantities permit object-specific calibration of the blending factor $C_{\rm blend}$ in Eq.~\eqref{eq:snline}. COSMOS provides two square degrees of well-characterized sky for testing joint scene reconstruction on a statistically useful galaxy population.

Deep photometry extends below the ELG detection threshold and measures the faint-end completeness that the wide survey lacks. The resulting constraint improves the count model in Section~\ref{sec:elgdepth}. JWST and MOSDEF spectra calibrate [O\,\textsc{iii}]/H$\beta$, [O\,\textsc{ii}]/H$\alpha$, and the relation between star-formation rate and line luminosity. The calibrated relations convert observed line strengths into physical quantities.

The deep and ultra-deep tiers occupy fields with existing multi-wavelength coverage. The same caps overlap the Euclid and Roman deep tiers, Rubin/LSST deep-drilling fields, and the HSC-SSP UltraDeep footprint in COSMOS and SXDS. Slitless redshifts and external imaging and grism catalogues on the same sky permit direct cross-calibration. The field placement integrates the wide survey into a coordinated multi-facility program.

\subsection{Ancillary Data and Synergy in the Two Reference Fields}\label{sec:anchorsynergy}

High Galactic latitude is not the only criterion for the two reference fields. Each field contains extensive panchromatic imaging and spectroscopy in its hemisphere. The northern emission-line-selected data and southern continuum-selected data provide complementary tests of slitless redshift selection. Table~\ref{tab:anchorancillary} lists the principal data sets and their calibration roles.

\begin{table}[htbp]
\centering
\sffamily\footnotesize
\setlength{\tabcolsep}{4pt}
\renewcommand{\arraystretch}{0.92}
\begin{tabularx}{\textwidth}{@{}p{2.1cm}p{2.8cm}p{3.2cm}X@{}}
\toprule
\textbf{Field} & \textbf{Survey}\newline\textbf{(facility)} & \textbf{Data} & \textbf{Role for the ELG survey}\\
\midrule
\multicolumn{4}{@{}l}{\textit{North reference field, NDWFS/HETDEX Bo\"otes at $b\simeq+67^\circ$ over $\sim$9.3 deg$^2$}}\\
 & NDWFS (Mayall 4\,m) \cite{jannuzi1999} & Deep $B_wRI$ optical $+$ $K$ & Source detection, morphology, photometric baseline\\
 & SDWFS (Spitzer) \cite{ashby2009} & IRAC 3.6--8\,$\mu$m & Stellar masses, photo-$z$ priors, mid-IR AGN flag\\
 & XBo\"otes (Chandra) \cite{murray2005} & 5\,ks X-ray mosaic & AGN identification and interloper flagging\\
 & AGES (Hectospec) \cite{kochanek2012} & $\sim$23{,}000 redshifts & Bright-end spectroscopic reference redshifts and selection function\\
 & HETDEX (HET/VIRUS) \cite{hetdex2021} & Blind emission-line IFU & External emission-line-selected reference sample\\
 & LoTSS Deep (LOFAR) \cite{tasse2021} & 150\,MHz radio & Radio-AGN and obscured star-formation cross-ID\\
\midrule
\multicolumn{4}{@{}l}{\textit{South reference field, SXDS/UDS/XMM-LSS at $b\simeq-59^\circ$}}\\
 & SXDS (Subaru) \cite{furusawa2008} & Deep $BVRi'z'$, 1.22 deg$^2$ & Deep optical detection and colors\\
 & UKIDSS UDS (UKIRT) \cite{lawrence2007} & Deepest wide $JHK$ & Stellar continuum, mass, high-$z$ photo-$z$\\
 & VIDEO (VISTA) \cite{jarvis2013} & $ZYJHK_s$ over $\sim$12 deg$^2$ & Near-IR over the medium-tier area\\
 & XMM-SERVS (XMM) \cite{chen2018} & X-ray over $\sim$5.3 deg$^2$ & AGN identification over the wide tier\\
 & HSC-SSP UltraDeep \cite{aihara2018} & Deepest HSC optical $+$ shapes & Weak-lensing shapes, deep photometry\\
 & VANDELS (VLT) \cite{mclure2018} & Deep VIMOS spectroscopy & Faint spectroscopic redshifts below the ELG limit\\
\bottomrule
\end{tabularx}
\caption{Principal ancillary data in the two survey reference fields and the calibration role each plays for the slitless ELG program. The northern field adds a blind emission-line survey (HETDEX) which no other deep field offers. The southern field adds the deepest near-infrared imaging (UDS) and weak-lensing-quality optical (HSC UltraDeep). The forthcoming Subaru PFS \cite{takada2014} and MOONS deep pointings target the southern field. Massively multiplexed continuum spectroscopy will arrive there during the mission.}
\label{tab:anchorancillary}
\end{table}

HETDEX has conducted a blind emission-line survey over the northern reference field without imaging preselection \cite{hetdex2021}. HETDEX detects Ly$\alpha$ emitters at $1.9\lesssim z\lesssim3.5$ and [O\,\textsc{ii}] emitters at $z\lesssim0.5$. The common emission-line selection directly tests catalogue completeness, purity, and [O\,\textsc{ii}]--Ly$\alpha$ confusion. VIRUS covers only $3500$--$5500$\,\AA, so the two surveys measure largely disjoint lines. For example, a HETDEX Ly$\alpha$ emitter at $z\simeq2.3$ has [O\,\textsc{iii}] at $1.65\,\mu$m and H$\alpha$ at $2.17\,\mu$m within the proposed near-infrared settings. Rest-frame optical lines from the proposed survey resolve the dominant HETDEX Ly$\alpha$ and [O\,\textsc{ii}] ambiguity while HETDEX tests blue-end single-line redshifts. Cross-correlation of the two tracers over $2\lesssim z\lesssim3.5$ suppresses cosmic variance in the multi-tracer limit \cite{mcdonaldseljak2009} and reveals selection errors that are absent from either autocorrelation alone. Combining HETDEX Ly$\alpha$ fluxes with Balmer-line fluxes gives object-specific Ly$\alpha$ escape fractions across Bo\"otes. NDWFS \cite{jannuzi1999}, SDWFS \cite{ashby2009}, and approximately 23{,}000 AGES redshifts \cite{kochanek2012} provide photometric-redshift and stellar-mass priors. XBo\"otes \cite{murray2005} and the LOFAR deep map \cite{tasse2021} identify active nuclei and obscured star formation that could contaminate line-flux statistics.

The southern reference field supplies deep near-infrared photometry and accurate galaxy shapes. UKIDSS UDS constrains stellar continua and photometric redshifts below the wide-tier line-flux limit \cite{lawrence2007}. VIDEO extends near-infrared coverage across the medium tier \cite{jarvis2013}. XMM-SERVS identifies active nuclei before their line emission biases the luminosity functions \cite{chen2018}. HSC-SSP UltraDeep provides weak-lensing-quality shapes and deep optical photometry for the diffuse-light and cluster-lensing measurements in Section~\ref{sec:lsb} \cite{aihara2018}. Slitless emission-line redshifts in turn calibrate photometric redshifts and weak-lensing source selection. VANDELS spectroscopy reaches below the ELG threshold and tests slitless redshifts with a continuum-selected sample \cite{mclure2018}. Planned Subaru Prime Focus Spectrograph \cite{takada2014} and MOONS observations will extend the spectroscopic reference sample during the mission.

Continuum selection in the southern field complements slitless line selection. Deep $JHK$ photometry detects massive, quiescent, and dusty galaxies with weak emission lines and measures population-dependent incompleteness in the emission-line census. Slitless spectra supply redshifts and line fluxes for continuum-faint dwarfs with high equivalent width. Combining a survey line flux with a UDS or VIDEO stellar mass places each common galaxy on the star-formation-rate versus stellar-mass plane. The joint property distribution $n(\boldsymbol{\theta}|z)$ and line completeness $C_i$ in Section~\ref{sec:elgdepth} follow from that comparison. Equivalent-width measurements similarly require the line flux from slitless spectroscopy and the continuum level from UDS or VIDEO.

Together, the reference fields provide three independent advantages. First, an emission-line-selected sample in the north and deep continuum redshifts in the south calibrate missed lines and misidentified interlopers against independent external data. Second, panchromatic imaging supplies the morphology and size priors that determine $C_{\rm blend}$ in Eq.~\eqref{eq:snline}. The same imaging constrains the line-ratio and continuum-magnitude conversions used by Tables~\ref{tab:desiFOVsum} and~\ref{tab:desiMcounts}. Measured inputs therefore determine the count model in Section~\ref{sec:elgdepth}. Third, the northern field near right ascension $14^{\rm h}30^{\rm m}$ and southern field near $2^{\rm h}20^{\rm m}$ support year-round scheduling, two statistically independent volumes, and a cross-hemisphere test of instrumental and selection systematics. Rubin, Euclid, and Roman also target both regions for deep observations, which places the survey within a coordinated multi-facility footprint from the first pointing.

\clearpage
\section{Companion Science Programs}

\subsection{Type Ia Supernovae}\label{sec:snia}

Both resolving-power candidates use the same image slicer. The broad supernova features require no additional information from $R=5000$. An $R=5000$ spectrum may be binned to the $R\simeq300$--500 analysis scale with the same photon-limited S/N as an $R=1000$ spectrum because the slicer accepts the same source area.

Type Ia supernovae (SNe Ia) provide a mature cosmological distance indicator and directly complement the dark-energy measurements of the flagship ELG survey. Spectra binned to $R\simeq300$--500 retain sufficient signal for classification, phase determination, dust and colour estimation, and population diagnostics. A dedicated cadence field is required because sparse wide-tier revisits do not sample maximum light reliably. One cadence field lies within each wide-tier footprint. The common footprint gives every SN Ia host a slitless redshift, local environment, large-scale density, and peculiar-velocity estimate from the same BAO and RSD volume. Short frequent visits follow the Rubin/LSST deep-drilling strategy within the wider main survey \cite{ivezic2019lsst}. A shallow $3\times900$\,s discovery visit detects SNe Ia photometrically to $z\simeq3$ and maximizes the monitored area for a fixed cadence budget. Deeper $2$--$30$\,hr spectra are reserved for selected candidates. The \fovsmall\ field satisfies the cadence requirement while the \fovlarge\ field contains more active supernovae and more efficient host-galaxy templates. At the relevant redshifts, the $0.36$--$3.0\,\mu$m band also records the rest-frame near-ultraviolet where iron-group line blanketing traces intrinsic luminosity diversity.

\subsubsection{Standardizable Candles and the Systematic Floor}

SNe Ia arise from thermonuclear disruption of carbon--oxygen white dwarfs in binary systems. Their cosmological utility rests on the Phillips relation between peak luminosity and the post-maximum decline rate $\Delta m_{15}(B)$ (Figure~\ref{fig:phillips}) \cite{phillips1993}. The light-curve-shape correction standardizes the luminosity. Application to high-redshift events revealed cosmic acceleration \cite{riess1998,perlmutter1999} and established dark energy as a central problem in physics. Current standardized distances reach approximately $5$--$7\%$ statistical precision per object.

Systematic uncertainty now limits the distance scale. After Phillips and colour corrections, the residual Hubble scatter of $\sigma\simeq0.10$--$0.15$\,mag exceeds the photometric errors and reveals intrinsic luminosity diversity outside the standard parameters. Evolution in progenitor metallicity, age, or explosion-channel mixture would convert that diversity into redshift-dependent biases in $H_0$ and $w(z)$. Spectroscopic diagnostics of each explosion's physical state therefore target a dominant uncertainty in the SN Ia distance ladder.

\subsubsection{Physical Subclasses beyond the Phillips Relation}

Several observations show that the Phillips relation is not a single physical sequence.
\begin{itemize}
\item \textbf{Progenitor-channel diversity.} Single-degenerate accretion and double-degenerate merger channels predict different nucleosynthetic yields and spectra even at fixed light-curve shape.
\item \textbf{Residual Hubble scatter.} The $\sigma\simeq0.10$--$0.15$\,mag scatter that survives standardization exceeds photometric errors, requiring an intrinsic component.
\item \textbf{Spectroscopic diversity.} The Si\,\textsc{ii}\,$\lambda6355$ velocity splits ``normal'' SNe Ia into Normal and High-Velocity groups \cite{wang2009}, and high-velocity events have since been linked to more massive hosts and distinct Hubble residuals.
\item \textbf{Host-mass step.} Standardized luminosities correlate with host stellar mass and star-formation rate, an indicator of metallicity and age that already forces an empirical ``mass-step'' correction in modern analyses.
\end{itemize}
The empirical correlations do not identify the underlying explosion physics. Ultraviolet spectra directly probe the iron-group composition and radial distribution established by the explosion.

\subsubsection{Optical Twins and Near-Ultraviolet Divergence}

\begin{figure}[htbp]
\centering
\includegraphics[width=0.62\textwidth]{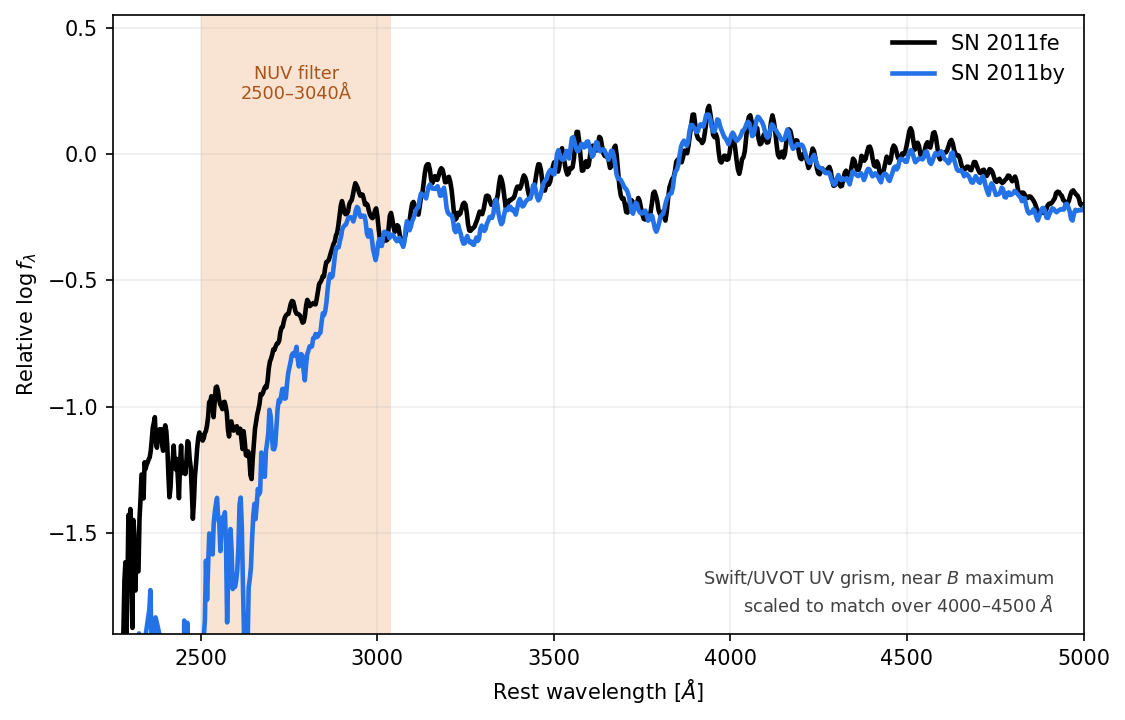}
\caption{Observed rest-frame UV--optical spectra of the optical twins SN\,2011fe (black) and SN\,2011by (blue), flux-calibrated Swift/UVOT UV-grism observations within half a day of $B$ maximum, from the Open Supernova Catalog archive \cite{guillochon2017}, deredshifted and scaled to match over $4000$--$4500$\,\AA. The shaded band marks the proposed rest-frame NUV filter ($2500$--$3040$\,\AA\ around the $2770$\,\AA\ diagnostic wavelength of Figure~\ref{fig:snuvlimit}). The optical region is nearly identical while the flux diverges strongly across that band, the near-ultraviolet metallicity signature of Foley \& Kirshner \cite{foleykirshner2013} predicted by Lentz et~al.\ \cite{lentz2000}. The optical twin condition isolates an intrinsic-luminosity difference visible \emph{only} in the rest-frame NUV \cite{graham2015}.}
\label{fig:uvtwins}
\end{figure}

The cleanest demonstration comes from ``optical twins'', which are SN Ia pairs with nearly identical optical spectra ($3500$--$7500$\,\AA), $\Delta m_{15}$, and colors that nonetheless differ measurably in peak luminosity. The canonical case is SN\,2011fe versus SN\,2011by (Figure~\ref{fig:uvtwins}). The two have near-identical optical spectra, yet a $\sim0.3$\,mag peak offset and a pronounced flux divergence below $\sim3500$\,\AA. Foley \& Kirshner attributed the difference to progenitor metallicity inferred from the UV \cite{foleykirshner2013}, while Graham et~al.\ found near-identical nebular spectra, pointing to the radial iron-group distribution and metallicity rather than total $^{56}$Ni mass alone \cite{graham2015}.

The rest-frame NUV at $\lambda\lesssim3500$\,\AA\ contains a dense forest of Fe\,\textsc{ii}, Fe\,\textsc{iii}, Co\,\textsc{ii}, and Ni\,\textsc{ii} lines. The resulting blanketing depends on three physical quantities. The total $^{56}$Ni mass sets the peak luminosity through $^{56}$Ni\,$\to^{56}$Co\,$\to^{56}$Fe. The radial distribution of iron-group elements sets the ejecta opacity and emergent UV flux. Progenitor metallicity sets the initial abundance of stable $^{54}$Fe and $^{58}$Ni, which suppress UV flux without adding to the radioactive energy budget \cite{lentz2000}. Two SNe Ia may therefore share an optical spectrum dominated by Si, S, and Ca in the outer ejecta while differing in NUV flux set by inner iron-group material. Swift/UVOT photometry separates normal SNe Ia into NUV-red and NUV-blue groups with an offset of approximately $0.4$\,mag in NUV$-$optical colour. The relative abundance of the groups evolves with redshift \cite{milne2015}. An evolving subgroup mixture can bias $H_0$ and mimic or conceal evolution in the dark-energy equation of state.

\subsubsection{Rest-Frame NUV Access with a 0.36--3.0\,$\mu$m Slitless Survey}

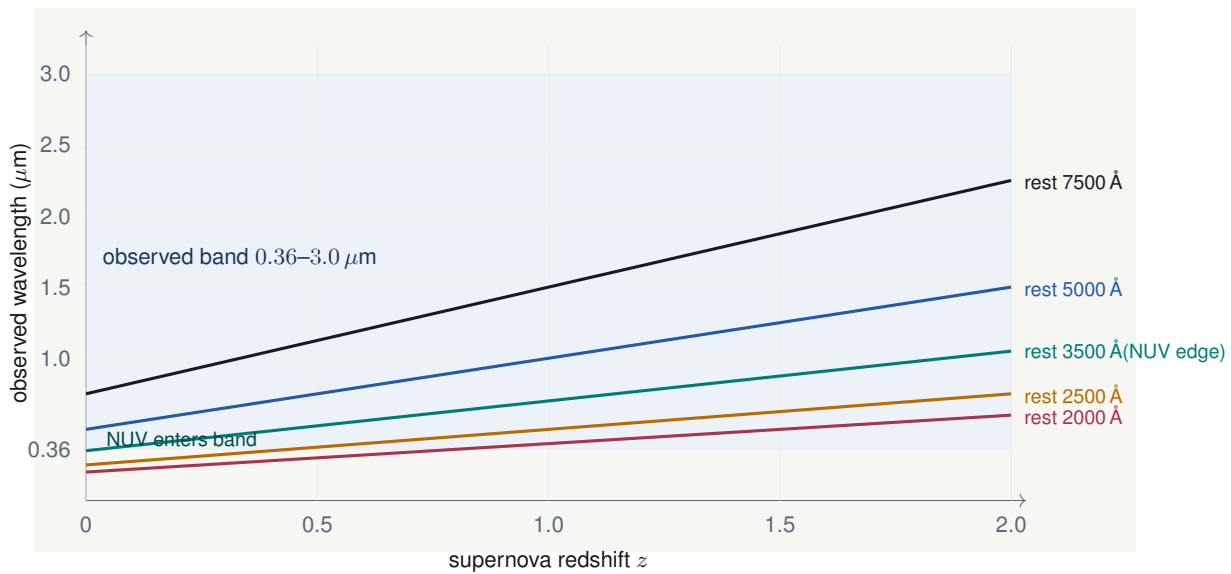
\begin{figure}[htbp]
\centering
\resizebox{0.96\textwidth}{!}{%
\begin{tikzpicture}[font=\sffamily\footnotesize]
  \def\w{12.6}\def\h{6.2}
  \def\zmax{2.0}\def\lmax{3.2}
  \fill[paper] (-0.7,-0.7) rectangle (\w+1.7,\h+0.5);
  \fill[blue!8] (0,{0.36/\lmax*\h}) rectangle (\w,{3.0/\lmax*\h});
  \node[anchor=west,color=blue!60!black] at (0.1,{1.7/\lmax*\h}) {observed band $0.36$--$3.0\,\mu$m};
  \draw[->,color=muted] (0,0) -- (\w+0.2,0);
  \draw[->,color=muted] (0,0) -- (0,\h+0.2);
  \foreach \z in {0,0.5,1.0,1.5,2.0} {
    \pgfmathsetmacro\xx{\z/\zmax*\w}
    \draw[line!55] (\xx,0) -- (\xx,\h);
    \node[anchor=north,color=muted] at (\xx,-0.08) {\z};}
  \foreach \l/\lab in {0.36/0.36,1.0/1.0,1.5/1.5,2.0/2.0,2.5/2.5,3.0/3.0} {
    \pgfmathsetmacro\yy{\l/\lmax*\h}
    \draw[line!45] (0,\yy) -- (\w,\yy);
    \node[anchor=east,color=muted] at (-0.08,\yy) {\lab};}
  \foreach \lr/\col/\name/\ly in {0.2/rose/{rest 2000\,\AA}/0.0, 0.25/gold/{rest 2500\,\AA}/0.0, 0.35/teal/{rest 3500\,\AA (NUV edge)}/0.0, 0.5/blue/{rest 5000\,\AA}/0.0, 0.75/ink/{rest 7500\,\AA}/0.0} {
    \draw[\col,line width=1.2pt] plot[domain=0:\zmax,samples=40] ({\x/\zmax*\w},{min(\lr*(1+\x),\lmax)/\lmax*\h});
    \pgfmathsetmacro\yend{min(\lr*(1+\zmax),\lmax)/\lmax*\h}
    \node[anchor=west,color=\col,font=\sffamily\scriptsize] at (\w+0.05,\yend) {\name};}
  \node[anchor=north,color=ink] at (0.5*\w,-0.55) {supernova redshift $z$};
  \node[rotate=90,anchor=south,color=ink] at (-0.62,0.5*\h) {observed wavelength ($\mu$m)};
  \node[anchor=west,color=teal!60!black,font=\sffamily\scriptsize] at (0.15,{0.43/\lmax*\h}) {NUV enters band};
\end{tikzpicture}}
\caption{How redshift brings the rest-frame iron-group diagnostic into the observed band. Each colored curve is a fixed rest-frame wavelength shifted as $\lambda_{\rm obs}=\lambda_{\rm rest}(1+z)$. The shaded region is the proposed $0.36$--$3.0\,\mu$m coverage. The rest-frame NUV iron-blanketing window ($2000$--$3500$\,\AA) enters the band progressively with its $2500$\,\AA\ core crossing near $z\simeq0.44$ and the full window in band by $z\simeq0.8$, while the optical-twin baseline ($3500$--$7500$\,\AA) stays in band to $z\simeq3$. In the mission's working range $0.5\lesssim z\lesssim1.5$ a single multi-roll slitless spectrum records both the optical light-curve parameters and the rest-NUV iron-group color.}
\label{fig:nuvband}
\end{figure}

At $z=0$, the rest-frame NUV lies below the $0.36\,\mu$m cutoff and requires a dedicated ultraviolet observatory. Cosmological redshift moves the iron-blanketing region into the proposed band as shown in Figure~\ref{fig:nuvband}. Rest-frame $2500$\,\AA\ reaches the blue edge at $z\simeq0.44$. The $2500$--$3500$\,\AA\ core lies in band by $z\simeq0.5$ and the full $2000$--$3500$\,\AA\ interval lies in band by $z\simeq0.8$. The optical-twin baseline at $3500$--$7500$\,\AA\ remains observable to $z\simeq3$. The time-domain tier therefore obtains the NUV diagnostic over the same $0.5\lesssim z\lesssim1.5$ interval used for SN Ia classification.

The SN Ia diagnostic is limited by blue-end signal-to-noise rather than spectral resolution. Photospheric P\,Cygni features have velocity widths near $10^4$\,\kms, which correspond to $R\sim c/v\sim30$. Iron-group blanketing produces an even broader NUV pseudo-continuum suppression. Both candidate resolving powers oversample the relevant structure by more than an order of magnitude. Binning to $R\simeq100$--300 increases signal-to-noise without erasing the diagnostic. The exposure-time reach in Figure~\ref{fig:snreach} therefore sets the precision of $\Delta\mathrm{NUV}$ and $\theta_{\rm Fe}$. A multi-orientation slitless spectrum binned to $R\simeq300$--500 records optical light-curve parameters and rest-frame NUV colour in one observation and permits subclass-aware standardization,
\begin{equation}
M_B = M_0 + \alpha\,(\Delta m_{15}-1.1) + \beta\,(B-V) + \gamma\,\Delta\mathrm{NUV} + \delta\,\theta_{\rm Fe},
\label{eq:nuvstd}
\end{equation}
where $\Delta\mathrm{NUV}$ is the rest-frame NUV colour excess and $\theta_{\rm Fe}$ parameterizes iron-group blanketing. The design goal reduces the Hubble residual scatter from $\sigma\simeq0.12$\,mag toward $\sigma\lesssim0.08$\,mag and improves per-object distance precision by approximately $30\%$. No existing Hubble-flow sample has enough rest-frame NUV spectra to demonstrate that reduction. The true improvement may be smaller if several unmodelled variables add independent residual scatter. The measurement also tests whether progenitor-metallicity evolution imposes a redshift-dependent standardization bias. Because the same explosion physics produces both NUV diversity and the host-mass step, $\theta_{\rm Fe}$ offers a physical alternative to empirical host corrections.

\subsubsection{Observing Strategy and Survey Reach}

Host-galaxy light is the central observational systematic because the supernova and host spectra overlap at the same sky position. Three measurements constrain the contamination. Supernova-free epochs form a deep dispersed host template. Subtraction of the template isolates the transient spectrum and leaves noise dominated by the supernova epoch. The unresolved supernova produces a narrow trace while the resolved host produces a smoother background with lower effective resolution as described in Section~\ref{sec:effres}. Host surface brightness at the transient position therefore matters more than total host magnitude. Roman simulations also demonstrate three-dimensional host reconstruction from multi-orientation slitless spectra before host subtraction \cite{astraatmadja2026}. The flagship three-orientation strategy supplies the required data in the cadence field. HST provides flight precedent through ACS-grism classification of GOODS supernovae above $z=1$, including SN\,2002fw at $z=1.3$ near maximum light \cite{riess2004}. Roman plans slitless prism spectroscopy for approximately 12{,}000 SNe Ia \cite{rose2021roman}. The proposed observations therefore support high-redshift standardization and physical classification rather than a stand-alone sub-percent $H_0$ measurement.

For a normal SN Ia with $M_B\simeq-19.3$, the approximate peak magnitude is $m\simeq\mu(z)-19.3$ before bandpass, stretch, color, and host corrections. Here $\mu(z)$ is the distance modulus, set by the luminosity distance $d_L(z)$,
\begin{equation}
\mu(z)=5\log_{10}\!\left[\frac{d_L(z)}{10\,\mathrm{pc}}\right]=25+5\log_{10}\!\left[\frac{d_L(z)}{\mathrm{Mpc}}\right],
\qquad
d_L(z)=\frac{c\,(1+z)}{H_0}\int_0^z\frac{dz'}{E(z')},
\label{eq:distmod}
\end{equation}
with $E(z)=\big(\Omega_m(1+z)^3+(1-\Omega_m)(1+z)^{3(1+w)}\big)^{1/2}$ for a flat model with dark-energy equation of state $w$. For $\Lambda$CDM, the expression reduces to $E(z)=\big(\Omega_m(1+z)^3+\Omega_\Lambda\big)^{1/2}$. Because $d_L\propto H_0^{-1}$, the distance modulus contains an additive $-5\log_{10}H_0$ term. Raising $H_0$ shifts the Hubble diagram toward smaller $\mu$ by $5\log_{10}(H_0/H_0')$\,mag without altering its shape. The redshift dependence of $\mu(z)$ instead constrains $\Omega_m$ and dark-energy parameters. At low redshift, the expression becomes $\mu\simeq25+5\log_{10}\!\big(cz/H_0\,[\mathrm{Mpc}]\big)+1.086\,(1-q_0)\,z+\cdots$, where $q_0=\tfrac12\Omega_m(1+3w)+\cdots$ is the deceleration parameter. The planning sensitivity reaches $m_{\rm AB}\simeq24.5$ and $z\simeq0.9$ in 2 hr near maximum light. Exposures of 10 hr reach $m_{\rm AB}\simeq25.4$ and $z\simeq1.25$. Exposures of 30 hr reach $m_{\rm AB}\simeq26.0$ and $z\simeq1.5$ for favourable events (Figure~\ref{fig:snreach}). The values assume spectra binned to $R\simeq300$--500 after combining three orientations. Blue-end signal limits the rest-NUV diagnostic in Eq.~\eqref{eq:nuvstd}. Iron-group colour is most precise for the brighter and lower-redshift half of each tier while classification and host-redshift measurements extend across the full range.

\begin{figure}[htbp]
\centering
\includegraphics[width=0.72\textwidth]{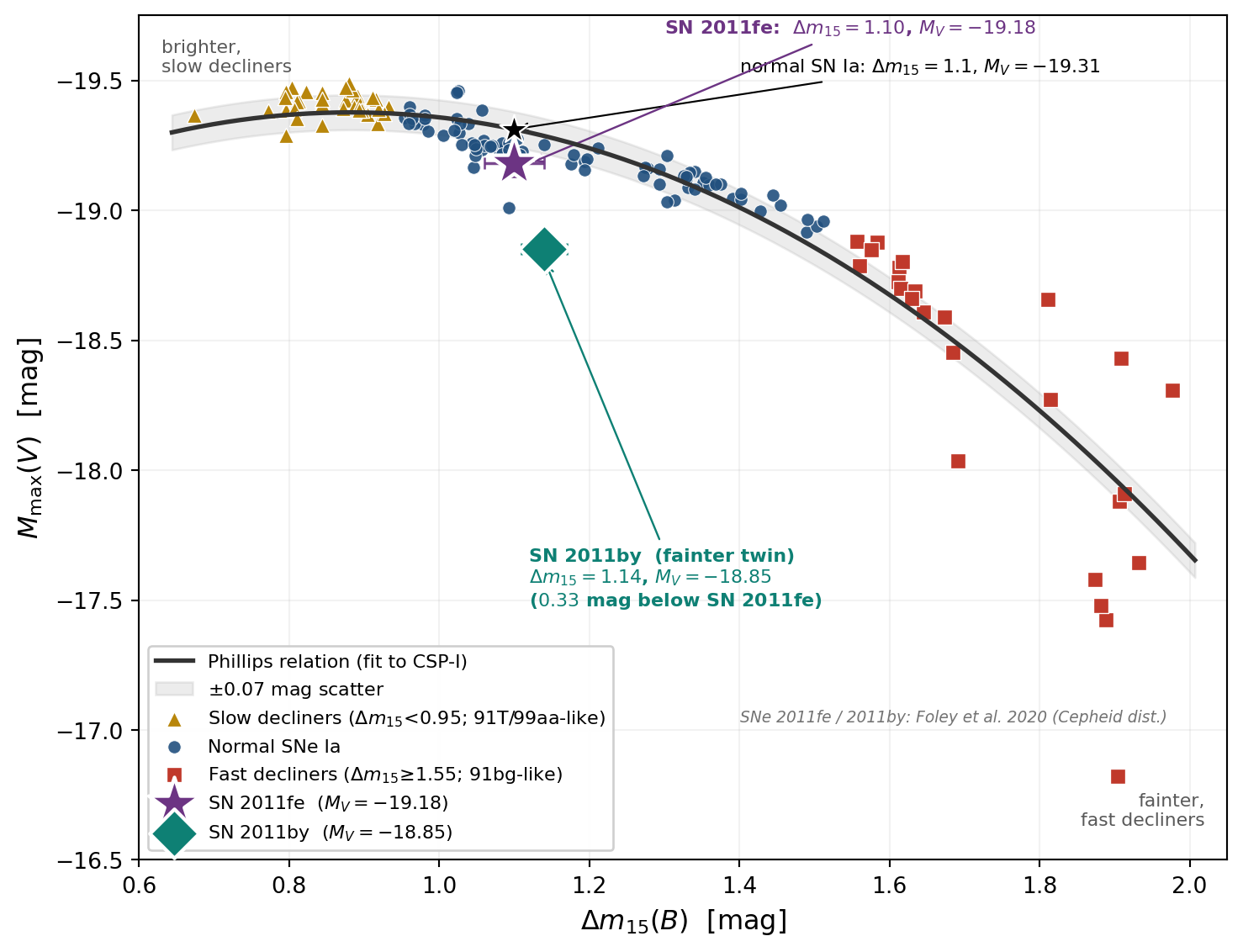}
\caption{The Phillips width--luminosity relation that underpins SN Ia standardization. The plot uses the 139 SNe Ia in the Carnegie Supernova Project\,I sample of Burns et al.\ \cite{burns2018csp} and follows the presentation of Figure 11 in Zeng et al.\ \cite{zeng2025sn2023ehl}. The peak absolute $V$ magnitude $M_{\max}(V)$ is corrected for host-galaxy extinction with the per-SN reddening and $R_V$ tabulated by the survey. Brighter supernovae decline more slowly as measured by $\Delta m_{15}(B)$. The solid curve is a quadratic fit that passes through $M_V=-19.31$ at $\Delta m_{15}=1.1$. The shaded band marks the residual scatter of $0.07$\,mag. Colours separate slow-declining overluminous 1991T/1999aa-like events, normal SNe Ia, and fast-declining subluminous 1991bg-like events. One light-curve-shape parameter explains most of the luminosity range but leaves residual structure for rest-frame NUV spectroscopy to diagnose. The CSP absolute scale adopts $H_0=72$\,\kms\,Mpc$^{-1}$. That choice fixes the zero point but not the shape of the relation.}
\label{fig:phillips}
\end{figure}

\begin{figure}[htbp]
\centering
\begin{tikzpicture}[font=\sffamily\footnotesize]
  \def\w{13.5}
  \def\h{5.0}
  \fill[paper] (-0.35,-0.35) rectangle (\w+0.45,\h+0.45);
  \draw[->,color=muted] (0,0) -- (\w+0.1,0);
  \draw[->,color=muted] (0,0) -- (0,\h+0.1);
  \foreach \x/\lab in {0/0.5,2.53/0.8,4.22/1.0,5.91/1.2,8.44/1.5,10.97/1.8,12.66/2.0} {
    \draw[line!65] (\x,0) -- (\x,\h);
    \node[anchor=north,color=muted] at (\x,-0.08) {\lab};
  }
  \foreach \y/\lab in {0/23.0,1.25/24.0,2.5/25.0,3.75/26.0,5.0/27.0} {
    \draw[line!65] (0,\y) -- (\w,\y);
    \node[anchor=east,color=muted] at (-0.08,\y) {\lab};
  }
  \draw[blue,line width=1.45pt,smooth]
    plot coordinates {(0,0.0) (2.53,1.50) (4.22,2.25) (5.91,2.86) (8.44,3.61) (10.97,4.23) (12.66,4.57)};
  \draw[teal,line width=0.9pt,dashed] (0,1.875) -- (\w,1.875);
  \draw[gold,line width=0.9pt,dashed] (0,3.0) -- (\w,3.0);
  \draw[rose,line width=0.9pt,dashed] (0,3.75) -- (\w,3.75);
  \node[anchor=west,color=teal] at (9.2,1.98) {2 hr gives $m\simeq24.5$};
  \node[anchor=west,color=gold] at (9.2,3.12) {10 hr gives $m\simeq25.4$};
  \node[anchor=west,color=rose] at (9.2,3.88) {30 hr gives $m\simeq26.0$};
  \node[anchor=north,color=ink] at (0.5*\w,-0.55) {redshift};
  \node[rotate=90,anchor=south,color=ink] at (-0.92,0.5*\h) {SN Ia peak magnitude};
  \node[anchor=west,color=blue,font=\bfseries] at (0.45,4.65) {$M_B=-19.3,\ m=\mu(z)+M_B$};
\end{tikzpicture}
\caption{Planning reach for binned $R\simeq300$--500 SN Ia slitless spectra. The exposure labels are total science time after combining the three required orientations.}
\label{fig:snreach}
\end{figure}
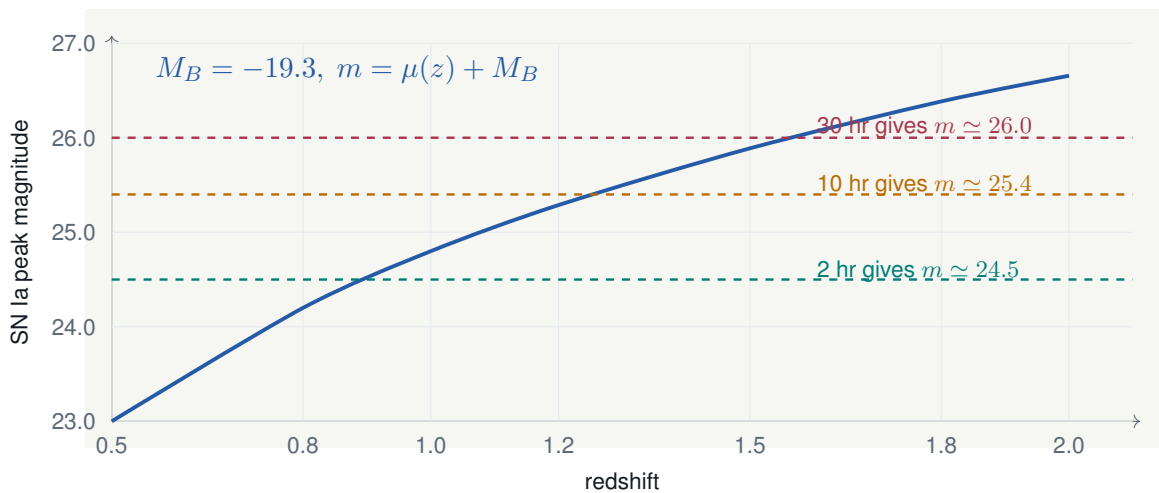

\begin{figure}[htbp]
\centering
\includegraphics[width=\textwidth]{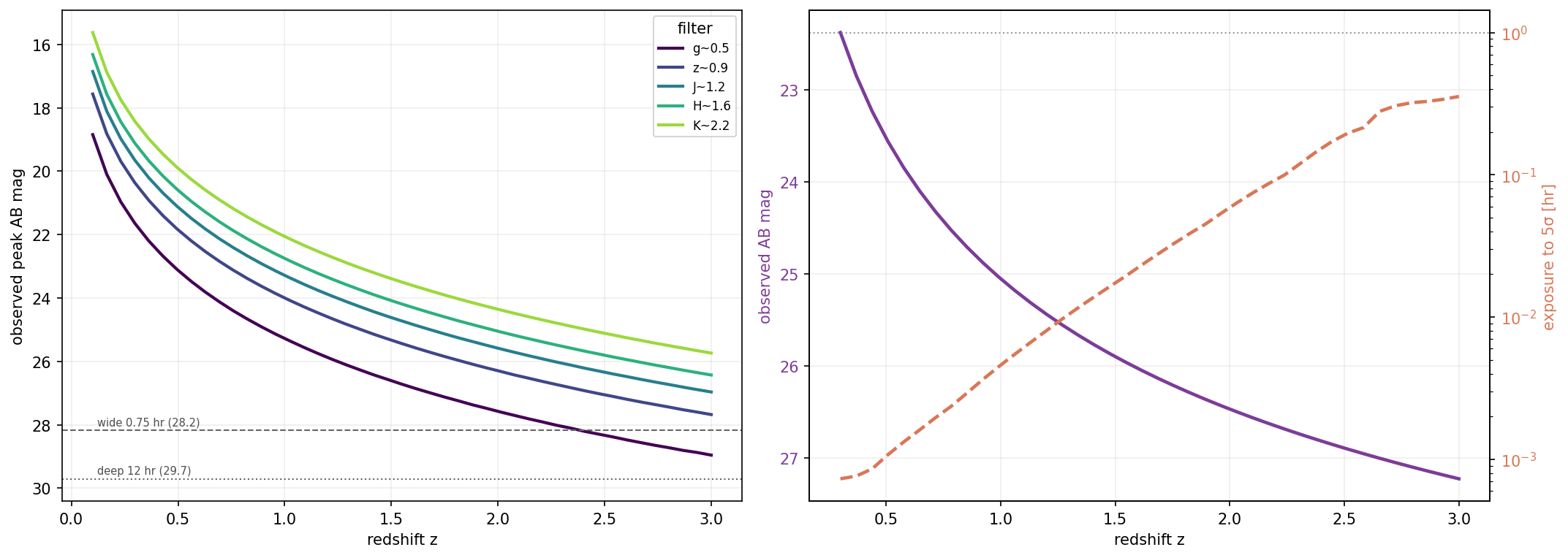}
\caption{The left panel gives observed AB magnitude by band as a function of redshift for a peak Type Ia SN with $M_B=-19.3$ and a Nugent template. Wide $0.75$\,hr and deep $12$\,hr $H$-band point-source depths are marked for the $3.5$\,m concept. Redder bands retain the peak above the wide-tier limit to $z\gtrsim2$. The right panel follows rest-frame near-ultraviolet emission at $2770$\,\AA\ as it redshifts through the observed band. Its left axis gives observed AB magnitude and its right axis gives the exposure required for a $5\sigma$ detection. The rest-NUV twin diagnostic requires minutes to $z\simeq2$.}
\label{fig:sniamag}
\label{fig:sniamuv}
\end{figure}

The exposure-time calculator independently evaluates the redshift reach and rest-ultraviolet sensitivity. Figure~\ref{fig:sniamag} gives observed magnitudes from the Nugent peak Type Ia template normalized to $M_B=-19.3$. Figure~\ref{fig:sniamuv} follows the rest-frame $2770$\,\AA\ diagnostic after the wavelength enters the band at $z\gtrsim0.3$. The idealized photon budget detects the rest-frame NUV of a peak Type Ia in minutes to $z\simeq2$. The result places the optical-twin NUV divergence within the redshift interval used for the distance scale.

\subsubsection{The S/N-Limited Redshift of the Twin \texorpdfstring{$U$}{U} and NUV Photometry}

The $5\sigma$ detection reach of Figure~\ref{fig:sniamuv} is not the right requirement for the twin program because the twin diagnostic is a magnitude rather than a detection. The NUV-red and NUV-blue subgroups are offset by $\sim0.4$\,mag in NUV$-$optical colour at matched optical light curves \cite{milne2015}. The optical side of that colour comes essentially free since the optical bands are one to two magnitudes brighter. The colour error is the NUV magnitude error. Assigning a single supernova to a subgroup with the $0.4$\,mag offset resolved at $4\sigma$ therefore requires $\sigma_m\simeq0.10$\,mag in the rest-frame band, which through $\sigma_m=2.5/(\ln 10\cdot {\rm S/N})$ means ${\rm S/N}\simeq11$. Using $\Delta{\rm NUV}$ as the continuous standardization covariate of Eq.~\eqref{eq:nuvstd} at the $\sigma\lesssim0.08$\,mag Hubble-residual target requires $\sigma_m\simeq0.05$\,mag, or ${\rm S/N}\simeq22$.

\begin{figure}[htbp]
\centering
\includegraphics[width=\textwidth]{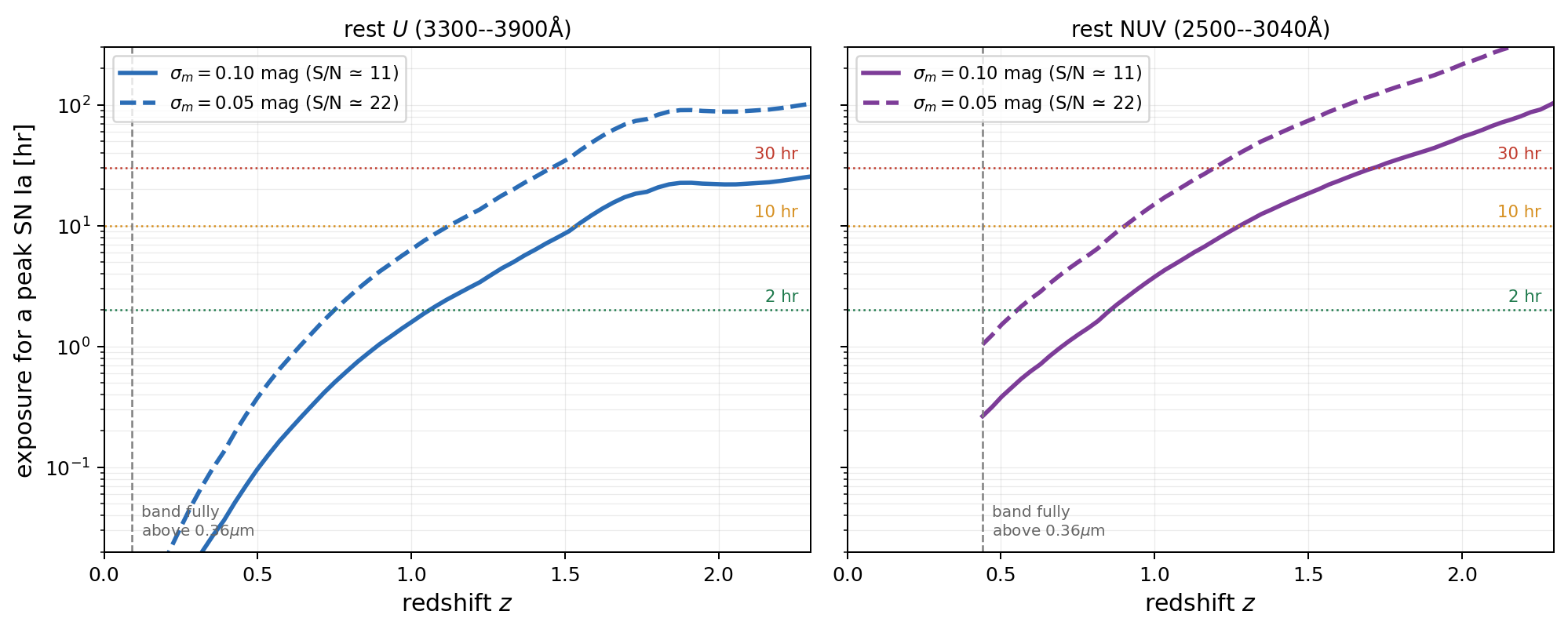}
\caption{The S/N-limited redshift of the SN Ia rest-frame $U$ (left) and NUV (right) filter photometry from the slitless exposure-time calculator for a peak normal SN Ia ($M_B=-19.3$, Nugent template). Each curve gives the exposure at which the synthetic filter magnitude reaches $\sigma_m=0.10$\,mag (solid, the $4\sigma$ NUV-red/NUV-blue subgroup assignment \cite{milne2015}) and $\sigma_m=0.05$\,mag (dashed, covariate-grade photometry, Eq.~\eqref{eq:nuvstd}). Dotted lines mark the $2$, $10$, and $30$\,hr planning exposures. The vertical dashed line marks where each filter clears the $0.36\,\mu$m blue edge. Subgroup classification reaches $z\simeq0.86$ (NUV) and $1.06$ ($U$) in $2$\,hr and $z\simeq1.28$ and $1.54$ in $10$\,hr.}
\label{fig:snuvlimit}
\end{figure}

Figure~\ref{fig:snuvlimit} evaluates the S/N thresholds with the slitless ETC and the Nugent peak template at $M_B=-19.3$. Synthetic rest-frame top-hat filters cover $3300$--$3900$\,\AA\ in $U$ and $2500$--$3040$\,\AA\ in the NUV. The calculation bins the combined slitless spectrum over each redshifted band and includes dispersed zodiacal emission, cirrus, telescope thermal emission, dark current, and read noise. The NUV band clears the $0.36\,\mu$m blue edge at $z\geq0.44$ while the $U$ band clears the edge at $z\geq0.09$. In 2 hr, subgroup assignment with $\sigma_m=0.10$ reaches $z\simeq0.86$ in the NUV and $z\simeq1.06$ in $U$. Covariate-grade photometry with $\sigma_m=0.05$ reaches $z\simeq0.55$ and $0.75$, respectively. A 10 hr exposure extends subgroup assignment to $z\simeq1.28$ in the NUV and $1.54$ in $U$. The corresponding covariate-grade limits are $z\simeq0.90$ and $1.12$. A 30 hr stack reaches $z\simeq1.71$ in the NUV. The flattening near $z\simeq1.8$ occurs when the redshifted $U$ band crosses the $1\,\mu$m dichroic into the near-infrared arm where zodiacal emission is lowest. Standard exposures therefore provide twin photometry to $z\simeq0.9$--$1.3$ and cover the central part of the $0.5\lesssim z\lesssim1.5$ classification interval. The limits come from the idealized slitless ETC without calibration to an achieved survey depth. Section~\ref{sec:etc} shows that the values are upper bounds and require validation with the flight spectral extraction.

\subsubsection{Cosmological Constraints on \texorpdfstring{$H_0$}{H0} and Dynamic Dark Energy}

\begin{figure}[htbp]
\centering
\includegraphics[width=0.76\textwidth]{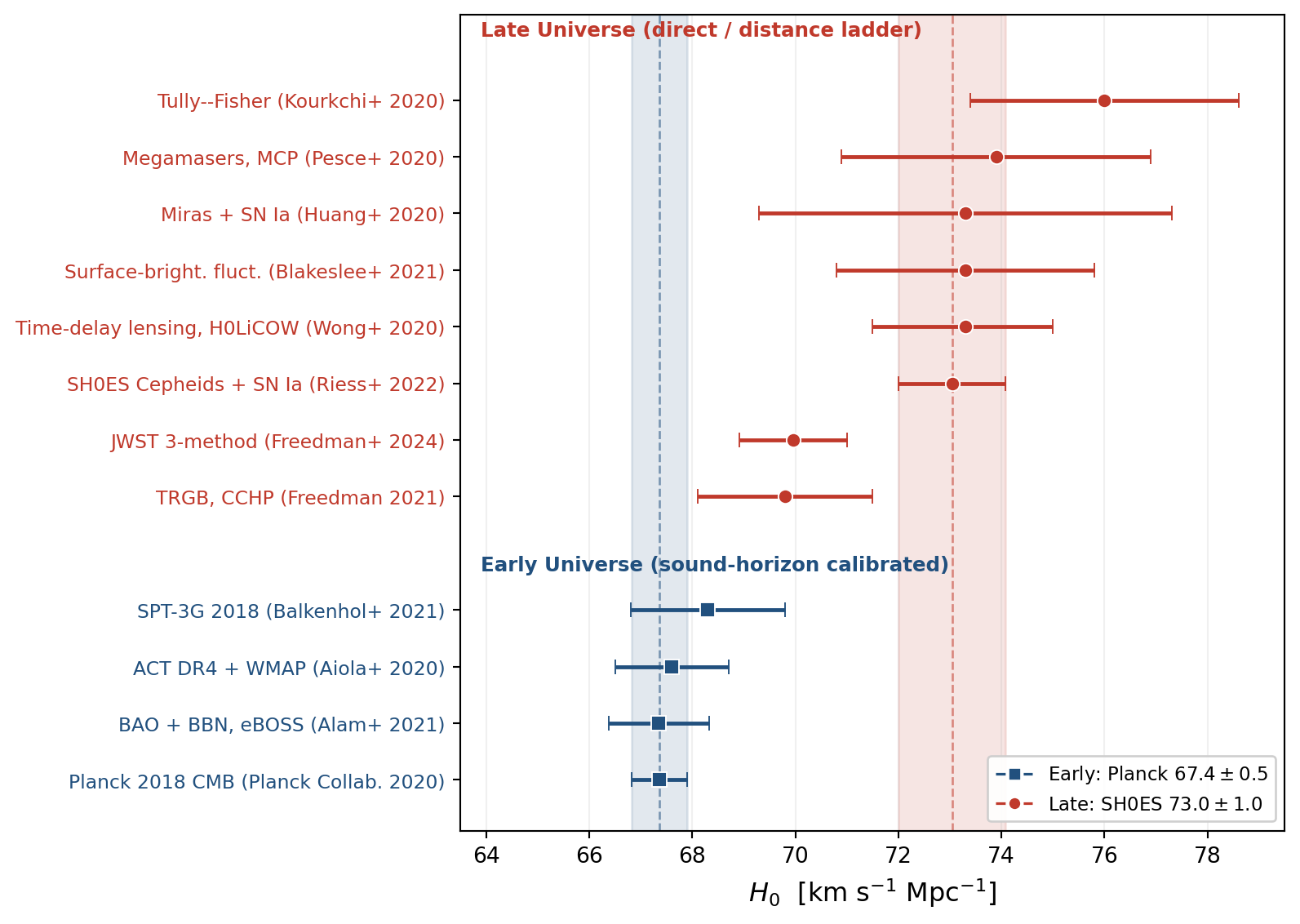}
\caption{Compilation of $H_0$ determinations with measured value on the horizontal axis and one method per row. Blue squares mark early-Universe probes calibrated by the sound horizon. The methods are Planck CMB \cite{planck2018}, ACT$+$WMAP, SPT-3G, and BAO$+$BBN, which cluster near $67.4\pm0.5$\,\kms\,Mpc$^{-1}$. Red circles mark late-Universe direct and distance-ladder probes. The methods are SH0ES Cepheids$+$SN\,Ia \cite{riess2022}, TRGB, JWST multi-method, surface-brightness fluctuations, megamasers, Miras, Tully--Fisher, and time-delay lensing, which cluster near $73\pm1$\,\kms\,Mpc$^{-1}$. The shaded Planck and SH0ES bands differ by $5.7$\,\kms\,Mpc$^{-1}$. The difference is approximately $4.8\sigma$ for the two quoted values and approximately $5\sigma$ for the full early-versus-late comparison. Rest-frame NUV spectroscopy permits a subclass-homogeneous and iron-group-aware SN\,Ia distance ladder that directly tests the dominant astrophysical uncertainty in the late-Universe measurements.}
\label{fig:h0tension}
\end{figure}

\begin{figure}[htbp]
\centering
\includegraphics[width=0.76\textwidth]{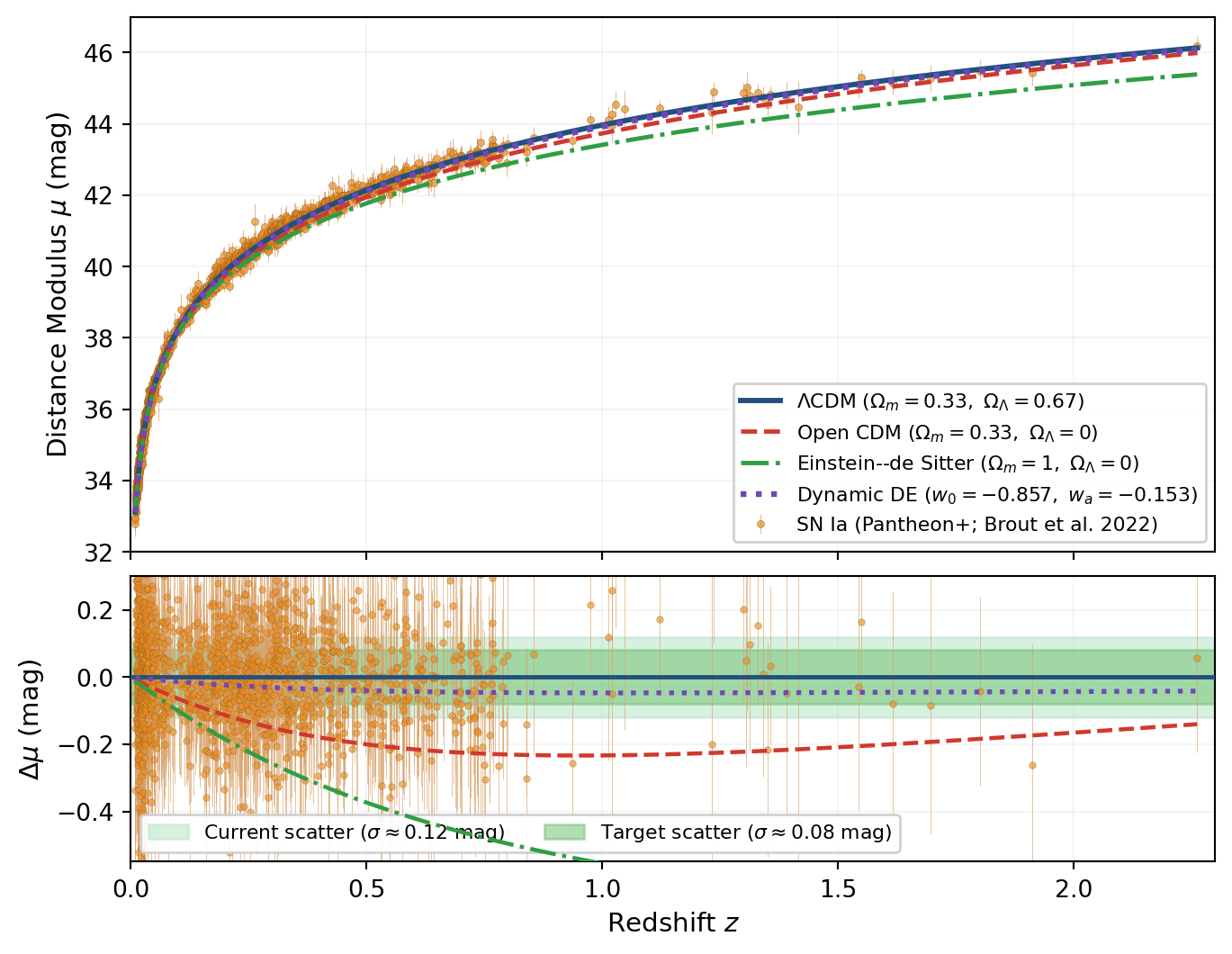}
\caption{SN Ia Hubble diagram and the precision required for dark-energy science. The upper panel gives distance modulus $\mu$ versus redshift for 1580 Hubble-flow SNe Ia in the Pantheon$+$SH0ES sample after removing Cepheid calibrators \cite{brout2022}. Curves show $\Lambda$CDM with $\Omega_m=0.33$ and $\Omega_\Lambda=0.67$, open CDM, Einstein--de\,Sitter, and a CPL dynamic-dark-energy model with $w_0=-0.857$ and $w_a=-0.153$. The reference uses the Pantheon$+$ flat-$\Lambda$CDM best fit with $H_0=73.04$ and $\Omega_m=0.334$. The lower panel gives residuals $\Delta\mu$ relative to $\Lambda$CDM. Shaded bands compare the current per-SN scatter of $\sigma\approx0.12$\,mag with the target $\sigma\approx0.08$\,mag after the NUV-augmented calibration in Eq.~\eqref{eq:nuvstd}. Distinguishing $\Lambda$CDM from dynamic dark energy requires systematic control below $0.1$\,mag.}
\label{fig:sniacosmo}
\end{figure}

The SN Ia Hubble diagram tests two precision cosmology questions (Figure~\ref{fig:sniacosmo}). The first is the approximately $5\sigma$ discrepancy between the local SH0ES value of $H_0=73.04\pm1.04$\,\kms\,Mpc$^{-1}$ \cite{riess2022} and the Planck CMB value of $67.4\pm0.5$\,\kms\,Mpc$^{-1}$ \cite{planck2018}. Figure~\ref{fig:h0tension} shows the broader set of measurements. Unequal mixtures of NUV-red and NUV-blue subclasses in the calibrator and Hubble-flow samples would bias the distance ladder. Rest-frame NUV spectra identify the subgroup and measure $\theta_{\rm Fe}$ for each event. A subgroup-aware standardization directly tests and corrects the associated astrophysical bias.

The second question tests whether the dark-energy equation of state evolves with redshift. Unrecognized evolution in the SN population can mimic or conceal dynamic dark energy. Figure~\ref{fig:sniacosmo} shows that separating $\Lambda$CDM from the illustrated CPL model requires residual control below $0.1$\,mag. The required precision is below the present per-SN scatter and motivates the NUV-augmented calibration in Eq.~\eqref{eq:nuvstd}. Combining the SN Ia distances, flagship BAO and RSD measurements, and external Planck or ACT CMB constraints yields a joint posterior on $H_0$, $\Omega_m$, and $(w_0,w_a)$ with an explicit physical model for the SN systematic.

The SN Ia program extends beyond classification and host-redshift measurement. Rest-frame NUV spectra over $0.5\lesssim z\lesssim1.5$ diagnose iron-group physics, test a physical interpretation of the host-mass step, and provide a subclass-controlled distance sample for joint analysis with the flagship BAO and RSD survey.

\subsection{Strong-Lens Cosmography}\label{sec:stronglens}

Strong gravitational lensing provides a geometric cosmological probe. Ratios of angular-diameter distances to the deflector and source determine the lens configuration and constrain the expansion history with systematics distinct from BAO and supernova measurements. Three effects couple the cosmological signal to the mass distribution. The mass-sheet degeneracy rescales convergence and adds a uniform sheet while preserving all observables except the absolute time delay. Stellar dynamics introduce a mass--anisotropy degeneracy when breaking the mass-sheet degeneracy. Mass along the line of sight contributes external convergence $\kappa_{\rm ext}$. Control of the three effects rather than the raw lens count determines the cosmological precision.

Diffraction-limited and photometrically stable direct imaging at approximately $0.1$\,arcsec provides the primary data for lens identification and reconstruction of Einstein rings and arcs. Spectroscopy supplies deflector and source redshifts, stellar velocity dispersions, and redshifts of line-of-sight galaxies. Weak-lensing shapes provide an additional mass constraint at larger radius. Forecasts predict approximately $1.7\times10^5$ discoverable galaxy--galaxy lenses in \textit{Euclid} and $1.2\times10^5$ in Rubin/LSST \cite{collett2015}. The \textit{Roman} High-Latitude Survey is expected to contain approximately $1.7\times10^4$ lenses \cite{weiner2020}. \textit{Euclid} has already published hundreds of candidates over its first 63\,\degti\ and projects more than $10^5$ over the full survey \cite{walmsley2025}. Redshift confirmation \cite{shu2025} and individual mass reconstruction remain the limiting measurements.

\subsubsection{Double-Source-Plane Lenses for \texorpdfstring{$w$}{w} and \texorpdfstring{$\Omega_m$}{Omega_m}}

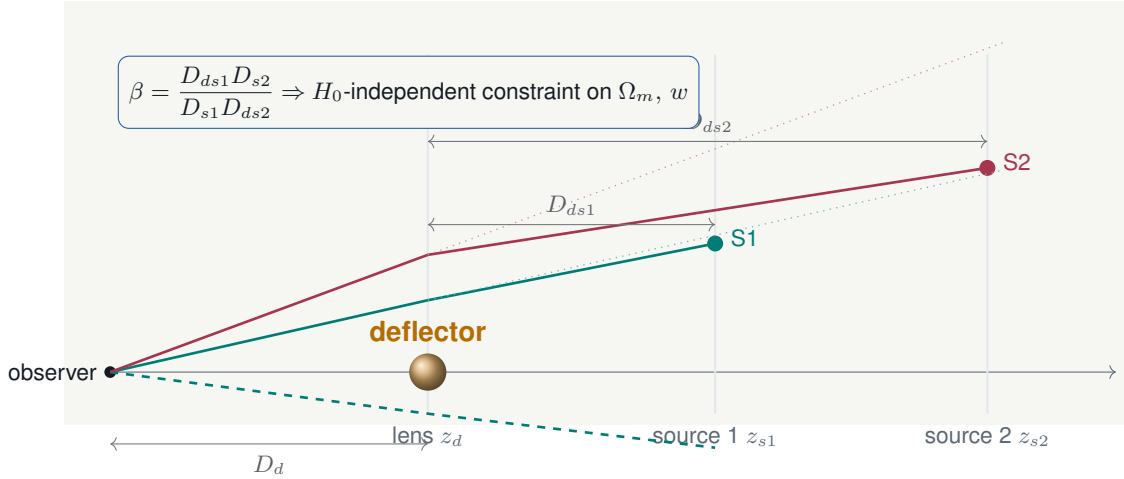
\begin{figure}[htbp]
\centering
\begin{tikzpicture}[font=\sffamily\footnotesize]
  \def\xo{0}\def\xl{4.2}\def\xa{8.0}\def\xb{11.6}
  \def\h{4.2}
  \fill[paper] (-0.6,-0.7) rectangle (\xb+1.9,\h+0.7);
  \draw[muted,->] (\xo,0) -- (\xb+1.7,0);
  \draw[line!70,line width=0.8pt] (\xl,-0.55) -- (\xl,\h);
  \draw[line!70,line width=0.8pt] (\xa,-0.55) -- (\xa,\h);
  \draw[line!70,line width=0.8pt] (\xb,-0.55) -- (\xb,\h);
  \node[anchor=north,color=muted] at (\xl,-0.6) {lens $z_d$};
  \node[anchor=north,color=muted] at (\xa,-0.6) {source 1 $z_{s1}$};
  \node[anchor=north,color=muted] at (\xb,-0.6) {source 2 $z_{s2}$};
  \fill[ink] (\xo,0) circle (2.2pt);
  \node[anchor=east,color=ink] at (\xo-0.05,0) {observer};
  \shade[ball color=gold!60] (\xl,0) circle (7pt);
  \node[anchor=south,color=gold,font=\bfseries] at (\xl,0.28) {deflector};
  \fill[teal] (\xa,1.7) circle (3pt); \node[anchor=west,color=teal] at (\xa+0.08,1.78) {S1};
  \fill[rose] (\xb,2.7) circle (3pt); \node[anchor=west,color=rose] at (\xb+0.08,2.78) {S2};
  \draw[teal,line width=1.0pt] (\xo,0) -- (\xl,0.95) -- (\xa,1.7);
  \draw[teal,line width=1.0pt,dashed] (\xo,0) -- (\xl,-0.55) -- (\xa,-1.0) coordinate(s1m);
  \draw[rose,line width=1.0pt] (\xo,0) -- (\xl,1.55) -- (\xb,2.7);
  \draw[teal!55,line width=0.5pt,dotted] (\xl,0.95) -- (\xb+0.2,{0.95/\xl*(\xb+0.2)});
  \draw[rose!55,line width=0.5pt,dotted] (\xl,1.55) -- (\xb+0.2,{1.55/\xl*(\xb+0.2)});
  \draw[ink!60,<->] (\xo,-0.95) -- (\xl,-0.95); \node[anchor=north,color=ink!70] at ({0.5*(\xo+\xl)},-0.97) {$D_d$};
  \draw[ink!60,<->] (\xl,1.95) -- (\xa,1.95); \node[anchor=south,color=ink!70] at ({0.5*(\xl+\xa)},1.95) {$D_{ds1}$};
  \draw[ink!60,<->] (\xl,3.05) -- (\xb,3.05); \node[anchor=south,color=ink!70] at ({0.5*(\xl+\xb)},3.05) {$D_{ds2}$};
  \node[anchor=west,draw=blue,fill=paper,rounded corners,inner sep=4pt,text=ink] at (0.1,3.7)
     {$\displaystyle \beta=\frac{D_{ds1}D_{s2}}{D_{s1}D_{ds2}}$ $\Rightarrow$ $H_0$-independent constraint on $\Omega_m,\,w$};
\end{tikzpicture}
\caption{Double-source-plane lens geometry. One deflector at $z_d$ lenses two sources at $z_{s1}$ (teal) and $z_{s2}$ (rose). The ratio of their Einstein radii fixes the distance ratio $\beta$ of Eq.~\eqref{eq:dsplbeta}, which constrains $\Omega_m$ and $w$ nearly independently of $H_0$ and of the absolute lens mass. Stable imaging measures the two ring radii, while slitless spectroscopy supplies $z_d$, $z_{s1}$, and $z_{s2}$ in the same pointing.}
\label{fig:dspl}
\end{figure}

The cleanest strong-lensing cosmology available to a spectroscopic-imaging mission is the \emph{double-source-plane lens} (DSPL), a single deflector lensing two background sources at distinct redshifts $z_{s1}<z_{s2}$ (Figure~\ref{fig:dspl}). The ratio of the two Einstein radii fixes the dimensionless distance ratio
\begin{equation}
\beta \;=\; \frac{D_{ds1}\,D_{s2}}{D_{s1}\,D_{ds2}},
\label{eq:dsplbeta}
\end{equation}
where $D_{s}$ and $D_{ds}$ are angular-diameter distances from the observer and deflector to each source. The common-lens distance ratio is nearly independent of $H_0$ and absolute mass normalization and instead constrains $\Omega_m$ and dark-energy parameters. The ``Jackpot'' lens SDSS\,J0946$+$1006 gives $\beta^{-1}=1.404\pm0.016$. Combined with CMB constraints in flat $w$CDM, the measurement gives $w=-1.17\pm0.20$ and improves the CMB-only precision by approximately $30\%$ \cite{collettauger2014}. Forecasts give $\sigma(w)\approx0.15$ from a small sample of accurately modelled DSPLs \cite{collett2012}. Combined AGEL systems have recently improved CMB-only $w$ precision by approximately $30$--40\% \cite{sahu2025,bowden2025}.

DSPLs are rare because imaging alone seldom identifies two source redshifts behind one deflector. Stable imaging resolves the two ring or arc systems and measures their radii. Slitless spectroscopy measures both source redshifts and the deflector redshift in the same pointing. The combined observation supports a systematically selected DSPL sample. DSPL distance-ratio cosmography is the primary strong-lensing goal because wide-field spectroscopy naturally supplies multiple secure source-plane redshifts and because $\beta$ avoids absolute $H_0$ calibration and reduces sensitivity to the mass-sheet degeneracy.

\subsubsection{Mass-Density Profiles and the Degeneracy-Breaking Program}

Massive early-type lens galaxies have nearly isothermal total density profiles with $\rho\propto r^{-\gamma'}$ and $\gamma'\simeq2$. SLACS dynamics gives $\langle\gamma'\rangle=2.085^{+0.025}_{-0.018}$ \cite{koopmans2009}. Joint SLACS lensing and dynamics gives $2.078\pm0.027$ with intrinsic scatter $0.16$ \cite{auger2010}, while extended-source modelling gives $2.075^{+0.023}_{-0.024}$ \cite{etherington2023}. At fixed mass and size, the slope evolves as $\partial\gamma'/\partial z=-0.31\pm0.10$ \cite{sonnenfeld2013}. A homogeneous measurement of $\gamma'(z,M_\ast)$ traces the balance between dissipative growth and dry merging in massive-galaxy assembly. The same measurement supplies density-slope and orbital-anisotropy constraints required to reduce the mass-sheet degeneracy in time-delay cosmography. Hierarchical inference across the lens population requires such a large and consistently selected sample \cite{birrer2020}.

Diffraction-limited imaging constrains the Einstein radius $\theta_E$, enclosed projected mass, and radial slope $\gamma'$ from the reconstructed ring. An image-sliced spectrum adds an aperture-integrated stellar velocity dispersion $\sigma$ for the deflector. The $R=1000$ candidate has a point-source width near $300$\,\kms\ and is insensitive to lower dispersions. Lens-selected deflectors are commonly massive elliptical galaxies with $\sigma\gtrsim250$\,\kms. Phase A simulations must determine the precision of integrated dispersions recovered by template fitting and morphology deconvolution across three orientations. The $R=5000$ candidate reduces the point-source width to approximately $60$\,\kms\ and provides stronger dispersion constraints without a photon-limited S/N penalty after integration over the same absorption feature. Image slicing does not guarantee spatially resolved kinematics. Flagship systems that require two-dimensional stellar-velocity fields need external integral-field spectroscopy \cite{tdcosmo2025}.

\subsubsection{Supporting Role in Time-Delay Cosmography}

Time-delay cosmography measures $H_0$ from the relative arrival times of multiply imaged variable sources. H0LiCOW obtained $H_0=73.3^{+1.7}_{-1.8}$ \kms\,Mpc$^{-1}$ with 2.4\% precision from six lensed quasars under power-law or composite mass models \cite{wong2020}. Allowing full mass-sheet freedom increases the uncertainty to approximately $8\%$ \cite{birrer2020}. The latest TDCOSMO analysis combines integrated and spatially resolved stellar kinematics and gives $H_0=71.6^{+3.9}_{-3.3}$ \kms\,Mpc$^{-1}$ \cite{birrer2025}. A competitive independent time-delay measurement requires dedicated cadence and resolved kinematics outside the baseline mission. The proposed survey instead supplies essential redshifts, environmental data, and mass-profile constraints for lensed quasars and supernovae discovered by Rubin and \textit{Roman}.

Spectroscopy first provides the deflector and source redshifts and the stellar velocity dispersion required by each time-delay analysis. The wide field also measures redshifts of faint galaxies along the sight line and constrains external convergence $\kappa_{\rm ext}$. Because $H_0\propto(1-\kappa_{\rm ext})$, uncertainty in external convergence directly propagates into $H_0$. For B1608$+$656, $\kappa_{\rm ext}=0.10^{+0.08}_{-0.05}$ dominated the error budget \cite{suyu2010}. Typical sight lines have convergence of a few percent with approximately $2.5\%$ uncertainty after weighted galaxy counts and field spectroscopy \cite{rusu2017,wells2023}. Dense redshifts of line-of-sight galaxies from overlapping flagship fields therefore improve both $\kappa_{\rm ext}$ and $H_0$ \cite{greene2013}.

\subsubsection{Automated Lens Identification and Mass Modelling}

\begin{figure}[htbp]
\centering
\resizebox{0.97\textwidth}{!}{%
\begin{tikzpicture}[font=\sffamily\footnotesize,
  img/.style={rounded corners,draw=blue,line width=0.9pt,fill=blue!8,inner sep=4pt,align=center,text width=2.5cm,minimum height=1.05cm},
  spec/.style={rounded corners,draw=teal,line width=0.9pt,fill=teal!8,inner sep=4pt,align=center,text width=2.5cm,minimum height=1.05cm},
  ai/.style={rounded corners,draw=rose,line width=0.9pt,fill=rose!10,inner sep=4pt,align=center,text width=2.5cm,minimum height=1.05cm},
  outbox/.style={rounded corners,draw=gold,line width=1.1pt,fill=gold!12,inner sep=4pt,align=center,text width=2.6cm,minimum height=1.05cm},
  ar/.style={-{Stealth[length=2.4mm]},line width=0.8pt,color=ink!75},
  arp/.style={-{Stealth[length=2.4mm]},line width=0.8pt,color=teal,dashed}]
  \fill[paper] (-1.85,-2.65) rectangle (12.6,2.55);
  \node[img]  (im)   at (0,1.25)    {Stable-PSF\\direct imaging};
  \node[spec] (sp)   at (0,-1.25)   {Wide-field slitless\\spectroscopy};
  \node[ai]   (find) at (3.7,1.25)  {CNN/transformer\\lens finder};
  \node[spec] (zz)   at (3.7,-1.25) {$z_d,z_{s1},z_{s2},$\\$\sigma$, LOS galaxies};
  \node[ai]   (sbi)  at (7.4,1.25)  {SBI mass model\\$\theta_E,\gamma',\kappa$ / lens};
  \node[spec] (kext) at (7.4,-1.25) {$\kappa_{\rm ext}$ from\\weighted counts};
  \node[outbox] (hier) at (10.9,0.0) {Hierarchical\\Bayesian inference};
  \node[anchor=north,text=gold,text width=3.0cm,align=center,font=\bfseries\footnotesize] at (10.9,-0.85)
       {$\Omega_m,w$ (DSPL)\\$\gamma'(z)$ and TDC priors};
  \draw[ar] (im)   -- (find);
  \draw[ar] (find) -- (sbi);
  \draw[ar] (sp)   -- (zz);
  \draw[ar] (zz)   -- (kext);
  \draw[ar] (sbi.east)  -- (hier.north west);
  \draw[ar] (kext.east) -- (hier.south west);
  \draw[arp] (zz.north) -- ++(0,0.55) -| (sbi.south);
  \node[text=teal,font=\scriptsize,fill=paper,inner sep=1pt] at (5.55,0.15) {$z,\sigma$ priors};
\end{tikzpicture}}
\caption{Strong-lensing analysis sequence. Imaging in blue supplies lens identification and simulation-based mass estimates. Spectroscopy in teal supplies redshifts, velocity dispersions, and line-of-sight galaxy counts for $\kappa_{\rm ext}$. Hierarchical Bayesian inference in gold jointly constrains cosmology and population-level nuisance distributions while marginalizing individual lenses.}
\label{fig:lenspipe}
\end{figure}
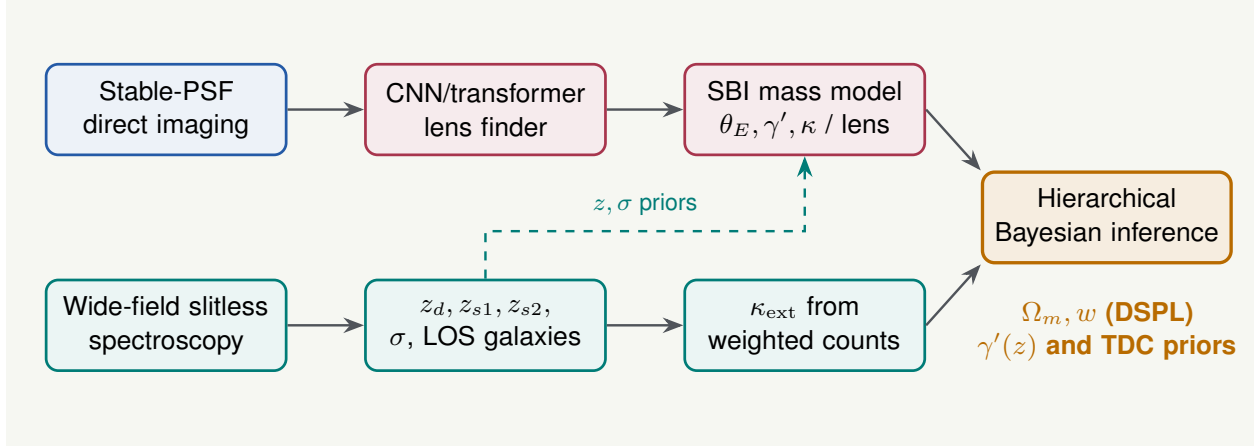

Samples approaching $10^5$ systems require automated lens identification and mass reconstruction (Figure~\ref{fig:lenspipe}). Convolutional and transformer classifiers examine stable-PSF imaging. Controlled tests show higher classification efficiency than visual inspection \cite{metcalf2019}. Similar classifiers have selected hundreds of candidates from millions of ground-based catalogue sources \cite{jacobs2019} and now support space-based lens catalogues \cite{walmsley2025}. Slitless spectra provide an independent selection through detection of background emission lines superposed on a foreground red galaxy and supply a source redshift at discovery.

Classical Markov-chain lens reconstruction requires days to weeks per system and does not scale to $10^5$ lenses. Convolutional estimators evaluate lens parameters approximately $10^7$ times faster than conventional maximum-likelihood modelling \cite{hezaveh2017}. Bayesian neural estimators return parameter posteriors \cite{perreaultlevasseur2017}, and population studies combine per-lens posteriors with approximately $9\%$ precision into sub-percent $H_0$ constraints over several hundred simulated lenses without detected bias \cite{park2021}. The proposed analysis uses neural posterior estimation trained on mission-specific PSF and noise realizations to infer $\theta_E$, $\gamma'$, ellipticity, and $\kappa$ for each lens. A hierarchical Bayesian model jointly constrains cosmology and population distributions of density slope, anisotropy, and mass-sheet terms while marginalizing individual lenses \cite{wagnercarena2021,legin2023}. Spectroscopic redshifts and dispersions supply informative constraints on individual mass models and the distance scale. DSPL $\beta$ measurements and the $\gamma'(z)$ distribution enter the same hierarchical inference.

Mass reconstruction requires stricter validation than source identification. The cited classifiers have been tested on real ground-based and space-based images while the neural mass estimator must initially learn from simulated mission PSFs and noise. Simulation-based posteriors become overconfident when the training distribution differs from observed lenses. Before inclusion in the cosmology sample, individual and hierarchical posteriors must reproduce confirmed precursor lenses and pass injection-recovery tests with withheld orientations and PSF realizations. Section~\ref{sec:deblending} applies the same principle to spectral deblending priors. Automated inference supplies the mass estimates while independent recovery tests determine whether the posteriors satisfy the cosmology requirements.

\subsubsection{Weak-Lensing Cross-Checks and Program Scope}

Competitive cosmic-shear measurements require the much larger imaging areas of Rubin, \textit{Euclid}, and \textit{Roman} and are not a primary mission product. Targeted weak lensing nevertheless provides an internal mass-profile test. Stacked galaxy--galaxy and group or cluster shear around strong-lens deflectors constrains mass beyond the Einstein radius independently of stellar dynamics. Joint strong and weak lensing tightens the total mass profile for cluster-scale deflectors. The measurements use the same imaging acquired for lens identification and require no additional observations.

The Years~2--5 strong-lensing program uses flagship imaging and fields shared with external time-delay campaigns. Targeted spectroscopy assigns $3\times2$--$3\times4$ hr over three orientations to the highest-value deflectors and all DSPL candidates. Deliverables include a spectroscopically confirmed lens and DSPL catalogue, posterior mass models, a population measurement of $\gamma'(z,M_\ast)$, and line-of-sight constraints on $\kappa_{\rm ext}$ for external time-delay studies. DSPL distance ratios provide an independent constraint on $(\Omega_m,w)$. Stable imaging, multiplexed spectroscopy, and validated automated inference therefore support a population-level cosmology program rather than a single-object $H_0$ claim.

\subsection{Lyman-\texorpdfstring{$\alpha$}{alpha} Forest, Absorption Systems, and Reionization}\label{sec:lyaforest}

The Lyman-$\alpha$ forest is the dense set of H\,\textsc{i} absorption lines imprinted on the spectrum of a background source by neutral hydrogen in the intergalactic medium (IGM) along the line of sight. Each absorber records the density, temperature, and peculiar velocity of diffuse gas at the redshift where Ly$\alpha$ ($\lambda_0=1215.67$\,\AA) comes into resonance. The forest is therefore a continuous tracer of large-scale structure across $2\lesssim z\lesssim4$, where few other tracers exist, and it underpins several distinct cosmological measurements. At higher column density the same absorbers steepen into Lyman-limit systems and damped Ly$\alpha$ absorbers, discrete tracers of circumgalactic and interstellar gas that this section surveys alongside the diffuse forest. We first set out what the forest delivers in principle and then quantify the conservative $R=1000$ reference case. The final instrument may adopt $R=5000$ after the cosmology trade.

\subsubsection{The Forest as a Cosmological Probe}

Four measurements give the forest its cosmological weight. The first is the one-dimensional transmitted-flux power spectrum $P_{\rm 1D}(k)$, which probes the linear matter power spectrum on the smallest scales accessible to large-scale structure and is therefore sensitive to the summed neutrino mass and to the free-streaming of warm dark matter. The eBOSS measurement used about 44{,}000 quasars at $z>2.1$ \cite{chabanier2019}. The DESI DR1 analyses now reach beyond 300{,}000 Ly$\alpha$ quasars \cite{karacayli2025,ravoux2025}. Forest data set the strongest astrophysical lower bounds on the warm-dark-matter particle mass with $m_{\rm WDM}\gtrsim3.3$--$5.3$\,keV for an early-decoupled thermal relic \cite{viel2013,irsic2017}.

The second measurement is the thermal history of the IGM, encoded in the temperature-density relation. Power-spectrum and line-width analyses trace the gas temperature $T_0$ rising to $\sim1.4\times10^4$\,K near $z\approx3$ as He\,\textsc{ii} reionizes, and then cooling toward lower redshift \cite{walther2019,gaikwad2021}. The third is the baryon acoustic oscillation scale at $z\simeq2.3$, obtained from the three-dimensional correlation of Ly$\alpha$ absorption across many sightlines. This is the only BAO measurement deep in the matter-dominated era. eBOSS measured it from more than 210{,}000 quasars \cite{dumasdesbourboux2020}. DESI now reaches about 2\% precision from over 420{,}000 Ly$\alpha$ spectra \cite{desi2024lya}. The fourth is Ly$\alpha$ forest tomography. When sightlines are dense enough, the forest can be inverted into a three-dimensional map of IGM density that resolves cosmic-web filaments, voids, and protoclusters at $z>2$ \cite{lee2018clamato,newman2020latis}.

\subsubsection{Spectral Resolution and Detectable H\,I Columns}\label{sec:lyaresolution}

The interpretation of the forest depends critically on spectral resolution. Individual forest absorbers have Doppler parameters of order $b\simeq20$--$30$\,km\,s$^{-1}$, set by the IGM temperature and turbulence \cite{rudie2012,rudie2013}. Resolving such a line requires a velocity resolution well below its width. Echelle spectrographs reach the required $R\gtrsim30{,}000$--$40{,}000$. The $R=1000$ candidate has a width of $300$\,km\,s$^{-1}$ and the $R=5000$ candidate has a width of $60$\,km\,s$^{-1}$. Neither candidate resolves the narrowest forest components into individual Voigt profiles. The $R=5000$ design nevertheless preserves more flux structure and improves statistical forest measurements without reducing integrated photon-limited S/N.

\begin{figure}[htbp]
\centering
\includegraphics[width=0.85\textwidth]{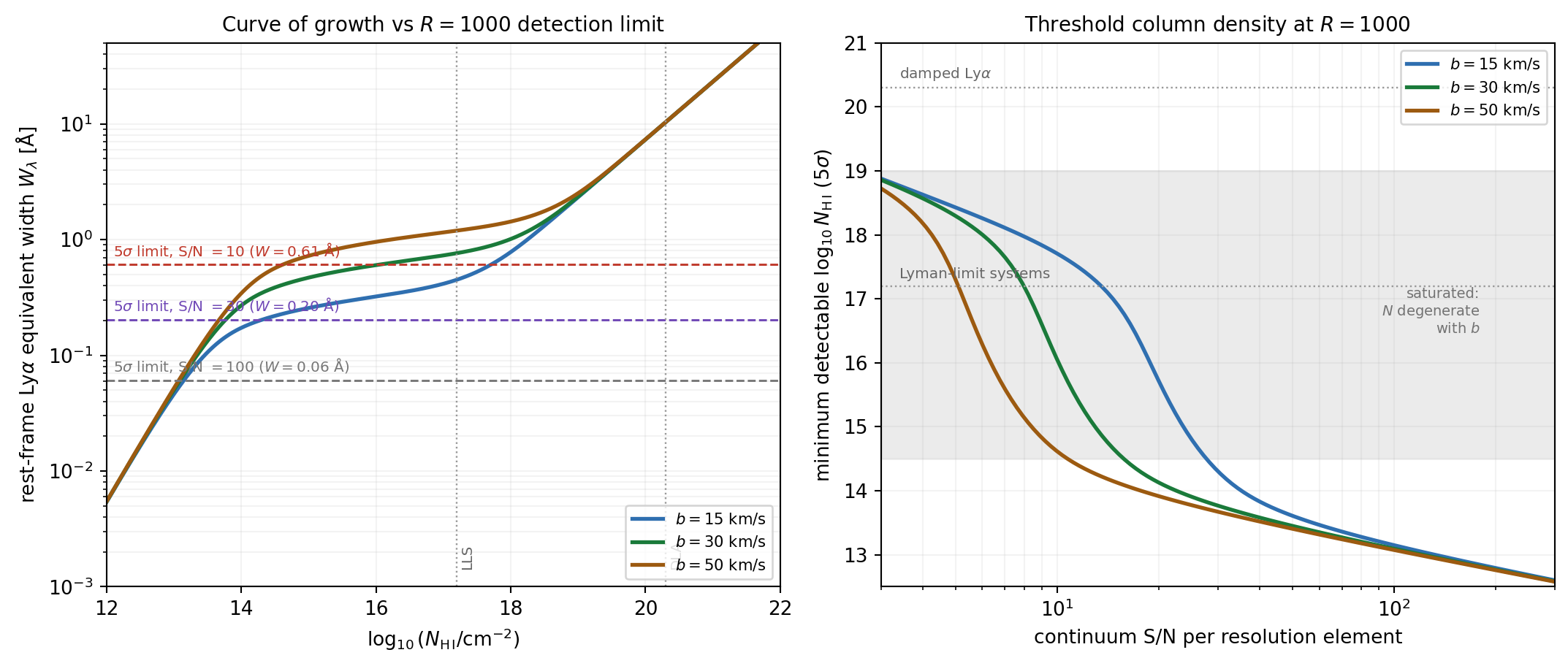}
\caption{H\,\textsc{i} column-density sensitivity of the $R=1000$ Ly$\alpha$ spectrum. The left panel gives the Ly$\alpha$ curve of growth from Voigt profiles with Doppler parameters $b=15$, 30, and 50\,km\,s$^{-1}$. Dashed lines give the $5\sigma$ equivalent-width limits from Eq.~\eqref{eq:wmin} for several continuum S/N values. The saturated plateau between the linear and damping branches gives a column density degenerate with $b$. The right panel gives the minimum detectable $\log N_{\rm H\,I}$ as a function of continuum S/N. Damped Ly$\alpha$ systems with $N_{\rm H\,I}\geq2\times10^{20}$\,cm$^{-2}$ lie on the damping branch and permit accurate column-density measurements at $R=1000$.}
\label{fig:lyathreshold}
\end{figure}

The spectrum detects absorption above a flux-decrement threshold. Integration of a Voigt profile gives the Ly$\alpha$ curve of growth and relates rest-frame equivalent width $W_\lambda$ to H\,\textsc{i} column density $N_{\rm H\,I}$ in Figure~\ref{fig:lyathreshold}. For an unresolved line, the minimum equivalent width detected at significance $N_\sigma$ in a spectrum of resolution $R$ and continuum signal-to-noise ${\rm S/N}$ per resolution element is
\begin{equation}
W_{\min}^{\rm rest} \;=\; \frac{N_\sigma\,\lambda_0}{R\,({\rm S/N})},
\label{eq:wmin}
\end{equation}
where the $(1+z)$ factor of the observed-frame width and the rest-frame conversion cancel and, therefore, the threshold is independent of redshift. At $R=1000$ and a $5\sigma$ threshold, Eq.~\eqref{eq:wmin} gives $W_{\min}^{\rm rest}=0.61$\,\AA\ at ${\rm S/N}=10$ and $0.20$\,\AA\ at ${\rm S/N}=30$.

Inverting the curve of growth gives the minimum detectable column density in Figure~\ref{fig:lyathreshold} and Table~\ref{tab:lyathresh}. At ${\rm S/N}=10$, the mission detects $N_{\rm H\,I}\gtrsim10^{16}$\,cm$^{-2}$. Diffuse-forest absorbers near $N_{\rm H\,I}\sim10^{14}$\,cm$^{-2}$ require ${\rm S/N}\gtrsim30$. Column densities between approximately $10^{14.5}$ and $10^{19}$\,cm$^{-2}$ occupy the saturated part of the curve of growth. Equivalent width then grows only logarithmically with $N_{\rm H\,I}$ and remains degenerate with the Doppler parameter $b$. An $R=1000$ spectrum detects absorption in that regime but does not determine column density from Ly$\alpha$ alone. The thresholds use the idealized continuum S/N from Section~\ref{sec:etc} and require validation with flight spectral extraction.

\begin{table}[htbp]
\centering
\sffamily\small
\begin{tabular}{@{}ccccc@{}}
\toprule
\textbf{Continuum S/N} & \textbf{$W_{\min}^{\rm rest}$} & \multicolumn{3}{c}{\textbf{minimum detectable $\log_{10}(N_{\rm H\,I}/{\rm cm^{-2}})$}}\\
(per res.\ element) & (\AA) & $b=15$ & $b=30$ & $b=50$ \\
\midrule
5 & 1.22 & 18.4 & 18.3 & 17.3\\
10 & 0.61 & 17.7 & 16.1 & 14.6\\
30 & 0.20 & 14.3 & 13.8 & 13.7\\
100 & 0.06 & 13.2 & 13.1 & 13.1\\
\bottomrule
\end{tabular}
\caption{Minimum detectable H\,\textsc{i} column density at $R=1000$ ($5\sigma$, Eq.~\eqref{eq:wmin}) for three Doppler parameters $b$ (km\,s$^{-1}$). Values with $\log N_{\rm H\,I}\gtrsim14.5$ fall on the saturated plateau of the curve of growth, where the column density is degenerate with $b$ and is not recoverable from the Ly$\alpha$ line alone.}
\label{tab:lyathresh}
\end{table}

Two regimes therefore remain robust at $R=1000$. The diffuse forest is measured statistically through the mean transmitted flux and the large-scale power spectrum rather than line by line. Damped Ly$\alpha$ absorbers, with $N_{\rm H\,I}\geq2\times10^{20}$\,cm$^{-2}$, develop damping wings tens of \AA\ wide in the observed frame, which are fully resolved at $R=1000$ and yield accurate column densities. The mission is thus a statistical-forest and damped-absorber instrument, not an echelle line-fitter.

\subsubsection{Tomographic Mapping with the Wide Survey}

\begin{figure}[htbp]
\centering
\includegraphics[width=0.85\textwidth]{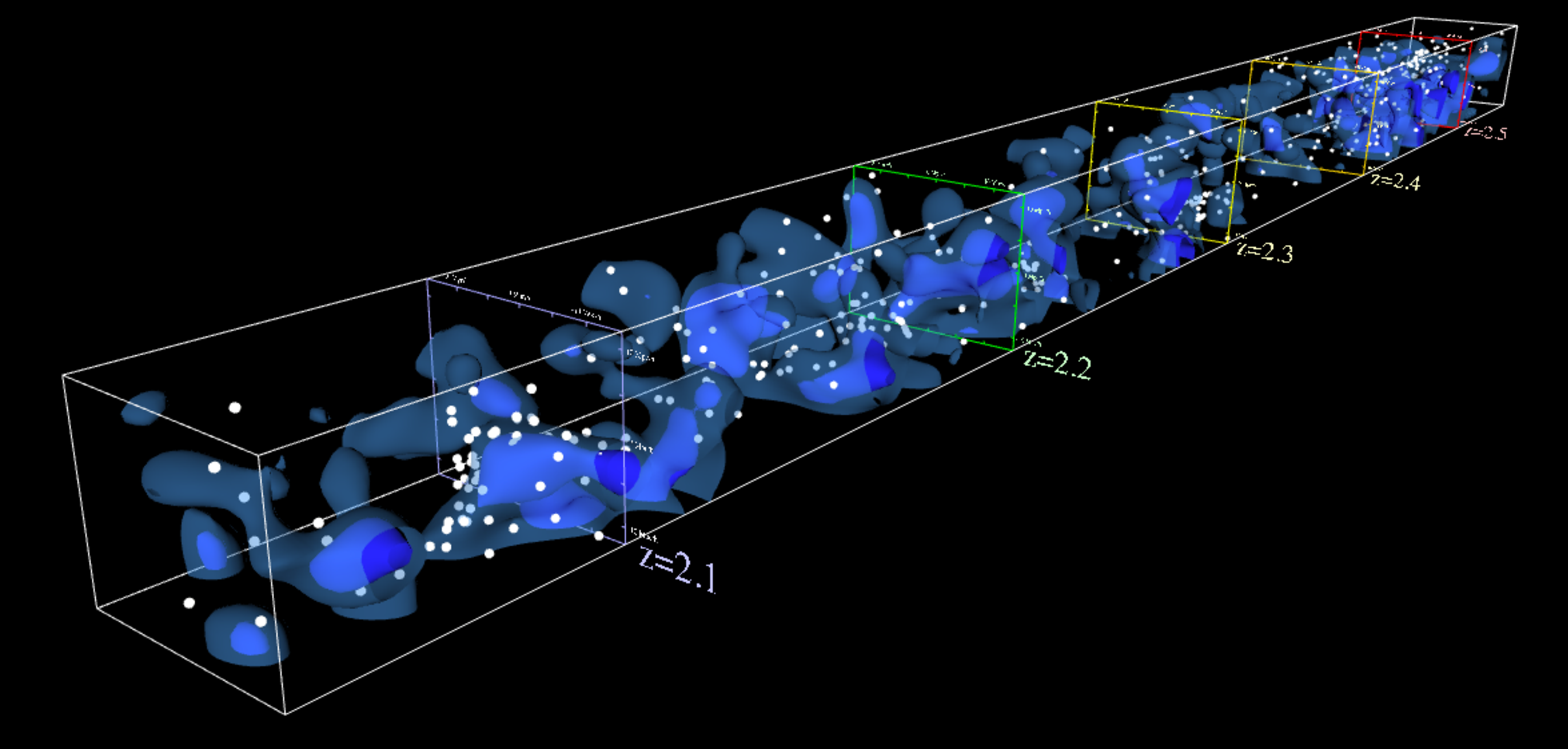}
\caption{External three-dimensional Ly$\alpha$ forest tomography from the CLAMATO DR1 map. Blue isodensity surfaces show reconstructed IGM absorption structures and points mark coeval galaxies. Multiple Ly$\alpha$ forest sightlines jointly determine the three-dimensional gas-density field. The image is from Lee et al. 2018, CLAMATO DR1, arXiv source figure \texttt{x3d\_screenshot\_v0.pdf} \cite{lee2018clamato}.}
\label{fig:lya3dexternal}
\end{figure}

The statistical forest matches the requirements of tomographic mapping, which needs many adjacent sightlines at moderate signal-to-noise rather than high resolution on a few. The achievable map resolution is set by the transverse sightline separation, which in turn depends on the surface density of usable background sources. At $z\simeq2.3$ the density of background galaxies and quasars usable as Ly$\alpha$ backlights rises steeply with depth from $\sim$360 to $\sim$3300\,deg$^{-2}$ between $g=24.0$ and $g=25.0$, giving mean transverse separations of $\sim3.6$ down to $\sim1.2\,h^{-1}$\,Mpc \cite{lee2014tomo}. CLAMATO realized this at $2.05<z<2.55$ with 240 sources over 0.157\,deg$^2$, a mean separation of $2.37\,h^{-1}$\,Mpc, and a map resolution of $2.5\,h^{-1}$\,Mpc (Figure~\ref{fig:lya3dexternal}) \cite{lee2018clamato}. LATIS extended this to 1.7\,deg$^2$ with about 3800 galaxy spectra at comparable resolution \cite{newman2020latis}.

The wide multi-orientation survey obtains faint star-forming galaxy and quasar continua across contiguous fields. Selected fields of approximately 1\,deg$^2$ with S/N of a few per resolution element for $g\lesssim24.5$ background sources provide transverse sampling near $2\,h^{-1}$\,Mpc. The sampling is comparable to CLAMATO and LATIS over a larger area. The reconstructed maps trace protoclusters and the cosmic web at $2\lesssim z\lesssim3$ and measure the density environment of the mission galaxy and quasar samples.

\subsubsection{Damped Ly$\alpha$ and Lyman-Limit System Census}

Damped Ly$\alpha$ absorbers (DLAs, $N_{\rm H\,I}\geq2\times10^{20}$\,cm$^{-2}$) and Lyman-limit systems (LLS, $N_{\rm H\,I}\gtrsim1.6\times10^{17}$\,cm$^{-2}$) trace the neutral-gas reservoir that fuels star formation. Section~\ref{sec:lyaresolution} shows that $R=1000$ measures broad DLA damping wings accurately. SDSS-III catalogued more than 12{,}000 DLAs from ground-based spectra \cite{noterdaeme2012}. Convolutional classifiers identify DLAs with approximately $97\%$ reliability for $\log N_{\rm H\,I}>19.5$ \cite{parks2018}. Application to homogeneous space-based spectra avoids telluric absorption and sky-subtraction residuals. High-resolution measurements at $\langle z\rangle\simeq2.4$ calibrate the column-density distribution \cite{rudie2013}. The wide slitless sample then measures its evolution and the neutral-gas mass density while associated metal lines trace chemical enrichment.

\subsubsection{Redshift Reach for Reionization}

The $0.36$--$3.0\,\mu$m band and quasar sightlines also reach the epoch of reionization. The CMB constrains the integrated ionization history through a Thomson optical depth of $\tau=0.0544\pm0.0073$ and a midpoint near $z_{\rm re}=7.67\pm0.73$ \cite{planck2018}. The redshift evolution and spatial variation of the volume-averaged neutral fraction $\bar{x}_{\rm H\,I}$ remain uncertain. Reionization diagnostics are broad spectral structures and are well matched to the candidate resolving powers. Ly$\alpha$ enters the blue edge at $z\simeq2.0$ and remains in band to $z\simeq23.7$. The Lyman break remains in band from $z\simeq2.9$ to $z\simeq32$. The spectra cover Ly$\alpha$ damping wings, Ly$\alpha$ emission, and rest-frame ultraviolet continua throughout reionization. Damping wings, Gunn-Peterson troughs, and proximity zones span tens to hundreds of observed-frame \AA\ and are resolved by a $300$\,km\,s$^{-1}$ element.

\subsubsection{Quasar Damping Wings and the Neutral Fraction}

The cleanest single-object probe of $\bar{x}_{\rm H\,I}$ is the IGM damping wing redward of Ly$\alpha$ in a luminous $z>7$ quasar. A neutral IGM produces a smooth absorption wing whose shape and depth scale with the neutral fraction. The feature is broad enough that $R=1000$ measures it accurately. Since the discovery of ULAS\,J1120+0641 at $z=7.085$ \cite{mortlock2011}, damping-wing modeling of the growing $z>7$ quasar sample has returned neutral fractions that rise steeply through the EoR. Two quasars at $z=7.09$ and $z=7.54$ give $\bar{x}_{\rm H\,I}\approx0.48$ and $\approx0.60$ \cite{davies2018}, J0252$-$0503 at $z=7.0$ gives $\approx0.70$ \cite{wang2020}, and independent analyses of J1120 and J1342 span $\bar{x}_{\rm H\,I}\approx0.4$ at $z\approx7.1$ to $\approx0.2$--$0.4$ at $z\approx7.5$ \cite{greig2017,greig2019}, with a third quasar at $z\approx7.5$ giving a consistent value \cite{yang2020}. The spread across objects is itself informative because it quantifies the patchiness of reionization.

The end of the process is pinned by the Gunn-Peterson trough in $z\simeq6$ quasars, where the Ly$\alpha$ optical depth steepens sharply and the transmitted flux vanishes \cite{fan2006}. Long dark troughs persisting below $z\simeq6$ require large-scale fluctuations in the neutral fraction \cite{becker2015}. The XQR-30 sample shows that a uniform ionizing background is excluded until $z\approx5.3$. Therefore, reionization ends later than once assumed \cite{bosman2022}. Model-independent dark-pixel counts bound the neutral fraction from above with $\bar{x}_{\rm H\,I}\lesssim0.06$ by $z\simeq5.9$ \cite{mcgreer2015}. Bright $z>6.5$ quasars from Euclid, Roman, and Rubin/LSST will supply the targets, although the subset bright enough for high-quality damping-wing work is modest \cite{mortlock2024}.

\subsubsection{Ly$\alpha$ Emission from Galaxies}

Galaxies provide a complementary and far more numerous probe through their Ly$\alpha$ emission, which is resonantly scattered by neutral hydrogen and therefore suppressed when the surrounding IGM is neutral. The fraction of Lyman-break galaxies showing strong Ly$\alpha$ emission rises toward $z\simeq6$, where roughly half of faint galaxies have rest-frame equivalent widths above $25$\,\AA\ \cite{stark2010,stark2011}, and then drops at $z\simeq7$ in a spatially patchy way \cite{pentericci2014}. Modeling this decline yields neutral fractions of $\bar{x}_{\rm H\,I}\approx0.59$ at $z\sim7$ \cite{mason2018}, $\approx0.49$ at $z\simeq7.6$ \cite{jung2020}, and as high as $\approx0.88$ at $z\simeq7.6$ in one analysis \cite{hoag2019}. The Ly$\alpha$ equivalent-width distribution and its evolution with redshift are the key observables. Both are directly measured from the mission's spectra.

JWST now detects Ly$\alpha$ deep in the epoch of reionization. GN-z11 shows Ly$\alpha$ at $z=10.6$ \cite{bunker2023}. Four galaxies at $z=10.3$--$13.2$ show Lyman-edge damping wings consistent with a largely neutral IGM \cite{curtislake2023}, and Ly$\alpha$ emission at $z\simeq13$ indicates an early ionized region around a luminous galaxy \cite{witstok2025}. A wide-field slitless survey extends the small JWST samples into uniform Ly$\alpha$ emitter and Lyman-break galaxy samples across independent sightlines. Multiple sightlines are required to measure the spatial variation of reionization.

\subsubsection{Synergy with 21\,cm and the CMB}

The redshifted 21\,cm line directly traces neutral gas and complements Ly$\alpha$ measurements. HERA Phase I constrains the 21\,cm power spectrum and requires IGM heating above the adiabatic-cooling floor by $z\approx10.4$ \cite{hera2023}. SKA will provide tomographic maps. Galaxy Ly$\alpha$ emission and quasar damping wings localize ionized and neutral regions for cross-correlation with 21\,cm data. The CMB optical depth \cite{planck2018} supplies an integral constraint on the reconstructed ionization history.

\subsubsection{Program Scope and Synergy}

The Ly$\alpha$ program emphasizes calibration, tomography, and an absorber census rather than DESI-scale BAO. DESI already measures Ly$\alpha$ BAO and $P_{\rm 1D}$ from hundreds of thousands of sightlines \cite{desi2024lya,karacayli2025}. The proposed mission does not match that statistical power, and forest $P_{\rm 1D}$ alone does not tighten the summed-neutrino-mass bound after combination with other probes \cite{chaves2026}. The forest sample contains 3000--5000 quasars at $2.1<z<3.5$. Exposures of 1--2\,hr reach $m_{\rm AB}\lesssim21.5$ while 4--6\,hr calibration fields reach $m_{\rm AB}\simeq22.5$ over 1000--3000\,\degti\ of parent imaging. A separate targeted reionization program observes 50--100 quasars at $z>6.5$ with $3\times2$--$3\times5$\,hr spectra. Damping wings then constrain the neutral fraction. Integrations of $3\times5$--$3\times10$\,hr target 100--200 bright Ly$\alpha$ emitters and Lyman-break galaxies in known overdensities. Continuum S/N and radiative-transfer modelling limit the galaxy measurements. Deliverables include a homogeneous quasar reference sample calibrated against DESI and high-resolution echelle data, megaparsec-resolution tomography fields, a DLA and LLS catalogue with metal-line measurements, damping-wing neutral fractions over $6.5\lesssim z\lesssim8$, the redshift evolution of the Ly$\alpha$ equivalent-width distribution, and ionized-region maps for 21\,cm cross-correlation. High-resolution ground-based spectra determine individual $b$ and $N_{\rm H\,I}$ values for a subset. The mission contribution is uniform space-based coverage of quasar sightlines over a wide area.

\subsection{Low-Surface-Brightness Galaxies and the Dark Matter of Clusters}\label{sec:lsb}

Low-surface-brightness (LSB) galaxies, faint stellar envelopes, and intracluster light lie below the night-sky background and are incomplete in ordinary flux-limited surveys. In clusters, bright member galaxies and X-ray gas are spatially biased relative to the total mass. Diffuse stellar light more closely follows the projected dark-matter distribution. Two observing components recover the missing populations and diffuse emission. Direct images from the wide, medium, and deep flagship tiers provide a blind census of field LSB galaxies and ultra-diffuse galaxies (UDGs) at no additional observing cost. The census also identifies candidates for the void program in Section~\ref{sec:dwarfvoid}. Long integrations at several orientations target approximately 20--40 nearby clusters and map UDG populations and intracluster light as tracers of cluster dark matter.

The three target classes require different confirmation data. Observed surface brightness alone defines an LSB galaxy candidate. A UDG additionally requires a physical effective radius in kiloparsecs and therefore a distance. Emission lines provide slitless redshifts for gas-rich systems while a known cluster redshift supplies the distance for quiescent cluster members. Void membership requires a three-dimensional position because a sky projection mixes foreground, void, and background galaxies. Section~\ref{sec:voidphenom} describes the redshift requirement. Imaging therefore discovers LSB galaxies while spectroscopy confirms physical UDG sizes and void membership.

\subsubsection{Space-Based Requirements for Low Surface Brightness}

\begin{figure}[htbp]
\centering
\includegraphics[width=\textwidth]{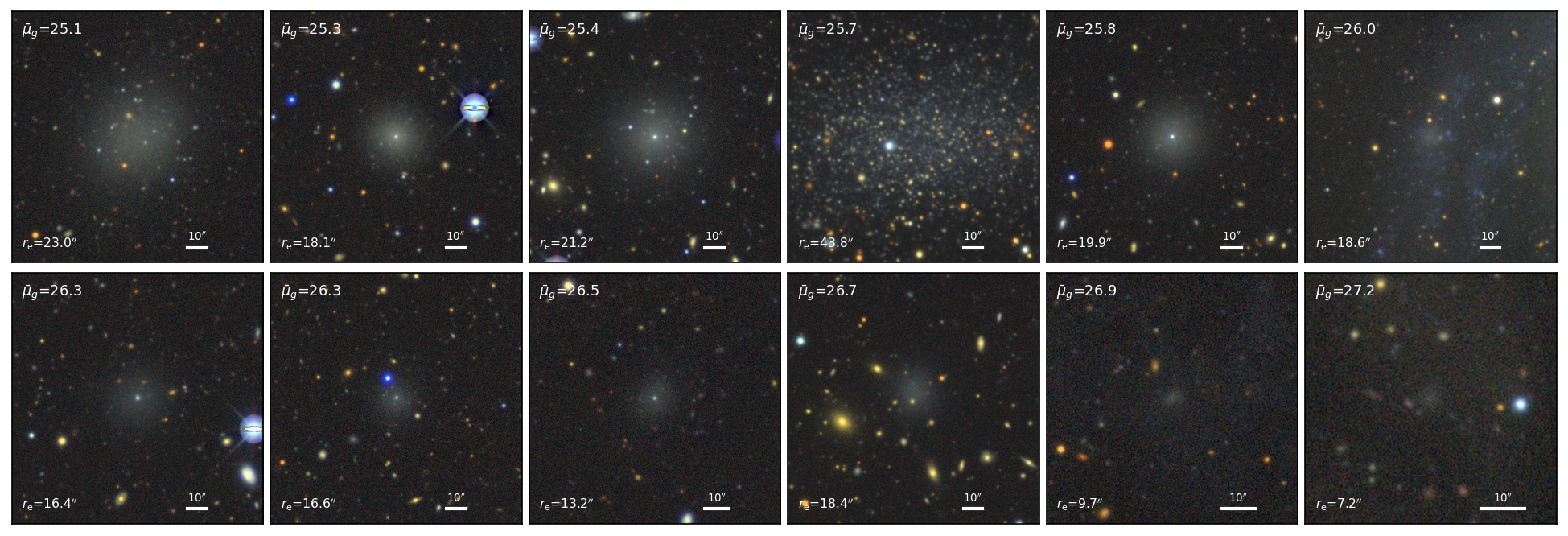}
\caption{Twelve low-surface-brightness galaxies from the published DES DR1 LSBG catalog of 23{,}790 objects \cite{tanoglidis2021}, in DECam Legacy Surveys DR10 colour imaging \cite{dey2019}. Panels are ordered by mean $g$-band surface brightness across twelve equal bins spanning $\bar\mu_g=25.0$--$27.2$ mag\,arcsec$^{-2}$, the range the published catalog populates, each showing the angularly largest object with its $\bar\mu_g$, effective radius, and a $10''$ scale bar. The brightest examples are obvious, the $\bar\mu_g\simeq26$ objects are already marginal, and the faintest nearly vanish in imaging that represents the state of the art for wide ground-based surveys. The parameter space beyond $\bar\mu_g\simeq27$, where the fuzzy-dark-matter candidate Nube lies (Figure~\ref{fig:nube}), is essentially unexplored over wide areas, exactly the regime the mission's $\mu\sim30$--$32$ mag\,arcsec$^{-2}$ reach opens.}
\label{fig:lsbggallery}
\end{figure}

Surface brightness is distance independent in a static Universe but cosmological dimming follows $\mu(z)=\mu_0+10\log_{10}(1+z)$ mag\,arcsec$^{-2}$. A UDG with $\mu_0\simeq24$ mag\,arcsec$^{-2}$ dims by $1.1$ mag at $z=0.3$ and by $3$ mag at $z=1$. Ground-based limits are set by variable airglow, scattered light, and flat-field structures on the same angular scales as diffuse sources. Reliable wide-area limits reach $\mu\sim28$--$29$ mag\,arcsec$^{-2}$ while the purpose-built Dragonfly Telephoto Array reaches $\mu\sim30$--$31$ mag\,arcsec$^{-2}$ \cite{abraham2014dragonfly}. A space platform removes airglow and stabilizes the point-spread function and photometric response. Integration time, zodiacal emission, and Galactic cirrus then set the depth. The \fovlarge\ field combines a wide contiguous area with long integrations. Figure~\ref{fig:lsbggallery} shows twelve galaxies from the DES LSB catalogue \cite{tanoglidis2021} spanning $\bar\mu_g=25.0$--$27.2$ mag\,arcsec$^{-2}$. The faintest examples approach the visibility limit of deep wide-area ground imaging.

\subsubsection{Ultra-Diffuse Galaxies as a Cluster Population}

\begin{figure}[htbp]
\centering
\includegraphics[width=\textwidth]{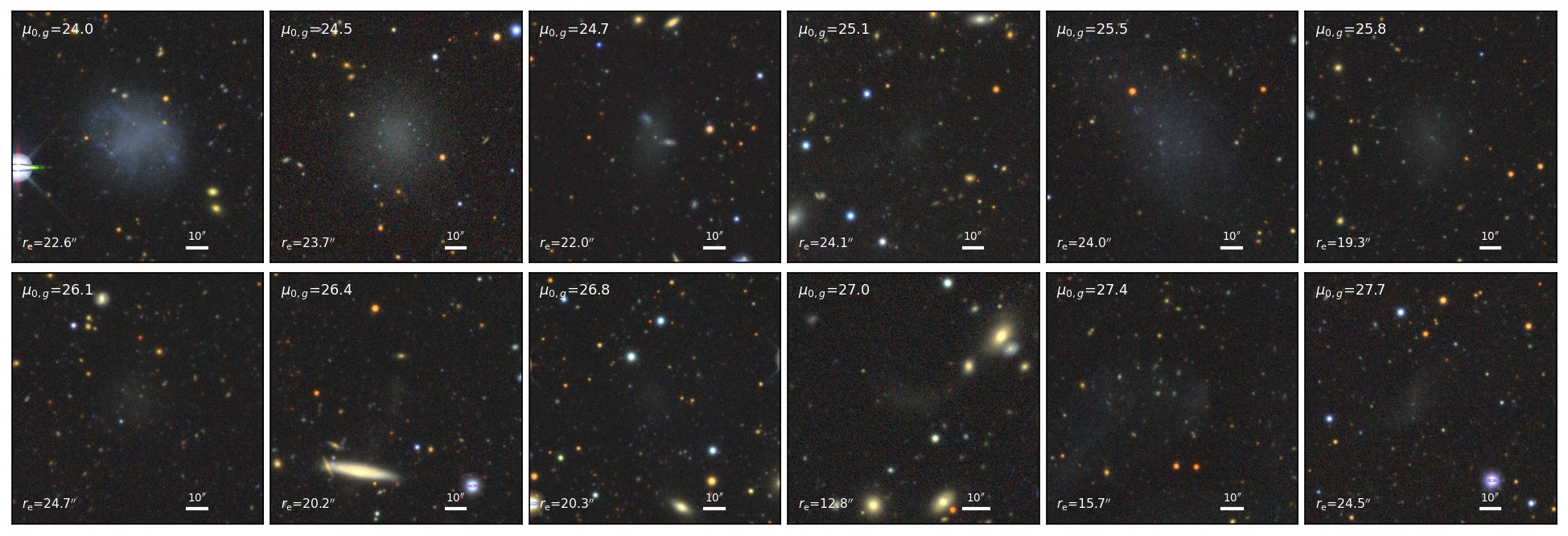}
\caption{Twelve ultra-diffuse-galaxy candidates over the DES southern sky (Dec $<+5^\circ$) from the published SMUDGes catalog of $7{,}070$ candidates selected over the DECam Legacy Surveys footprint with $r_{\rm e}\geq5.3''$ and $\mu_{0,g}\geq24$ mag\,arcsec$^{-2}$ \cite{zaritsky2023}, shown in Legacy Surveys DR10 colour imaging \cite{dey2019}. The panels are ordered across twelve equal bins of central surface brightness spanning $\mu_{0,g}=24.0$--$27.7$ mag\,arcsec$^{-2}$, showing the angularly largest candidate per bin among those with $r_{\rm e}\leq25''$, with a $10''$ scale bar. As in Figure~\ref{fig:lsbggallery} the faint end of the published population sits at the edge of visibility in the deepest wide-area ground imaging. The UDG census, its cluster-to-field comparison, and the dark-matter statistics built on them (Section~\ref{sec:lsb}) therefore require the space-based surface-brightness reach of this mission.}
\label{fig:udggallery}
\end{figure}

The clearest recent demonstration of the hidden LSB population is the discovery of ultra-diffuse galaxies (UDGs). Deep imaging of the Coma cluster with Dragonfly revealed 47 Milky-Way-sized but extremely diffuse galaxies, objects with effective radii $r_{\rm e}\gtrsim1.5$\,kpc yet central surface brightness $\mu_0(g)\gtrsim24$ mag\,arcsec$^{-2}$ and stellar masses one hundred to one thousand times below the Milky Way \cite{vandokkum2015coma}. Follow-up with the wider field of Subaru/Suprime-Cam expanded the Coma census to roughly a thousand such systems \cite{koda2015,yagi2016}. Comparable populations were found in Virgo and other clusters \cite{mihos2015}. That these fragile, low-density systems survive in the cluster tidal field at all implies that many sit inside massive dark matter halos. Beyond clusters, the SMUDGes program has extended the census to the field, selecting $7{,}070$ UDG candidates over the full DECam Legacy Surveys footprint \cite{zaritsky2023}. Figure~\ref{fig:udggallery} shows twelve of its candidates over the DES southern sky, spanning central surface brightness $\mu_{0,g}=24.0$--$27.7$ mag\,arcsec$^{-2}$.

The dark matter content of UDGs is, however, strikingly bimodal. That bimodality is what makes them interesting for the nature of dark matter rather than merely for galaxy formation. At one extreme sits Dragonfly\,44 in Coma. Its stellar velocity dispersion of $\sigma\simeq47$ km\,s$^{-1}$ and its roughly one hundred globular clusters imply a dynamical mass within the half-light radius of order $10^{10}\,M_\odot$ and a dark-matter fraction near unity, comparable to that of a galaxy a hundred times more luminous \cite{vandokkum2016df44}. At the other extreme lie NGC\,1052--DF2 and NGC\,1052--DF4. The globular-cluster and stellar kinematics of these two UDGs give dispersions of only $\sigma\simeq3$--$4$ km\,s$^{-1}$, consistent with the baryonic mass alone and apparently \emph{lacking} dark matter \cite{vandokkum2018df2,vandokkum2019df4}. Dark-matter-free galaxies are not predicted by collisionless cold dark matter without an external stripping or collisional formation channel. Confirming, counting, and locating such systems within their host environments is a direct test of structure formation. The internal stellar dispersions themselves, a few to a few tens of km\,s$^{-1}$, lie below the $300$ km\,s$^{-1}$ resolution element of an $R=1000$ spectrograph and remain the domain of high-resolution follow-up. What the mission supplies is the step on which all of this work depends, a deep, wide, uniform photometric and slitless census. The census finds the systems. It determines cluster membership through the redshift of any emission or strong absorption features and through the radial velocities of associated compact sources. It characterizes their structure and stellar populations across an entire cluster in a single pointing.

\subsubsection{A Field Census of LSBGs and UDGs from the Wide Survey}

The cluster program above is targeted, but the mission produces a second low-surface-brightness census at no additional observing cost because the direct imaging taken for the flagship ELG survey already covers the $100$--$300$\,deg$^2$ wide footprint and its nested medium and deep tiers. Table~\ref{tab:lsbsurvey} gives the surface-brightness reach of that imaging from the exposure-time calculator. Even the wide tier goes roughly three magnitudes deeper than the DES imaging from which the published catalogs of Figures~\ref{fig:lsbggallery} and \ref{fig:udggallery} were selected. The guaranteed floor is therefore set by the published surface densities themselves. The DES DR1 catalog contains $23{,}790$ LSBGs over $\sim$5{,}000\,deg$^2$, about $4.8$ per deg$^2$ to its $\bar\mu_g\simeq27.2$ limit \cite{tanoglidis2021}. The wide footprint alone contains five hundred to fourteen hundred such galaxies. Beyond them the so far unexplored population at $\bar\mu_g\gtrsim27$, the regime of the almost-dark galaxy Nube (Figure~\ref{fig:nube}), is opened over hundreds of square degrees for the first time.

\begin{table}[htbp]
\centering
\sffamily\small
\setlength{\tabcolsep}{4pt}
\begin{tabularx}{\textwidth}{@{}p{2.9cm}p{1.7cm}p{2.35cm}p{2.35cm}X@{}}
\toprule
\textbf{Tier} & \textbf{Exposure} & \textbf{$\mu_{5\sigma}(10'')$, $g$} & \textbf{$\mu_{5\sigma}(10'')$, $H$} & \textbf{Role for the LSB census}\\
\midrule
Wide ($\sim$150\,deg$^2$/cap) & $0.75$ hr & $29.8$ & $29.5$ & Discovery over the full footprint, three magnitudes past the DES selection\\
Medium (30\,deg$^2$) & $3$ hr & $30.6$ & $30.3$ & Structural parameters and colour gradients for the faint tail\\
Deep (1--3\,deg$^2$) & $12$ hr & $31.3$ & $31.0$ & The $\bar\mu\gtrsim28$ regime, unexplored beyond individual objects\\
\bottomrule
\end{tabularx}
\caption{Surface-brightness reach of the flagship-survey direct imaging for the field LSBG and UDG census from the exposure-time calculator over a $10\times10$\,arcsec$^2$ bin. The census comes free with the wide survey. The published DES surface density of $4.8$ LSBGs per deg$^2$ \cite{tanoglidis2021} sets its guaranteed floor.}
\label{tab:lsbsurvey}
\end{table}

Simultaneous spectroscopy converts the imaging catalogue into a physical population measurement. UDG classification requires a distance to convert angular effective radius into kiloparsecs. Published field candidates often rely on assumed cluster association or individual follow-up \cite{zaritsky2023}. Gas-rich LSB galaxies have high line-to-continuum contrast and are favourable slitless targets. Emission-line redshifts therefore confirm or reject the UDG classification and place each source within the three-dimensional ELG density field. The resulting catalogue has uniform selection, physical sizes for the emission-line subset, and environments from voids to groups. Section~\ref{sec:dwarfvoid} and Table~\ref{tab:dmmodels} use those population measurements for dark-matter tests.

\begin{figure}[htbp]
\centering
\includegraphics[width=\textwidth]{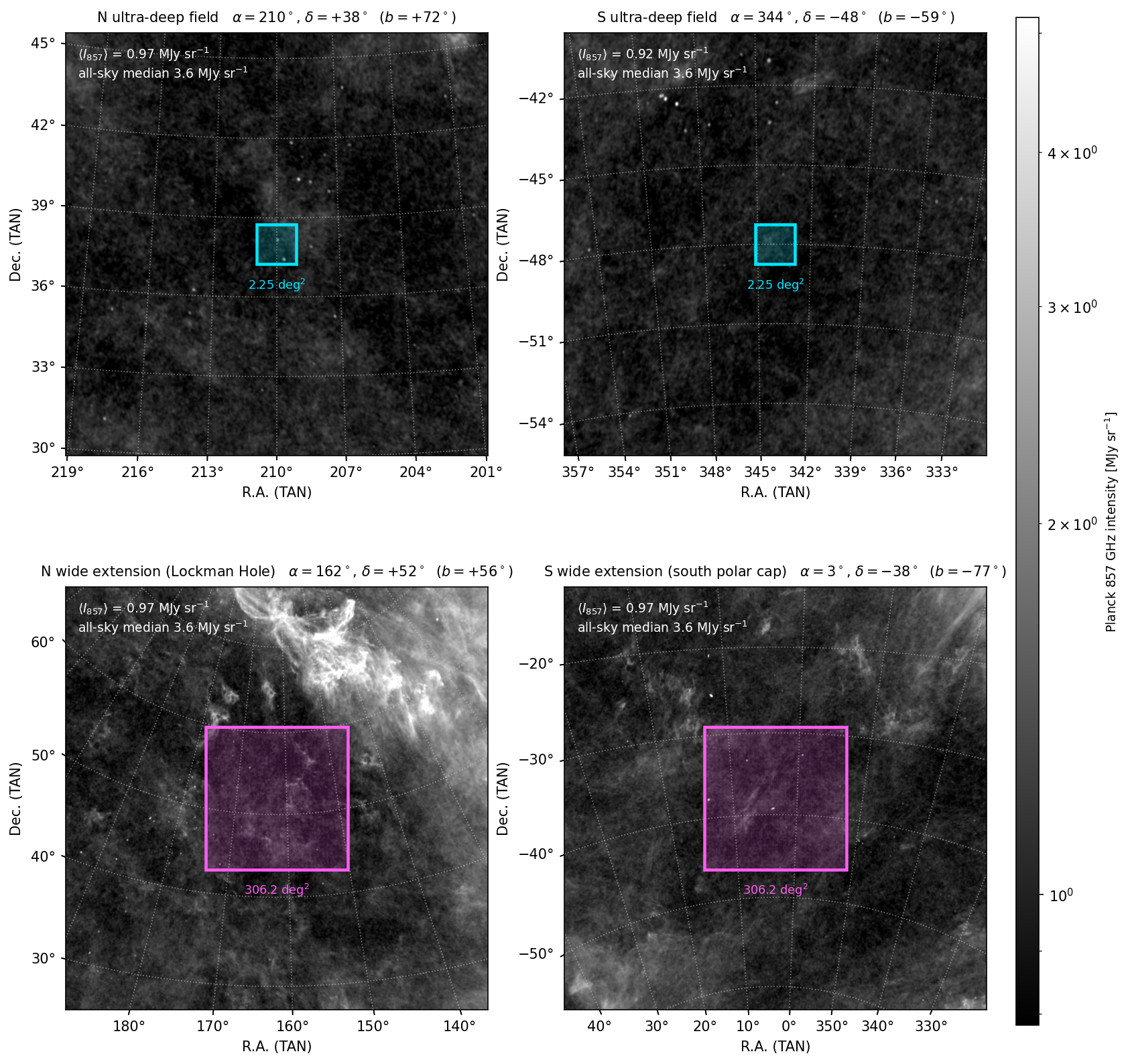}
\caption{Four dedicated imaging fields on the Planck 857\,GHz thermal-dust map \cite{planck2018hfi} with the same gnomonic projection and graticule as Figure~\ref{fig:deepzoom}. The top row shows two ultra-deep fields of $3\times3$ tiles and $2.25$\,deg$^2$ each. The fields minimize mean 857\,GHz intensity within a $7^\circ$ aperture over the sky at $|b|>48^\circ$. The bottom row shows two imaging-only wide extensions of $35\times35$ tiles and approximately $306$\,deg$^2$ each. A matching aperture selects minima at $|b|>55^\circ$ beside the Lockman Hole in the north and the south Galactic pole in the south. The measured mean intensity in each footprint is approximately one quarter of the all-sky median. Figure~\ref{fig:deepzoom} shows the SN Ia cadence fields that lie inside the wide tier.}
\label{fig:imgfields}
\end{figure}

For the faintest diffuse work the program adds two dedicated ultra-deep imaging fields, deliberately far smaller than the blind census, which keeps riding the full survey tiers. Each field is a $3\times3$ block of $30'$ tiles, $1.5^\circ$ on a side and $2.25$\,deg$^2$, sited on the cleanest sky available, found by minimising the mean Planck 857\,GHz intensity over a $7^\circ$ aperture across all $|b|>48^\circ$ sky \cite{planck2018hfi}. The minima fall at $(\alpha,\delta)=(210^\circ,+38^\circ)$, $b=+72^\circ$, in the north, just outside the corner of the flagship Bo\"otes wide tier, and $(344^\circ,-48^\circ)$, $b=-59^\circ$, in the south. Their field-mean $857$\,GHz intensities are $0.97$ and $0.92$\,MJy\,sr$^{-1}$, about a quarter of the all-sky median of $3.6$\,MJy\,sr$^{-1}$ and $\sim30\%$ below the SXDS window ($1.30$\,MJy\,sr$^{-1}$). The top row of Figure~\ref{fig:imgfields} zooms on both fields, confirming directly on the dust map that the footprints are free of cirrus structure. Their locations are marked on the all-sky view of Figure~\ref{fig:skymap}. The fields are observed in the five survey bands, $g$ ($0.5\,\mu$m) and $z$ ($0.9\,\mu$m) on the CCD arm and $J$, $H$, and $K$ ($1.2$, $1.6$, $2.2\,\mu$m) on the HgCdTe arm (Figure~\ref{fig:compareimg}). Because the dichroic feeds both arms simultaneously, the deep $g$ and $H$ pair is collected at once. Sixteen hours per tile in that pair reach $\mu_{5\sigma}(10'')=31.5$ in $g$ and $31.2$ in $H$ from the exposure-time calculator, the regime of the almost-dark galaxy Nube (Figure~\ref{fig:nube}), for about $290$\,hr over both caps. Two shallower passes add $z$, $J$, and $K$ for the colours, stellar-population diagnostics, and photometric redshifts of the catalog.

In the opposite direction the imaging also extends far beyond the spectroscopic footprint since imaging-only tiles are cheap. An imaging-only wide extension adds a $35\times35$-tile field, $17.5^\circ$ on a side and $\sim$306\,deg$^2$, per cap, sited by the same cirrus-minimisation criterion with a matching aperture. The northern extension field, at $(\alpha,\delta)=(162^\circ,+52^\circ)$, $b=+56^\circ$, sits beside the Lockman Hole, the classic low-column window of the northern sky. The southern field, at $(3^\circ,-38^\circ)$, $b=-77^\circ$, lies near the south Galactic pole. Both fields have footprint-mean $857$\,GHz intensities of $0.97$\,MJy\,sr$^{-1}$. A quarter of an hour per tile in the simultaneous $g$+$H$ pair reaches $\mu_{5\sigma}(10'')=29.2$ in $g$ and $28.9$ in $H$, still about two magnitudes past the DES imaging behind Figures~\ref{fig:lsbggallery} and \ref{fig:udggallery}, for about $610$\,hr in total. The extension roughly triples the area of the blind census, raising its guaranteed floor at the published DES surface density \cite{tanoglidis2021} to over three thousand LSBGs. Because the imaging-only wide extension and the two ultra-deep fields have no spectroscopy, they extend the LSBG catalog alone. A candidate found there enters the confirmed UDG sample only once an external facility supplies its distance, and it plays no part in the void-membership test of Section~\ref{sec:voidphenom}, which needs that same redshift-based three-dimensional position. Both extension fields are marked in magenta on Figure~\ref{fig:skymap} and zoomed in the bottom row of Figure~\ref{fig:imgfields}. The flat-field and cirrus-foreground control that this diffuse photometry requires is common to every imaging program of the mission and is treated in Section~\ref{sec:flatcirrus}.

\subsubsection{The Missing Population and the Stellar Mass Function as a Feedback Test}

\begin{figure}[htbp]
\centering
\includegraphics[width=\textwidth]{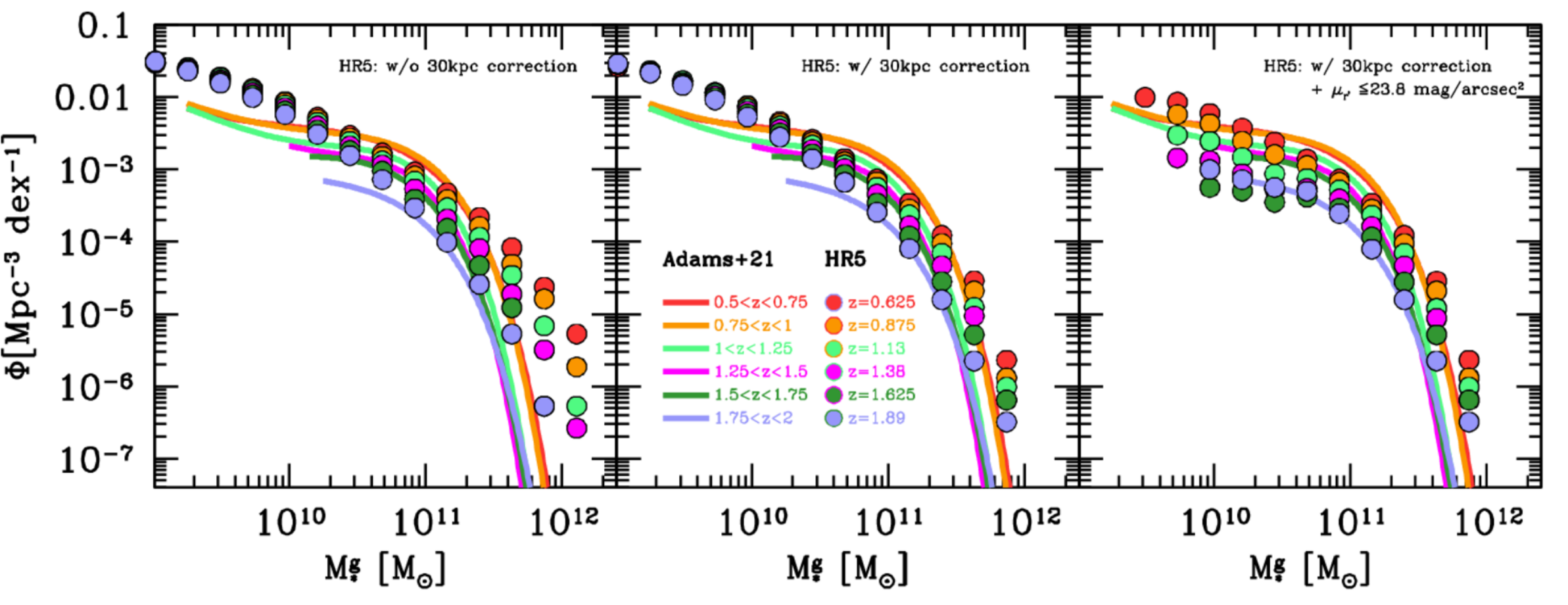}
\caption{Low-surface-brightness galaxies missing from the observed stellar mass function in the mission team's Horizon Run~5 simulation \cite{kim2023gsmf}. Symbols give simulated mass functions at $z=0.625$--$1.89$, and lines give the observed functions from Adams et~al.\ \cite{adams2021} in matching bins. The left panel shows that the unmodified simulation overpredicts low-mass galaxy counts. The middle panel measures simulated stellar masses within the observational $30$\,kpc aperture and corrects the massive end. The right panel additionally applies the observational surface-brightness limit of $\mu_{r'}\leq23.8$ mag\,arcsec$^{-2}$. The low-mass excess then disappears, which identifies undetected galaxies rather than feedback physics as the source of the disagreement. Reproduced from Kim et~al.\ \cite{kim2023gsmf}.}
\label{fig:hr5gsmf}
\end{figure}

The census also matters directly for galaxy formation and for the cosmological galaxy--halo connection through the galaxy stellar mass function (GSMF). The low-mass slope of the GSMF is the primary observable against which the stellar-feedback prescriptions of modern galaxy-formation simulations are calibrated because supernova feedback is what suppresses the steep low-mass rise of the halo mass function down to the shallow observed slope. The flagship simulations tune their feedback until the observed function is reproduced. The mission team's own Horizon Run 5 simulation shows why that calibration rests on shaky ground. Its mock surveys apparently overproduce low-mass galaxies relative to the observed GSMFs of Adams et~al.\ \cite{adams2021} at $0.625\le z\le2$, yet imposing the surface-brightness limit of the observational fits, $\mu_{r'}\leq23.8$ mag\,arcsec$^{-2}$, on the simulated galaxies removes most of the discrepancy (Figure~\ref{fig:hr5gsmf}) \cite{kim2023gsmf}. The observed GSMF is missing its low-surface-brightness population at exactly the masses where feedback is calibrated. Therefore, a feedback model verified against the observed function inherits the truncation. The entanglement is not resolved by simulation alone. What resolves it is a census whose selection function reaches the missing population. The survey provides precisely that, diffuse-light depth three or more magnitudes past the current wide surveys (Table~\ref{tab:lsbsurvey}) with slitless emission-line redshifts that convert detections into stellar masses over the same $0.6\lesssim z\lesssim1.6$ window where the Horizon Run 5 mock surveys locate the bias. The GSMF measurement lives on the spectroscopic survey tiers, where the direct imaging and the slitless spectra share the same $100$--$300$\,deg$^2$ footprint by construction. The imaging-only wide extension and the ultra-deep fields include no spectroscopy. They extend the discovery counts and the surface-brightness reach of the census but enter the mass function only through photometric redshifts. The resulting selection-controlled faint end of the GSMF becomes a direct test of stellar feedback, complementing the feedback-versus-dark-sector degeneracy of Section~\ref{sec:dwarfvoid}. Because the GSMF sets the abundance-matching relation, it also removes a selection systematic from the small-scale galaxy--halo inferences built on it. For the gas-poor quiescent subset the emission-line redshifts are unavailable even on the spectroscopic tiers. The census falls back on the five-band photometric redshifts, a limitation shared with every imaging census.

\subsubsection{Intracluster Light as a Luminous Tracer of Dark Matter}

The second and complementary LSB tracer is the intracluster light (ICL). The ICL is the diffuse stellar component built from stars stripped from galaxies during the hierarchical assembly of the cluster. Those stars are bound to the cluster potential rather than to any single galaxy. Because they are collisionless test particles that have phase-mixed in the global potential, their spatial distribution should follow the total mass. Stacked-cluster photometry first established the ICL as a measurable component at the $\sim10$--$20\%$ level of the total cluster light \cite{zibetti2005}. The decisive result for dark matter is that the projected shape of the ICL closely follows the mass distribution reconstructed from gravitational lensing. In the Hubble Frontier Fields clusters, the ICL isocontours trace the lensing mass contours out to radii of $\sim140$\,kpc more faithfully than the X-ray-emitting gas does \cite{montes2019icl}. That makes the ICL the most accessible luminous tracer of the two-dimensional dark matter distribution in clusters. It offers a cheap, purely photometric route to mapping cluster dark matter on scales and for cluster samples far larger than strong lensing can reach, provided the diffuse light can be measured to $\mu\gtrsim29$ mag\,arcsec$^{-2}$ over the full cluster extent.

A space-based wide-field instrument adds two things to ICL science beyond imaging depth. First, the stable PSF and the absence of airglow allow the diffuse light to be separated cleanly from the wings of bright cluster members, the dominant systematic in ICL photometry. Second, slitless spectroscopy provides discrete dynamical tracers of the same diffuse potential. These are the intracluster planetary nebulae, identifiable by their unresolved [O\,\textsc{iii}]\,$\lambda5007$ emission, and the brighter intracluster globular clusters. Individual line-of-sight velocities at $\sim300$ km\,s$^{-1}$ resolution do not resolve the internal motions of these tracers. The \emph{ensemble} velocity field of hundreds of intracluster planetary nebulae across a cluster, however, constrains the diffuse-light kinematics. It tests whether the ICL and the dark matter share the same dynamical state \cite{montes2022review}.

\subsubsection{Why Diffuse Light Follows the Dark Matter}

The reason the intracluster light is a good mass tracer, while the X-ray gas is not, is dynamical. Intracluster stars are stripped from their parent galaxies during infall and tidal interactions. They then move as collisionless test particles in the global cluster potential. Violent relaxation and phase mixing distribute them according to the total gravitational field, dark matter included. The hot intracluster gas, by contrast, is collisional and pressure-supported. Its distribution is therefore shaped by shocks, sloshing, cooling, and active-galactic-nucleus feedback. During a merger it lags and offsets from the collisionless mass, as the Bullet Cluster makes vivid. The empirical result that the ICL isophotes track the lensing mass more tightly than the X-ray isophotes out to $\sim140$\,kpc is the macroscopic expression of this difference \cite{montes2019icl}. The diffuse light therefore offers a two-dimensional, purely photometric reconstruction of the projected dark-matter shape that can be obtained for far larger cluster samples than strong lensing permits and at higher spatial fidelity than X-ray or Sunyaev--Zel'dovich maps.

The diffuse light encodes assembly history as well as shape. The ICL fraction rises with cluster mass and with dynamical age. The radial color and metallicity gradients of the ICL record the masses and infall times of the galaxies that were disrupted to build it. Deep multi-band photometry of the diffuse component therefore constrains when and how the cluster assembled its mass \cite{montes2022review}. A second, discrete tracer is available in the globular-cluster systems of the cluster galaxies and of the UDGs themselves. Because the number of globular clusters correlates almost linearly with host halo mass, globular-cluster counts provide a photometric halo-mass estimate independent of stellar luminosity. This is exactly the diagnostic that flagged Dragonfly\,44 as unusually massive (about a hundred globular clusters) and NGC\,1052--DF2 and DF4 as anomalous. Over a full cluster it yields a halo-mass-weighted map of the substructure that complements the smooth ICL.

\subsubsection{Ultra-Diffuse Galaxy Formation and What the Census Decides}

The UDGs are not a single phenomenon. The competing formation channels make distinct, testable predictions that a uniform cluster census can separate. In the ``failed galaxy'' picture, UDGs are objects that formed in relatively massive halos but quenched early, retaining large sizes, old red stellar populations, rich globular-cluster systems, and high dark-matter fractions. In the feedback picture, they are ordinary dwarfs whose stellar bodies were puffed up by episodic gas outflows, predicting younger populations, lower dark-matter fractions, and a link to the dwarf mass--size relation \cite{dicintio2017}. In the high-spin picture, they form in dwarf-mass halos with anomalously high angular momentum, which spreads the disk to low surface brightness without changing the halo mass \cite{amorisco2016}. Tidal interactions in the cluster can additionally transform and further diffuse infalling dwarfs. These channels differ in the predicted abundance of UDGs as a function of cluster mass, their radial distribution within the cluster, their globular-cluster richness, their stellar ages and metallicities, and their dark-matter content, including the existence of the dark-matter-deficient tail. No single object decides between them. A homogeneous sample spanning many clusters of different mass, with structural parameters, stellar-population estimates, and globular-cluster counts measured uniformly, does.

The same wide, deep fields that measure the diffuse light and the UDGs also contain a dense background of faint emission-line galaxies whose shapes can be used for weak-lensing mass reconstruction of the foreground cluster. Because the slitless data supply spectroscopic or grism redshifts for a fraction of these background sources, the lensing kernel is calibrated internally. The lensing mass map can be compared directly, in the same field and on the same astrometric frame, with the ICL-derived luminous tracer. This internal cross-check between an independent gravitational mass map and the diffuse-light tracer is what turns ``the ICL follows the mass'' from a calibrated assumption into a measurement for each individual cluster.

\subsubsection{Cluster-Scale Tests of Dark-Matter Self-Interaction}

Clusters are the highest-velocity ``particle colliders'' available for dark matter. They complement the low-velocity dwarfs of Section~\ref{sec:dwarfvoid} and probe the \emph{velocity dependence} of any self-interaction. Two cluster observables bear directly on the self-interaction cross section. Both work by comparing collisionless tracers with the collisional gas. The first is the spatial offset between components in merging clusters. In a collision the galaxies and the dark matter, being effectively collisionless, pass through one another while the gas is shocked and lags behind. The Bullet Cluster showed the lensing mass leading the X-ray gas, a direct demonstration that the dominant mass is collisionless \cite{clowe2006}. Measuring the small residual offset between the collisionless stellar component and the lensing mass across a sample of merging clusters bounds the momentum exchange of dark matter, currently giving $\sigma/m\lesssim0.5$ cm$^2$\,g$^{-1}$ at cluster collision velocities \cite{harvey2015}. The second observable is the central shape and density of relaxed cluster halos. Self-interactions make the inner halo rounder and reduce its central density. The ellipticity and core size of the mass distribution constrain $\sigma/m$ \cite{peter2013}.

\begin{figure}[htbp]
\centering
\includegraphics[width=0.76\textwidth]{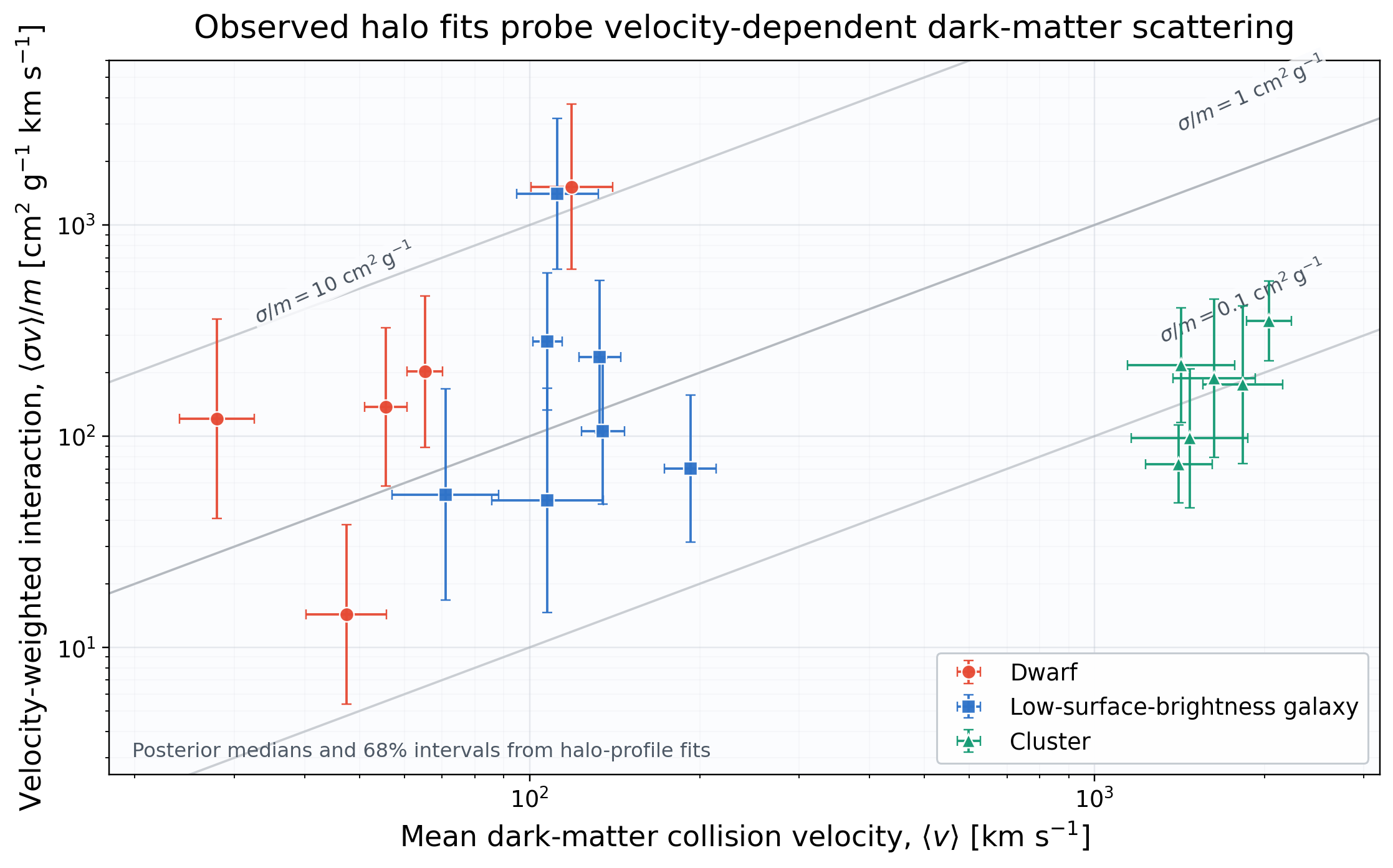}
\caption{Velocity-weighted dark-matter self-interaction rates inferred from observed halo profiles by Kaplinghat, Tulin, and Yu \cite{kaplinghat2016}. The points give posterior medians and 68\% intervals for five dwarf galaxies, seven low-surface-brightness galaxies, and six galaxy clusters. The diagonal lines mark constant $\sigma/m$. Simulation particles and the particle-physics fit from the published figure are omitted because the comparison here concerns the observational inference. The numerical plotting table was reconstructed from the vector coordinates in the authors' arXiv figure because the individual posterior values were not published as a table. The mission replaces the heterogeneous input samples with uniform measurements at the galaxy and cluster velocity scales.}
\label{fig:sidmsigmav}
\end{figure}

The mission supplies the collisionless luminous tracers that both tests require, at higher fidelity than before. The intracluster light and the member-galaxy distribution are the collisionless components whose centroids and shapes should be compared with the lensing mass and the X-ray gas. The deep, space-stable diffuse-light measurement pins the stellar centroid and the halo ellipticity far more accurately than the handful of bright galaxies used in early offset studies. Combined with the dwarf-scale constraints of Section~\ref{sec:dwarfvoid}, a cluster-scale measurement of the same cross section maps $\sigma/m$ across roughly two decades in collision velocity from tens of km\,s$^{-1}$ in dwarfs to thousands in clusters. The velocity span tests a velocity-dependent interaction rather than a single cross section. Figure~\ref{fig:sidmsigmav} shows the observational inference from halo-profile fits. The dwarf and low-surface-brightness samples favor larger effective cross sections than the cluster sample. The characteristic value decreases from order $1$\,cm$^2$\,g$^{-1}$ at galaxy velocities to order $0.1$\,cm$^2$\,g$^{-1}$ at cluster velocities although the object-level posteriors remain broad \cite{kaplinghat2016}. The predicted halo shapes, central densities, and merger offsets for each cross section are produced by the mission team's companion self-interacting-dark-matter simulations and confronted directly with the ICL- and lensing-derived maps.

\subsubsection{Depth, Cadence, and the Cluster Sample}

The defining advantage of the platform for this program is that the surface-brightness limit improves with integration time essentially without the systematic floor that halts ground-based diffuse photometry. With no airglow and a stable photometric response, the noise in a stacked image continues to fall as the square root of exposure until it reaches the astrophysical foreground set by Galactic cirrus and zodiacal light. Reaching $\mu\sim29$--$30$ mag\,arcsec$^{-2}$ is a matter of accumulating exposure rather than of defeating a fluctuating sky. The multi-orientation observing strategy adopted for the flagship survey is doubly valuable here. Rotating the field between visits averages down flat-field residuals and the azimuthal structure of the scattered-light wings of bright stars and galaxies, the dominant systematics in any diffuse-light measurement. A single deep stare does not do that. A target list of $\sim20$--$40$ clusters at $0.02\lesssim z\lesssim0.10$ matches the field of view to the cluster virial extent. It places the UDG population at sizes that are cleanly resolved and keeps the ICL within reach of the surface-brightness limit out to several hundred kiloparsecs. Spread over the mission lifetime as a deep companion to the wide survey, each cluster receives long multi-roll integrations. The sample spans the cluster-mass range needed to test the predicted scaling of UDG abundance and ICL fraction with halo mass.

\subsubsection{Intracluster Planetary Nebulae and Globular Clusters as Discrete Tracers}

The slitless mode turns the diffuse light into a dynamical tracer through two populations of compact objects that a grism survey detects naturally. Because intracluster planetary nebulae radiate almost all of their light in the [O\,\textsc{iii}]\,$\lambda5007$ line, with negligible continuum, they appear in a slitless survey as pure emission-line point sources, exactly the kind of object the instrument is optimized to find. Hundreds to thousands of them populate a nearby cluster. Although the $300$\,km\,s$^{-1}$ resolution element does not resolve the motion of a single planetary nebula precisely, the relevant dynamical signal here is the cluster-scale velocity field, whose dispersion of $500$--$1000$\,km\,s$^{-1}$ is comfortably larger than the resolution. With adequate signal-to-noise on the bright line the per-object velocity is recovered to well within the cluster dispersion. The \emph{ensemble} line-of-sight velocity distribution of the intracluster planetary nebulae measures the kinematics of the diffuse stellar component directly. This tests whether the intracluster light and the dark matter share the same dynamical state and not merely the same projected shape. It extends the mass profile to large radii where lensing and X-ray data weaken. The planetary-nebula luminosity function additionally provides an independent distance estimate for nearby cluster fields \cite{ciardullo2012}.

The second population is the globular clusters. The total number of globular clusters in a galaxy correlates nearly linearly with its host halo mass over several decades, making globular-cluster counts a photometric halo-mass estimator that is independent of the diffuse stellar luminosity \cite{harris2013}. Applied across a cluster, this yields a halo-mass-weighted map of the surviving substructure. Applied to individual UDGs it provides the halo masses that, in combination with the stellar kinematics measured by high-resolution follow-up, flag the dark-matter-rich and dark-matter-deficient extremes. The intracluster globular clusters, which have been stripped into the general cluster potential, trace the same diffuse halo as the intracluster light, giving a discrete spatial tracer to complement the continuous surface-brightness map. Both populations are by-products of the same deep imaging and slitless spectroscopy taken for the diffuse-light and UDG science. They add dynamical information at no extra observational cost.

\subsubsection{The Diffuse Light to the Halo Edge and the Splashback Radius}

The dark-matter content of a cluster is encoded not only in the shape of the inner mass distribution but in the extent and slope of its outskirts. Here also the diffuse light is the accessible tracer. Cold dark matter predicts a physical halo boundary, the splashback radius, marking the apocenter of material on its first orbit after collapse, beyond which the mean density profile steepens abruptly. The location of this feature is set by the recent mass-accretion rate and therefore by the growth of structure \cite{more2015}. The splashback radius has been detected statistically in the radial profile of satellite galaxies around massive clusters \cite{more2016}. Because the intracluster light and the satellite galaxies are both collisionless tracers of the same potential, mapping their profiles to large radius locates the feature and measures the cluster accretion rate. This is a measurement of the dark-matter distribution at the halo edge that complements the core-focused lensing and ICL-shape analyses. It is sensitive to the dark-matter model. Warm, fuzzy, and self-interacting variants modify the abundance and survival of the infalling substructure that defines the splashback transition.

Reaching this regime requires diffuse-light photometry to $\mu\sim30$ mag\,arcsec$^{-2}$ over the full virial extent and into the infall region at several megaparsecs, a combination of depth and contiguous area that ground-based facilities achieve only with great difficulty against their fluctuating sky. The platform's stable response and airglow-free background, together with the wide field tiled across the cluster outskirts, make the outer ICL profile and the radial distribution of faint member galaxies measurable in the same data set. Hence, the splashback feature and the cluster assembly history come essentially for free alongside the core mapping. Combined with the inner ICL shape, the merger offsets, and the globular-cluster and planetary-nebula dynamics, the result is a dark-matter map of each cluster spanning from the core to the accretion boundary.

\subsubsection{Observing Strategy and Survey Reach}

The LSB cluster program is a deep, targeted use of the wide field rather than a blind survey. The adopted plan targets a sample of $\sim20$--$40$ nearby clusters at $0.02\lesssim z\lesssim0.10$, where UDGs are spatially resolved and the ICL fills a large fraction of the field. Long multi-orientation integrations per cluster reach a $1\sigma$ surface-brightness limit of $\mu\sim29$--$30$ mag\,arcsec$^{-2}$ in the stacked image. Table~\ref{tab:lsbreach} summarizes the planning reach. The same exposures yield the slitless spectra of compact emission-line tracers (intracluster planetary nebulae, background emission-line galaxies for weak-lensing shape calibration) and of any star-forming LSB dwarfs. Table~\ref{tab:lsbanchors} places the published measurements that reference each scientific claim of the program next to the capability the mission adds to it.

\begin{table}[htbp]
\centering
\sffamily\small
\begin{tabularx}{\textwidth}{@{}p{3.6cm}p{3.2cm}X@{}}
\toprule
\textbf{Product} & \textbf{Planning depth} & \textbf{Dark-matter use}\\
\midrule
UDG photometric census & $\mu(g)\sim30$ mag\,arcsec$^{-2}$, $r_{\rm e}\gtrsim1$\,kpc & Abundance, radial distribution, structural diversity in the host potential, and the DM-rich versus DM-free fraction\\
Intracluster light map & $\mu\sim29$--$30$ mag\,arcsec$^{-2}$ to $\gtrsim300$\,kpc & Two-dimensional luminous tracer of the projected total-mass (dark matter) distribution\\
Compact dynamical tracers & [O\,\textsc{iii}] PNe, intracluster GCs & Ensemble velocity field of the diffuse component and consistency of light and mass\\
\bottomrule
\end{tabularx}
\caption{Planning reach of the low-surface-brightness cluster program. Internal stellar velocity dispersions of individual UDGs (a few to tens of km\,s$^{-1}$) are below the $R=1000$ resolution element and require high-resolution follow-up. The mission supplies the wide-field deep census, the ICL map, and the ensemble kinematics of compact tracers.}
\label{tab:lsbreach}
\end{table}

\begin{table}[htbp]
\centering
\sffamily\small
\begin{tabularx}{\textwidth}{@{}p{2.6cm}p{5.6cm}X@{}}
\toprule
\textbf{Observable} & \textbf{Published reference} & \textbf{What the mission adds}\\
\midrule
UDG abundance & 47 Coma UDGs from Dragonfly \cite{vandokkum2015coma}, expanded to $\sim\!10^3$ by Subaru imaging \cite{koda2015,yagi2016} & One uniform census across $\sim$20--40 clusters to $\mu(g)\sim30$ mag\,arcsec$^{-2}$\\
UDG dark-matter extremes & DF44 with $\sigma\simeq47$ km\,s$^{-1}$ and $\sim$100 globular clusters \cite{vandokkum2016df44} against DF2/DF4 with $\sigma\simeq3$--$4$ km\,s$^{-1}$ \cite{vandokkum2018df2,vandokkum2019df4} & Globular-cluster counts and membership over whole clusters that select both extremes for kinematic follow-up\\
ICL fraction & 10--20\% of the total cluster light in stacked photometry \cite{zibetti2005} & Per-cluster ICL maps to $\mu\sim29$--$30$ mag\,arcsec$^{-2}$ beyond $300$ kpc\\
ICL as mass tracer & ICL isocontours follow the lensing mass to $\sim$140 kpc in the Frontier Fields \cite{montes2019icl} & The same test repeated cluster by cluster against internally calibrated weak lensing\\
SIDM cross section & $\sigma/m\lesssim0.5$ cm$^2$\,g$^{-1}$ from stellar--lensing offsets in merging clusters \cite{harvey2015} & Deep ICL centroids and halo ellipticities for much larger merger samples\\
Halo edge & Splashback radius detected statistically in stacked satellite profiles \cite{more2016} & Diffuse light and satellites traced to the boundary of individual clusters\\
Depth ceiling & Ground limits $\mu\sim28$--$29$, Dragonfly $\mu\sim30$--$31$ mag\,arcsec$^{-2}$ \cite{abraham2014dragonfly} & ETC reach $\mu_{5\sigma}(10'')=28.9$--$31.9$ for $0.25$--$64$ hr (Figure~\ref{fig:iclimg})\\
\bottomrule
\end{tabularx}
\caption{Published measurements that define the cluster low-surface-brightness program, placed next to the step the mission adds to each. Every quantitative claim of the program traces back to one of these references.}
\label{tab:lsbanchors}
\end{table}

\begin{figure}[htbp]
\centering
\includegraphics[width=0.7\textwidth]{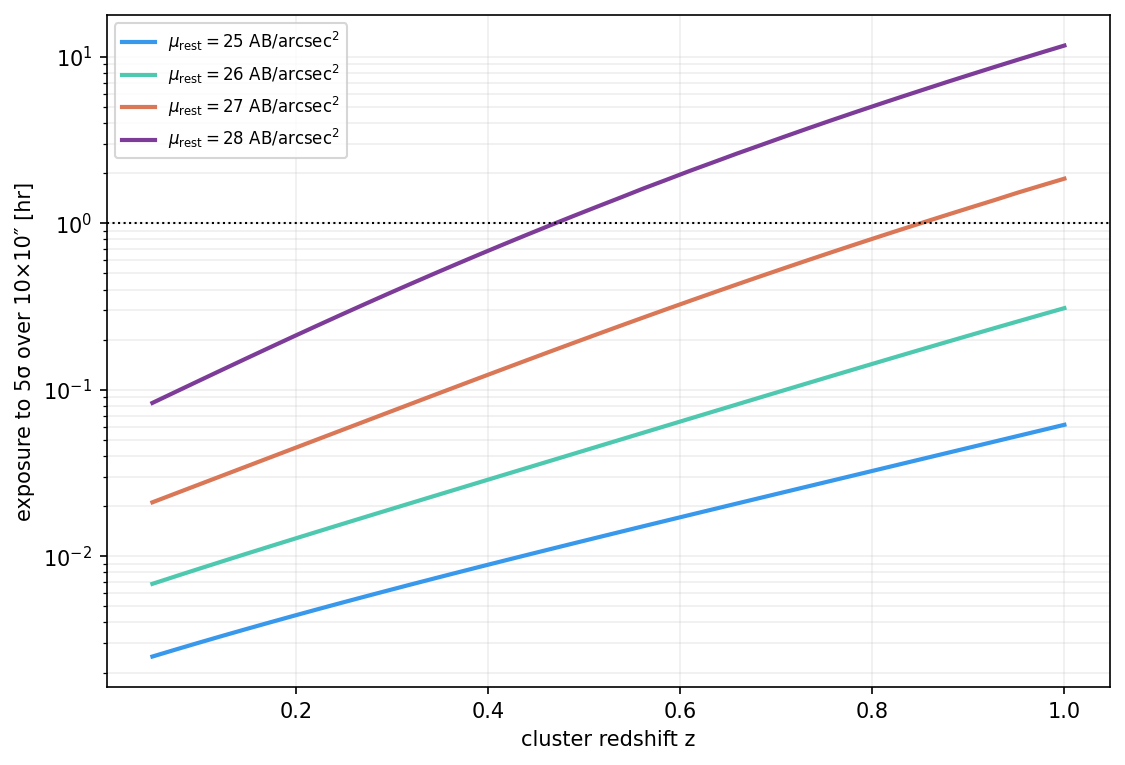}
\caption{Exposure to a $5\sigma$ detection of intracluster light over a $10\times10$\,arcsec$^2$ bin versus cluster redshift in the $H$ band for four rest-frame surface brightnesses, including the $(1+z)^4$ dimming. The $1$\,hr level is marked.}
\label{fig:icl}
\end{figure}

The exposure-time calculator sets the surface-brightness reach directly. Figure~\ref{fig:icl} gives the time to a $5\sigma$ detection of intracluster light binned over a $10\times10$\,arcsec$^2$ region as a function of cluster redshift, including the $(1+z)^4$ cosmological surface-brightness dimming that dominates the trend. A rest-frame $\mu\simeq26$\,AB\,arcsec$^{-2}$ envelope is reached in about a minute to $z\simeq0.3$ and the fainter $\mu\simeq28$ outskirts in about an hour. The diffuse light can be traced toward the cluster edge across the sample.

\begin{figure}[htbp]
\centering
\includegraphics[width=\textwidth]{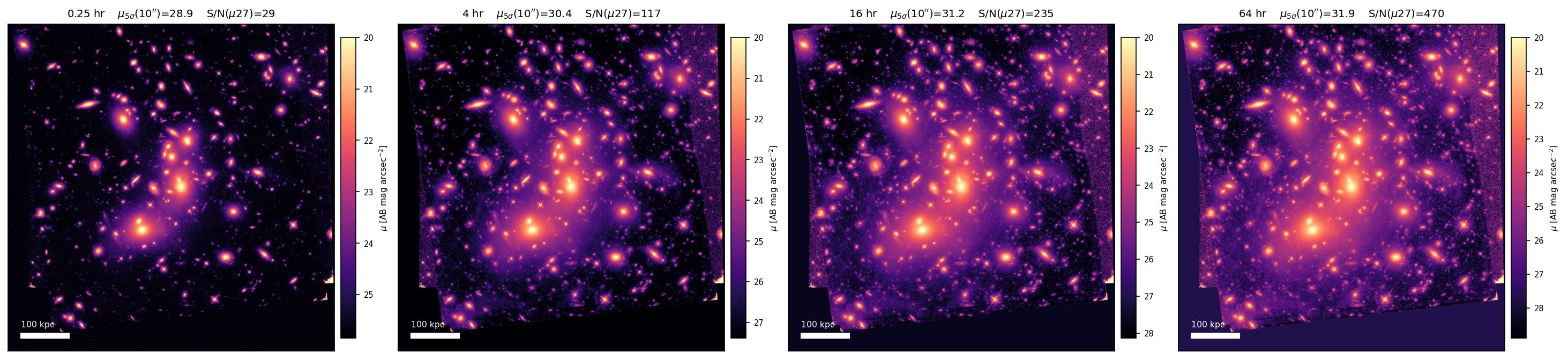}
\caption{Simulated appearance of the Abell~2744 intracluster light to the 3.5\,m ST vs total exposure time, as a dedicated deep program rather than a wide-survey tier. The deep HST Frontier Fields WFC3/IR F160W mosaic \cite{lotz2017} is the true scene ($z=0.308$, observed $H$ band). For each exposure the ETC source, sky, dark, and read noise are added in quadrature to the noise already present in the truth image. The displayed per-pixel noise therefore equals the ETC prediction. The annotated signal-to-noise is the ETC value for a $\mu=27$\,AB\,arcsec$^{-2}$ patch over a $10\times10$\,arcsec$^2$ box. Each panel is stretched between a fixed bright limit and its own $1\sigma$ per-pixel noise floor. The growth of traceable diffuse light with exposure is directly visible. Each panel has its own colour bar and a $100$\,kpc scale bar.}
\label{fig:iclimg}
\end{figure}

A dedicated deep pointing on a single cluster turns this reach into a direct image. Figure~\ref{fig:iclimg} takes the deep HST Frontier Fields WFC3/IR F160W mosaic of Abell~2744 \cite{lotz2017} as the true scene and passes it through the same exposure-time calculator, adding the source, sky, dark, and read noise in quadrature to the noise already present in the true scene so that each simulated frame reproduces exactly the ETC per-pixel noise. Four total integrations of a dedicated intracluster-light program are shown, at a quarter of an hour, four, sixteen, and sixty-four hours, reaching $5\sigma$ surface-brightness limits over a $10\times10$\,arcsec$^2$ bin of $\mu\simeq28.9$, $30.4$, $31.2$, and $31.9$\,AB\,arcsec$^{-2}$. Each panel is stretched between a fixed bright limit and its own per-pixel noise floor. The background tone is common across the four and the eye reads directly how much fainter the diffuse light is traced as the exposure grows. The intracluster envelope that is buried in the noise at a quarter of an hour is mapped smoothly between the cluster galaxies and out to the faint tidal features by sixteen to sixty-four hours. At the longest integrations the reach passes the depth of the space-based Frontier Fields imaging that itself defines the state of the art for cluster diffuse light. That is the concrete sense in which the platform advances intracluster-light science.

\subsubsection{Program Scope and Synergy}

The scope of the program is cluster-dark-matter mapping and an LSB census. The deliverables are a uniform UDG catalog with structural parameters and stellar-population estimates across the target clusters, deep ICL maps that serve as luminous dark-matter tracers for cross-comparison with weak- and strong-lensing mass reconstructions, and a sample of intracluster compact emission-line tracers for ensemble dynamics. The natural external synergies are with Euclid and Rubin/LSST, which provide the wide but shallower imaging that selects target clusters, feeding weak-lensing mass maps. Internally the strong-lens program of Section~\ref{sec:stronglens} supplies the mass models that serve as an independent reference for the ICL-as-tracer hypothesis. High-resolution spectroscopy from large ground-based telescopes remains the source of individual UDG velocity dispersions. The mission's role is the deep, wide, space-stable diffuse-light measurement that no ground-based facility can match. Taken together, the inner ICL shape compared against lensing, the merger offsets of the collisionless stellar component, the globular-cluster and planetary-nebula dynamics, and the outer profile to the splashback radius constitute a dark-matter map of each target cluster from the core to the accretion boundary. The self-interaction cross section at cluster velocities comes with it, complementing the dwarf-scale measurement. The concrete deliverables are the following.
\begin{itemize}[leftmargin=1.4em,itemsep=1pt,topsep=2pt]
\item A uniform ultra-diffuse-galaxy catalog across $\sim20$--$40$ clusters with structural parameters, stellar-population estimates, and globular-cluster counts.
\item A field LSBG and UDG catalog over the full $100$--$300$\,deg$^2$ wide footprint to $\mu_{5\sigma}(10'')\simeq29.8$--$31.3$ in $g$ (Table~\ref{tab:lsbsurvey}) with slitless redshifts, physical sizes, and survey-internal environments for the emission-line subset.
\item Deep intracluster-light maps to $\mu\sim29$--$30$ mag\,arcsec$^{-2}$ serving as luminous tracers of the projected dark-matter distribution from the core to the splashback radius.
\item Collisionless-tracer centroids and shapes for merger-offset and halo-ellipticity constraints on $\sigma/m$ at cluster velocities.
\item Ensemble kinematics of intracluster planetary nebulae and globular clusters for the diffuse-component dynamics.
\item Internal cross-checks against weak- and strong-lensing mass maps built from background sources in the same fields.
\end{itemize}

\subsection{Dwarf and Void Galaxies as Tests of Dark-Matter Physics}\label{sec:dwarfvoid}

Dwarf galaxies are the smallest known gravitationally bound stellar systems and the most dark-matter-dominated. Dynamical mass-to-light ratios reach hundreds to thousands in the faintest Milky Way satellites. Their shallow potentials and low star-formation efficiencies amplify signatures of dark-matter microphysics as well as the baryonic feedback that may produce similar density profiles. Cosmic voids sharpen the comparison because their dwarf populations evolve with fewer mergers, less ram-pressure stripping, and weaker tidal perturbations than dwarfs in groups and clusters. The program will construct one uniformly selected census across void, field, group, and cluster environments. The abundance, central density, stellar population, and gas content of that census will test cold, self-interacting, and fuzzy dark-matter predictions under a common selection function.

\subsubsection{Dwarfs as Dark-Matter Laboratories}

The dark-matter content of a pressure-supported dwarf is measured from the line-of-sight stellar velocity dispersion $\sigma_{\rm los}$ and the projected half-light radius $r_{1/2}$ through the mass estimator
\begin{equation}
M_{1/2}=\frac{4\,\sigma_{\rm los}^2\,r_{1/2}}{G}\simeq930\left(\frac{\sigma_{\rm los}}{\rm km\,s^{-1}}\right)^{2}\!\left(\frac{r_{1/2}}{\rm pc}\right)M_\odot,
\label{eq:wolf}
\end{equation}
which is robust to the unknown velocity anisotropy because it is evaluated at the half-light radius \cite{wolf2010}. Applied to the classical and ultra-faint dwarf spheroidals \cite{mateo1998}, Eq.~\eqref{eq:wolf} yields masses that exceed the stellar mass by one to four orders of magnitude. The ultra-faint dwarfs in particular, with luminosities as low as a few hundred $L_\odot$ and velocity dispersions of $2$--$10$ km\,s$^{-1}$, are the cleanest probes of the halo mass function and inner density structure at the smallest accessible scales \cite{simon2019}. Their abundance, radial distribution, and internal densities each encode a different aspect of the dark-matter model.

\subsubsection{Small-Scale Challenges to Cold Dark Matter}

Collisionless cold dark matter (CDM) reproduces the observed Universe on large scales but faces a set of persistent tensions on dwarf-galaxy scales, reviewed comprehensively by \cite{bullock2017}. The \emph{core--cusp} problem is the mismatch between the steep central density cusps, $\rho\propto r^{-1}$, predicted by dissipationless CDM simulations and the shallower, cored profiles inferred from the rotation curves and stellar kinematics of many dwarfs \cite{deblok2010}. The \emph{missing-satellites} problem is the order-of-magnitude excess of predicted subhalos over observed satellites \cite{moore1999}. The \emph{too-big-to-fail} problem is the observation that the densest predicted subhalos are too centrally dense to host the observed bright satellites \cite{boylankolchin2011}. Most sharply, the \emph{diversity} problem is the wide scatter in the inner slopes of dwarf rotation curves at fixed maximum circular velocity, with galaxies ranging from strongly cored to cuspy where CDM predicts near-uniformity \cite{oman2015}. Baryonic feedback can plausibly erase cusps in gas-rich dwarfs with extended star formation, but it struggles to explain the gas-poor and ultra-faint regime and the full breadth of the diversity, which keeps open the possibility that the resolution lies in the dark sector itself.

\subsubsection{Self-Interacting and Fuzzy Dark Matter}

\begin{figure}[htbp]
\centering
\includegraphics[width=\textwidth]{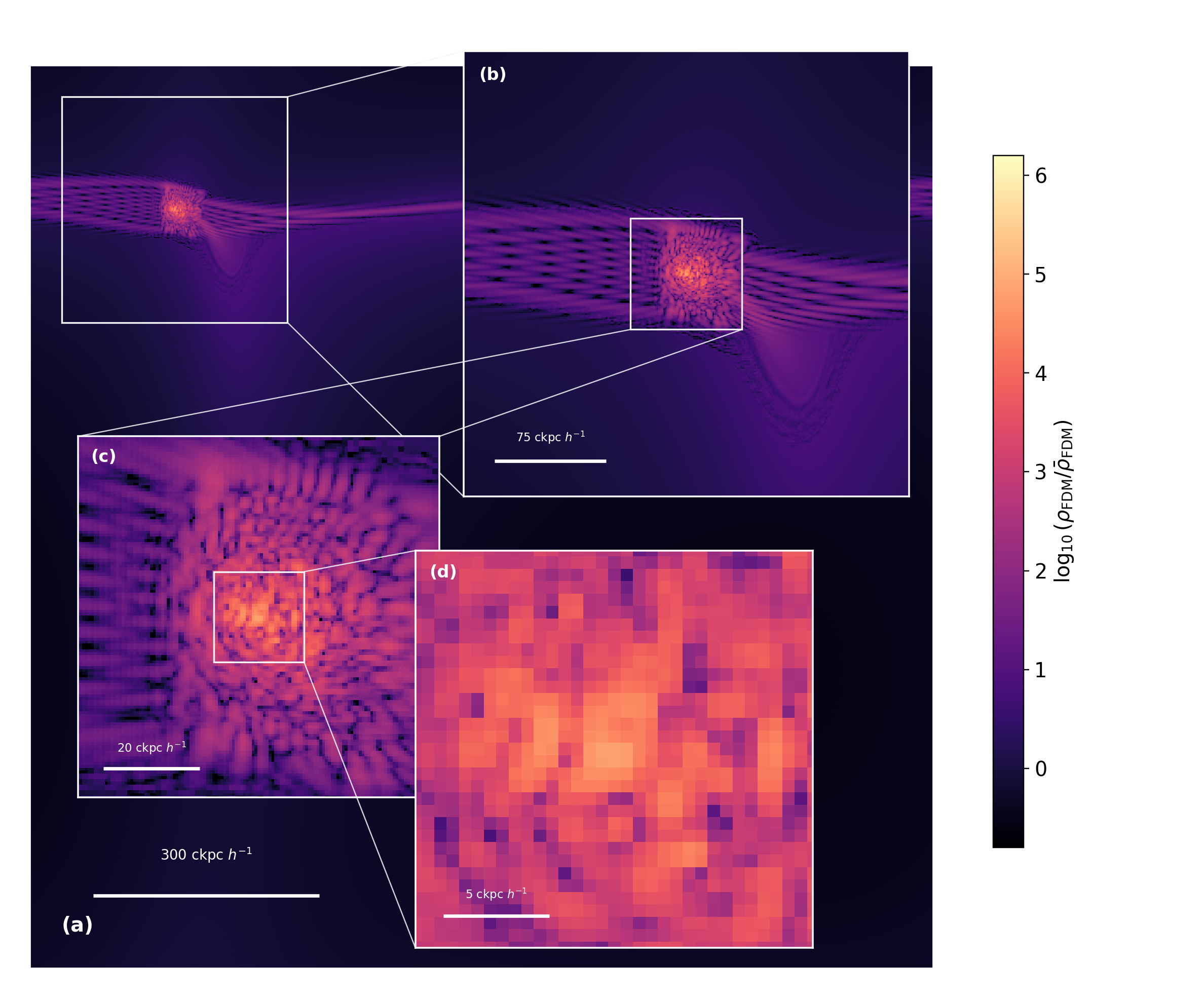}
\caption{Successive zoom into a halo from the mission team's cuRAMSES wave-dark-matter simulation at $z=1.49$ with $m=10^{-22}$\,eV. Panel (a) shows the complete $1.2$\,cMpc\,$h^{-1}$ periodic simulation box. The periodic origin is translated to place the selected halo near the upper-left corner without changing the density field. Panels (b)--(d) follow the marked region through widths of $300$, $75$, and $18.75$\,ckpc\,$h^{-1}$. The finest leaf-cell scale present in the output is $0.59$\,ckpc\,$h^{-1}$. Density is computed as $|\psi|^2$ and is divided by the volume-averaged wave-dark-matter density. A common colour scale follows the resolved interference field from the cosmic environment to the central density enhancement expected around a collapsed FDM halo \cite{schive2014}. A soliton classification is not imposed without a profile fit.}
\label{fig:fdmslice}
\end{figure}

Self-interacting dark matter (SIDM) reduces these tensions by giving dark-matter particles an elastic self-scattering cross section \cite{spergel2000}. Scattering transports heat into the halo center. The rate per particle, $\Gamma\simeq(\sigma/m)\,\rho\,v_{\rm rms}$, becomes of order one over a Hubble time when $\sigma/m\sim1$ cm$^2$\,g$^{-1}$ at dwarf-halo densities and velocities. The energy transport then produces a constant-density core \cite{tulinyu2018}. Inferred core sizes and central densities constrain the cross section at the corresponding relative velocity. Measurements from dwarf to cluster scales may therefore reconstruct a velocity-dependent $\sigma/m$ \cite{kaplinghat2016}. SIDM also predicts gravothermal core collapse to a high central density. The same scattering physics can consequently produce unusually diffuse dwarfs and unusually dense dwarfs at fixed halo mass \cite{ren2019}. An ultralight boson with $m\sim10^{-22}$\,eV predicts a different mechanism. Its kiloparsec-scale de~Broglie wavelength suppresses small-scale structure and replaces central cusps with solitonic cores \cite{hu2000,schive2014}. The soliton core radius follows the core--halo relation
\begin{equation}
r_{\rm c}\simeq1.6\;{\rm kpc}\left(\frac{m}{10^{-22}\,\rm eV}\right)^{-1}\!\left(\frac{M_{\rm halo}}{10^{9}\,M_\odot}\right)^{-1/3},
\label{eq:soliton}
\end{equation}
Dwarf galaxies host the largest predicted cores because $r_{\rm c}$ increases toward lower halo mass. They therefore provide the strongest constraints on the boson mass \cite{schive2014}. Figure~\ref{fig:fdmslice} shows a halo selected from the mission team's cosmological wave-dark-matter calculation. The direct density slice contains interference minima and maxima from the complex field together with a compact central enhancement. A radial profile fit and a numerical-convergence test are required before identifying the enhancement as a soliton. SIDM and FDM predict population distributions rather than a unique signature in one galaxy. Their tests therefore require central-density and core-size distributions across a uniformly selected sample spanning halo mass and environment. The wide-field census will measure those distributions for direct comparison with the mission team's SIDM and wave-dark-matter simulations.

\begin{figure}[htbp]
\centering
\includegraphics[width=0.72\textwidth]{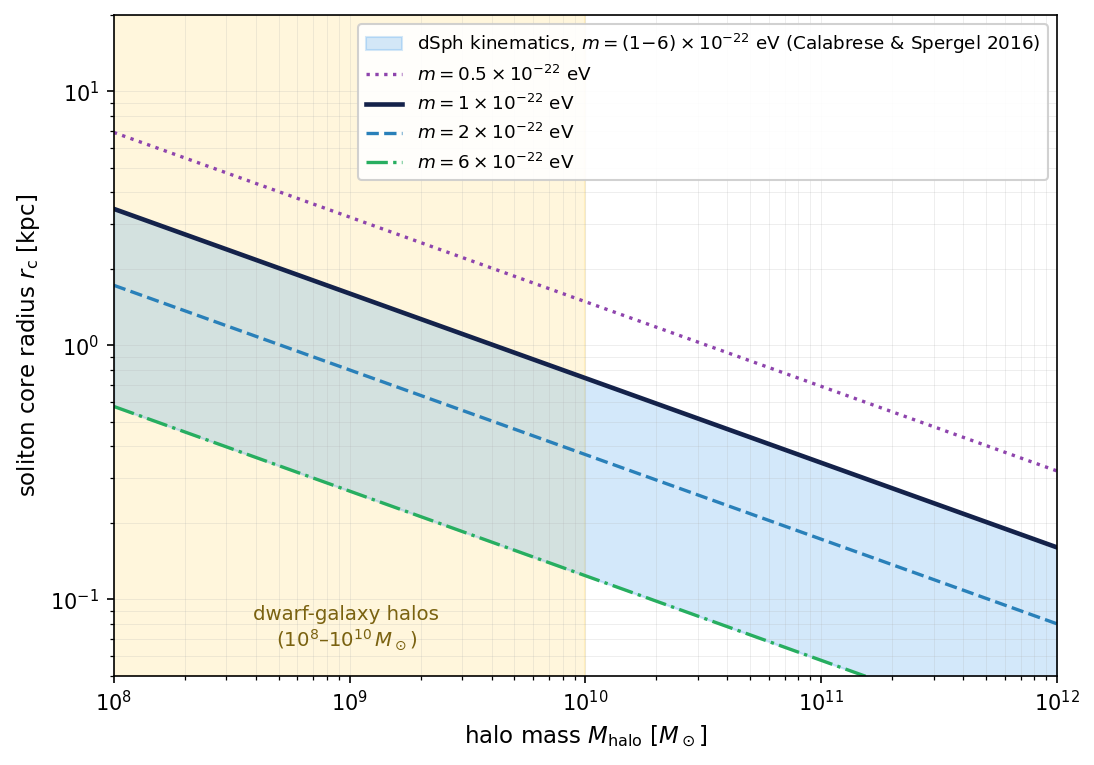}
\caption{The published fuzzy-dark-matter core--halo relation of Eq.~\eqref{eq:soliton} \cite{schive2014} at $z=0$ for four boson masses with the band preferred by dwarf-spheroidal kinematics \cite{calabrese2016}. In the dwarf halo-mass range (shaded) the predicted soliton cores reach a substantial fraction of the galaxy size. The size and central-surface-brightness distributions of a uniform dwarf census therefore constrain the boson mass at the population level.}
\label{fig:soliton}
\end{figure}

\begin{figure}[htbp]
\centering
\includegraphics[width=0.78\textwidth]{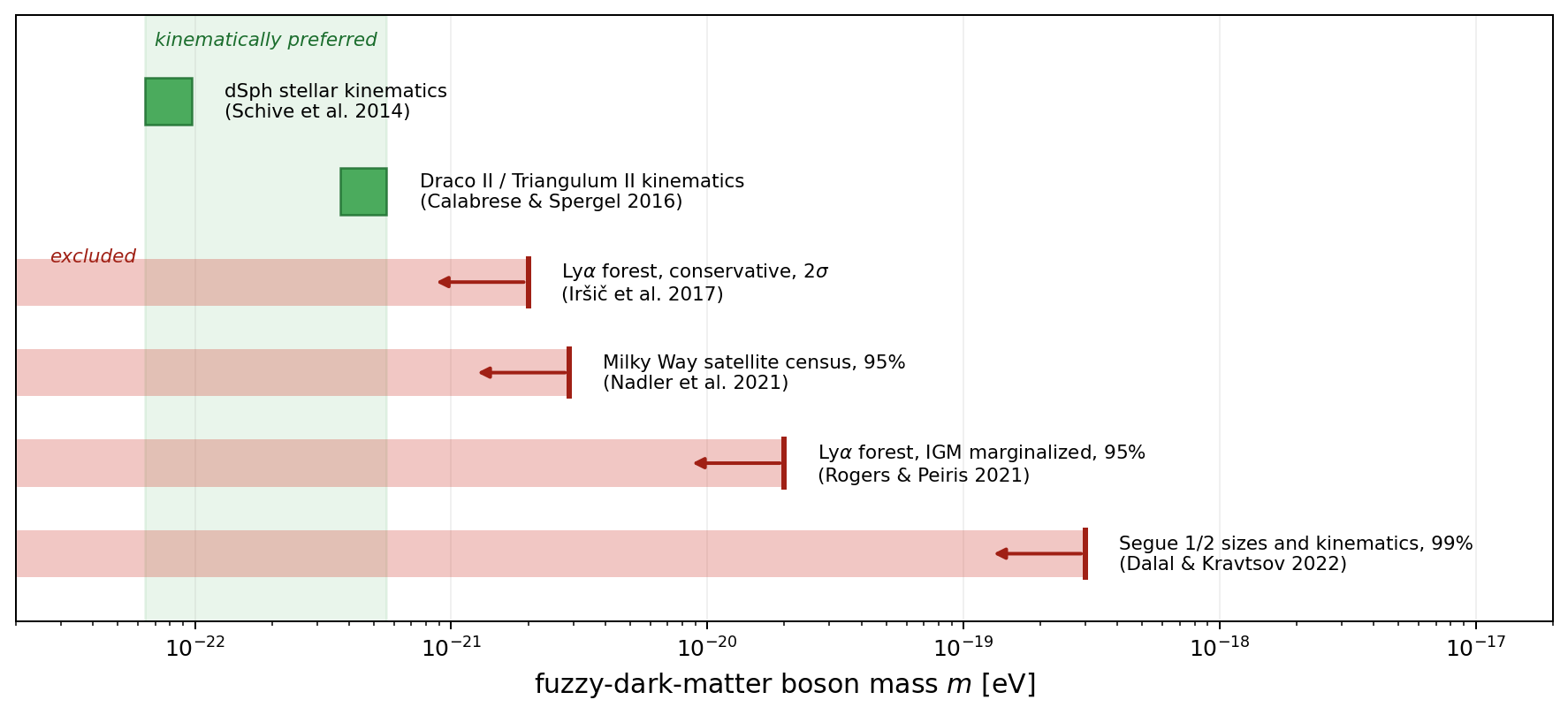}
\caption{Published constraints on the fuzzy-dark-matter boson mass, drawn from the quoted papers. Green bands mark the masses preferred by stellar kinematics of dwarf spheroidals \cite{schive2014} and of the ultra-faint dwarfs Draco~II and Triangulum~II \cite{calabrese2016}. Red regions are excluded by the Ly$\alpha$ forest under conservative thermal-history assumptions \cite{irsic2017fdm}, by the Milky Way satellite census \cite{nadler2021}, by the Ly$\alpha$ forest marginalized over the intergalactic thermal history \cite{rogers2021}, and by the sizes and stellar kinematics of Segue~1 and 2 \cite{dalal2022}. The direct conflict between the kinematically preferred band and the exclusion limits is the current state of the field. The uniform core-size census of this survey tests the preferred band on the very population that produces it.}
\label{fig:fdmconstraints}
\end{figure}

Stellar-kinematic analyses of individual dwarfs give boson masses clustering near $m\sim(1$--$6)\times10^{-22}$\,eV \cite{calabrese2016}, shown against the core--halo relation of Eq.~\eqref{eq:soliton} in Figure~\ref{fig:soliton}. These kinematically preferred masses are under pressure from independent probes, compiled in Figure~\ref{fig:fdmconstraints}. The Ly$\alpha$ forest excludes $m<2\times10^{-21}$\,eV even with a conservative thermal history \cite{irsic2017fdm} and $m<2\times10^{-20}$\,eV once the intergalactic thermal history is marginalized \cite{rogers2021}, the Milky Way satellite census requires $m>2.9\times10^{-21}$\,eV \cite{nadler2021}, and the sizes and stellar kinematics of Segue~1 and Segue~2 exclude everything below $3\times10^{-19}$\,eV \cite{dalal2022}. Taken at face value these bounds close the window in which the boson explains the dwarf cores, but they rest on different astrophysical assumptions and different systems from the kinematic preferences. What settles the question is a single uniform dwarf census whose core-size and central-density distributions confront the preferred band directly on the very population that produces it.

\subsubsection{The Halo Mass Function and the Void Phenomenon}\label{sec:voidphenom}

Beyond the inner density structure, the sheer \emph{abundance} of low-mass halos is a sharp discriminant between dark-matter models. Cold dark matter predicts a halo mass function that continues as a power law far below the dwarf scale, whereas warm dark matter truncates it through the free-streaming of a thermal relic and fuzzy dark matter truncates it through the de~Broglie wavelength that suppresses power below a half-mode scale. Both cutoffs land at dwarf and satellite masses for the parameter ranges of current interest. The census of Milky Way satellites already turns this into a quantitative bound, using the observed number and luminosity function of satellites discovered by the Dark Energy Survey and Pan-STARRS, corrected for completeness \cite{nadler2021}. The satellite measurement is made in the tidally active environment of a massive host. Self-interacting dark matter also accelerates the tidal evaporation and core collapse of subhalos in this environment. The two processes couple subhalo counts and central densities to the host environment.

Voids provide the least tidally processed environment for isolating low-mass halo abundance. Cold dark matter predicts numerous low-mass halos inside voids, yet the observed population remains sparse and is dominated by faint late-type dwarfs. The discrepancy constitutes the long-standing ``void phenomenon'' \cite{peebles2001}. One explanation retains the low-mass halos but suppresses their star formation through the ultraviolet background and inefficient gas cooling below a halo-mass threshold of a few $\times10^{9}\,M_\odot$ \cite{hoeft2010}. Under that explanation, voids should contain gas-bearing optically dark or ultra-faint systems. Warm or fuzzy dark matter instead reduces the number of low-mass halos. The faint-end slopes of the void galaxy luminosity function and H\,\textsc{i} mass function therefore distinguish inefficient star formation from a physical cutoff in the halo mass function. The measurement complements satellite constraints because void halos have not undergone repeated stripping by a massive host. Void abundance, size, shape, and galaxy outflow velocity also constrain structure growth, the dark-energy equation of state, and the summed neutrino mass \cite{hamaus2016}. Accurate redshifts for faint galaxies will improve both the three-dimensional void boundaries and the interior tracer density.

\begin{figure}[htbp]
\centering
\includegraphics[width=\textwidth]{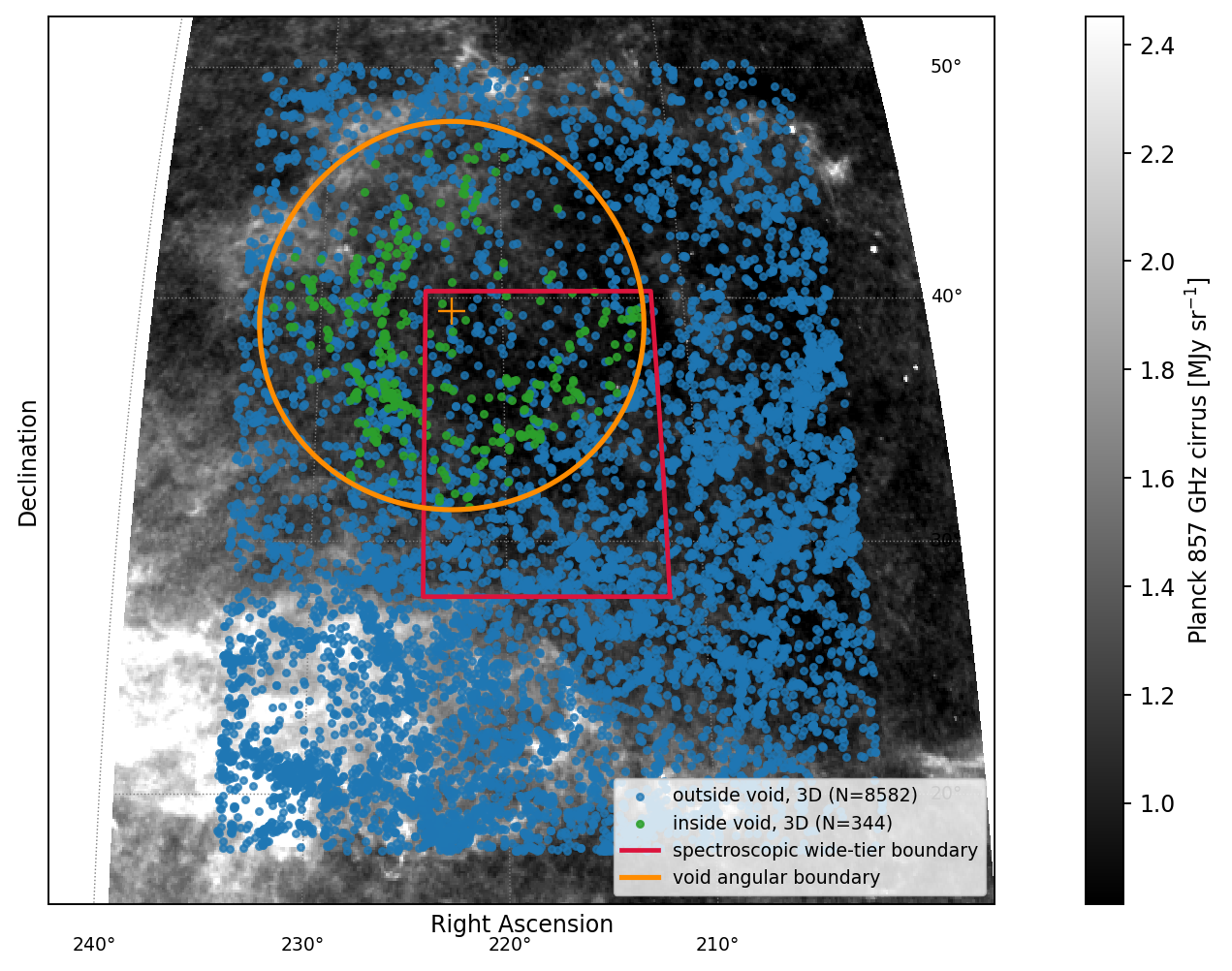}
\caption{Sky projection of the one void of the independent Bayesian catalog of Malandrino et al.\ \cite{malandrino2026} whose center falls inside the flagship Bo\"otes wide spectroscopic tier (Section~\ref{sec:deepfields}), matching the position, redshift, and size of the classic Bo\"otes Void \cite{kirshner1981,kirshner1987}, plotted at the void's own cataloged position and redshift with its radius shown at one-half of the catalog's quoted mean value, restricting the plotted interior to galaxies well away from the wall population that a void-finder radius can otherwise include. Points are real SDSS-I/II Main Galaxy Sample spectroscopic redshifts (\texttt{survey='sdss'}, excluding BOSS/eBOSS), queried live from SkyServer DR17 over the void's own radial extent, and classified by their true three-dimensional comoving distance from the void center. Green points lie inside the void's sphere. Blue points lie outside it, at the same sky position and the same distance shell, but not within the sphere itself since the sky projection also includes galaxies in front of and behind the void along the line of sight. The grey background is the Planck 857~GHz cirrus map of the same field, reprojected through the same transform (the flat-fielding and cirrus control common to every imaging program of the mission is treated in Section~\ref{sec:flatcirrus}). Red marks the spectroscopic wide-tier boundary. Orange marks the void's angular boundary. The projection is Mollweide, mirrored to match the view looking up at the sky rather than down at a map, with right ascension increasing to the left.}
\label{fig:voidwedge}
\end{figure}

Figure~\ref{fig:voidwedge} demonstrates the target selection with observed redshifts and a published void catalog. The independent Bayesian catalog of Malandrino et al.\ \cite{malandrino2026} contains one void whose center lies inside the flagship Bo\"otes wide spectroscopic tier of Section~\ref{sec:deepfields}. Its published position, redshift, and size agree with the Bo\"otes Void identified by Kirshner et al.\ \cite{kirshner1981,kirshner1987}. The plotted selection adopts one-half of the catalog mean radius to exclude most of the wall population. Void membership requires the same redshift-based three-dimensional position needed to assign a physical size to a UDG in Section~\ref{sec:lsb}. The imaging-only wide extension and the ultra-deep fields lack that information, whereas the wide, medium, and deep spectroscopic tiers provide it. SDSS-I/II Main Galaxy Sample redshifts within the corresponding radial shell classify each galaxy by comoving distance from the void center. A sky-only selection would mix interior galaxies with foreground and background systems. The survey will apply the same three-dimensional cross-match to published void catalogs throughout its spectroscopic tiers before tiling the selected interiors.

\subsubsection{Disentangling Baryonic Feedback from the Dark Sector}

Supernova feedback also transforms cusps into cores when repeated gas outflows transfer energy to the dark matter. The resulting density profile may resemble a dark-sector signature. The competing mechanisms predict different correlations. A feedback-driven core requires sufficient integrated star formation, so its radius should correlate with stellar mass and star-formation duration. Feedback becomes energetically inefficient in gas-poor ultra-faint dwarfs. A large core in such a system would favor dark-sector physics \cite{deblok2010,bullock2017}. SIDM relates core size to halo density, velocity, and scattering cross section. The model also predicts diffuse cored dwarfs and dense gravothermally collapsed dwarfs at the same halo mass \cite{kaplinghat2016,ren2019}. FDM predicts the solitonic core scaling of Eq.~\eqref{eq:soliton}. The decisive observable is therefore the joint distribution of central density, stellar mass, star-formation history, and environment rather than one rotation curve. Integrated spectra and multiband photometry will measure stellar age, metallicity, and star-formation history for the uniformly selected sample. Reionization quenching in an otherwise abundant low-mass halo population predicts surviving ultra-faint systems that are predominantly old and metal-poor. A warm- or fuzzy-dark-matter cutoff predicts fewer halos rather than only older stellar populations. The age and metallicity distributions thus provide an independent distinction \cite{kirby2013}. Table~\ref{tab:dmmodels} links each model to its defining scale, current constraints, and census observable. The mission team's cosmological simulations will predict the corresponding multivariate distributions.

\begin{table}[htbp]
\centering
\sffamily\small
\begin{tabularx}{\textwidth}{@{}p{1.3cm}p{4.6cm}p{4.4cm}X@{}}
\toprule
\textbf{Model} & \textbf{Defining scale and constraint} & \textbf{Dwarf-scale signature} & \textbf{Census observable}\\
\midrule
CDM & Collisionless at all scales \cite{bullock2017} & $\rho\propto r^{-1}$ cusps and an unbroken low-mass halo mass function \cite{deblok2010,moore1999} & Faint-end abundance and a near-uniform inner-structure population in tension with the observed diversity \cite{oman2015}\\
WDM & Thermal relic of a few keV, half-mode mass $M_{\rm hm}\sim10^{8}$--$10^{10}\,M_\odot$, bounded from below by the Milky Way satellite census \cite{nadler2021} & Turnover and deficit at the faint end of the halo and galaxy mass functions & Faint-end luminosity functions in field and void environments with one selection function\\
SIDM & $\sigma/m\sim1$ cm$^2$\,g$^{-1}$ at dwarf velocities \cite{spergel2000,tulinyu2018}, $\lesssim0.5$ cm$^2$\,g$^{-1}$ at cluster velocities \cite{harvey2015} & Coexisting diffuse cored dwarfs and gravothermally collapsed dense dwarfs at fixed mass \cite{kaplinghat2016,ren2019} & Distribution of central density and size across the census, plus the outlier target list for kinematic follow-up\\
FDM & Ultralight boson with $m\sim10^{-22}$ eV, kinematic estimates $(1$--$6)\times10^{-22}$ eV \cite{hu2000,schive2014,calabrese2016} & Soliton cores largest in the smallest halos (Eq.~\eqref{eq:soliton}, Figure~\ref{fig:soliton}) plus a mass-function cutoff & Core-size trend with halo mass jointly with the faint-end abundance\\
\bottomrule
\end{tabularx}
\caption{The competing dark-matter models, their published defining parameters and current constraints, the dwarf-scale signatures they predict, and the observable of the uniform census that tests each. The quantitative predictions for every row are supplied by the mission team's companion simulations.}
\label{tab:dmmodels}
\end{table}

\subsubsection{Dark and Almost-Dark Void Galaxies}

\begin{figure}[htbp]
\centering
\includegraphics[width=\textwidth]{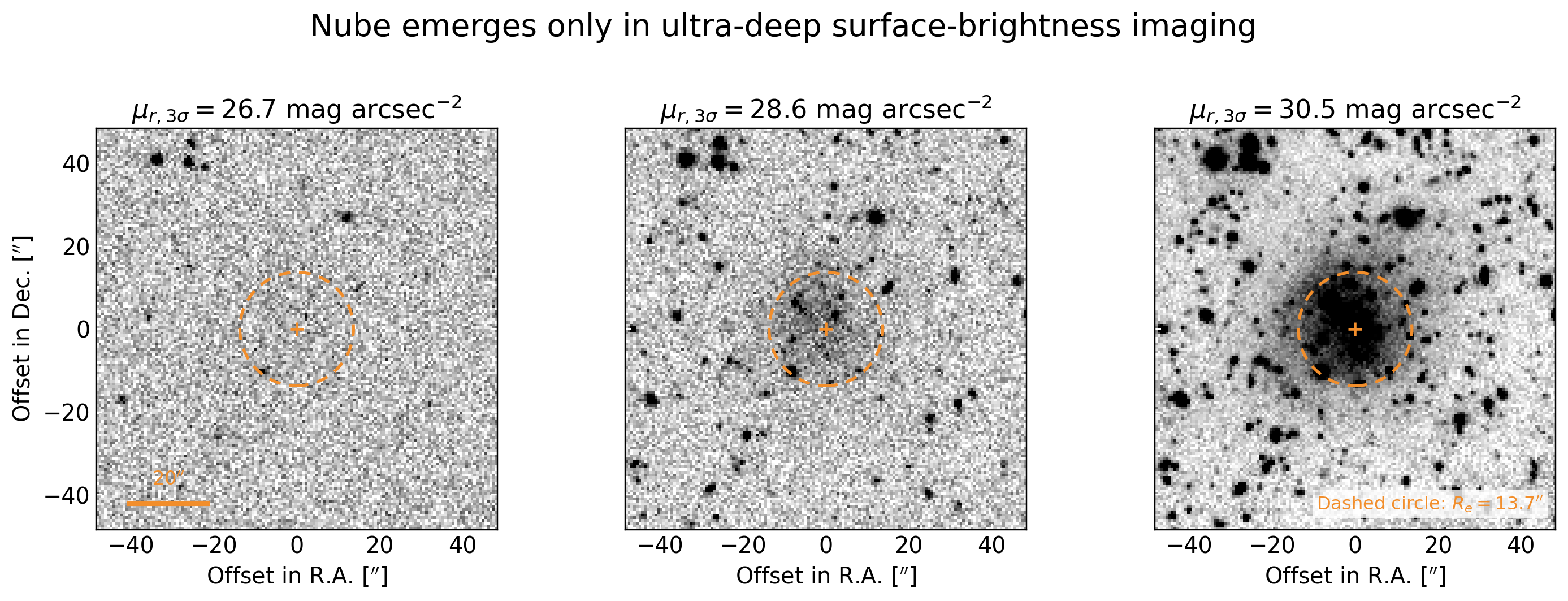}
\caption{The almost-dark galaxy Nube in a $100''\times100''$ cutout from the public HiPERCAM $r$-band mosaic of Montes et al.\ \cite{montes2024nube}. The right panel is the observed image at the measured $3\sigma$ limit of $30.5$\,mag\,arcsec$^{-2}$ in $10''\times10''$ boxes. The middle and left panels use the same observation after calibrated noise was added in flux units to reach $28.6$ and $26.7$\,mag\,arcsec$^{-2}$. The two limits reproduce the Stripe~82 and SDSS depths quoted by the discovery study without replacing the observed Nube morphology. Every panel is rebinned to $0.645''$ pixels and shown with the same signal-to-noise stretch. The dashed circle gives the measured effective radius of $13.7''$. Nube becomes secure only in the deepest image.}
\label{fig:nube}
\end{figure}

A dark-galaxy candidate is a dark-matter halo containing gas but few or no stars. VIRGOHI\,21 illustrates the principal false positive. The H\,\textsc{i} cloud showed apparent rotation and no optical counterpart, leading to an initial dark-object interpretation \cite{minchin2005}, but later observations favored tidal debris \cite{duc2008}. Blind H\,\textsc{i} surveys such as ALFALFA have systematically selected the small fraction of sources without obvious optical counterparts. Coma\,P (HI1232+20) is a metal-poor H\,\textsc{i} cloud with $M_{\rm HI}/M_\star>200$ and only a trace low-surface-brightness stellar component \cite{janowiecki2015}. ALFALFA also identified $115$ isolated gas-rich ultra-diffuse galaxies. The systems are bluer and more irregular than cluster UDGs and may occupy high-spin dwarf halos with very low star-formation efficiency \cite{leisman2017}. Nube contains approximately the stellar mass of the Small Magellanic Cloud distributed over such a large area that conventional imaging missed the galaxy. Its structure has motivated an FDM interpretation \cite{montes2024nube}. Figure~\ref{fig:nube} applies three calibrated surface-brightness limits to one public HiPERCAM observation. Nube remains undetected at $\mu_r=26.7$\,mag\,arcsec$^{-2}$, begins to emerge near the Stripe~82 depth of $28.6$\,mag\,arcsec$^{-2}$, and becomes unambiguous at $30.5$\,mag\,arcsec$^{-2}$. Gas-rich systems with even weak star formation may remain faint in the continuum while producing detectable nebular lines. Table~\ref{tab:darkbench} therefore orders the benchmarks from line-emitting H\,\textsc{i}-bearing UDGs through almost-dark clouds to the hypothetical population of star-free gas halos.

\begin{table}[htbp]
\centering
\sffamily\small
\begin{tabularx}{\textwidth}{@{}p{3.2cm}p{6.4cm}X@{}}
\toprule
\textbf{System} & \textbf{Published properties} & \textbf{Place in the census}\\
\midrule
H\,\textsc{i}-bearing UDGs & 115 isolated, gas-rich, extremely diffuse field systems in high-spin dwarf halos with very low star-formation efficiency \cite{leisman2017} & Line-emitting gas-rich dwarfs, detected directly through H$\alpha$ and [O\,\textsc{iii}]\\
Coma\,P (HI1232$+$20) & Metal-poor H\,\textsc{i} cloud with $M_{\rm HI}/M_\star>200$ and only a trace very-low-surface-brightness stellar body \cite{janowiecki2015} & Almost-dark class, weak lines with an H\,\textsc{i} counterpart\\
Nube & Stellar mass of the Small Magellanic Cloud spread so thinly it evaded standard imaging, discussed as a possible fuzzy-dark-matter signature \cite{montes2024nube} & Almost-dark class, recovered by the deep-imaging stack (Figure~\ref{fig:nube})\\
VIRGOHI\,21 & Rotating H\,\textsc{i} structure with no optical counterpart, later discussed as a candidate tidal feature \cite{minchin2005,duc2008} & The false positive that isolation and dynamical-mass vetting removes\\
\bottomrule
\end{tabularx}
\caption{Benchmark systems along the almost-dark sequence and the class each occupies in the three-way classification of the census. The genuinely dark end of the sequence, gas halos with neither lines nor continuum, is the population the census is designed to isolate.}
\label{tab:darkbench}
\end{table}

\subsubsection{Environmental Control across Field, Group, Cluster, and Void Samples}

\begin{figure}[htbp]
\centering
\includegraphics[width=0.72\textwidth]{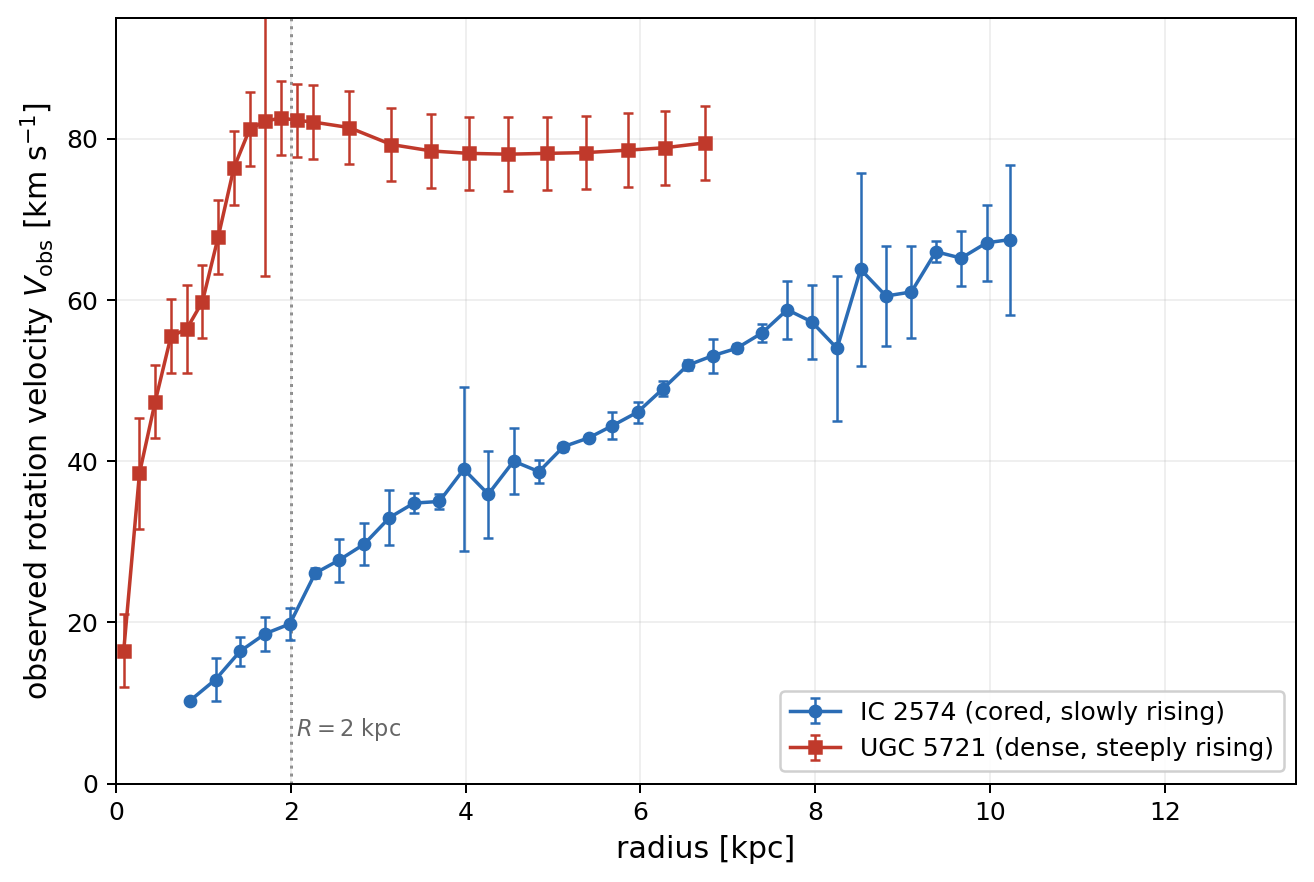}
\caption{The rotation-curve diversity problem in the observed data, drawn directly from the public SPARC mass-model files \cite{lelli2016}. IC~2574 and UGC~5721 reach nearly the same outermost rotation speed, $67$ and $79$\,km\,s$^{-1}$. They occupy halos of comparable mass, yet at $2$\,kpc one rotates at $20$ and the other at $82$\,km\,s$^{-1}$, a factor of four in velocity and more than an order of magnitude in enclosed mass. Reproducing both extremes at fixed halo mass is the benchmark for the dark-matter models of Table~\ref{tab:dmmodels}. Self-interacting dark matter fits both galaxies with a single cross section \cite{kaplinghat2016,ren2019}.}
\label{fig:sparcdiv}
\end{figure}

The sharpest current statement of the small-scale tension comes from the rotation curves of gas-rich field dwarfs and low-surface-brightness disks, for which homogeneous compilations such as the SPARC database of mass models provide spatially resolved inner kinematics for 175 galaxies \cite{lelli2016}. Figure~\ref{fig:sparcdiv} shows the diversity problem in the observed data, two SPARC dwarfs of nearly the same outermost rotation speed, and hence comparable halo mass, whose inner curves differ by a factor of four in velocity. The interpretation of such comparisons is limited by the heterogeneity of existing samples, drawn from many surveys with different selection criteria, distance scales, and environments, so that intrinsic scatter is entangled with selection and systematic effects.

Environment separates dark-sector signatures from tidal and hydrodynamic processing. Cluster dwarfs experience tidal stripping, ram pressure, and repeated high-speed encounters. Group dwarfs are more likely to have interacted with neighboring galaxies. Tidal mass loss may reduce central density and imitate an SIDM core, so any dark-matter inference must model those processes. Void galaxies provide a low-density comparison rather than an interaction-free control. The Void Galaxy Survey finds predominantly gas-rich low-luminosity blue disks, including systems with evidence of continued accretion or interactions \cite{kreckel2012}. A MaNGA comparison finds no statistically significant environmental difference in dark-matter-to-stellar-mass ratio or in its relation to gas-phase metallicity \cite{douglass2019}. The available observations therefore do not support an assumption that void dwarfs have uniformly larger dark-matter fractions. Their gas-rich disks nonetheless provide rotation curves for the core--cusp test and the diversity comparison in Figure~\ref{fig:sparcdiv}. One instrument and one selection function across void, field, group, and cluster samples will measure environmental trends without cross-survey calibration differences. The same census will constrain the steepening and scatter of the stellar-to-halo mass relation near the threshold of galaxy formation and test whether that threshold corresponds to a minimum halo mass \cite{simon2019,bullock2017}.

\subsubsection{A Starlight-Independent Halo Census from Strong Lensing}

Galaxy counts include only halos that formed detectable stars. Gravitational lensing instead selects halos through mass perturbations. Gravitational imaging of extended Einstein rings has detected a dark satellite of mass $\sim2\times10^{8}\,M_\odot$ in a lens galaxy \cite{vegetti2012}, and high-resolution interferometry has localized a comparable subhalo in SDP.81 \cite{hezaveh2016}. Flux-ratio and astrometric anomalies in multiply imaged quasars provide a complementary statistical measurement of the same low-mass population. The strong-lens survey of Section~\ref{sec:stronglens} supplies the lens sample and automated mass-reconstruction methods for substructure detection. The dwarf and void survey measures the luminous halo population with a controlled selection function as well as stellar-population diagnostics of when each galaxy formed its stars. Comparable lensing and luminous halo mass functions would indicate that star-formation efficiency dominates the faint-galaxy deficit. A deficit in both populations would instead favor a physical suppression of low-mass halos by warm or fuzzy dark matter. The two censuses therefore separate halo abundance from the probability that a halo forms a visible galaxy.

\subsubsection{Detection with Slitless Spectroscopy, Imaging, and H\,\textsc{i}}

\begin{figure}[htbp]
\centering
\includegraphics[width=0.55\textwidth]{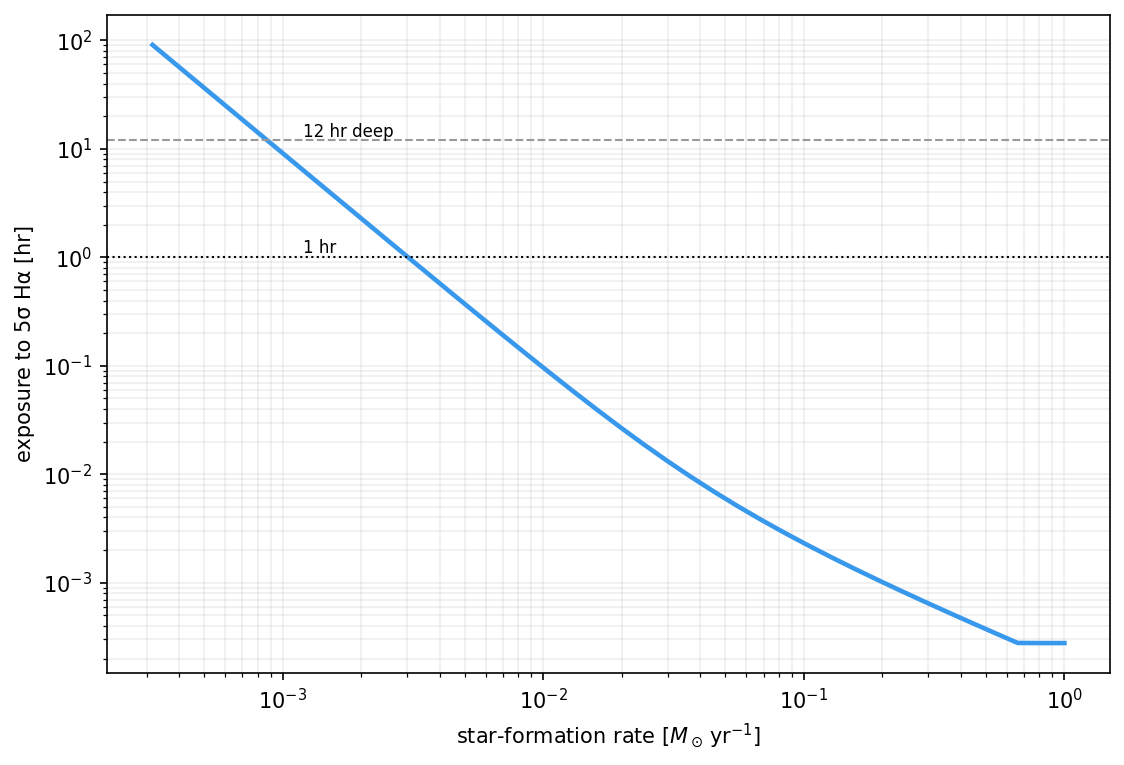}
\caption{Exposure to a $5\sigma$ H$\alpha$ detection versus star-formation rate for a void galaxy at $z=0.05$, from the Kennicutt SFR calibration. The $1$ and $12$\,hr levels are marked.}
\label{fig:darkgal}
\end{figure}

A slitless survey at $R=1000$ does not resolve the few-km\,s$^{-1}$ stellar velocity dispersions of dwarf galaxies. Those dispersions are approximately thirty times narrower than one resolution element and require high-resolution multi-object follow-up. The mission instead discovers systems and measures the quantities needed to select that follow-up. Continuum-selected surveys miss gas-rich star-poor systems when their stellar surface brightness falls below the imaging limit. Blind H\,\textsc{i} surveys have consequently supplied most dark-galaxy candidates. Slitless spectroscopy provides an independent selection through H$\alpha$, [O\,\textsc{iii}], and [O\,\textsc{ii}] emission. Even weak ongoing star formation in an optically faint halo may produce a large line-to-continuum ratio. Figure~\ref{fig:darkgal} gives the exposure required for a $5\sigma$ H$\alpha$ detection from a galaxy at $z=0.05$. The calculation uses ${\rm SFR}\simeq7.9\times10^{-42}\,(L_{\rm H\alpha}/{\rm erg\,s^{-1}})\,M_\odot\,{\rm yr^{-1}}$ \cite{kennicutt1998}. The idealized line-flux calculation reaches $\sim3\times10^{-3}\,M_\odot\,{\rm yr}^{-1}$ in approximately one hour and $\sim10^{-3}\,M_\odot\,{\rm yr}^{-1}$ in a deep pointing. Realistic source morphology, overlap, detector noise, and recovery completeness will reduce those limits by the margin quantified in Section~\ref{sec:etc}.

Cross-matching the emission-line catalog with H\,\textsc{i} maps from ALFALFA, Apertif, MeerKAT, and ultimately the SKA will measure gas and stellar emission together. Line and H\,\textsc{i} detections identify gas-rich star-forming dwarfs. H\,\textsc{i} detections with weak line emission define the almost-dark population. H\,\textsc{i} detections without lines or continuum provide candidate star-free gas halos. The last classification requires more than an H\,\textsc{i} signal. Isolation within a void reduces the probability of tidal debris. A dynamical mass substantially above the combined gas and stellar mass indicates a non-baryonic mass component. Deep airglow-free imaging provides individual and stacked surface-brightness limits on any stellar counterpart. Joint application of the three tests will convert individual candidates into a statistical measurement of failed galaxy formation at low halo mass.

The void search is built around nearby voids selected from catalogs constructed from SDSS DR7 and DESI DR1 \cite{sutter2012,rincon2025}. The sample is restricted to $z\lesssim0.05$ to keep the diagnostic nebular lines within the most sensitive part of the band and to spatially resolve even very low-mass dwarfs. Tiling the interiors of a sample of large voids to the wide- and medium-tier depths yields a volume-complete emission-line census whose faint end measures the void galaxy luminosity and star-formation-rate functions with a uniform, space-based selection that heterogeneous, atmosphere-limited ground-based void surveys do not match. Because the search exploits the same emission-line sensitivity, multi-roll depth, and wide field that the flagship survey already delivers, the void, field, group, and cluster volumes are all covered at marginal additional cost. Table~\ref{tab:dwarfvoidreach} summarizes the resulting reach. Figure~\ref{fig:voidwedge} of Section~\ref{sec:voidphenom} makes the underlying target selection concrete with real archival data.

\begin{table}[htbp]
\centering
\sffamily\small
\begin{tabularx}{\textwidth}{@{}p{3.4cm}p{3.6cm}X@{}}
\toprule
\textbf{Target class} & \textbf{Selection} & \textbf{Dark-matter use}\\
\midrule
Dwarf/UDG census (field, group, cluster) & Deep imaging $+$ grism, wide field & Halo mass function and abundance versus CDM, WDM, and FDM suppression. Membership, distances, and stellar populations from emission and absorption redshifts, SBF, and TRGB\\
Structural diversity & Sizes, central surface brightness, across environments & Distribution of inner densities (SIDM core/collapse, FDM solitons) and the environmental control of Section~\ref{sec:dwarfvoid}\\
Star-forming void dwarfs & Blind emission lines (H$\alpha$, [O\,\textsc{iii}]) & Faint end of the void luminosity function and low-mass halo abundance versus WDM and FDM\\
Almost-dark / H\,\textsc{i}-bearing UDGs & Lines $+$ H\,\textsc{i} cross-match & High gas-to-star systems and the star-formation threshold in low-mass halos\\
Optically dark halos & H\,\textsc{i} detection, no line emission & Census of star-free gas halos, the void phenomenon\\
\bottomrule
\end{tabularx}
\caption{Planning reach of the combined dwarf and void dark-matter program. The mission delivers wide-field discovery, membership, and stellar-population characterization with a uniform selection function across environments. Internal stellar velocity dispersions lie below both candidate resolutions and require high-resolution follow-up while lensing substructure (Section~\ref{sec:stronglens}) provides a starlight-independent halo census.}
\label{tab:dwarfvoidreach}
\end{table}

\subsubsection{Program Scope}

The program combines flagship imaging and slitless spectra with deeper observations of selected nearby hosts, groups, and void interiors. Archival and forthcoming H\,\textsc{i} maps add the neutral-gas measurement. Dwarf abundance and radial distributions constrain the low-mass halo mass function. Void interiors provide the least tidally processed measurement of its cutoff. Structural and chemical diagnostics distinguish baryonic core formation from SIDM and FDM predictions. Strong-lens substructure counts halos without requiring starlight. A common selection function across void, field, group, and cluster samples controls the environmental and cross-survey differences that limit current comparisons. The mission team's SIDM and wave-dark-matter simulations will predict the joint distributions for direct comparison with the catalog. Most of the census uses exposures already required by the flagship survey. The concrete deliverables are the following.
\begin{itemize}[leftmargin=1.4em,itemsep=1pt,topsep=2pt]
\item A uniform dwarf and ultra-diffuse-galaxy catalog with structural parameters, stellar metallicities, ages, and a controlled selection function across field, group, host, and void environments.
\item The faint-end galaxy stellar-mass function, satellite radial distributions, and void galaxy and H\,\textsc{i} mass functions that jointly constrain the halo mass function and its warm or fuzzy cutoff.
\item The joint distribution of central density, stellar mass, star-formation history, and environment that separates baryon-driven from SIDM and FDM core formation.
\item A catalog of almost-dark and H\,\textsc{i}-bearing ultra-diffuse systems, and a prioritized target list of central-density, metallicity, and isolation/dynamics-vetted dark-halo-candidate outliers for high-resolution kinematic and deep H\,\textsc{i} follow-up with the SKA and its precursors.
\item A starlight-independent halo census from the strong-lens substructure program of Section~\ref{sec:stronglens}, confronted against the luminous census within the same dark-matter models.
\item Improved interior tracer densities for void-dynamics cosmology, as a by-product of the same redshift survey.
\end{itemize}

\clearpage
\section{Program Triage}

Table~\ref{tab:triage} translates the preceding science cases into program priority, observing mode, area, depth, and decision status. The ELG BAO and RSD survey defines the flagship requirements. Companion programs use compatible spectroscopic or imaging observations but do not set the flagship performance claims.

\begin{table}[htbp]
\centering
\sffamily\footnotesize
\setlength{\tabcolsep}{2.4pt}
\renewcommand{\arraystretch}{0.90}
\begin{tabularx}{\textwidth}{@{}>{\raggedright\arraybackslash}p{2.7cm}>{\raggedright\arraybackslash}p{3.0cm}>{\raggedright\arraybackslash}p{3.2cm}>{\raggedright\arraybackslash}p{3.2cm}Y@{}}
\toprule
\textbf{Program and priority} & \textbf{Observing mode} & \textbf{Area or sample} & \textbf{Depth or exposure} & \textbf{Decision}\\
\midrule
\textbf{ELG BAO/RSD}\newline Flagship & Imaging and 3-roll image-sliced spectroscopy at the selected $R$ & 100--300 \degti\ with $10^6$--$3\times10^6$ reference redshifts & $F_{5\sigma}=10^{-16}$ wide and $2.5\times10^{-17}$ deep & \textcolor{green}{Core mission}\\
\textbf{SN Ia}\newline Major companion & Cadenced wide-tier field and targeted slitless typing & 5--10 \degti\ with 1000--3000 discoveries & 2--30 hr spectra with useful typing to $z\simeq1.5$ & \textcolor{green}{Proceed}\\
\textbf{LSB blind census}\newline Survey by-product & Imaging during flagship tiers & 100--300 \degti\ with $\gtrsim$500--1400 LSBGs and UDGs & $\mu_{5\sigma}(10'',g)=29.8$--$31.3$ & \textcolor{green}{Proceed}\\
\textbf{LSB ultra-deep fields}\newline Major companion & Imaging only & Two $2.25$ \degti\ low-cirrus fields & 16 hr per tile in simultaneous $g+H$ with $\mu_{5\sigma}(10'')=31.5/31.2$ & \textcolor{green}{Proceed}\\
\textbf{Imaging wide extension}\newline Major companion & Imaging only & 612 \degti\ across two low-cirrus fields & 0.25 hr per tile in $g+H$ with $\mu_{5\sigma}(10'',g)=29.2$ & \textcolor{green}{Proceed}\\
\textbf{ICL clusters}\newline Major companion & Targeted deep imaging stacks & 20--40 clusters at $0.02\lesssim z\lesssim0.10$ & Multi-roll 4--64 hr per cluster with $\mu_{5\sigma}(10'')\simeq30.4$--$31.9$ & \textcolor{green}{Proceed}\\
\midrule
\textbf{ELG line profiles}\newline Calibration & Both resolution candidates & Compact ELGs across redshift, size, and line-flux bins & Matched ETC and recovery tests & \textcolor{gold}{Phase A validation}\\
\textbf{High-$z$ quasars}\newline Major companion & Image-sliced spectra at the selected $R$ & 50--100 quasars at $z>6.5$ & $3\times2$--$3\times5$ hr per target & \textcolor{green}{Proceed}\\
\textbf{Ly$\alpha$ forest}\newline Calibration & Image-sliced spectra at the selected $R$ & 3000--5000 quasars at $2.1<z<3.5$ & 1--6 hr for $m_{\rm AB}\lesssim22.5$ & \textcolor{gold}{Moderate claims}\\
\textbf{Strong lenses}\newline Pathfinder & Imaging and spectra at the selected $R$ & 20--30 clean systems & External IFU support where required & \textcolor{gold}{Ancillary}\\
\textbf{LBG topology}\newline Ancillary & Image-sliced spectra at the selected $R$ & 100--200 bright systems & $3\times5$--$3\times10$ hr & \textcolor{rose}{Do not flagship}\\
\bottomrule
\end{tabularx}
\caption{Program-level numerical scope. Spectroscopic entries use the required image slicer and three orientations but exclude operational overhead. The numerical ELG yield comes from the $R=1000$ reference calculation. The final resolving power requires the matched cosmology comparison of Table~\ref{tab:etcmodes}. The observing column excludes serendipitous supernova discoveries from repeat visits in other imaging tiers because the proposal has not forecast that rate.}
\label{tab:triage}
\end{table}

\clearpage
\section{Observing Architecture}

\begin{figure}[htbp]
\centering
\begin{tikzpicture}[font=\sffamily]
  \fill[paper] (-0.35,-0.25) rectangle (16.0,5.7);
  \node[anchor=west,font=\bfseries\large,color=ink] at (0,5.35) {Operational flow for the robotic survey};
  \foreach \x/\title/\body/\col in {
    0.25/{Direct image}/{catalogs, morphology, astrometry}/blue,
    4.15/{0/45/90 deg}/{three slitless views with distinct overlaps}/teal,
    8.05/{Two arms}/{dichroic optical + near-IR spectra}/gold,
    11.95/{Scene model}/{redshifts, masks, clustering products}/green} {
    \fill[white,draw=\col,rounded corners=2pt,line width=0.65pt] (\x,2.15) rectangle ++(3.25,2.15);
    \node[anchor=west,font=\bfseries,color=\col] at (\x+0.22,3.72) {\title};
    \node[anchor=west,text width=2.75cm,color=softink,font=\footnotesize] at (\x+0.22,2.95) {\body};
  }
  \foreach \xa/\xb in {3.58/4.08,7.48/7.98,11.38/11.88}
    \draw[-{Stealth[length=2.6mm]},line width=0.8pt,color=muted] (\xa,3.22) -- (\xb,3.22);
  \draw[teal,line width=0.9pt] (0.45,1.25) -- (15.0,1.25);
  \foreach \x/\lab in {0.45/{simulate},3.7/{schedule},7.0/{extract},10.3/{validate},13.6/{release}}
    {
    \fill[teal] (\x,1.25) circle (2pt);
    \node[anchor=north,color=muted,font=\footnotesize] at (\x,1.05) {\lab};
    }
\end{tikzpicture}
\caption{Operational sequence for the robotic survey. Direct imaging constrains source position and morphology before three slitless orientations provide overlap diversity. Joint scene modelling then yields spectra, redshift probabilities, masks, and clustering products.}
\label{fig:ops}
\end{figure}
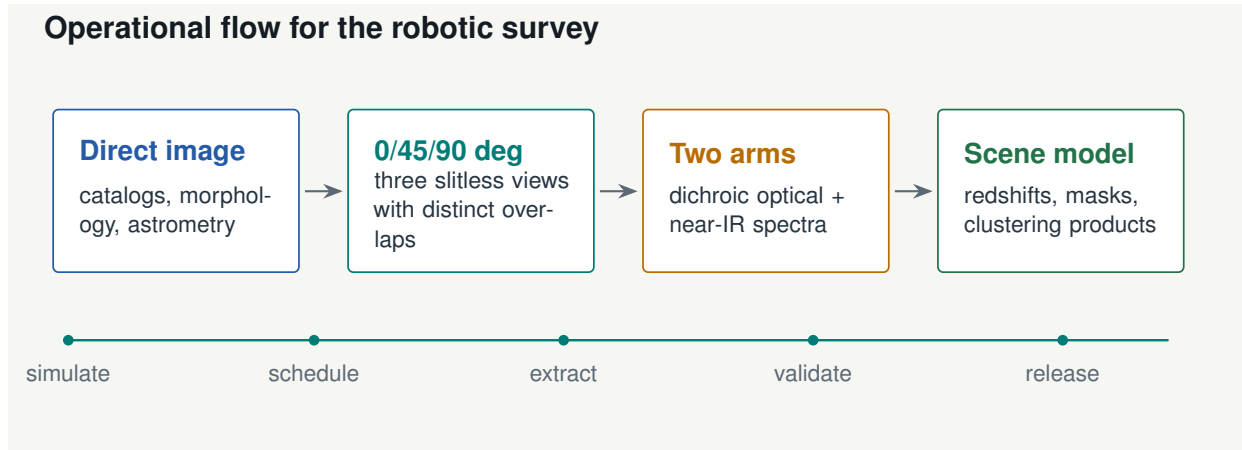

\subsection{Wide Survey Mode}

Each detector-level observing bundle contains a direct image, dispersion views at $0^\circ$, $45^\circ$, and $90^\circ$, and at least two subpixel dithers per orientation. The validation baseline adopts two 900~s integrations at each orientation. Splitting the exposure permits bad-pixel rejection and cosmic-ray identification while preserving the quoted total integration. The final number of nondestructive reads and exposure segments will follow the selected ramp-read mode and the cosmic-ray rate measured in orbit.

The wide survey uses the \fovlarge\ field, two band-limited grism settings per wavelength channel, and three orientations. A complete visit therefore includes direct imaging, grism exposures at $0^\circ$, $45^\circ$, and $90^\circ$, dithers, and calibration overhead. Every survey-time estimate includes the complete visit rather than one grism exposure.

Table~\ref{tab:timeline} gives the current database-driven reference allocation. The scheduler places every Volume II program within the same 260-week capacity model used for the other science volumes. The resulting 6650~hr total corresponds to the midpoint of the previous 4700--8600~hr planning envelope and equals 15.17 per cent of the five-year wall clock.

\begin{table}[htbp]
\centering
\sffamily\small
\begin{tabularx}{\textwidth}{@{}
  >{\raggedright\arraybackslash}p{3.7cm}
  >{\raggedright\arraybackslash}p{2.25cm}
  >{\raggedright\arraybackslash}p{2.2cm}
  >{\raggedright\arraybackslash}X@{}}
\toprule
\textbf{Mission phase} & \textbf{Nominal period} & \textbf{Science time} & \textbf{Main deliverable}\\
 & mission year & hr & \\
\midrule
Commissioning and reference fields & Year 1 & 400 hr & Wavelength solution, point-spread-function and trace model, detector calibration package, multi-orientation blind-recovery validation, image-slicer optical validation, and ultra-deep reference fields.\\
Wide ELG BAO/RSD & Years 1--4 & 1850 hr & 100--300 \degti ELG redshift map with survey masks and random catalogs.\\
Medium/deep ELG tiers & Years 2--5 & 900 hr & Completeness, purity, luminosity-function, overlap-calibration transfer functions, and tier-specific $P_{\rm secure}$.\\
Imaging-only wide extension & Year 5 & 610 hr & LSBG and UDG census beyond the spectroscopic footprint with approximately 306 \degti\ per cap in $g$ and $H$.\\
Ultra-deep imaging fields & Year 5 & 290 hr & Nube-depth LSBG and UDG imaging in two dedicated fields with $\mu_{5\sigma}\sim31.5$ in $g$ and 31.2 in $H$.\\
SN Ia time-domain cadence & Years 2--5 & 412 hr & Repeated visits that establish the cadence and photometric reference sequence.\\
SN Ia time-domain field & Years 4--5 & 788 hr & Spectral time series and host-scene models for $z\lesssim1.5$ SNe Ia in a field nested inside the wide tier.\\
Quasar and reionization targets & Years 4--5 & 900 hr & Ly$\alpha$ calibration spectra and a $z>6.5$ damping-wing sample.\\
Pathfinders and contingency & Year 5 & 500 hr & Resolving-power calibration with image-sliced ELGs, strong-lens targets, LBG spectroscopy, ICL cluster imaging, and special-field programs.\\
\midrule
\textbf{Volume II total} & \textbf{Years 1--5} & \textbf{6650 hr} & \textbf{15.17 per cent of the five-year wall clock.}\\
\bottomrule
\end{tabularx}
\caption{Database-driven five-year reference allocation for Volume II. The shared scheduler applies a $90^\circ$--$180^\circ$ solar-elongation field of regard, common maintenance and calibration periods, and protected Director weeks. The allocation demonstrates seasonal and shared-capacity feasibility. It does not replace tile-level optimization with the final instrument overheads.}
\label{tab:timeline}
\end{table}

The wide survey targets two high-Galactic-latitude caps that avoid the Milky Way and overlap the existing wide surveys, chosen together with the legacy deep fields they enclose as described in Section~\ref{sec:deepfields}.

\subsection{Deep and Time-Domain Mode}

The \fovsmall\ mode serves programs whose sensitivity depends more strongly on exposure depth or cadence than on solid angle. Applications include supernova cadence fields, intracluster-light imaging from Section~\ref{sec:lsb}, deep calibration fields, quasar monitoring, and recovery tests in fields with existing spectroscopy. Adopting the smaller field for the RSD survey would reduce the cosmological footprint at fixed mission time.

\subsection{Resolving-Power Selection}

The image slicer is mandatory at both candidate resolving powers. A direct image measures source position, size, and orientation. The slicer reformats the field into a pseudo-slit before the disperser records one or more band-limited settings. Compact ELGs provide the primary calibration sample because external spectra resolve their [O\,\textsc{ii}], H$\beta$, [O\,\textsc{iii}], and H$\alpha$ profiles. At least two dithers move sources across slice boundaries. An orientation is repeated when source morphology would otherwise dominate the measured line-spread function.

Selection between $R=1000$ and $R=5000$ requires four matched tests. Optical models for both candidates must cover the same wavelength interval and report usable field area on the adopted detector format. Calibration exposures must constrain wavelength and line-spread-function variation across slices. Source-injection tests must recover line flux and redshift without significant bias in crowded fields. The image-sliced ETC must verify the expected $R^0$ scaling of photon-limited integrated S/N over a fixed physical line interval and quantify departures caused by detector noise. The cosmology forecast must then propagate completeness, catastrophic-redshift fraction, usable area, and $n(z)$ into the BAO and RSD covariance. The current $R=1000$ yield is a reference calculation rather than a selected flight specification.

\clearpage
\section{Flat Fielding and the Cirrus Foreground in the Imaging Surveys}\label{sec:flatcirrus}

Deep surface photometry imposes common calibration requirements on the wide ELG tiers, imaging-only extensions, two ultra-deep fields, and targeted cluster program of Section~\ref{sec:lsb}. Multiplicative response errors and additive diffuse foregrounds dominate the error budget. The ultra-deep fields set the most stringent limits considered below.

Surface photometry at $\mu\gtrsim29$ mag\,arcsec$^{-2}$ requires simultaneous control of the multiplicative flat field and additive diffuse foregrounds. A one-percent residual illumination error on a zodiacal background of $\mu_g\simeq22$ mag\,arcsec$^{-2}$ produces a spurious diffuse signal at $\mu_g\simeq27$. That artifact is more than four magnitudes brighter than the ultra-deep limit of $31.5$. The low-spatial-frequency flat-field residual must therefore remain below $10^{-4}$ across the field.

Deep surveys establish the required calibration methods. The sky and repeated stellar measurements provide the low-spatial-frequency reference because an internal lamp illuminates the optical train differently and may introduce its own scattered-light pattern. A star flat solves the position-dependent photometric response from repeated measurements of the same stars at different focal-plane positions \cite{manfroid1996}. Dark Energy Camera processing applies star-flat solutions above dome flats to separate quantum efficiency, pixel-area variation, and stray light \cite{bernstein2017pasp}. SDSS extended the method to overlapping scans through photometric self-calibration, reaching one-percent relative calibration in $griz$ over $8500$\,deg$^2$ \cite{padmanabhan2008}. Survey simulations show that a stable global solution requires repeated placement of each source on widely separated detector regions \cite{holmes2012}. Diffuse-light measurements additionally use dark-sky flats from masked and dithered science exposures. The 10.4-m GTC reached $\mu_r=31.5$ mag\,arcsec$^{-2}$ at $3\sigma$ in $10\times10$\,arcsec$^2$ apertures with that method \cite{trujillofliri2016}. Tunable monochromatic illumination measured by calibrated photodiodes provides an independent transfer of laboratory throughput standards to the telescope and detector \cite{stubbs2006}.

The mission combines all three approaches. Internal illumination calibrates high-spatial-frequency pixel response. Masked and dithered science exposures form a dark-sky superflat for low-spatial-frequency illumination. The method is practical in space because zodiacal light is smooth across one field and lacks the rapid temporal variation of atmospheric airglow. Large dithers and three roll angles separate detector-fixed response from sky-fixed diffuse structure. Overlapping tiles then support a joint least-squares solution for multiplicative response, additive sky, and relative photometric zero point \cite{padmanabhan2008}. Repeated stellar measurements at different rolls and offsets connect every detector region to the same calibration system.

Galactic cirrus is a sky-fixed additive foreground rather than a detector-calibration error. At the mean $857$\,GHz intensities of the selected ultra-deep fields, the scattered-light scaling in Section~\ref{sec:etc} predicts $\mu_g\simeq27.2$ mag\,arcsec$^{-2}$ from the measured optical $b(\lambda)$ ratios \cite{ienaka2013}. The corresponding near-infrared estimate is $\mu_H\simeq26.7$ mag\,arcsec$^{-2}$. Both values are several magnitudes brighter than the survey limit, so spatial fluctuations set a stronger confusion floor than the mean intensity alone. Figure~\ref{fig:imgfields} maps thermal-dust structure across the candidate footprints, and Figure~\ref{fig:skymap} places those fields within the wider survey. The maps inform field selection but do not predict optical scattered light at each pixel.

Cirrus control proceeds in three stages. Field selection provides the largest reduction because cirrus fluctuation power scales approximately with the cube of mean brightness \cite{gautier1992}. Selecting fields at one quarter of the median sky brightness suppresses the fluctuation power by a factor of approximately fifty relative to a typical high-latitude field. Planck $857$\,GHz and IRAS $100\,\mu$m maps provide a low-order spatial template. The fit allows its amplitude and wavelength dependence to vary because dust temperature, grain properties, and scattering geometry differ across the sky. Measured $b(\lambda)$ ratios provide informative priors \cite{ienaka2013}. Five-band photometry then assigns each low-surface-brightness candidate a cirrus-contamination probability from far-infrared correlation, morphology, and spectral energy distribution. Cirrus and stellar populations occupy different preferred optical colour ranges \cite{roman2020}, but colour enters as a likelihood rather than a hard cut. Injection and recovery on the observed backgrounds calibrate completeness and false-positive rate. The procedure includes cirrus knots known to contaminate ground-based low-surface-brightness catalogs \cite{zaritsky2023} without assuming perfect separation by one colour.

\clearpage
\section{Expected Deliverables}

The mission will release the following science-ready products in addition to calibrated images.
\begin{itemize}
  \item Direct-image source catalogs with morphological parameters and astrometric covariance.
  \item Calibrated two-dimensional slitless exposures for each filter, orientation, and dispersion direction.
  \item Extracted one-dimensional spectra with contamination estimates and spectral covariance.
  \item Probabilistic redshift catalogs with emission-line identification likelihoods.
  \item Survey masks, random catalogs, and completeness maps for clustering analyses.
  \item Supernova host-galaxy models and binned spectral time series.
  \item High-redshift quasar continuum fits and damping-wing likelihood functions.
  \item Public detector-level simulations reproducing the final survey selection function.
\end{itemize}

\clearpage
\section{Conclusion}

The \facility\ derives its cosmological capability from dense and homogeneous redshift mapping over a large solid angle. The \fovlarge\ mode therefore enables the ELG BAO and RSD survey, whereas the \fovsmall\ mode supports deep calibration, supernova cadence, and targeted quasar observations. Standard two-hour visits measure the rest-frame $U$ and NUV magnitudes that distinguish optically similar SN~Ia subgroups to $z\simeq0.9$--$1.1$. Ten-hour stacks extend the measurement to $z\simeq1.3$--$1.5$ as shown in Figure~\ref{fig:snuvlimit}. Deep imaging within the spectroscopic tiers uses emission-line redshifts to recover low-surface-brightness galaxies that are absent from the observed stellar mass function in the team's Horizon Run~5 mock surveys at $0.625\le z\le2$ as shown in Figure~\ref{fig:hr5gsmf}. The resulting faint-end stellar mass function tests feedback rather than only calibrating an incomplete sample. A common selection function across void, field, group, and cluster environments also permits direct comparison of the low-mass halo abundance and inner density structure with cold, self-interacting, and fuzzy dark-matter predictions.

Roman, Euclid, DESI, and Rubin will map larger volumes and obtain larger samples than the proposed mission. The distinctive dark-matter measurement instead comes from a matched-depth sample selected with one instrument across field, group, cluster, and void environments. That controlled environmental comparison separates SIDM or wave-dark-matter signatures in low-mass halo structure from tidal processing and cross-survey selection differences. Heterogeneous samples currently limit that distinction as discussed in Section~\ref{sec:dwarfvoid}. The proposed survey will therefore test whether dark matter remains cold and collisionless on dwarf-galaxy scales or requires self-interaction or wave-like behavior.

The revised mission statement follows.
\begin{leadbox}[title=Mission Statement]{teal}
\textbf{The \facility\ will conduct a band-limited multi-roll survey with an image-sliced spectrograph. The instrument study will select $R=1000$ or $R=5000$ from matched forecasts of secure ELG density, usable survey area, and BAO and RSD covariance. The image slicer preserves the integrated photon-limited S/N at either resolving power. Supernova, high-redshift quasar, low-surface-brightness, and dwarf-galaxy programs exploit the same stable robotic observatory.}
\end{leadbox}

\begin{figure}[htbp]
\centering
\includegraphics[width=0.98\textwidth]{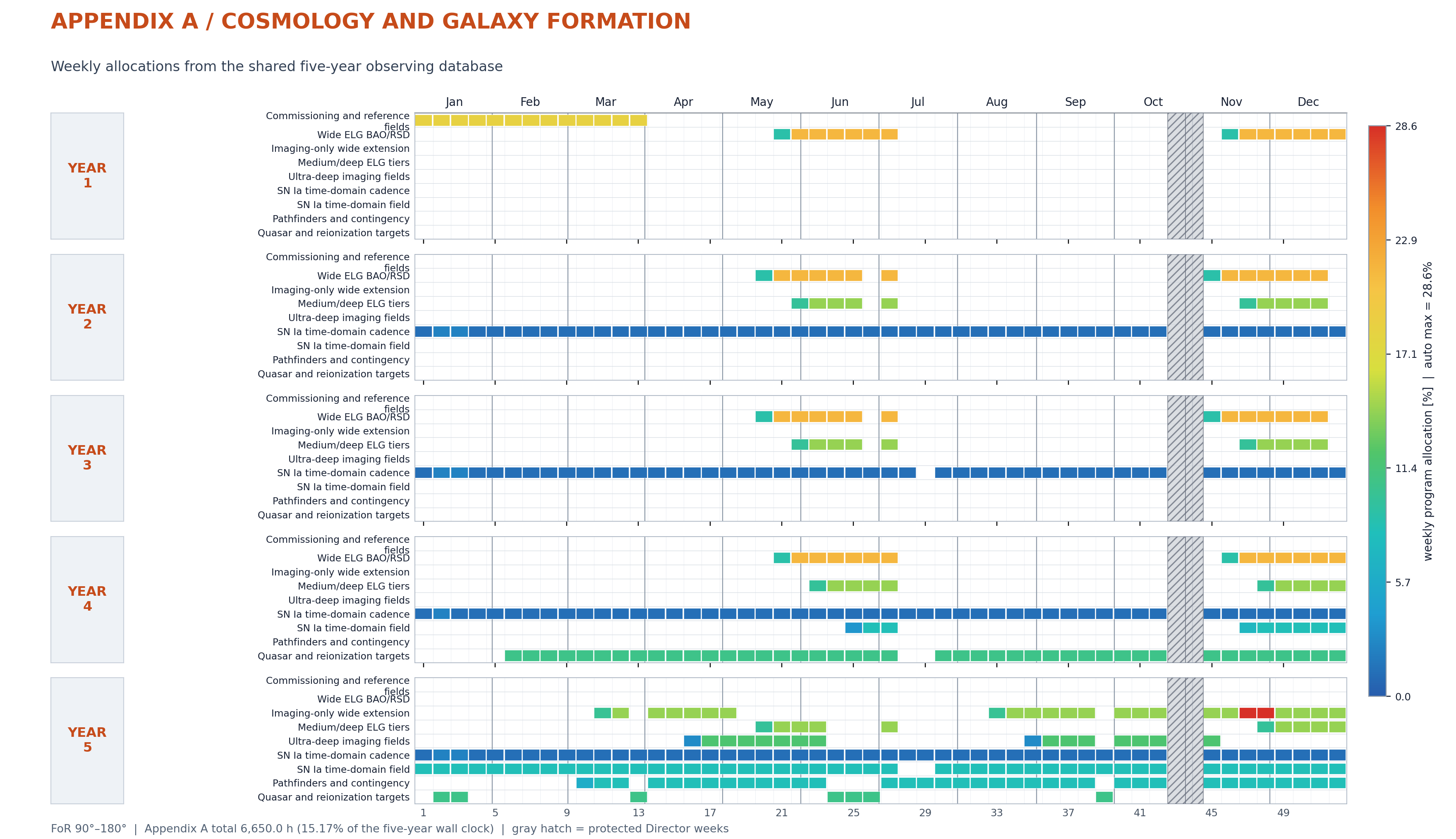}
\caption{Weekly Volume II allocation generated from the shared observing database. Color gives the fraction of one 168-hr week assigned to each program. The color scale uses the largest Volume II allocation. Gray hatching marks Director weeks 43 and 44 in every mission year. The calculation uses a $90^\circ$--$180^\circ$ field of regard and includes the common maintenance, calibration, Volume III, Volume IV, and Volume V demands when enforcing weekly capacity.}
\label{fig:surveytimeline}
\end{figure}

Figure~\ref{fig:surveytimeline} shows the weekly assignments associated with Table~\ref{tab:timeline}. Routine science begins in Year~1 week~14 after commissioning and performance acceptance. Fixed fields are scheduled only when at least three days of the week satisfy the adopted solar-elongation limit. Flexible target lists occupy the remaining eligible weeks under the same mission-wide capacity constraint.

\section*{Figures and Sources}
Most schematic figures in this document are generated directly in \LaTeX\ using TikZ. Figure~\ref{fig:crowded} is a slitless scene simulation generated locally at the baseline instrument parameters with the tracked script \texttt{make\_slitless\_scene\_sim.py}, using the repository's FSPS galaxy templates as source spectra. Figure~\ref{fig:slicer} is generated locally with the tracked script \texttt{make\_hsc\_slicer\_demo.py}, using the same instrument parameters and FSPS templates as Figure~\ref{fig:crowded}, but with source positions, brightnesses, sizes, ellipticities, and position angles measured from a real Subaru HSC-SSP deep image of the SXDS/UDS/XMM-LSS reference \cite{aihara2018} instead of a synthetic population. Figure~\ref{fig:elgspectrum} uses an observed bright SDSS DR17 low-redshift emission-line galaxy spectrum converted to a local plotting table. Figure~\ref{fig:elgobswavez} stacks actual SDSS/BOSS and DESI DR1 spectra retrieved through SPARCL together with HST WFC3/G141 3D-HST COSMOS grism spectra (v4.1.5 1D extractions) and JWST/NIRSpec PRISM star-forming galaxy spectra (msaexp 1D extractions from the DAWN JWST Archive) on an approximately uniform display grid in redshift, spanning the full $0.36$--$3.0\,\mu$m proposed range. Figures~\ref{fig:elgmagdepth} and \ref{fig:desiMlf} use the DESI DR1 LSS ELG catalog with an observed-frame absolute magnitude computed locally from the downloaded FITS table. In Figure~\ref{fig:desiMlf} the binned observed-frame luminosity function is additionally fit with a Schechter function on the bright side of each redshift bin and shown as smooth curves. Figure~\ref{fig:phillips} plots the real Carnegie Supernova Project\,I sample of 139 SNe Ia (Burns et~al.\ \cite{burns2018csp}, VizieR catalog J/ApJ/869/56), computing the host-extinction-corrected peak absolute magnitude $M_V=V_{\max}-\mu_{\rm CV}-R_V\,E(B\!-\!V)$ from the tabulated distance modulus and reddening, plotting it against the tabulated $\Delta m_{15}(B)$, and overlaying a quadratic Phillips relation fitted locally to those points. The figure layout follows Figure 11 of Zeng et~al.\ \cite{zeng2025sn2023ehl}. Figure~\ref{fig:uvtwins} plots the observed flux-calibrated Swift/UVOT UV-grism spectra of SN\,2011fe and SN\,2011by near maximum, retrieved from the Open Supernova Catalog archive \cite{guillochon2017}, deredshifted and scaled over the optical. Figure~\ref{fig:sniacosmo} (the multi-cosmology SN Ia Hubble diagram with current and target Hubble-residual scatter) is generated locally with \texttt{astropy} distance moduli for the four cosmologies against the real Pantheon$+$SH0ES sample \cite{brout2022}, retaining the 1580 Hubble-flow SNe Ia ($z>0.01$, Cepheid calibrators removed). Figure~\ref{fig:nuvband} is generated in TikZ and shows the redshifting of fixed rest-frame wavelengths into the proposed observed band. Figure~\ref{fig:lyathreshold} is computed locally, integrating a Voigt profile to obtain the Ly$\alpha$ curve of growth and inverting it against the $R=1000$ detection limit of Eq.~\eqref{eq:wmin} to give the minimum detectable H\,\textsc{i} column density. Figure~\ref{fig:lya3dexternal} is an external CLAMATO 3D Ly$\alpha$ forest tomography rendering from the arXiv source package. Figure~\ref{fig:hr4xi} combines an observed BOSS CMASS-North pair-count measurement (obscorr pipeline, the same source data behind \texttt{plot\_bao\_jk.py}) with the author's Horizon Run 4 mock $\xi(\sigma,\pi)$ pair-count measurement in redshift space, rather than reproducing any published figure. Figure~\ref{fig:sidmsigmav} is drawn locally with \texttt{make\_sidm\_observational\_inference.py} from the observed halo-fit points of Kaplinghat, Tulin, and Yu \cite{kaplinghat2016}. The accompanying table records the posterior medians and intervals reconstructed from the vector arXiv figure. Figure~\ref{fig:fdmslice} is drawn locally with \texttt{make\_fdm\_halo\_slice.py} from the mission team's cuRAMSES wave-dark-matter output. The script reads the AMR leaf geometry and the complex field and computes $\rho=|\psi|^2$ without using a published image. Figure~\ref{fig:nube} is drawn locally with \texttt{make\_nube\_depth\_comparison.py} from the public HiPERCAM $r$-band FITS mosaic in CDS catalog J/A+A/681/A15 \cite{montes2024nube}. The shallower panels add calibrated noise to the same observed frame. Figure~\ref{fig:sparcdiv} is drawn locally from the public SPARC mass-model files \cite{lelli2016}. Figure~\ref{fig:fdmconstraints} is drawn locally from the published bounds quoted in its caption. Figure~\ref{fig:hr5gsmf} is the original figure from the arXiv source package of the team's own Horizon Run 5 stellar-mass-function study \cite{kim2023gsmf}, reproduced as published. Figure~\ref{fig:surveytimeline} is generated locally with the tracked script \texttt{make\_five\_year\_optimized\_schedule.py}. The script reads program times, target coordinates, cadence constraints, maintenance periods, Director weeks, and the $90^\circ$--$180^\circ$ field of regard from \texttt{observing\_schedule\_inputs.sqlite}. It solves the shared weekly allocation before plotting the Volume II rows. Literature and documentation sources are listed below.